\documentclass{ieeeaccess}
\usepackage{cite}
\usepackage{amsmath,amssymb,amsfonts}
\usepackage{algorithmic}
\usepackage{graphicx}
\usepackage{textcomp}
\usepackage{subfigure}
\usepackage{algorithm}
\usepackage{color}
\usepackage{tabularx}
\usepackage{multirow} 
\usepackage{paralist}
\usepackage{array}
 \usepackage{multirow}
\usepackage{enumitem}
\usepackage{wrapfig}
\usepackage{url}
\usepackage{flushend}

\definecolor{purple}{RGB}{128, 0, 128}
\definecolor{red}{rgb}{0.9, 0.0, 0.0}
\definecolor{blue}{rgb}{0.0, 0.0, 0.9}
\definecolor{green}{rgb}{0.0, 0.9, 0.0}

\newcommand{\be}{\begin{equation}}
\newcommand{\ee}{\end{equation}}
\newcommand{\ba}{\begin{array}}
\newcommand{\ea}{\end{array}}
\newcommand{\bea}{\begin{eqnarray}}
\newcommand{\eea}{\end{eqnarray}}

\newcommand{\vbar}{\raisebox{.17ex}{\rule{.04em}{1.35ex}}}
\newcommand{\vbarind}{\raisebox{.01ex}{\rule{.04em}{1.1ex}}}
\newcommand{\R}{\ifmmode {\rm I}\hspace{-.2em}{\rm R} \else ${\rm I}\hspace{-.2em}{\rm R}$ \fi}
\newcommand{\T}{\ifmmode {\rm I}\hspace{-.2em}{\rm T} \else ${\rm I}\hspace{-.2em}{\rm T}$ \fi}
\newcommand{\N}{\ifmmode {\rm I}\hspace{-.2em}{\rm N} \else \mbox{${\rm I}\hspace{-.2em}{\rm N}$} \fi}
\newcommand{\B}{\ifmmode {\rm I}\hspace{-.2em}{\rm B} \else \mbox{${\rm I}\hspace{-.2em}{\rm B}$} \fi}
\newcommand{\Hil}{\ifmmode {\rm I}\hspace{-.2em}{\rm H} \else \mbox{${\rm I}\hspace{-.2em}{\rm H}$} \fi}
\newcommand{\C}{\ifmmode \hspace{.2em}\vbar\hspace{-.31em}{\rm C} \else \mbox{$\hspace{.2em}\vbar\hspace{-.31em}{\rm C}$} \fi}
\newcommand{\Cind}{\ifmmode \hspace{.2em}\vbarind\hspace{-.25em}{\rm C} \else \mbox{$\hspace{.2em}\vbarind\hspace{-.25em}{\rm C}$} \fi}
\newcommand{\Q}{\ifmmode \hspace{.2em}\vbar\hspace{-.31em}{\rm Q} \else \mbox{$\hspace{.2em}\vbar\hspace{-.31em}{\rm Q}$} \fi}
\newcommand{\Z}{\ifmmode {\rm Z}\hspace{-.28em}{\rm Z} \else ${\rm Z}\hspace{-.28em}{\rm Z}$ \fi}

\def\BibTeX{{\rm B\kern-.05em{\sc i\kern-.025em b}\kern-.08em
    T\kern-.1667em\lower.7ex\hbox{E}\kern-.125emX}}
 
\begin{document}
\history{Received 29 November 2024, accepted 18 December 2024, date of publication 00 xxxx 0000, date of current version 00 xxxx 0000.}
\doi{10.1109/ACCESS.2024.3521579}

\title{6G: The Intelligent Network of Everything}

\author{\uppercase{Harri Pennanen}\authorrefmark{1}, \IEEEmembership{Member, IEEE},
\uppercase{Tuomo H\"anninen}\authorrefmark{1}, \uppercase{Oskari Tervo}\authorrefmark{2}, \uppercase{Antti T\"olli}\authorrefmark{1}, \IEEEmembership{Senior Member, IEEE}, and \uppercase{Matti Latva-aho}\authorrefmark{1},
\IEEEmembership{Fellow, IEEE}}
\address[1]{Centre for Wireless Communications, University of Oulu, Finland (e-mail: harri.pennanen@oulu.fi, tuomo.hanninen@oulu.fi, antti.tolli@oulu.fi, matti.latva-aho@oulu.fi)}
\address[2]{Unaffiliated (e-mail: okeketervo@gmail.com)}
\tfootnote{This research was supported by the Research Council of Finland (former Academy of Finland) 6G Flagship Programme (Grant Number: 346208). This article has been accepted for publication in IEEE Access. This is the author's version which has not been fully edited and content may change prior to final publication. Citation information: DOI 10.1109/ACCESS.2024.3521579. This work is licensed under a Creative Commons Attribution 4.0 License. For more information, see https://creativecommons.org/licenses/by/4.0/}

\markboth
{Pennanen \headeretal: 6G: The Intelligent Network of Everything}
{Pennanen \headeretal: 6G: The Intelligent Network of Everything}

\corresp{Corresponding author: Harri Pennanen (e-mail: harri.pennanen@oulu.fi).}

\begin{abstract} 
The global 6G vision has taken its shape after years of international research and development efforts. This work culminated in ITU-R Recommendation on "IMT-2030 Framework". While the definition phase of technological requirements is currently ongoing, 3GPP standardization process on 6G networks is expected to start in 2025 and worldwide commercialization around 2029--2030. It is timely to present an up-to-date overview of the entire 6G research field. This article serves as a comprehensive guide to 6G by providing an overall vision, a contemporary survey of the main literature, and an informative tutorial-type presentation style. In our vision, 6G will be based on three fundamental elements: wireless, artificial intelligence, and Internet of Everything. Consequently, 6G can ultimately become \textit{the Intelligent Network of Everything} while serving as an enabling platform for the next major disruption in mobile communication, called \textit{mobile intelligence}. The potential of mobile intelligence is that anything can be made connected, intelligent, and aware of its environment. This will revolutionize the way how devices, systems, and applications are designed; how they operate and interact with humans and each other; and how they can be used for the benefit of people, society, and the world in general. After high-level visioning, the main details of 6G are discussed, including fundamental elements, disruptive applications, key use cases, main performance requirements, potential technologies, and defining features. A special focus is given to a comprehensive set of potential 6G technologies, each of which is introduced in a tutorial manner, with a discussion on the vision, introduction, past, present, opportunities, challenges, literature, and future research directions. Finally, we speculate on what comes after 6G and sketch the first high-level vision of 7G. All in all, the objective of this article is to provide a thorough guide to 6G in order to serve as a source of knowledge and inspiration for further research and development work in academia, industry, and standardization bodies. \end{abstract}


\titlepgskip=-15pt

\maketitle

\section{INTRODUCTION}
\label{Intro}
\begin{figure}[!htb]
\center{\includegraphics[width=0.92\columnwidth]
{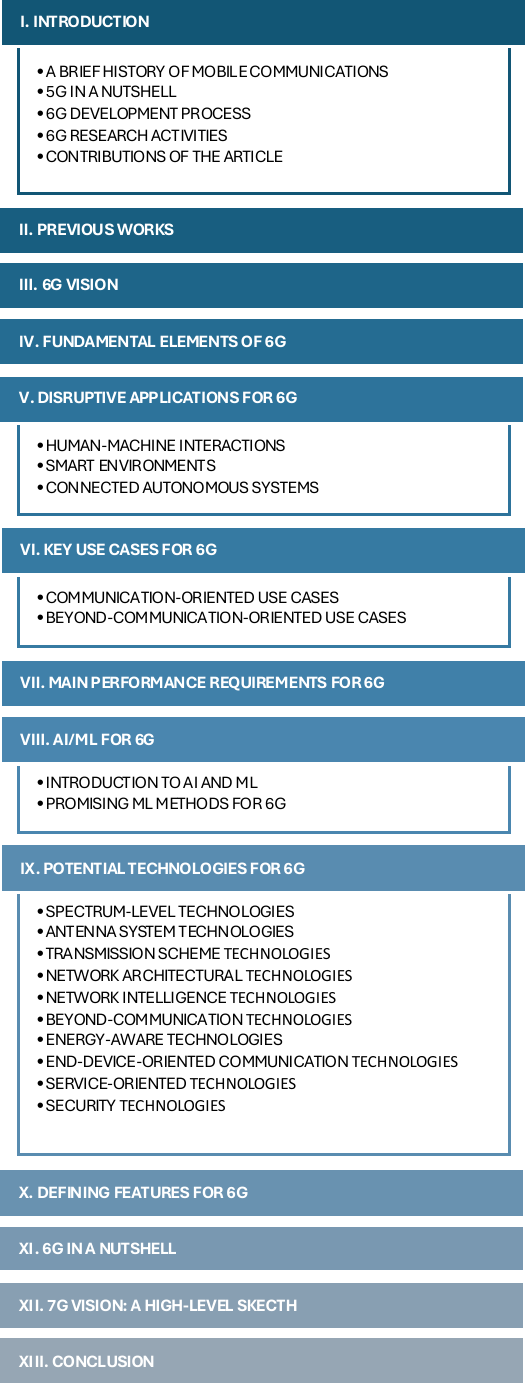}}
\caption{\label{Fig_Content}Content of the article.}
\end{figure}

This section takes a glance at the past, present, and future of mobile communications by introducing the evolution from 1G to 6G. After that, a special attention is given on 5G by providing a brief overview on its fundamentals, background, and standardization. The focus is then shifted to 6G. A generic development process and its timeline is discussed first. Then, an up-to-date review is given on the worldwide research activities toward 6G. Finally, the contributions of the article are presented. For the convenience of the readers, the content of this article is introduced in Figure \ref{Fig_Content}. 

\begin{figure*}[!htb]
\center{\includegraphics[width=\textwidth]
{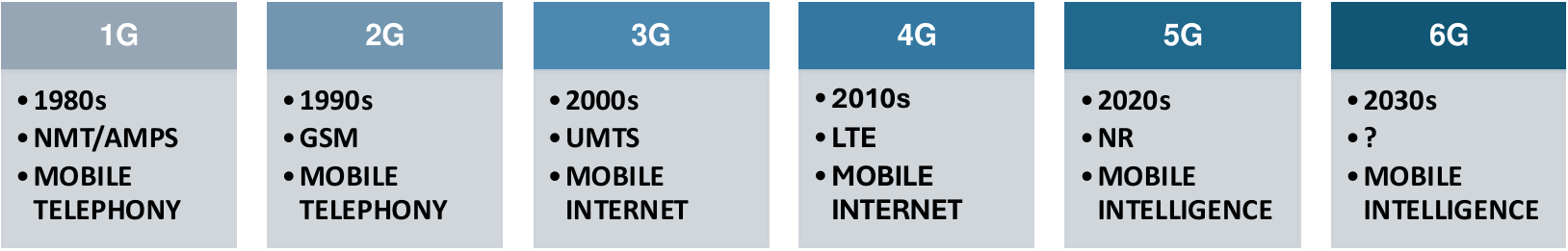}}
\caption{\label{Fig_Evolution}Evolution of mobile networks from 1G to 6G.}
\end{figure*}

\subsection{A BRIEF HISTORY OF MOBILE COMMUNICATIONS}
The generations from 1G to 6G are briefly reviewed, with a focus on the disruptive nature of each generation. This evolution is summarized in Figure \ref{Fig_Evolution}. 

\subsubsection{1G: MOBILE MEETS TELEPHONY}
Mobile communications began over four decades ago in the 1980s, when {1G networks} were introduced. 1G was based on analog communication with voice-only services. There were numerous regional mobile systems under the umbrella of 1G, such as the Advanced Mobile Phone System (AMPS) in North America, the Nordic Mobile Telephone (NMT) in Nordic countries, and the Total Access Communication System (TACS) in the United Kingdom (UK). 1G started the era of mobile telephony. The disruptive feature of mobile telephony was that one could make a phone call anywhere to anybody (in the coverage area). In practice, however, the number of users was rather limited because mobile phones were expensive, large in size (they could not fit in a pocket), and heavy to carry. Consequently, the disruptive nature of 1G and mobile telephony remained modest during the 1980s. 

\subsubsection{2G: MOBILE TELEPHONY}
The real disruption occurred in the 1990s, when {2G networks} were introduced. 2G was based on digital communication, allowing calling, texting, and limited data services. Due to digital signal processing technology, transceivers became more complex, powerful, and energy efficient, making phones fit in pockets and having a great voice quality. The dominant 2G system was the Global System for Mobile Communications (GSM), which was first used in Europe and later around the world. Other 2G systems were Digital AMPS and cdmaOne in North America and Personal Digital Cellular (PDC) in Japan. In the 2G era, the game changer was the evolution from analog to digital technology, making mobile phones affordable, small, and easy to carry. People were able to call and text anyone anywhere due to the high penetration rate and broad coverage of 2G. 2G freed the potential of mobile telephony and became a huge success, changing everyday life and interactions between people.

\subsubsection{3G: MOBILE MEETS INTERNET}
In the 2000s, {3G networks} were introduced, with data-centric communication as a key feature. Calling and texting were also included. The radio access technology used in 3G was code-division multiple access (CDMA), allowing for more efficient spectrum use. Turbo coding enabled 3G communication links to approach the Shannon capacity limit in practice. High data rates were possible due to the significantly increased system bandwidth. CDMA2000 was used in North America, while the dominant 3G system elsewhere in the world was the Universal Mobile Telecommunications System (UMTS). 3G provided a platform for the next major disruption, called mobile internet (i.e., mobile broadband). However, mobile internet did not boom in the early phase of 3G since mobile phones did not have touch screens, and their operating systems were not properly optimized for mobile internet usage. In addition, the majority of Internet content was not optimized for mobile phone usage. In other words, the ecosystem for mobile internet was immature at that time. This led to a clumsy user experience of mobile internet. However, all started to change after the introduction of the iPhone in 2007 and the iPhone 3G in 2008. The iPhones had touch screens, and their operating systems were properly optimized for mobile internet. Touch screens and mobile applications disrupted the whole mobile phone industry. The golden age of mobile internet, touch screen devices, and mobile applications started at the verge of the 4G era in the late 2000s. 

\subsubsection{4G: MOBILE INTERNET} 
The first commercial {4G networks} were launched in 2009. 4G was designed for broadband mobile internet with higher data rates. The air interface technology was changed from CDMA to orthogonal frequency-division multiplexing (OFDM). OFDM was chosen due to its robustness, flexibility, and straightforward support for multiple-input multiple-output (MIMO) and lower-complexity receivers. 4G was the first large-scale wireless system to exploit full-dimension MIMO channels. 4G was able to adapt to the prevailing channel and interference conditions. This with the increased system bandwidth and spectrum re-farming enabled high-rate communications and flexible use of broadband mobile internet. Mobile internet boomed in the 4G era and became an essential part of everyday life. The 3rd Generation Partnership Project (3GPP) developed Long-Term Evolution (LTE) (Release 8) was the dominant 4G system commonly used around the world. The evolution of LTE included LTE-Advanced (LTE-A) (Release 10) and LTE-A Pro (Release 13). The main technological components of the evolution were carrier aggregation, enhanced MIMO, relaying, coordinated multi-point (CoMP) transmission/reception, dual connectivity, and licensed-assisted access. LTE evolution also introduced support for vehicle-to-everything (V2X) communications and Machine-Type Communications (MTC). 

\subsubsection{5G: MOBILE MEETS INTELLIGENCE} 
In 2019, the first {5G networks}, based on the 3GPP 5G System technology (Release 15 \cite{Rel-15}), were rolled out. 5G was designed for software-native networking, service-oriented architecture, enhanced mobile internet, mission-critical communications, and massive Internet of Things (IoT) connectivity. In general, 5G is a flexible network with greatly expanded capabilities, providing a robust platform for novel wireless services and applications, particularly for vertical industries. The 5G System introduced a new core network (i.e., 5G Core) and a new radio interface (i.e., 5G New Radio (NR)) as part of the access network. The 5G Core supports service-based architecture (SBA), software-defined networking (SDN), network function virtualization (NFV), and network slicing. The key technologies of 5G radio access network (RAN) include millimeter wave (mmWave) communications, massive MIMO, small cells, flexible OFDM-based transmission scheme, cellular/industrial IoT, edge computing, and private networks. 5G also introduced support for non-terrestrial networks (NTNs), integrated access and backhaul (IAB), NR position, unlicensed NR, unmanned aerial systems (UAS), and NR V2X. Currently, there are four 5G Releases published in 3GPP, i.e., Releases 15 (2018), 16 (2020), 17 (2022), and 18 (2023). The most disruptive feature of Release 18 (i.e., 5G-Advanced) was the introduction of artificial intelligence/machine learning (AI/ML) to mobile networks. This was the turning point when mobile met intelligence in the network design. 

\subsubsection{6G: MOBILE INTELLIGENCE} 
Commercial 6G networks are expected to be launched around 2029--2030. Currently, 6G is in the development phase and standardization will start in 2025. Key technologies under development include terahertz (THz) communications, ultra-massive MIMO, reconfigurable intelligent surfaces (RISs), ultra-dense networks (UDNs), network/edge intelligence, integrated non-terrestrial and terrestrial networks (INTNs), integrated sensing and communication (ISAC), 6G V2X communications, cellular-connected unmanned aerial vehicle (UAV) communications, and THz IAB. Ultimately, 6G aims to be "the Intelligent Network of Everything", with extreme performance (capacity, rates, latency, reliability, connection density, coverage, and mobility), pervasive AI/ML (core, edge, and air interface), ubiquitous Internet of Everything (IoE) (sensors, devices, machines, vehicles, drones, robots, etc.), and beyond-communication capabilities (computation, caching, sensing, positioning, and energy). The 6G ecosystem is expected to provide a fruitful platform for the next major disruption, known as mobile intelligence (also known as wireless intelligence and connected intelligence). Mobile intelligence will revolutionize the design, operation, and interactions of devices, systems, and applications. In the big picture, mobile intelligence will be deeply integrated into the future society, providing diverse benefits at all its levels. 

\begin{figure}[!htb]
\center{\includegraphics[width=0.65\columnwidth]
{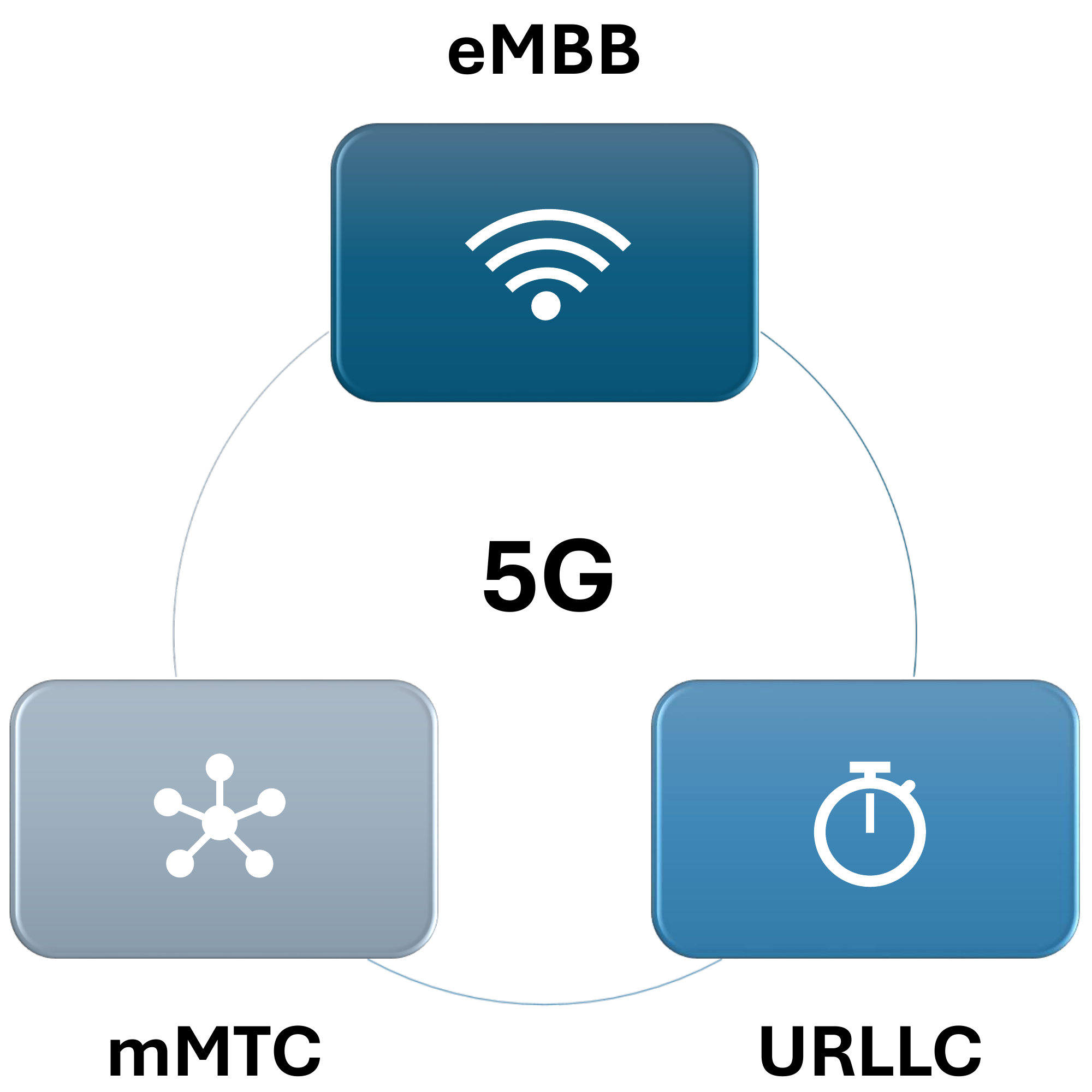}}
\caption{\label{Fig_5G}Key use cases of 5G.}
\end{figure}

\subsection{5G IN A NUTSHELL}
This section presents a brief overview of 5G. The main aspects are reviewed, including introduction, background, and standardization. More comprehensive overviews of 5G have been presented in the following publications: 5G NR \cite{Lin-19}, 5G Evolution \cite{Kim-19, Ghosh-19, Parkvall-20}, and 5G-Advanced \cite{Lin-22, Chen-23, Liberg-24}. 

\subsubsection{INTRODUCTION} 
5G refers to mobile communication networks that meet the performance and service requirements of International Mobile Telecommunications for 2020 and beyond (IMT-2020) systems set by the International Telecommunication Union Radiocommunication Sector (ITU-R). The main performance requirements of 5G include {20 Gbit/s peak rate, 100 Mbit/s user-experienced rate, 1 ms latency, 1-10$^{\text{-}5}$ reliability, and 1 million/km$^2$ connection density} \cite{ITU-M.2410-0}. 5G supports a variety of use cases, from enhanced Mobile Broadband (eMBB) to Ultra-Reliable Low-Latency Communications (URLLC) and massive MTC (mMTC) \cite{ITU-M.2083-0}, as illustrated in Figure \ref{Fig_5G}. Consequently, 5G enables diverse services across a wide range of vertical domains. The 5G System, standardized by 3GPP, is the dominant 5G technology used worldwide. 

\begin{table*}[htb!]
\begin{center}
\caption{Summary of 3GPP 5G Releases 15-19}
\label{Table_5G}
\centering
\begin{tabularx}{\textwidth}{| >{\centering\arraybackslash}X | >{\centering\arraybackslash}X |
>{\centering\arraybackslash}X | 
>{\centering\arraybackslash}X |
>{\centering\arraybackslash}X |
>{\centering\arraybackslash}X |
>{\centering\arraybackslash}X |}
\hline
\centering
\vspace{3mm} \vspace{3mm} & \centering \textbf{Release 15} & \centering \textbf{Release 16} & \centering \textbf{Release 17} & \centering \textbf{Release 18}  & \vspace{1.5mm} \begin{center} \textbf{Release 19} \end{center} \\
\hline
\vspace{3mm} Timeline \vspace{3mm}  & 07/2017--12/2018 & 01/2019--06/2020 & 07/2020--03/2022 & 04/2022--12/2023 & 01/2024--06/2025 \\
\hline
\vspace{3mm} Evolution type \vspace{3mm}  & 5G Phase 1 & 5G Phase 2 & 5G Evolution & 5G-Advanced & 5G-Advanced Evolution \\
\hline
\vspace{3mm} Main features \vspace{3mm}  & \vspace{3mm} SDN - SBA - NFV - network slicing - edge computing - mmWave - mMIMO \vspace{3mm} & Private networks - IoT - unlicensed spectrum - positioning - SON & NTNs - RAN slicing - mmWave (higher) - NR light - IAB - UAS & AI/ML - mobile IAB - dual multi-SIM - mobile small data & AI/ML (air interface) - network energy savings - ambient IoT \\
\hline
\end{tabularx}
\end{center}
\end{table*}

\subsubsection{BACKGROUND}
The research on 5G began in the early 2010s in academia and industry. At that time, ITU-R also started its development work toward IMT-2020. The role of ITU-R was to build an overall vision, define performance/service requirements, evaluate candidate radio interface technologies (RITs), and select IMT-2020 compliant (5G) systems by the end of 2020. In 2014, a landmark 5G paper, "\textit{What Will 5G Be?}", was published \cite{Andrews-14}. In this early phase, this study identified the key elements of 5G, providing important guidelines for further research and development work. The first standardization efforts by 3GPP were taken in 2016. At the end of 2018, 3GPP completed the first standard for {the 5G System}, i.e., Release 15 \cite{Rel-15}. Release 15 was the first phase of 5G. The second phase, Release 16 \cite{Rel-16}, was completed in mid-2020. Release 17, a continuation of 5G evolution, was finalized in the first quarter of 2022. Release 18, i.e., 5G-Advanced, was completed by the end of 2023. Currently, Release 19 is in preparation and expected to be finalized by mid-2025. Release 20 will be the last 5G dominant Release, along with the first studies on 6G. The estimated timeline of Release 20 is 7/2025--12/2026. The main features of Releases 15--19 are summarized in Table \ref{Table_5G}. 

Releases 15 and 16 were submitted to the ITU-R evaluation process as a potential RIT candidate to meet the IMT-2020 requirements. The evaluation and selection process culminated in February 2021, when ITU-R published Recommendation ITU-R M.2150-0 \textit{"Detailed specifications of the radio interfaces of IMT-2020"} \cite{ITU-M.2150}. This document presented three approved IMT-2020 compliant systems, i.e., 3GPP 5G set of RITs (SRITs), 3GPP 5G RIT, and 5G RIT for India (5Gi). The first technology refers to 5G Non-Stand Alone (NSA), whereas the second represents 5G Stand Alone (SA). 5G NSA is dependent on 4G LTE and cannot work independently. The NSA version enabled a faster and smoother launch of 5G due to its easier implementation compared to SA. Unlike NSA, 5G SA operates independently without the aid of 4G LTE. The first commercial 5G networks were launched during the first half of 2019. Most of the first 5G deployments were based on the NSA version, operating at 3.5 gigahertz (GHz) carrier frequency.  

\subsubsection{RELEASE 15: 5G PHASE 1}
{Release 15} (7/2017--12/2018) introduced the first phase of 5G. The corresponding 5G System comprised three main parts: 5G Core, 5G RAN, and 5G user equipment (UE) \cite{Rel-15}. The 5G Core was designed to support flexible functionalities via its cloud-native service-oriented architecture, enabling novel services and business opportunities. The main features of 5G Core were SDN, SBA, NFV, and network slicing \cite{Ghosh-19}. The 5G RAN was designed to support high-speed mobile internet, mission-critical communication, and massive IoT connectivity. The main features of 5G RAN included mmWave communications (24.25--52.6 GHz), massive MIMO, OFDM with flexible numerology, ultra-lean transmission, dynamic time-division duplex (TDD), forward compatibility, and edge computing \cite{Dahlman-20}. In addition to a new set of features, each release specifies enhancements to a set of existing features. In Release 15, the main enhanced features were MTC and V2X communications. A comprehensive review of the Release 15 work items can be found in \cite{Rel-15}. 

\subsubsection{RELEASE 16: 5G PHASE 2}
Release 16 (1/2019--6/2020) introduced the second phase of 5G. The main new features were private networks, industrial/cellular IoT, NR unlicensed, NR positioning, self-organized network (SON) properties, local area network services, wireless-wireline convergence, high-speed train scenario, and maritime communication services \cite{Rel-16}. The main enhanced features were URLLC, V2X, mission-critical services, location services, network slicing, and SBA \cite{Rel-16}. Further details on the Release 16 work items can be found in \cite{Rel-16}. 

\subsubsection{RELEASE 17: 5G EVOLUTION}
Release 17 (7/2020--3/2022) continued the work on 5G evolution. The main new features were operation at higher mmWave frequencies (52.7--71 GHz), RAN slicing, NR light, IAB, UAS, and NTNs \cite{Ghosh-19}. The existing features that were further enhanced included massive MIMO, cellular/industrial IoT, URLLC, V2X, multicast/broadcast, positioning, network slicing, edge computing, network automation, non-public networks, and wireless-wireline convergence \cite{Ghosh-19}. More information on the work items of Release 17 can be found in \cite{Rel-17}. 

\begin{figure*}[!htb]
\center{\includegraphics[width=\textwidth]
{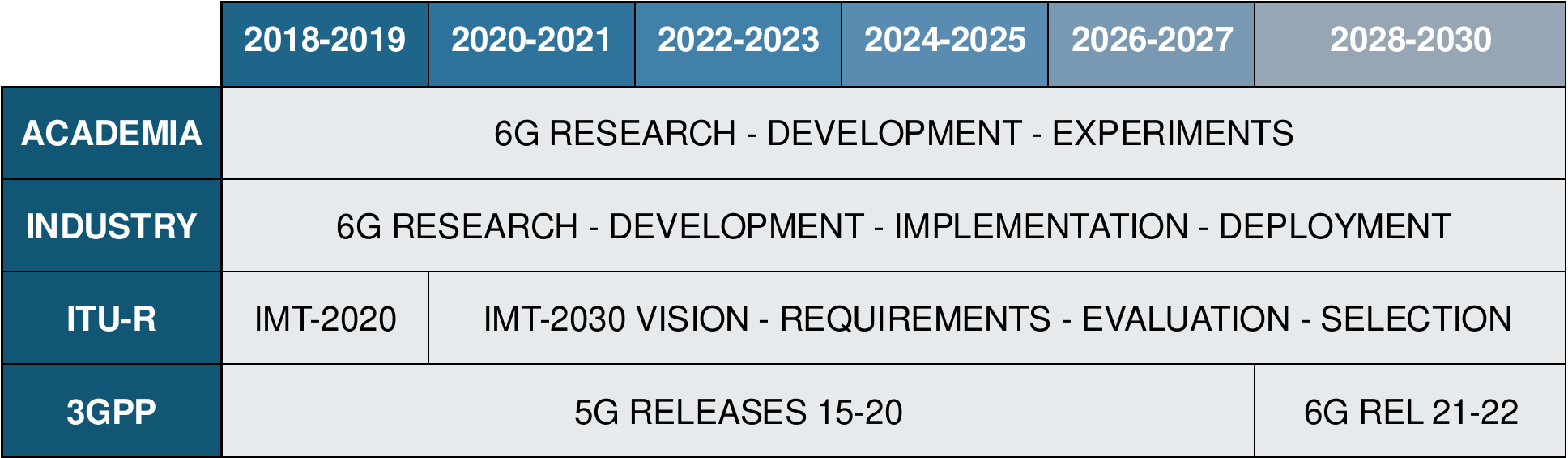}}
\caption{\label{Fig_Timeline}Estimated timeline for 6G development process.}
\end{figure*}

\subsubsection{RELEASE 18: 5G-ADVANCED}
Release 18 (4/2022--12/2023) introduced the first standard of 5G-Advanced, representing a significant evolution step for 5G \cite{Lin-22}. A major milestone was the introduction of AI/ML technology, which provides intelligent network features for 5G. Other novel features were mobile IAB, mobile-terminated small data, and a dual multi-subscriber identity module (SIM) \cite{Lin-22}. Enhancements were provided for coverage, mobility, MIMO, multicasting/broadcasting, spectrum sharing, multi-carrier, sidelink, and NTNs \cite{Lin-22}. An overview of the Release 18 work items can be found in \cite{Rel-18}. 

\subsubsection{RELEASE 19: 5G-ADVANCED EVOLUTION}
Release 19 (1/2024--6/2025) will be the first evolution of 5G-Advanced. 3GPP agreed on its content in December 2023, and the work began in the beginning of 2024. Release 19 continues the 5G evolution toward 6G, and can be seen largely as the preparation studies toward 6G, while still continuing the development based on commercial deployment needs. One focus will be on developing further use cases, such as the NTN evolution for NR and IoT, femto cells, wireless access/backhaul, low-power wake-up signals, ambient IoT, and multi-hop relays. Another focus area is to further improve the eMBB experience, such as extended reality (XR), mobility, MIMO operation, and duplex. Some interesting starting studies toward 6G technologies include channel modeling for 7-24GHz, which is one of the important new 6G bands in the 6G roadmap and channel modeling for sensing use cases. In addition, improvements of the other key performance indicators will continue in terms of network energy saving enhancements, AI/ML for air interface, and next-generation RAN (NG-RAN), as well as the enhancements on SON and minimization of drive tests (MDT).

\subsection{6G DEVELOPMENT PROCESS}
In this section, the generic development process of 6G is discussed, covering the big picture and the roles of ITU-R and 3GPP. Generally, ITU-R and 3GPP are vital consortia in 6G standardization, gathering together a large number of stakeholders to develop consistent, interoperable, and global 6G standards. The estimated timeline for the development process of 6G is summarized in Figure \ref{Fig_Timeline}. 

\subsubsection{BIG PICTURE}
The development of 6G follows a process similar to that of 5G. In short, the mobile network industry carries out the process from research and development to testing and implementation. Academia also participates in the research and development work, often cooperating with industry. Standardization and regulations are handled by the international and national standardization bodies and regulatory organizations. In these consortia, there are a vast variety of cooperating stakeholders involved, ranging from companies and universities to alliances and associations. Mobile network operators, service providers, and network vendors carry out the deployment and commercialization phases. 

\subsubsection{ROLE OF ITU-R}
\label{ITU}
ITU acts as a specialized agency of the United Nations for information and communication technologies (ICT) \cite{ITU}. Its slogan "committed to connecting the world" reflects the aim of advancing global connectivity and access to digital technologies \cite{ITU}. In ITU, there are 194 member states and over 1000 companies, universities, and organizations \cite{ITU}. ITU consists of three sectors, i.e., Radiocommunication (ITU-R), Standardization (ITU-T), and Development (ITU-D) \cite{ITU}. ITU-R is responsible for the worldwide management of radiocommunication and satellite systems. ITU-T standardizes international ICT systems through its Study Groups full of experts around the world. ITU-D aims to close the digital divide by providing digital services for the least developed countries. From these three sectors, ITU-R plays a key role in 6G standardization. Specifically, the role of ITU-R includes three main phases \cite{ITU-IMT}: phase 1) build a vision of IMT-2030 (2020--2023), phase 2) define performance/service requirements and evaluation methodology/criteria (2024--2026), and phase 3) evaluate candidate RITs and select the IMT-2030 compliant (i.e., 6G) ones by the end of 2030 (2027--2030). 

ITU-R began its phase 1 visioning work of IMT-2030 in early 2020 and completed it by the end of 2023. The developed IMT-2030 vision was based on two ITU-R documents. In November 2022, ITU-R published its first technological report on IMT-2030, i.e., Report ITU-R M.2516-0, titled "\textit{Future technology trends of terrestrial IMT systems toward 2030 and beyond}" \cite{ITU-M2516}. The report focused on emerging services and applications, technology trends and enablers, and enhanced radio interface and network technologies. In November 2023, the visioning phase culminated in the publication of Recommendation ITU-R M.2160-0, titled "\textit{Framework and overall objectives of the future development of IMT for 2030 and beyond}" \cite{ITU-M2160}. This document discussed the trends, usage scenarios, capabilities, and ongoing development of IMT-2030. 

The phase 2 of the IMT-2030 development process began in the beginning of 2024 and is expected to be finalized by the end of 2026. Phase 2 is two-fold. First, ITU-R defines the performance and service requirements that IMT-2030 systems need to meet. These requirements will be presented in a report on "Technical Performance Requirements". Second, ITU-R develops a comprehensive framework for the evaluation of candidate IMT-2030 systems by defining the evaluation criteria and methodology. This evaluation framework will be introduced in a report on "Evaluation Methodology". In phase 2, there will be also a third report, "Submission Template", for the candidate RIT proposals. 

Phase 3 is planned to begin in early 2027 and be completed by the end of 2030. In phase 3, ITU-R evaluates candidate RITs, developed by 3GPP and other organizations, against the IMT-2030 requirements, relying on the evaluation framework defined in phase 2. Finally, ITU-R selects and approves the RITs that meet the IMT-2030 requirements. These results will be reported in an ITU-R Recommendation document, with a working title of "Detailed specifications on the terrestrial radio interfaces of international mobile telecommunications-2030 (IMT-2030)" \cite{ITU-IMT}. This document will finalize the work on IMT-2030 systems. The latest details on the IMT-2030 development can be found in \cite{ITU-IMT}. 

ITU-R plays also another key role in the 6G development process by defining the frequency bands for IMT-2030 through World Radio Conferences (WRCs), organized around every four years. At WRC 2023, ITU-R already began the work toward the candidate spectrum bands for IMT-2030. In Resolution 256, ITU-R defined three frequency ranges from the mid-band spectrum to be studied for IMT-2030 \cite{ITU-WRC}, i.e., 4.4--4.8 GHz, 7.125--8.4 GHz, and 14.8--15.35 GHz. These spectrum bands are particularly important for achieving wide coverage since they have favorable propagation characteristics. The plan is to conduct a thorough study of these bands so that they can be approved at WRC 2027. 

In addition to the mid-band frequencies, ITU-R defined five spectrum bands from above 100 GHz frequencies for preliminary studies toward possible approval at WRC 2031, i.e., 102--109.5 GHz, 151.5--164 GHz, 167--174.8 GHz, 209--226 GHz and 252--275 GHz, as presented in Resolution 255 \cite{ITU-WRC}. The main promise of these frequency bands is to enable massive network capacity and ultra-broadband communication links in dense hotspot areas. Other potential use cases include ISAC, D2D, and wireless backhaul/fronthaul. For the feasibility of mobile communications at THz frequencies, ITU-R has published a report on the topic in May 2024, i.e., Report ITU-R M.2541-0, titled "\textit{Technical feasibility of IMT in bands above 100 GHz}" \cite{ITU-M2541}. The conclusion of the report was that it is technically feasible for the IMT-2030 systems to operate at frequencies above 100 GHz, provided that the corresponding challenges are properly resolved. While ITU-R identifies high path, atmospheric, and blocking losses as the main challenges, potential technical solutions include antenna, semiconductor, material, and MIMO technologies. 

\subsubsection{ROLE OF 3GPP}
3GPP is a large consortium that develops international standards for mobile telecommunication networks. Since its establishment in 1998, 3GPP has defined standards for 3G, 4G, and 5G networks. This work will continue for 6G as well. 3GPP comprises seven standard development organizations (SDOs) as organizational partners, 26 associations, alliances, and forums as market representation partners, and over 850 companies and organizations as individual members \cite{3GPP}. Whereas organizational partners have the authority to define/publish 3GPP standards and determine the general policies of 3GPP, market representation partners are able to give market advice and guidance to 3GPP. Individual members have the right to contribute to the 3GPP specifications. 

Organizational Partners include seven SDOs around the world \cite{3GPP}, i.e., the Alliance for Telecommunications Industry Solutions (ATIS) from North America, the European Telecommunications Standards Institute (ETSI) from Europe, China Communications Standards Association (CCSA) from China, Telecommunications Technology Association (TTA) from South Korea, the Association of Radio Industries and Businesses (ARIB) from Japan, Telecommunication Technology Committee (TTC) from Japan, and Telecommunications Standards Development Society, India (TSDSI) from India. These SDOs play a key role in 3GPP and 6G standardization. 

3GPP cellular standards are developed through Releases, each of which comprises a major package of new technological features for mobile networks, covering the core network, radio access, terminal, and service capabilities. Every mobile generation contains a set of Releases, each providing an evolutionary step forward. For example, 3G and its evolution was defined in five Releases (99, 4--7), 4G in seven Releases (8--14), and 5G will include six Releases (15--20). Note that there is some overlapping in Releases when the generations are changing, as the current and new/old generation features are developed at the same time. In each Release, new technological features are studied and defined in Technical Reports (TRs) and Technical Specifications (TSs). 

Specifically, TRs and TSs are the outputs of study items and work items, respectively. A study item is an exploratory phase to examine the potential and feasibility of new features and technologies, with no commitment to immediate standardization. A work item is a phase to develop exact specifications for new features and technologies, with the intention of producing formal standards. The creation of TRs/TSs is handled by the Technical Specification Groups (TSGs) and their dedicated Working Groups (WGs). In 3GPP, there are three TSGs \cite{3GPP}: Core Network and Terminals (CT), Radio Access Networks (RAN), and Service and System Aspects (SA). RAN and SA TSGs have six WGs each, whereas CT has four. Overall, TSGs are responsible for preparing, approving, and maintaining TRs/TSs. 

When 5G evolution approaches its limits, the standardization efforts begin to shift toward 6G. The 6G standardization process is estimated to begin in 2025. If Release 19 will be completed by mid-2025, as planned, the preparation of Release 20 will begin in July 2025. It is planned that Release 20 will begin the 6G standardization process by being the first Release to study 6G features. Assuming the average 18 months preparation time for a Release, Release 20 is anticipated to be finalized by the end of 2026 and the preparation of Release 21 to begin in early 2027. Release 21 is expected to be the first Release to define 6G technologies. Thus, being the first official 6G standard. Most likely Release 21 will be included in the 3GPP submission to ITU-R as a potential 6G radio interface candidate to meet the IMT-2030 requirements. It remains to be seen if Release 22 will complement the 3GPP submission, similarly as Release 16 complemented Release 15 in the IMT-2020 submission. If Release 21 will be completed by mid-2028, as assumed, the first commercial 6G networks could be launched already around late 2028 to early 2029. Thus, the worldwide commercialization phase of 6G networks could begin around 2029--2030. 

\subsection{6G RESEARCH ACTIVITIES}
\begin{figure}[!tb]
\center{\includegraphics[width=\columnwidth]
{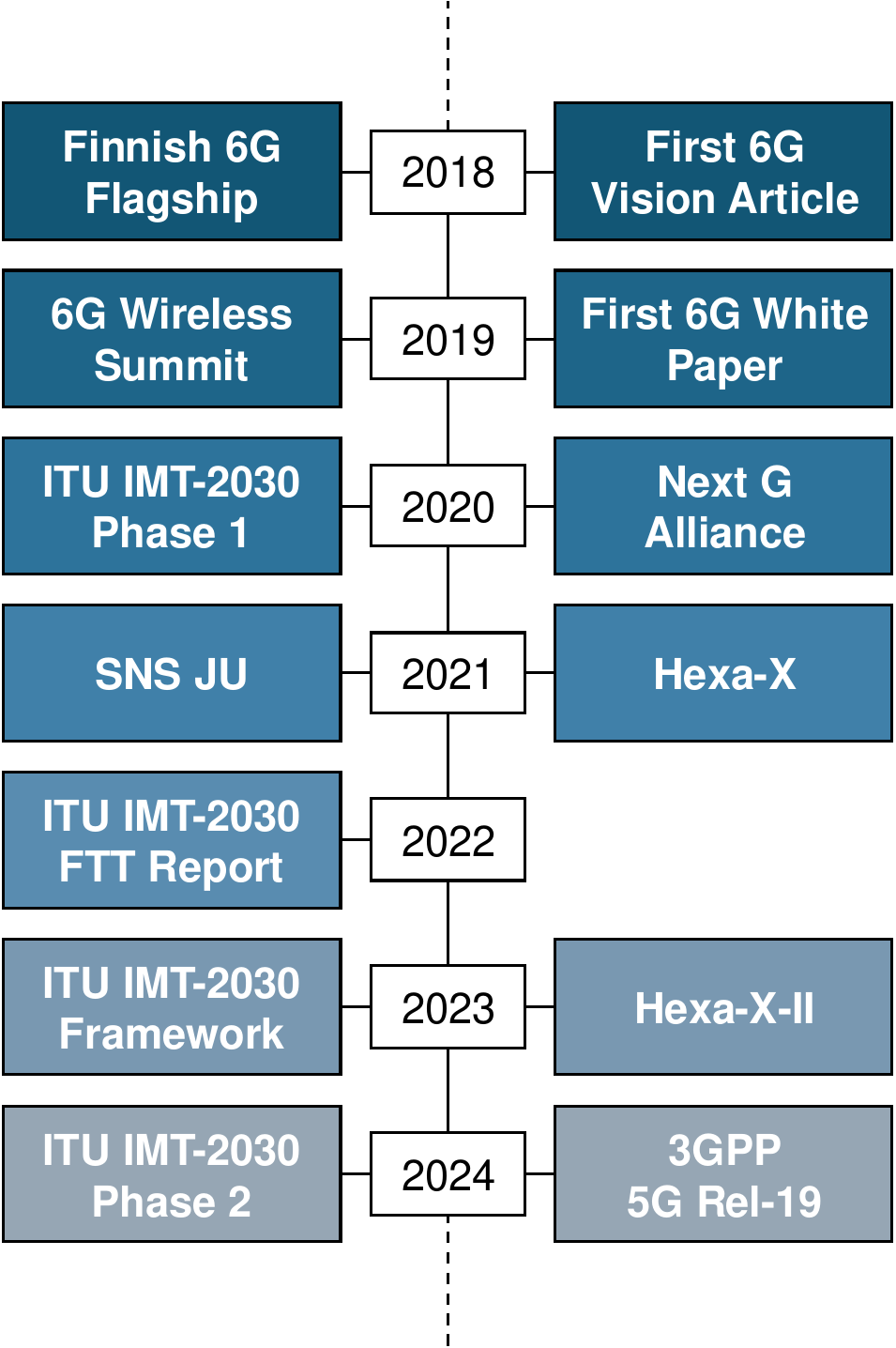}}
\caption{\label{Fig_Activities}Main benchmarks of 6G research.}
\end{figure}

This section highlights the primary research steps taken worldwide since 2018. Many different types of activities are introduced, including initiatives, projects, collaborations, funding, experiments, events, and publications. For clarity, the activities are introduced in a chronological order. Further details on the 6G activities presented in this section can be found in \cite{RCRWireless}, where a comprehensive list of 6G research progress is provided. The main benchmarks of 6G research are summarized in Figure \ref{Fig_Activities}. The identification of the main research activities and benchmarks is based on the extensive search of the 6G literature. 

\subsubsection{ACTIVITIES IN 2018}
To the best of our knowledge, the first research activities on 6G, which focused solely on 6G and used the term "6G", were conducted in 2018 in terms of projects, events, and publications. The first large-scale 6G project \cite{6GFlagship}, with a total volume of 249 million euros, was launched in April in Finland by a consortium of the Finnish leading academic and industrial organizations. The 6G Flagship Program is an eight-year research initiative that aims to develop and test 6G enabling technologies. In September, the first 6G vision paper was published in the Institute of Electrical and Electronics Engineers (IEEE) Vehicular Technology Magazine \cite{David-18b}, under the title "\textit{6G Vision and Requirements: Is There Any Need for Beyond 5G?}" \cite{David-18}. The paper contemplated the need for 6G, with a positive conclusion by reviewing the strengths and deficiencies of each generation. In November, another 6G vision paper was published by the Finnish 6G Flagship Program, sketching its tentative vision for 6G and emphasizing the importance of interdisciplinary research in wireless communications, computer engineering, electronics, and material science \cite{Katz-18}. In December, Carleton University organized the first workshop series on 6G, covering physical (PHY) layer technologies, network architectures, and AI/ML methods. 

\subsubsection{ACTIVITIES IN 2019}
6G research activities began to spread around the world in 2019. In January, South Korean LG Electronics opened a center for 6G research at the Korean Advanced Institute of Science and Technology (KAIST). Samsung also established a new 6G research unit called the Advanced Communications Research Center. In March, the Federal Communications Commission announced that it will open the spectrum from 95 GHz to 3 THz for experimental usage and 6G research in the United States (US). In April, the UK launched a five-year project to study optical wireless technologies for 6G and beyond. China Mobile and Tsinghua University reported their cooperation to study 6G technologies in May. In June, SK Telecom partnered with Ericsson and Nokia to conduct 6G research. South Korea and Finland reported joint cooperation in 6G development. 

In August, Keysight Technologies joined the Finnish 6G Flagship Program, strengthening its testing and experimental capabilities. Huawei announced that it had already begun to study 6G at its Canadian research center in Ottawa. In November, the Chinese Ministry of Science and Technology stated that it is going to establish two national 6G groups. One will promote 6G research and development work, consisting of governmental departments, while the other will be formed by academic and industrial organizations, focusing on the technical part of 6G. Rohde $\&$ Schwarz reported collaboration with a group of German research institutions to study communications at THz frequencies. The Indian company Tech Mahindra and Finnish government-led Business Finland agency signed a Memorandum of Understanding (MoU) for the joint exploration of 5G/6G, with an agreement to establish an innovation lab in Finland. 

A handful of 6G events were arranged around the world in 2019. The first conference, fully dedicated on 6G, was held in March, i.e., the annual 6G Wireless Summit organized by the Finnish 6G Flagship Program \cite{6GSummit}. As an outcome of the Summit, the first 6G white paper, "\textit{Key Drivers and Research Challenges for 6G Ubiquitous Wireless Intelligence}" \cite{Latva-aho-20}, was published. A few 6G panels/workshops were organized at major conferences. Different research groups around the world began to propose their own visions of what 6G might be in the 2030s \cite{Zhang-19, Yang-19, Strinati-19, Zong-19, Letaief-19, Zhang-19c}. These 6G vision papers discussed many aspects of 6G, such as key trends, main requirements, system architectures, potential technologies, possible applications, opportunities, and challenges. In September, IEEE Vehicular Technology Magazine released a special issue on 6G. 

\subsubsection{ACTIVITIES IN 2020}
6G research continued to spread around the world in 2020. In January, the government of Japan organized a 6G panel with academic and industrial representatives to plan a comprehensive strategy for the Japanese 6G research. In February, the ITU-R Working Party 5D (WP 5D), in charge of IMT systems, decided to begin the visioning work toward IMT-2030. In May, the Chinese technology vendor ZTE and mobile network operator China Unicom agreed to collaborate on 6G research, including the joint development, verification, and trials of 6G enabling technologies. Later in May, the Alliance for Telecommunications Industry Solutions (ATIS) released a call to action to advance the US toward a global leadership in 6G and beyond. The aim was to build close collaboration between the government, academia, and industry to provide a comprehensive path from research and development to standardization and commercialization. In June, the Finnish 6G Flagship Program published eleven 6G white papers, providing a comprehensive study on different aspects of 6G \cite{Rajatheva-20, Ali-20, Mahmood-20, Taleb-20, Bourdoux-20, Peltonen-20, Matinmikko-20, Saarnisaari-20, Pouttu-20, Ylianttila-20, Yrjola-20}. Later in autumn, the 6G Flagship Program released an online webinar series, in which each white paper was introduced and thoroughly discussed.

In July, Samsung published a white paper, revealing its 6G vision on the next hyper-connected experience for all \cite{Samsung-6G}. In August, the South Korean government announced that it is planning to allocate 169 million dollars for the development of 6G technologies during the period from 2021 to 2026, and then start a pilot project for 6G services, with five focus areas: immersive content, digital healthcare, autonomous vehicles, smart factories, and smart cities. In September, Nokia released a white paper on its 6G vision, merging digital, physical, and human worlds \cite{Nokia-6G}. In October, the Next G Alliance was established to promote 6G in North America \cite{NextG}. The founding members included numerous major players in the mobile industry. The target of the Next G Alliance is to ensure that North America is a global leader in the 6G development. ITU-R WP 5D released a Liaison statement, inviting external organizations to contribute to the IMT-2030 visioning work by providing views on different technological trends and aspects of 6G. In November, China launched the first experimental "6G" satellite to test THz communication in space. The University of Surrey opened a 6G Innovation Center and published a white paper, introducing its 6G vision \cite{Tafazolli-20}. 

In 2020, a few 6G conferences were organized virtually. The 6G theme spread to all major telecommunication conferences in the forms of keynote speeches, plenary talks, panels, workshops, and forums. Many major technology companies published white papers to reveal their initial 6G visions. IEEE Vehicular Technology Magazine and IEEE Access released special issues on the 6G-specific topics. Numerous 6G visions \cite{Dang-20, Tomkos-20, Chen-20, Shafin-20, Viswanathan-20, Giordani-20, Alsharif-20, Saad-20, Chowdhury-20, Akyildiz-20, Khan-20, Tariq-20, Bariah-20, Ziegler-20, Liu-20b, Gui-20, Yang-20, Akhtar-20} and one survey \cite{Lu-20} were also published in 2020. 

\subsubsection{ACTIVITIES IN 2021}
At the beginning of 2021, many European Union (EU)-funded 6G projects were launched, including also Hexa-X. The Hexa-X initiative is the European 6G flagship project that aims to develop 6G technology enablers to connect human, physical, and digital worlds \cite{HexaX}. 
Also in January, a team of five German Fraunhofer research institutes established the 6G Sentinel project to develop key technologies for 6G, with the main focus on THz communications and flexible network solutions. The Japanese government reported that it is planning to invest 482 million dollars for its 6G research and development work, i.e., 60 $\%$ of the funding for public-private sectors to develop 6G technologies and 40 $\%$ to set up a research facility to test the developed technologies. 

In March, ITU-R WP 5D established the 6G Vision Group and started the development work toward "IMT-2030 Framework". The one6G association was established to promote 6G development by bringing together industrial and academic players around the world \cite{one6G}. In April, the German government announced that it will fund German 6G research into key technologies and innovative products with 700 million euros by 2025. LG reported a recent collaboration with Keysight Technologies and KAIST to develop 6G enabling technologies, especially for THz communications. A few months later, LG announced a successful demonstration of wireless THz transmission over a 100 meter long outdoor link. Furthermore, LG demonstrated 6G RF front-end technology, jointly developed with Fraunhofer Research Institute, for sub-THz frequencies using the Keysight testbed equipment at the Korean Science and Technology Exhibition 2021. 

In June, Vodafone Germany announced that it will open a new 5G/6G research and development center in Dresden. The Finnish 6G Flagship Program and Japanese Beyond 5G Promotion Consortium agreed on a collaboration for joint research and development on 6G technologies, aiming to contribute to the regulation and standardization of 6G. Samsung and the University of California, Santa Barbara demonstrated an end-to-end sub-THz link with fully digital beamforming at a 140 GHz carrier frequency at the IEEE International Conference on Communications (ICC) 2021. In July, the University of Texas partnered with a group of American and Korean technology companies to establish a joint 6G research and development center, with a special focus on wireless ML, advanced sensing, innovative networking, new spectrum technologies, and satellite communications. Ericsson and the Massachusetts Institute of Technology reported a collaboration on two 6G-related project, focusing on neuromorphic computing and "zero-energy" devices. The Chinese technology company OPPO announced the establishment of a research team to study 6G requirements, technologies, and system architectures. OPPO also released a 6G white paper on AI-based intelligent networking \cite{Oppo-21}. Mobile network operators LG UPlus and KDDI reported that they have signed an MoU to study novel business opportunities for 5G networks and develop 6G enabling technologies. 

According to the report of Nikkei in September \cite{Nikkei}, the distribution of 6G patents among major countries is as follows: China 40.3 $\%$, US 35.2 $\%$. Japan 9.9 $\%$, Europe 8.9 $\%$, and South Korea 4.2 $\%$. In late 2021, Huawei reported that it had demonstrated a 240 Gbit/s data rate over a 500 meter outdoor link operating at a 220 GHz carrier frequency with a bandwidth of 13.5 GHz. In November, the University of Oulu and the University of Tokyo agreed on a bilateral partnership in 6G research. Ericsson and King Abdullah University of Science and Technology announced a cooperation on 6G research with the main emphasis on ML-assisted massive MIMO, RISs, and THz communications. The EU founded the European Smart Networks and Services Joint Undertaking (SNS JU) targeting to guarantee the European industrial leadership in 5G/6G and advance green/digital transition \cite{SNSJU}. The funding budget for SNS JU is 900 million euros for the period 2021-2027. In December, SNS JU accepted its first Work Program 2021-2022 with a budget of 240 million euros for the research and innovation work on 5G evolution and 6G. The European 6G flagship project Hexa-X published a 6G article \cite{Uusitalo-21}, discussing its 6G vision and the value 6G is expected to bring. The Next G Alliance and South Korean 5G Forum signed an MoU to accelerate the development of 6G networks. 

Numerous, mostly virtual or hybrid, 6G events were arranged in 2021, including conferences, workshops, panels, forums, plenary talks, and keynote speeches. In terms of publications, many IEEE journals and magazines released the 6G-specific special issues. Furthermore, many technology companies, academic organizations, and research coalitions published 6G white papers. In addition, several 6G visions \cite{You-21, Bhat-21, Rajatheva-21, Tataria-21, Uusitalo-21} and surveys \cite{Mahmoud-21, Jiang-21, Alwis-21, Dogra-21, Alsabah-21} were published. A handful of 6G books were also released. 
 
\subsubsection{ACTIVITIES IN 2022}
Many different types of activities occurred in January. The project TiC6G was launched by the University of Glasgow, with a group of academic and industrial partners, aiming to test prototype devices at THz frequencies using the new cutting-edge laboratory equipment. The Japanese operator Nippon Telegraph and Telephone (NTT) reported that it will build a 6G pilot network for the Osaka World Expo 2025. The Chinese Purple Mountain Laboratories, together with Fudan University and China Mobile, reported that it had obtained a data rate of 206.25 Gbit/s for a wireless THz transmission in a laboratory environment. The Next G Alliance published a white paper titled "\textit{Green G: The Path toward Sustainable 6G}" \cite{NextGLib}. The University of Oulu announced a partnership with Jio Estonia to study 6G technologies, particularly for non-terrestrial communications, holographic radio, and three-dimensional (3D) connected intelligence. Huawei published its 6G vision in a white paper entitled "\textit{6G: The Next Horizon}", envisioning the transition from connected things to connected intelligence \cite{Huawei-6G}. 

In February, Keysight Technologies announced a partnership with Samsung to develop, test, and verify 6G technologies. The Next-Generation Mobile Networks (NGMN) Alliance published a white paper on 6G use cases, categorizing them into four main classes, i.e., enhanced human communications, enhanced machine communications, enabling services, and network evolution \cite{NGMNpub}. At the Mobile World Congress (MWC) 2022, LTE showcased its cutting-edge RIS technology, including four different prototypes, i.e., liquid crystal RIS, PIN diode RIS, transparent planar RIS, and transparent flexible RIS. At the same venue, the US technology company VMware reported its collaboration with German research institutes to accelerate 6G research, focusing on the fusion of cloud, networking, and AI technologies. 

In April, Kyocera, a Japanese technology firm, announced that it has developed a transmissive metasurface that can be used to extend coverage in 5G and 6G networks by redirecting radio signals to avoid obstacles. In May, an MoU was signed between the European 6G Smart Networks and Services Industry Association (6G-IA) and the Japanese Beyond 5G Promotion Consortium to accelerate the development of 6G networks and deepen the cooperation between Europe and Japan. Samsung released a new white paper, "\textit{6G Spectrum: Expanding the Frontier}" \cite{Samsung-22}, discussing spectrum policies and potential frequency bands for 6G. The Finnish leading academic and industrial organizations established a national 6G coalition, called 6G Finland, aiming to strengthen national 6G cooperation, form new international partnerships, and promote the Finnish 6G expertise on a global scale. LG demonstrated its 6G relevant full-duplex and power amplifier technologies at the IEEE ICC 2022. The Next G Alliance and the Japanese Beyond 5G Promotion Consortium agreed on 6G cooperation by signing an MoU in May. 

In June, the Next G Alliance published a white paper, titled "\textit{6G Applications and Use Cases}" \cite{NextGLib}, focusing on four categories, i.e., networked robotics, XR, distributed communication and sensing, and personalized user experiences. Nokia, NTT, and NTT Docomo formed a 6G partnership to demonstrate an AI/ML-enhanced air interface and sub-THz communication at 140 GHz spectrum. The Singaporean Nanyang Technological University announced the use of Keysight test and measurement solutions to verify its on-chip THz electronic-photonic 6G device technologies. 
The Japanese companies NTT Docomo, Fujitsu, and NEC reported to perform joint 6G indoor and outdoor trials. 6G-IA and the Chinese IMT-2030 Promotion Group signed an MoU for 6G cooperation. The Open RAN (O-RAN) Alliance reported a launch of its Next Generation Research Group, focusing on the utilization of O-RAN technologies in 6G. 

In July, LG Uplus and Nokia signed an MoU for collaboration in 5G/6G standardization. Nokia announced its leadership in a German 6G light house project 6G-ANNA, aiming to promote 6G development and standardization in Germany. The Next G Alliance published a white paper, entitled "\textit{6G Technologies}" \cite{NextGLib}, on the technologies required to meet the demands of its 6G vision. The UK government reported to grant 25 million pounds funding for the research and development work on 6G technologies. In August, the Next G Alliance and European 6G-IA signed an MoU for the agreement of 6G research collaboration, including joint workshops, webinars, and trials. While already supporting 6G research at the University of Texas at Austin, the Northeastern University, the University of Surrey, and Viavi Solutions announced a new funding program to advance worldwide 6G development in academia and industry. The Next G Alliance reported the establishment of a 6G research council, aiming to promote cooperation between academia and industry and align North American 6G strategy with the US and Canadian governments. 

In September, LG reported its final THz communication test at the Fraunhofer Heinrich-Hertz Institute in Berlin, with a successful information transfer over a 320 meter outdoor link at a frequency range between 155 and 175 GHz. The Finnish University of Oulu and Japanese National Institute of Communication and Technology (NICT) announced a cooperation in B5G/6G research. Rohde $\&$ Schwarz and China Mobile reported their partnership in the development of ISAC solutions for 6G. The Singaporean Infocom Media Development Authority and the Singapore University of Technology and Design revealed their collaboration to establish a 6G Research and Development (R$\&$D) laboratory in Singapore, being the first in kind in Southeast Asia. Nokia informed its 5G/6G collaboration with Vodafone New Zealand. 
In October, Keysight Technologies and the Indian Institute of Technology, Madras signed an MoU, expanding their cooperation from 5G to 6G design. Deutsche Telekom reported its leadership in the German government-funded project 6G-TakeOff, aiming to develop an architectural framework for the integration of terrestrial and non-terrestrial networks in 6G. Ericsson and the University of Texas at Austin deepen their collaboration into 6G-empowered XR. VMware announced that it is launching the Next G-AI Research and Innovation Center in Montreal, Canada, gathering multidisciplinary expertise to advance the technological development of 5G and 6G systems. Samsung Electronics reported the founding of a new 6G research group in the UK, which is part of the global 6G development strategy. The German government-funded project 6G NeXt was kicked-off, with Deutsche Telekom in lead, and the goal of developing a flexible infrastructure platform to study and test future XR applications. 

In November, ITU-R WP 5D completed its work on the first report on IMT-2030, i.e., "IMT-2030 Future Technology Trends" \cite{ITU-M2516}. It was reported that the Indian Department of Telecommunications suggests opening up the frequency range from 95 GHz to 3 THz for experimenting and testing of 6G technologies and products. Nokia announced its leadership in the German government-supported KOMSENS-6G project, which focuses on the integration of sensing and communication. The Japanese Beyond 5G Promotion Group and Northeastern University agreed on a 6G collaboration by signing an MoU to foster the development of 6G networks. NTT Docomo and SK Telecom announced their partnership in the fields of smart-life and metaverse, searching for opportunities to jointly produce immersive XR content. Ericsson reported its ten-year 6G research initiative in the UK, focusing on the areas of AI, cognitive networks, and network security. Nokia announced the launch of a new 5G/6G R$\&$D center in Portugal. The UK government reported a 110 million pound grant for 5G/6G research and development. 28 million is dedicated to the collaboration of academia (Universities of York, Bristol, and Surrey) and industry (Nokia, Ericsson, and Samsung), while 80 million is invested in a new 5G/6G R$\&$D laboratory. All types of physical, hybrid, and virtual 6G events were held in 2022. Numerous IEEE special issues on 6G were released. A handful of 6G visions and surveys were also published \cite{Wang-22, Hakeem-22, Alraih-22, Salameh-22, Asghar-22, Akbar-22}. 
Moreover, lots of 6G white papers and books were written. 

\subsubsection{ACTIVITIES IN 2023}
In January, the European Hexa-X-II flagship project (i.e., a continuation of Hexa-X) was launched \cite{HexaX}. As part of SNS JU Call 1, 35 other EU funded 5G/6G research and innovation projects were initiated, with a total funding of 250 million euros, in order to strengthen the European expertise in the development of future mobile networks \cite{SNSJU}. 6G-IA and the European Telecommunications Standards Institute (ETSI) agreed on a close 5G/6G cooperation with each other by signing an MoU, with the goal of advancing 6G pre-standardization activities. The ATIS and O-RAN Alliance agreed on a partnership via an MoU to foster the development of O-RAN technologies. Rohde $\&$ Schwarz announced its channel measurement campaign conducted in urban micro outdoor and indoor scenarios at frequencies of 158 GHz and 300 GHz, contributing to the ITU-R report titled "\textit{Technical feasibility of IMT in bands above 100 GHz}". The Finnish University of Oulu and South Korean Electronics and Telecommunications Research Institute announced the launch of their new collaboration project, 6GBRIDGE-6GCORE, aiming to develop a service-centric 6G system architecture. 

In February, the University of Sheffield announced the opening of the new UK Research and Innovation National 6G Radio Systems Facility, with top-notch 6G R$\&$D capabilities and support from more than 40 companies and academic organizations. Nokia, NTT Docomo, and NTT reported two major milestones achieved on the evolution road toward 6G, including sub-THz communication (25 Gbit/s at 144 GHz) and the integration of AI into air interface (ML-based waveform), implemented as proof of concepts at Nokia Bell Labs in Germany. The Finnish 6G Flagship and Brazilian research institute Inatel announced their partnership in developing advanced 6G solutions for rural and remote areas. At the MWC 2023, Nokia demonstrated network sensing capabilities using its prototype radio equipment. Also at MWC, Bosch reported its collaboration with Nokia on 6G-based IoT solutions for industry 4.0 applications. The Taiwanese operator Chunghwa Telecom and Ericsson signed an MoU at MWC, joining forces to advance 5G and 6G standardization. 

In March, the Indian government unveiled a national 6G initiative, targeting to launch 6G networks in India by 2030. Anritsu and Danish Aalborg University announced their collaborative 6G project on channel sounding for joint communication and sensing at mmWave and sub-THz frequencies. In April, the Spanish operator Telefonica, NEC, Bluespecs, and IMDEA Research Institute reported the launch of the ENABLE-6G project, funded by the EU and Spanish government. The project aims to develop mechanisms to tackle the challenges 6G will encounter in terms of performance, energy efficiency, network sensing, and privacy/security. Keysight Technologies, the University of Surrey, and the National Physical Laboratory demonstrated over 100 Gbit/s data rates at a 300 GHz carrier frequency using their new sub-THz 6G testbed. 

In May, the Next G Alliance released a white paper on the 6G roadmap for vertical industries \cite{NextGLib}. The French technology company Capgemini announced the opening of a new 6G research lab in India to enable testing and experimentation of novel 6G solutions. The Brazilian government reported the allocation of 36 million dollars to three research centers to advance the development of 5G, 6G, and O-RAN technologies. Rohde $\&$ Schwarz and the French Institute of Electronics, Microelectronics and Nanotechnology announced their cooperation in THz communication research, with successful communication over 645 meters using a THz backhaul link operating at a frequency of 300 GHz. 

In June, the Next G Alliance published a white paper on 6G technologies for the wide-area cloud evolution \cite{NextGLib}. InterDigital and the University of Surrey announced a 6G research partnership to study joint communication and sensing and RIS technologies. The Indian Department of Telecommunications reported the establishment of the Brahat 6G Alliance, consisting of 75 companies, with the goal of strengthening 6G expertise and facilitating market access of Indian telecommunication products and services. In July, Northeastern University launched an Open Testing and Integration center to develop and test advanced O-RAN technologies. The GSM Association and the European Space Agency joined forces by signing an MoU to advance the integration of terrestrial and satellite networks in 5G and 6G. NTT and Fujitsu demonstrated a data rate of 30 Gbit/s over a 0.5 meter link distance at 300 GHz operating frequency using beamforming. 

In August, the Next G Alliance published 6G white papers on social/economic opportunities and spectrum considerations \cite{NextGLib}. The Next G Alliance also signed an MoU with the Indian Brahat 6G Alliance on 6G collaboration. In September, the NGMN Alliance released a 6G report from the perspective of network operators, titled "\textit{6G Position Statement: An Operator View}" \cite{NGMNpub}. InterDigital and the Indian Institute of Technology, Kanpur agreed on a 6G cooperation with a special focus on extreme MIMO. LG Electronics and LG Uplus reported a record-breaking 6G spectrum test in an outdoor environment at LG Sciencepark in Seoul, sending data over 500 meters at THz frequencies. In October, Nokia opened a 6G laboratory at its Global R$\&$D center in Bangalore, India to promote 6G development, especially in the network sensing technology. InterDigital and the University Carlos III of Madrid agreed on a 6G research partnership, focusing on joint communication and sensing. Ericsson established a new 6G research team at its R$\&$D center in Chennai, India. 

In November, the government of South Korea revealed its 6G development plan, with a total budget of 325 million dollars. The University of Glasgow launched a 6G research laboratory, named "Terahertz On-chip Circuit Test Cluster for 6G Communications and Beyond". The 6G-SANDBOX initiative and Taiwanese Industrial Technology Research Institute signed an MoU to advance 6G cooperation between Europe and Taiwan. ITU-R completed its visioning work by publishing the first Recommendation document on IMT-2030, i.e., "IMT-2030 Framework" \cite{ITU-M2160}. In December, SNS JU and the Next G Alliance released a document titled "\textit{EU-US Beyond 5G/6G Roadmap}" to characterize their collaboration in the development of 6G networks for 2025 and beyond. 3GPP completed Release 18 and agreed on the content of Release 19. 

\subsubsection{ACTIVITIES IN 2024}
In January, ITU-R began the second phase of IMT-2030 development, i.e., defining its requirements and corresponding evaluation methodologies. 3GPP began the preparation of Release 19, continuing the development of 5G-Advanced toward 6G. SNS JU launched its Call 2 phase, including 27 6G research and innovation projects, with a total EU funding of 130 million euros \cite{SNSJU}. NTT Docomo introduced their forthcoming AI-empowered 5G/6G technologies, such as human-augmented mobile platform to express immersive sensory perceptions in the metaverse environments. 

In February, SK Telecom and Intel reported that they have developed an AI-empowered "Inline Service Mesh" technology for the 6G core network to reduce latency and increase service efficiency. China Mobile announced the launch of its 6G test satellite, claiming it to be the first of its kind in the world. Samsung and Princeton University announced their partnership in innovative 6G research through the Princeton NextG Initiative Corporate Affiliates Program, aiming to develop new technological innovations and foster cooperation between academia, industry, and regulators. Other collaborators in this program are Nokia Bell Labs, Ericsson, Vodafone, MediaTek, Intel, and Qualcomm Technologies. SK Telecom and Rohde $\&$ Schwarz joined a group of industrial players collaborating in 6G spectrum trials at sub-THz frequencies. The existing partners include NTT Docomo, NTT, NEC Corporation, Fujitsu, Keysight Technologies, Nokia, and Ericsson. Nokia and the Indian Institute of Science announced their 6G collaboration, focusing on three 6G research areas: network architecture, radio technologies, and AI/ML-aided air interface. At the MWC 2024, Ericsson and Turkcell partnered in 6G research to promote technological advances in Turkie. Also at the MWC, Ericsson and Turk Telecom agreed on 6G cooperation by signing an MoU. 

In March, Nvidia announced that its 6G research cloud platform, based on a digital twin technology, is available, allowing simulations and testing in the realistic 6G environments. In April, Viavi Solutions reported its progress with Northeastern University in the development of a large-scale digital twin of a 6G network. The 6G Non-Terrestrial Networks project, funded by SNS JU, published a white paper on its vision of NTNs in 6G networks, highlighting the importance of non-terrestrial access and discussing the markets, coverage, and design of NTNs. The University of Glasgow reported that it has designed a digitally controlled metasurface antenna for the mmWave frequencies, paving the way toward advanced 6G antenna technologies. TDRA, the telecommunication regulator of the United Arab Emirates, unveiled its 6G roadmap, which promotes the leadership in the 6G development. 

In May, the South Korean mobile operator KT Corporation and Nokia announced their partnership in developing O-RAN technologies for 6G networks. Ericsson published a white paper on 6G spectrum considerations, discussing the importance of mid and sub-THz bands to enable mobile life in the 2030s \cite{Ericsson-6G}. The National Telecommunications and Information Administration of the US released a request for comment on 6G, asking for opinions of experts from academia and industry on the timing of trials/commercialization, societal benefits, and disaster resilience of 6G. A Japanese consortium, NTT Docomo, NICT, Panasonic, and SKY Perfect JSAT Corporation, informed their successful field test in the 38 GHz spectrum band to simulate the usage of high-altitude platform stations (HAPSs) at an altitude of 4 kilometers using a small plane Cessna. The test was claimed to be the first in kind, providing practical progress toward the use of NTNs at high altitudes in 6G. In June, Nokia and the Gati Shakti Vishwavidyalaya university announced that they have formed a partnership to conduct 5G/6G research on transportation and smart factory use cases. Nokia and Nordic mobile operator Telia reported their successful field trial at the upper 6 GHz spectrum, agreed on at the WRC 2023. These results pave the way toward the needed capacity enhancements in the 6G era using the mid-band spectrum. 

\begin{table*}[htb!]
\begin{center}
\caption{Comparison of comprehensive 6G surveys}
\label{Table_Comparison}
\centering
\begin{tabularx}{\textwidth}{| >{\centering\arraybackslash}X | >{\centering\arraybackslash}X |
>{\centering\arraybackslash}X | 
>{\centering\arraybackslash}X |
>{\centering\arraybackslash}X |
>{\centering\arraybackslash}X |
>{\centering\arraybackslash}X |
>{\centering\arraybackslash}X | 
>{\centering\arraybackslash}X |
>{\centering\arraybackslash}X |
>{\centering\arraybackslash}X |
>{\centering\arraybackslash}X |}
\hline
\centering
\vspace{3mm} \textbf{Ref.} \vspace{3mm} & \centering \textbf{Evolution  1G-6G} & \centering \textbf{5G Overview} & \centering \textbf{Develop. Process} & \centering \textbf{Activities} & \centering \textbf{Vision} & \centering \textbf{Appli-cations} & \centering \textbf{Use Cases} & \centering \textbf{Require-ments} & \centering \textbf{AI/ML} & \centering \textbf{Techno-logies} & \vspace{1.5mm} \begin{center} \textbf{7G Vision} \end{center} \\
\hline
Alsharif -20 \cite{Alsharif-20} & \textcolor{red}{\(\blacksquare\)} & \textcolor{green}{\(\blacksquare\)} & \textcolor{green}{\(\blacksquare\)} & \textcolor{blue}{\(\blacksquare\)}  & \textcolor{blue}{\(\blacksquare\)} & \textcolor{blue}{\(\blacksquare\)} & \textcolor{blue}{\(\blacksquare\)} &  & \textcolor{blue}{\(\blacksquare\)} & \textcolor{blue}{\(\blacksquare\)} &  \\
\hline
Akyildiz -20 \cite{Akyildiz-20} &  & \textcolor{blue}{\(\blacksquare\)} & \textcolor{green}{\(\blacksquare\)} &   & \textcolor{red}{\(\blacksquare\)} & \textcolor{blue}{\(\blacksquare\)} &  & \textcolor{blue}{\(\blacksquare\)} & \textcolor{red}{\(\blacksquare\)} & \textcolor{red}{\(\blacksquare\)} &  \\
\hline
Khan -20 \cite{Khan-20} & \textcolor{blue}{\(\blacksquare\)} & \textcolor{green}{\(\blacksquare\)} &  &  \textcolor{blue}{\(\blacksquare\)} & \textcolor{blue}{\(\blacksquare\)} & \textcolor{green}{\(\blacksquare\)} & \textcolor{green}{\(\blacksquare\)} & \textcolor{blue}{\(\blacksquare\)} & \textcolor{blue}{\(\blacksquare\)} & \textcolor{green}{\(\blacksquare\)} &  \\
\hline
Bariah -20 \cite{Bariah-20} &  & \textcolor{blue}{\(\blacksquare\)} &  &  & \textcolor{blue}{\(\blacksquare\)} & \textcolor{red}{\(\blacksquare\)} &  & \textcolor{blue}{\(\blacksquare\)} &  & \textcolor{red}{\(\blacksquare\)} &  \\
\hline
Akhtar -20 \cite{Akhtar-20} & \textcolor{blue}{\(\blacksquare\)} & \textcolor{blue}{\(\blacksquare\)} & \textcolor{green}{\(\blacksquare\)} &  \textcolor{green}{\(\blacksquare\)} & \textcolor{red}{\(\blacksquare\)} & \textcolor{blue}{\(\blacksquare\)} &  & \textcolor{red}{\(\blacksquare\)} & \textcolor{green}{\(\blacksquare\)} & \textcolor{blue}{\(\blacksquare\)} &  \\
\hline
You -21 \cite{You-21} &  & \textcolor{blue}{\(\blacksquare\)} &  &   & \textcolor{red}{\(\blacksquare\)} & \textcolor{red}{\(\blacksquare\)} & \textcolor{blue}{\(\blacksquare\)} & \textcolor{red}{\(\blacksquare\)} & \textcolor{red}{\(\blacksquare\)} & \textcolor{purple}{\(\blacksquare\)} &  \\
\hline
Rajatheva -21 \cite{Rajatheva-21} &  & \textcolor{red}{\(\blacksquare\)} &  &   & \textcolor{blue}{\(\blacksquare\)} & \textcolor{green}{\(\blacksquare\)} & \textcolor{blue}{\(\blacksquare\)} & \textcolor{red}{\(\blacksquare\)} & \textcolor{blue}{\(\blacksquare\)} & \textcolor{red}{\(\blacksquare\)} &  \\
\hline
Jiang -21 \cite{Jiang-21} & \textcolor{blue}{\(\blacksquare\)} & \textcolor{blue}{\(\blacksquare\)} & \textcolor{blue}{\(\blacksquare\)} & \textcolor{blue}{\(\blacksquare\)} & \textcolor{red}{\(\blacksquare\)} & \textcolor{blue}{\(\blacksquare\)} & \textcolor{blue}{\(\blacksquare\)} & \textcolor{red}{\(\blacksquare\)} & \textcolor{blue}{\(\blacksquare\)} & \textcolor{red}{\(\blacksquare\)} &  \\
\hline
Bhat -21 \cite{Bhat-21} &  & \textcolor{green}{\(\blacksquare\)} &  & \textcolor{blue}{\(\blacksquare\)}  & \textcolor{blue}{\(\blacksquare\)} & \textcolor{blue}{\(\blacksquare\)} & \textcolor{green}{\(\blacksquare\)} & \textcolor{blue}{\(\blacksquare\)} & \textcolor{red}{\(\blacksquare\)} & \textcolor{red}{\(\blacksquare\)} &  \\
\hline
Alwis -21 \cite{Alwis-21} & \textcolor{green}{\(\blacksquare\)} & \textcolor{blue}{\(\blacksquare\)} & \textcolor{blue}{\(\blacksquare\)} & \textcolor{red}{\(\blacksquare\)} & \textcolor{red}{\(\blacksquare\)} & \textcolor{red}{\(\blacksquare\)} & \textcolor{red}{\(\blacksquare\)} & \textcolor{blue}{\(\blacksquare\)} & \textcolor{blue}{\(\blacksquare\)} & \textcolor{red}{\(\blacksquare\)} &  \\
\hline
Tataria -21 \cite{Tataria-21} &  & \textcolor{green}{\(\blacksquare\)} &  &   & \textcolor{red}{\(\blacksquare\)} & \textcolor{red}{\(\blacksquare\)} &  & \textcolor{red}{\(\blacksquare\)} & \textcolor{blue}{\(\blacksquare\)} & \textcolor{red}{\(\blacksquare\)} &  \\
\hline
Alsabah -21 \cite{Alsabah-21} & \textcolor{red}{\(\blacksquare\)} & \textcolor{green}{\(\blacksquare\)} &  &   & \textcolor{red}{\(\blacksquare\)} & \textcolor{red}{\(\blacksquare\)} & \textcolor{green}{\(\blacksquare\)} & \textcolor{green}{\(\blacksquare\)} & \textcolor{red}{\(\blacksquare\)} & \textcolor{red}{\(\blacksquare\)} &  \\
\hline
Uusitalo -21 \cite{Uusitalo-21} &  & \textcolor{green}{\(\blacksquare\)} & \textcolor{green}{\(\blacksquare\)} &  \textcolor{blue}{\(\blacksquare\)} & \textcolor{red}{\(\blacksquare\)} & \textcolor{blue}{\(\blacksquare\)} &  & \textcolor{green}{\(\blacksquare\)} & \textcolor{blue}{\(\blacksquare\)} & \textcolor{blue}{\(\blacksquare\)} &  \\
\hline
Alraih -22 \cite{Alraih-22} &  \textcolor{red}{\(\blacksquare\)} &  \textcolor{blue}{\(\blacksquare\)} &  & \textcolor{green}{\(\blacksquare\)} &  \textcolor{blue}{\(\blacksquare\)} &  \textcolor{green}{\(\blacksquare\)} &  \textcolor{blue}{\(\blacksquare\)} &  \textcolor{red}{\(\blacksquare\)} &  \textcolor{green}{\(\blacksquare\)} &  \textcolor{red}{\(\blacksquare\)} &  \\
\hline
Hakeem -22 \cite{Hakeem-22} & \textcolor{blue}{\(\blacksquare\)} & \textcolor{red}{\(\blacksquare\)} &  & \textcolor{red}{\(\blacksquare\)}  & \textcolor{blue}{\(\blacksquare\)} & \textcolor{red}{\(\blacksquare\)} &  & \textcolor{red}{\(\blacksquare\)} & \textcolor{green}{\(\blacksquare\)} & \textcolor{green}{\(\blacksquare\)} &  \\
\hline
Salameh -22 \cite{Salameh-22} & \textcolor{blue}{\(\blacksquare\)} & \textcolor{red}{\(\blacksquare\)} & \textcolor{blue}{\(\blacksquare\)} &  \textcolor{red}{\(\blacksquare\)} & \textcolor{red}{\(\blacksquare\)} & \textcolor{green}{\(\blacksquare\)} &  & \textcolor{green}{\(\blacksquare\)} & \textcolor{blue}{\(\blacksquare\)} & \textcolor{blue}{\(\blacksquare\)} &  \\
\hline
Wang -22 \cite{Wang-22} &  &  &  &   & \textcolor{blue}{\(\blacksquare\)} & \textcolor{red}{\(\blacksquare\)} &  & \textcolor{green}{\(\blacksquare\)} & \textcolor{blue}{\(\blacksquare\)} & \textcolor{red}{\(\blacksquare\)} &  \\
\hline
Asghar -22 \cite{Asghar-22} & \textcolor{green}{\(\blacksquare\)} & \textcolor{red}{\(\blacksquare\)} &  &   & \textcolor{green}{\(\blacksquare\)} & \textcolor{blue}{\(\blacksquare\)} & \textcolor{green}{\(\blacksquare\)} & \textcolor{green}{\(\blacksquare\)} & \textcolor{blue}{\(\blacksquare\)} & \textcolor{blue}{\(\blacksquare\)} &  \\
\hline
Akbar -22 \cite{Akbar-22} & \textcolor{red}{\(\blacksquare\)} & \textcolor{blue}{\(\blacksquare\)} &  &  \textcolor{green}{\(\blacksquare\)} & \textcolor{green}{\(\blacksquare\)} & \textcolor{red}{\(\blacksquare\)}  & \textcolor{green}{\(\blacksquare\)}  &\textcolor{green}{\(\blacksquare\)}  &  \textcolor{blue}{\(\blacksquare\)} & \textcolor{green}{\(\blacksquare\)}  &  \\
\hline
Banafaa -23 \cite{Banafaa-23} & \textcolor{green}{\(\blacksquare\)} & \textcolor{blue}{\(\blacksquare\)} &  &  \textcolor{blue}{\(\blacksquare\)} & \textcolor{blue}{\(\blacksquare\)} & \textcolor{red}{\(\blacksquare\)} &  \textcolor{green}{\(\blacksquare\)} & \textcolor{blue}{\(\blacksquare\)} & \textcolor{blue}{\(\blacksquare\)} & \textcolor{red}{\(\blacksquare\)} &  \\
\hline
Shen -23 \cite{Shen-23} &  & \textcolor{green}{\(\blacksquare\)} & \textcolor{green}{\(\blacksquare\)} &  & \textcolor{blue}{\(\blacksquare\)} & \textcolor{green}{\(\blacksquare\)} &  \textcolor{green}{\(\blacksquare\)} & \textcolor{green}{\(\blacksquare\)} & \textcolor{red}{\(\blacksquare\)} & \textcolor{red}{\(\blacksquare\)} &  \\
\hline
Wang -23 \cite{Wang-23c} & \textcolor{green}{\(\blacksquare\)} & \textcolor{red}{\(\blacksquare\)} &  & \textcolor{blue}{\(\blacksquare\)} & \textcolor{red}{\(\blacksquare\)} & \textcolor{red}{\(\blacksquare\)} & \textcolor{red}{\(\blacksquare\)} & \textcolor{red}{\(\blacksquare\)} & \textcolor{red}{\(\blacksquare\)} & \textcolor{red}{\(\blacksquare\)} &  \\
\hline
Quy -23 \cite{Quy-23} & \textcolor{blue}{\(\blacksquare\)} & \textcolor{blue}{\(\blacksquare\)} &  & \textcolor{green}{\(\blacksquare\)} & \textcolor{red}{\(\blacksquare\)} & \textcolor{green}{\(\blacksquare\)} &  & \textcolor{green}{\(\blacksquare\)} & \textcolor{green}{\(\blacksquare\)} & \textcolor{blue}{\(\blacksquare\)} &  \\
\hline
Mohsan -23 \cite{Mohsan-23b} & \textcolor{blue}{\(\blacksquare\)} & \textcolor{blue}{\(\blacksquare\)} &  & \textcolor{red}{\(\blacksquare\)} & \textcolor{blue}{\(\blacksquare\)} & \textcolor{red}{\(\blacksquare\)} & \textcolor{red}{\(\blacksquare\)} & \textcolor{red}{\(\blacksquare\)} & \textcolor{blue}{\(\blacksquare\)} & \textcolor{blue}{\(\blacksquare\)} &  \\
\hline
Ishteyaq -24 \cite{Ishteyaq-24} & \textcolor{blue}{\(\blacksquare\)} & \textcolor{blue}{\(\blacksquare\)} &  & \textcolor{blue}{\(\blacksquare\)} & \textcolor{red}{\(\blacksquare\)} & \textcolor{blue}{\(\blacksquare\)} & \textcolor{red}{\(\blacksquare\)} & \textcolor{red}{\(\blacksquare\)} & \textcolor{green}{\(\blacksquare\)} & \textcolor{green}{\(\blacksquare\)} &   \\
\hline
Shafi -24 \cite{Shafi-24} &  & \textcolor{green}{\(\blacksquare\)} &  & \textcolor{green}{\(\blacksquare\)} & \textcolor{green}{\(\blacksquare\)} &  &  & \textcolor{green}{\(\blacksquare\)} & \textcolor{blue}{\(\blacksquare\)} & \textcolor{red}{\(\blacksquare\)} &  \\
\hline
Giuliano -24 \cite{Giuliano-24} & \textcolor{green}{\(\blacksquare\)} & \textcolor{red}{\(\blacksquare\)} &  &  & \textcolor{red}{\(\blacksquare\)} & \textcolor{red}{\(\blacksquare\)} & \textcolor{red}{\(\blacksquare\)} & \textcolor{red}{\(\blacksquare\)} & \textcolor{blue}{\(\blacksquare\)} & \textcolor{blue}{\(\blacksquare\)} &  \\
\hline
This survey & \textcolor{red}{\(\blacksquare\)}
& \textcolor{red}{\(\blacksquare\)} & \textcolor{red}{\(\blacksquare\)} & \textcolor{red}{\(\blacksquare\)} & \textcolor{red}{\(\blacksquare\)} & \textcolor{red}{\(\blacksquare\)} & \textcolor{red}{\(\blacksquare\)} & \textcolor{red}{\(\blacksquare\)} & \textcolor{red}{\(\blacksquare\)} & \textcolor{purple}{\(\blacksquare\)} & \textcolor{blue}{\(\blacksquare\)}  \\
\hline
\multicolumn{12}{|c|}{\multirow{2}{*}{\textcolor{purple}{\(\blacksquare\)} = \textbf{EXCEPTIONAL COVERAGE}, \textcolor{red}{\(\blacksquare\)} = \textbf{HIGH COVERAGE}, \textcolor{blue}{\(\blacksquare\)} = \textbf{MEDIUM COVERAGE}, \textcolor{green}{\(\blacksquare\)} = \textbf{LOW COVERAGE}}} \\
\multicolumn{12}{|c|}{} \\
\hline
\end{tabularx}
\end{center}
\end{table*}

\subsection{CONTRIBUTIONS OF THE ARTICLE}
In this section, the contributions of this article are summarized. In Table \ref{Table_Comparison}, we compare our article against other comprehensive 6G surveys in the literature. 

\begin{itemize}
\item \textbf{Comprehensive Vision, Survey, and Tutorial on 6G}:
The main contribution of this article is to provide a thorough vision, survey, and tutorial on 6G, all in the same package. This can be seen as a sum of the contributions of all individual sections. To the best of our knowledge, this is the most comprehensive 6G survey in the literature, as indicated in Table \ref{Table_Comparison}. The contributions of each section are introduced below. They are presented in the order of their potential value to the readers. 

\begin{itemize}
\item \textbf{Potential Technologies for 6G}: 
We identify and review 27 potential 6G technologies in 10 different technology categories. Each technology is reviewed in a tutorial manner, using the same template of vision, introduction, past and present, opportunities and challenges, and literature and future directions. To the best of our knowledge, this is the most comprehensive set of 6G technologies reviewed in a single article, as indicated in Table \ref{Table_Comparison}. In addition, our tutorial presentation approach is unique in the 6G literature. 

\item \textbf{6G Vision}: 
We introduce an insightful 6G vision of smart wireless world via 6G-enabled mobile intelligence. In this vision, 6G will evolve toward the Intelligent Network of Everything while serving as a fruitful platform for mobile intelligence to grow toward its potential. Note that our vision is well aligned with the IMT-2030 Framework of ITU-R, although more far-reaching. Based on our 6G vision, we can already see some major trends that evolve beyond 6G and sketch the first high-level vision for 7G. Our insightful and far-reaching 6G vision is novel in the literature. 

\item \textbf{7G Vision}: 
To place our 6G vision in a broader perspective, we discuss what comes after 6G and sketch a high-level vision for 7G. First, we briefly describe the evolution path from 1G to 7G, with the cycles of two generations. Then, we introduce our 7G vision, focusing on the main disruption, fundamental elements, and possible applications of 7G. Our 7G vision can be summarized in a few words: "Deeply Intelligent Network of Hyperverses". To the best of our knowledge, this is the first structured 7G vision in the literature. In addition, the introduced evolution road from 1G to 7G is also a novel view. 

\item \textbf{AI/ML for 6G}: 
Since AI/ML is considered the most revolutionizing technology for 6G, we dedicate a whole section for it. Specifically, we introduce the concepts of AI and ML, review three promising ML methods for 6G, and discuss the opportunities and challenges involved in integrating AI/ML into 6G networks. Most of the existing 6G surveys do not have a dedicated section for 6G AI/ML, with a generic introduction to AI/ML. 

\item \textbf{Disruptive Applications for 6G}: 
We define 12 disruptive applications for 6G in three categories: human-machine interactions, smart environments, and connected autonomous systems. To the best of our knowledge, this is a unique set of categories/applications in the 6G literature. 

\item \textbf{Key Use Cases for 6G}: 
Since 6G is expected to expand its capabilities beyond communication, we identify five communication-oriented and three beyond-communication-oriented use cases. To the best of our knowledge, this is a unique set of categories/use cases in the 6G literature. 

\item \textbf{Main Performance Requirements for 6G}: 
We review the main performance requirements of IMT-2030, introduced by ITU-R, and compare them to the typically proposed values in the 6G literature. To the best of our knowledge, this is the first time when IMT-2030 requirements are compared to the typical values in the 6G literature. 

\item \textbf{Fundamental Elements of 6G}: 
We envision that 6G will be based on three fundamental pillars, i.e., wireless, AI, and IoE. This is a simple and insightful view. 

\item \textbf{Defining Features for 6G}: 
We identify 12 main features that define the essence of 6G: extreme capacity $\&$ performance, ultra-flexible $\&$ agile, highly intelligent $\&$ aware, ubiquitously available $\&$ reliable, truly green $\&$ sustainable, and thoroughly secure $\&$ trustworthy. To the best of our knowledge, this is a unique set of features proposed in the 6G literature. 

\item \textbf{6G in a Nutshell}: 
We provide a compact summary of our 6G vision in bullet points, allowing readers to grasp the big picture of 6G at one glance. To the best of our knowledge, this style is unique in the 6G literature. 
 
\item \textbf{Development Process of 6G}:
We provide an up-to-date and thorough overview of the 6G development process. Specifically, we discuss the roles of ITU-R and 3GPP, as they are the most important standardization bodies in 6G development. The estimated timeline is also introduced. As indicated in Table \ref{Table_Comparison}, this is the most comprehensive overview of the 6G development process in the literature. 

\item \textbf{Research Activities toward 6G}: 
We review the main worldwide 6G research activities in a chronological order from 2018 to 2024. Various types of activities are considered, such as projects, collaborations, funding, standardization, experiments, events, and publications. This is one of the most thorough overviews of 6G research activities in the literature, as indicated in Table \ref{Table_Comparison}. 

\item \textbf{Previous Works}: 
We provide a thorough and up-to-date review of the main prior works in terms of 6G visions and surveys, comprising over 60 articles. To the best of our knowledge, this is one of the most comprehensive reviews of the existing 6G visions and surveys in the literature. 

\item \textbf{Overview of 5G}: 
We present an up-to-date overview of 5G in a compact form, focusing on the past, present, and future. In particular, we review the 3GPP 5G System and Releases 15, 16, 17, 18, and 19. 

\item \textbf{Evolution from 1G to 6G}: 
We provide an insightful introduction to the evolution of mobile communications by highlighting the essentials of each generation. In particular, we discuss how disruptive different generations have been and envision what will be the next major disruption enabled by 6G. To the best of our knowledge, this is a novel view on the evolution of mobile networks. 
\end{itemize} 
\end{itemize}

The remainder of this paper is organized as follows. Section \ref{Literature} reviews the previous works in the literature. In Section \ref{Vision}, our 6G vision is introduced. Three fundamental elements of 6G are identified in Section \ref{Elements}. Section \ref{Applications} defines a comprehensive set of disruptive 6G applications. Key use cases are proposed for 6G in Section \ref{Usecases}. Section \ref{Requirements} reviews the main performance requirements that 6G is expected to satisfy. AI/ML is discussed for 6G in Section \ref{AI}. In Section \ref{Technologies}, a comprehensive set of potential 6G technologies is reviewed. Twelve features that define the essence of 6G are identified in Section \ref{Features}. Section \ref{Summary} summarizes our 6G vision in the bullet points. In Section \ref{7G}, the post-6G era is speculated and a high-level vision is sketched for 7G. Finally, conclusions are drawn in Section \ref{Conclusion}. The list of abbreviations can be found in Appendix. The outline of the article is summarized in Figure \ref{Fig_Content}. 

\section{PREVIOUS WORKS}
\label{Literature}
In this section, we review the main prior works in terms of 6G vision and survey articles. At the end, we discuss the justification of our work with respect to these previous studies. 

\subsection{6G VISION ARTICLES} 
The first 6G vision articles were written in 2018. In \cite{David-18}, the authors speculated the need for 6G by discussing the strengths and weaknesses of each generation. In \cite{Katz-18}, the paper introduced the 6G vision of the Finnish 6G Flagship Program, focusing on interdisciplinary research between wireless communications, computer engineering, electronics, and material science. In 2019, a handful of 6G papers were published. In \cite{Zhang-19}, the authors introduced their 6G vision, discussing usage scenarios, target requirements, integrated space-air-ground-underwater networks, AI-based network design, and promising technologies. The paper \cite{Yang-19} presented a 6G vision from the perspectives of time-frequency-space resource usage, promising techniques toward 6G, ML-aided intelligent transmission, and key challenges. In \cite{Strinati-19}, 6G was envisioned as a key enabler on the road to a "global brain". The paper focused on new services, performance requirements, pervasive AI, THz and visible light communications (VLC), as well as the synergy between communication, computation, caching, and control. In \cite{Zong-19}, 6G was studied in terms of research activities, key drivers, main requirements, application scenarios, system architectures, and promising technologies. AI-empowered 6G was considered in \cite{Letaief-19}, covering the network architecture, AI-enabled technologies, 6G for AI, and hardware-aware communications. In \cite{Zhang-19c}, the proposed 6G vision emphasized mobile ultra-broadband, super IoT, and AI, with a discussion on THz communications, symbiotic radio, satellite-aided IoT, deep learning (DL), and reinforcement learning. 

The peak of 6G visions was in 2020. Numerous papers were published. The work \cite{Dang-20} contemplated what 6G should be in the 2030s, discussing potential applications, main challenges, key features, enabling communication technologies, and beyond technology impact. In \cite{Tomkos-20}, the authors envisioned the evolution toward the 6G era, focusing on the potential and challenges of AI/ML. The covered topics include the evolution from mobile edge computing to edge AI, distributed AI, communications for ML, and ML for communications. In the 6G vision of \cite{Chen-20}, the focus was on overcoming the challenges related to coverage, capacity, data rates, and mobility. In \cite{Shafin-20}, the authors explored AI-enabled wireless networks toward 6G, discussing AI for the PHY, medium access control (MAC), and network layers. The paper \cite{Viswanathan-20} identified six main requirements, fundamental design dimensions, and key technologies for 6G. The study \cite{Giordani-20} focused on 6G enabling technologies in three categories: disruptive communications, innovative network architectures, and integrated intelligence. 

In \cite{Alsharif-20}, 6G was studied in terms of key features, open challenges, possible solutions, and research activities. In \cite{Saad-20}, the most cited 6G vision paper discussed a comprehensive set of topics, including driving applications, key trends, performance metrics, new service classes, technological enablers, open issues, and future guidelines. The paper \cite{Chowdhury-20} focused on communication trends, service requirements, network features, applications, enabling technologies, research progress, technological challenges, and future research directions. In \cite{Akyildiz-20}, the authors introduced a vision for 6G and beyond, discussing THz communications, intelligent wireless environments, pervasive AI, network automation, reconfigurable transceiver front-ends, backscatter communications, internet of space things, cell-free massive MIMO, and beyond 6G technologies. The 6G vision in \cite{Khan-20} focused on communication, networking, and computing technologies. 

In \cite{Tariq-20}, a speculative study was presented on 6G, with a discussion on tentative vision, usage scenarios, key technologies, and main challenges. The article \cite{Bariah-20} provided an in-depth discussion on 6G enabling technologies. In \cite{Ziegler-20}, the authors envisioned that 6G will act as an enabling platform to connect the digital, physical, and biological worlds. In their 6G vision \cite{Liu-20b}, the authors focused on network architecture based on the convergence of information, communication, and computation technologies. The work \cite{Gui-20} discussed the integration of comfort, intelligence, and security in 6G networks. The paper \cite{Yang-20} proposed an AI-empowered architecture to enable their vision of intelligent 6G networks by introducing four layers, i.e., intelligent sensing layer, data mining and analytics layer, intelligent control layer, and smart application layer. In \cite{Akhtar-20}, 6G was considered in terms of the system architecture, network dimensions, possible technologies, promising applications, and performance indicators. 

In 2021-2024, a few 6G visions were introduced each year. In \cite{You-21}, the authors' vision was based on four new paradigm shifts, including integrated space-air-ground networks, spectrum operation at sub-6 GHz, mmWave, THz, and optical frequencies, AI-empowered mobile networks, and comprehensive network security. In \cite{Bhat-21}, the vision of the 6G ecosystem was built on the current research status and anticipated future trends. The article \cite{Rajatheva-21} studied broadband connectivity for 6G to score the target data rate of 1 Tbit/s. The authors introduced a comprehensive set of technological enablers at the spectrum, infrastructure, and algorithmic levels. 
In \cite{Tataria-21}, a top-down vision of the 6G ecosystem was provided, discussing lifestyle and societal changes, applications and their technical requirements, key challenges and opportunities for all layers, new spectrum bands and deployment scenarios, design principles and required changes in radio access and core network architectures, novel PHY layer solutions, propagation characteristics, real-time signal processing, and radio frequency (RF) transceiver design. 
The work \cite{Uusitalo-21} introduced the 6G vision of the European 6G flagship program Hexa-X. The five main contributions include a joint academic and industry perspective of 6G; review of concurrent 6G initiatives; identification of six key challenges; insightful discussion on application scenarios; and analysis of technological transformations in the radio, network, and orchestration domains. 

In \cite{Wang-22}, the proposed 6G vision focused on immersive, intelligent, and ubiquitous applications and the needed technologies. The paper \cite{Hakeem-22} introduced its 6G exploration with a discussion on the evolution of mobile networks, key features and the required capabilities, architectural perspectives, hardware-software evolution, European 6G projects, and main applications and their challenges. A vision of 6G hyper-connectivity, with nearly unlimited data rates, coverage, and computing, was given in \cite{Lee-23}. 
In \cite{Wang-23b}, the authors introduced a value-oriented 6G by defining multi-dimensional performance indicators, enabling technological elements, and a case study on intelligent multiple access. The work \cite{Erfanian-23} briefly presented the 6G vision of the NGMN Alliance, focusing on the key use cases. The paper \cite{Liu-23} introduced an evolutionary framework for 6G. The aspects explored are ITU-R activities, usage scenarios, required capabilities, and potential technologies. In \cite{Mohjazi-24}, the authors focused on building a 6G vision based on four use cases, i.e., Internet of Senses, connected intelligent machines, digitalized and programmable physical world, and connected sustainable world. In \cite{Mohr-24}, the European 6G vision was introduced, based on the perspective of the collaborative 6G research program SNS JU.  

\subsection{6G SURVEY ARTICLES}
The first 6G survey article \cite{Lu-20} was published in 2020, reviewing architectural framework, core technologies, application scenarios, and future challenges. The work \cite{Mahmoud-21} provided a 6G survey on network requirements, core features, potential applications, novel services, research progress, and open challenges. In \cite{Jiang-21}, the work provided a comprehensive picture of the road toward 6G by reviewing the key drivers, application scenarios, target requirements, development efforts, and technological enablers. The authors in \cite{Alwis-21} provided a thorough survey of the 6G frontiers, with a discussion on societal and technological trends, emerging applications and their requirements, research and standardization efforts, and technology enablers. In \cite{Dogra-21}, a survey was conducted on the evolution from 5G to 6G networks. In \cite{Alsabah-21}, a comprehensive review was presented on the potential 6G technologies for achieving high data rates, enhanced energy efficiency, full coverage, security and privacy, URLLC, and network intelligence. In \cite{Alraih-22}, 6G was reviewed in terms of use cases and applications, performance requirements, spectrum technologies, technical enablers, open challenges, and future research avenues. 

The authors of \cite{Salameh-22} studied the evolution from 5G to 6G by highlighting the shortcomings of 5G and the societal, economic, technological, and operational aspects of 6G. In \cite{Asghar-22}, the evolution of wireless networks to 6G was explored by discussing the evolution of network architecture, application areas, driving technologies, and ML techniques. In \cite{Akbar-22}, 6G was surveyed in terms of communication aspects, network architecture, and AI-enabled technologies. Recent advances and open problems in 5G/6G research were studied in \cite{Salahdine-23}. The work \cite{Banafaa-23} reviewed 6G communication technology from the aspects of usage scenarios, application opportunities, and potential technologies. In \cite{Yeh-23}, the authors surveyed key 6G technologies from the perspective of interactivity, connectivity, and intelligence. The paper \cite{Shen-23} explored the five facets of 6G, i.e., next-generation architectures, spectrum, and services; networking; IoT; wireless positioning and sensing; and applications of DL. A comprehensive survey of 6G and its recent research advances was provided in \cite{Wang-23c}. The paper reviewed the global 6G vision, network requirements, application scenarios, network architecture, core technologies, and existing testbeds. 

The authors of \cite{Chafii-23} identified 12 scientific challenges for 6G to rethink the theoretical foundations of traditional communication and to tackle the new paradigm of the convergence of communication and beyond-communication technologies. The work \cite{Quy-23} examined 6G architectures and technologies to form an enabling platform for innovative future applications. The survey \cite{Mohsan-23b} built a comprehensive picture of 6G by analyzing recent advances, key enablers, and technological challenges. In \cite{Ishteyaq-24}, the authors provided an insightful analysis of 6G enabling technologies and discussed the design of a comprehensive 6G network architecture. The article \cite{Shafi-24} reviewed the key technologies of 6G networks, focusing on the extreme performance that they can provide in diverse dimensions. In \cite{Singh-24}, 6G evolution was discussed, with a special focus on the usage scenarios of IMT-2030. The work \cite{Giuliano-24} explored the evolution from 5G-Advanced to 6G from the perspective of three novel services, including immersive communications, everything connected, and high-accuracy sensing. In \cite{Kerboeuf-24}, the authors designed and reviewed an end-to-end 6G system, based on the views of the European 6G flagship project Hexa-X-II. 

\subsection{DISCUSSION} 
Over the years, 6G has been reviewed from a broad range of aspects in the literature, such as the expected timeline, research activities, novel features, performance requirements, usage scenarios, driving applications, network architectures, and enabling technologies. Throughout the 6G literature, one can notice three common factors above all. First, the performance of 6G is expected to be significantly improved in all dimensions compared to that of 5G. Second, the number and variety of connected objects are expected to grow notable. This is seen as the evolution from IoT to IoE. Third, AI/ML is expected to play a key role in 6G, making it a highly intelligent entity. All these findings from the existing literature support our vision where 6G will be based on extreme wireless, pervasive AI/ML, and ubiquitous IoE, and evolve toward the Intelligent Network of Everything. 
Due to the years of global 6G research, ITU-R has completed the visioning phase of the IMT-2030 development process and moved to the second phase of defining the performance requirements and evaluation methodology. In the development of technical details, 3GPP is expected to begin the 6G standardization process after mid-2025. At this point, 6G literature is missing an up-to-date overview of the entire 6G research field, providing the latest details in all of the aforementioned 6G aspects. The time is also right to begin visioning beyond 6G. To close this gap in the literature, we provide a contemporary vision, survey, and tutorial on 6G, covering a wide range of timely topics and sketching the first high-level vision beyond 6G.

\begin{figure}[!htb]
\center{\includegraphics[width=\columnwidth]
{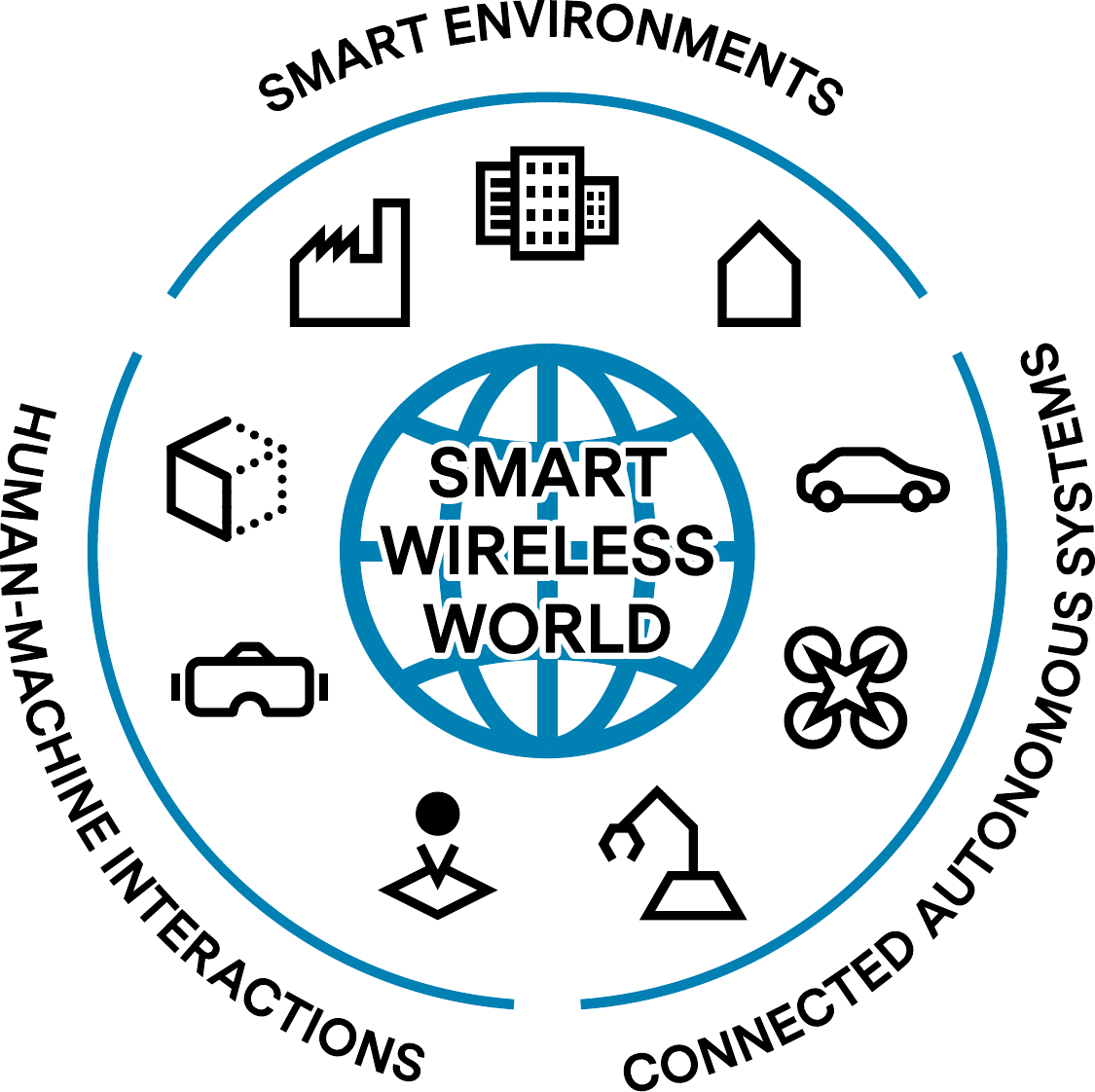}}
\caption{\label{Fig_Vision}6G vision: smart wireless world via 6G-enabled mobile intelligence.}
\end{figure}

\section{6G VISION}
\label{Vision}
In this section, our 6G vision is introduced. First, the big picture of 6G is discussed. Then, the focus is shifted to the main details. Specifically, a brief introduction is provided to the fundamental elements, disruptive applications, key use cases, target performance requirements, potential technologies, and defining features. These topics are discussed in detail in their dedicated sections later in the paper. It is worth mentioning that our 6G vision is well aligned with that of ITU-R, although more far-reaching with a broader perspective. 

\begin{itemize}
\item \textbf{Big Picture}: 
Based on the current trends, we envision that the next major disruption in mobile communications will be the 6G-enabled \textit{mobile intelligence}. Mobile intelligence refers to the fusion of wireless, AI, and IoE technologies at all levels of the society. Mobile intelligence has the potential to make anything connected, smart, and aware of the surrounding environment. Consequently, mobile intelligence will revolutionize the design, operation, interactions, and use of devices, systems, and applications. We envisage that 6G will evolve toward \textit{the Intelligent Network of Everything}, while serving as a fruitful platform for mobile intelligence to grow toward its potential. The 6G-enabled mobile intelligence will create a \textit{smart wireless world}, where there are unprecedented opportunities to produce greater value for the benefit of people, society, and the world in general. Eventually, mobile intelligence will penetrate all walks of life and become an essential part of the future society. 

\item \textbf{Fundamental Elements}: 
To provide a major disruption, 6G needs to be based on three profound elements: wireless, AI, and IoE. Wireless consists of communication, sensing, and energy dimensions. AI refers to the extensive use of AI at all levels of the 6G ecosystem, including the network core, network edges, air interface, devices, services, and applications. IoE refers to the massive number of network-connected objects, such as sensors, devices, machines, vehicles, drones, robots, etc. Further details on the fundamental elements are provided in Section \ref{Elements}.  

\item \textbf{Disruptive Applications}: 
6G is expected to support a wide range of game-changing applications. We divide the main ones into three categories: {human-machine interactions, smart environments, and connected autonomous systems}. The concept of human-machine interactions refers to the different ways in which humans interact with machines, devices, and smart entities. We define five key interactions: metaverse, XR, holographic-type communications, digital twins, and Tactile Internet. To support such data-hungry and time-sensitive applications, 6G must provide ultra-high data rates and extremely low latency. 
The concept of smart environment is defined as an entity that exploits wireless, AI, and IoE technologies to produce greater value. We specify four main environments, i.e., smart society, smart city, smart factory, and smart home. 

For smart environments, 6G needs to provide customized wireless network services with IoE communications, beyond-communication capabilities, and secure network solutions. Connected autonomous systems refer to entities that can operate independently without human involvement, relying on versatile technologies, such as AI, control, sensing, positioning, and connectivity. We define three systems: connected autonomous vehicle systems, connected autonomous aerial vehicle systems, and connected autonomous robotic systems. Such systems set stringent performance requirements for 6G, particularly in terms of reliability, availability, latency, and mobility. 
A detailed discussion of 6G applications is provided in Section \ref{Applications}. The 6G vision of the smart wireless world with the aforementioned applications is illustrated in Figure \ref{Fig_Vision}. 

\item \textbf{Key Use Cases}: 
Since 6G is expected to significantly expand the capabilities of mobile networks, we introduce five communication-oriented and three beyond-communication-oriented use cases accordingly. In the communication domain, there are six main performance dimensions that must be optimized for 6G, i.e., capacity, latency, reliability, density, coverage, and mobility. These performance dimensions are among the list of IMT-2030 capabilities introduced by ITU-R in \cite{ITU-M2160}. The corresponding use cases are ultra-broadband multimedia communications, extreme time-sensitive and mission-critical communications, ultra-massive communications, global-scale communications, and hyper-mobility communications. In addition to the enhanced communication features, 6G aims to extend its capabilities beyond communication. The corresponding use cases include network intelligence, network sensing, and network energy. Further details on these use cases are provided in Section \ref{Usecases}. 

\item \textbf{Performance Requirements}: 
To support diverse use cases and applications, the performance requirements of 6G need to be pushed to their limits. ITU-R has introduced example target values for nine main performance metrics: peak rate (50/100/200 Gbit/s), user experienced rate (300/500 Mbit/s), spectral efficiency (1.5X/3X), area traffic capacity (30/50 Mbit/s/m$^2$), latency (0.1--1 ms), reliability (1-10$^{\text{-}5}$--1-10$^{\text{-}7}$), connection density (10$^{6}$--10$^{8}$/km$^2$), mobility (500--1000 km/h), and positioning accuracy (1--10 cm). Compared to the targets proposed in the 6G literature, the values of ITU-R are rather moderate. A more detailed discussion on the target requirements is provided in Section \ref{Requirements}. 

\item \textbf{Potential Technologies}: 
A vast variety of advanced 6G technologies are required to support the expected requirements, use cases, and applications. A comprehensive set of key technologies is identified and reviewed in ten different network categories, including spectrum, antenna systems, transmission scheme, network architecture, network intelligence, beyond-communication, energy awareness, end-devices, services, and security. 
At the core of spectrum-level technologies is THz communications, providing extremely high data rates. Extreme antenna systems, such as ultra-massive MIMO and RISs, offer high spectral efficiency and extended coverage. An ultra-flexible transmission scheme based on a multi-waveform design, with flexible numerology, and fast grant-free access will form the basis for the 6G air interface. Integrating non-terrestrial access to the network design leads to a 3D space-air-ground architecture, with potential global-scale coverage. 

In the network intelligence domain, AI/ML plays a revolutionary role by empowering the network core, edges, and air interface. Integrating communication, computation, sensing, and energy greatly expands the capabilities of mobile networks. Green communication and networking provide significant ecological benefits, making 6G more energy-efficient and sustainable. Advanced device-to-device (D2D), V2X, and cellular-connected UAV communications extend the support for the IoE-type connectivity in the 6G era. 
The main service-oriented technologies are private networks, which significantly broaden the service and business opportunities of mobile networks in diverse vertical domains. A holistic network security architecture is vital for the success of 6G, providing trustworthy, secure, and privacy-protected use. In Section \ref{Technologies}, the potential 6G technologies are reviewed in a tutorial manner. 

\item \textbf{Defining Features}: 
We introduce 12 features that define the essence of 6G: {extreme capacity $\&$ performance}, {ultra-flexible $\&$ agile}, {highly intelligent $\&$ aware}, {ubiquitously available $\&$ reliable}, {truly green $\&$ sustainable}, and {thoroughly secure $\&$ trustworthy}. The core of 6G will be extreme performance, especially in terms of capacity, latency, reliability, density, coverage, and mobility. 6G needs to be flexible and agile to adapt to versatile wireless environments and application scenarios. A high level of intelligence and awareness requires pervasive AI and accurate sensing, providing many benefits from efficient network optimization and increased automation to object recognition and velocity estimation. To support diverse services, 6G must be broadly available and reliable. To obtain significant ecological benefits, sustainability and greenness need to be integral parts of 6G. Since 6G will be a pivotal part of the future society, it needs to be secure and trustworthy. Further discussion on the defining features can be found in Section \ref{Features}. 
\end{itemize}

\begin{figure}[!htb]
\center{\includegraphics[width=0.7\columnwidth]
{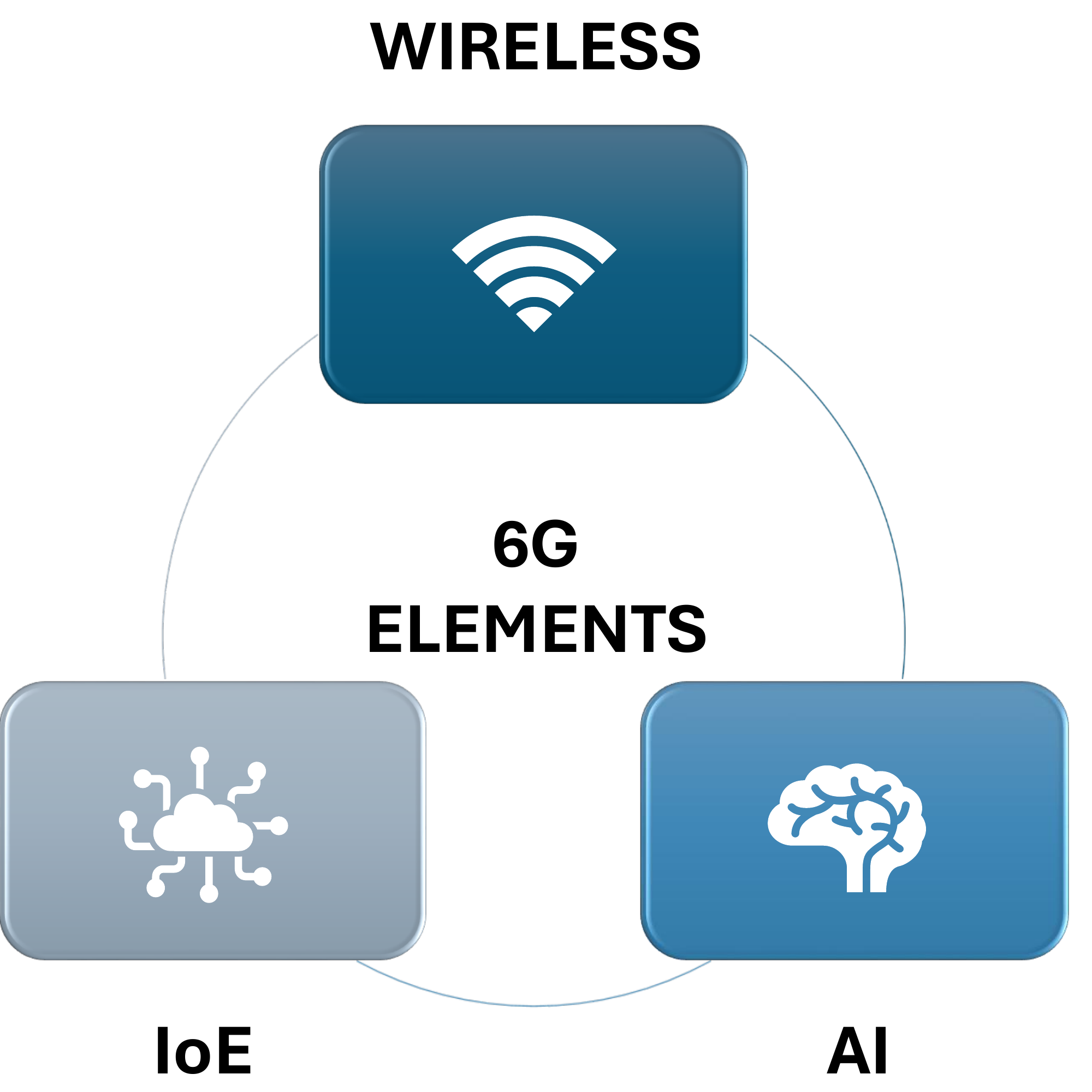}}
\caption{\label{Fig_Elements}Fundamental elements of 6G.}
\end{figure}

\section{FUNDAMENTAL ELEMENTS OF 6G} 
\label{Elements}
This section discusses the foundational building blocks of 6G. To become a highly disruptive network, 6G needs to be based on three fundamental elements, i.e., wireless, AI, and IoE, as illustrated in Figure \ref{Fig_Elements}. 

\begin{itemize}
\item \textbf{Wireless:}
In the context of 6G, wireless refers to a variety of technologies including communication, sensing, positioning, and energy. Extreme communication lays the foundation for 6G and its numerous applications. In the communication domain, there are six main performance dimensions that need to be pushed to the extreme levels, i.e., capacity, latency, reliability, density, coverage, and mobility. Consequently, 6G is expected to support a vast range of communication scenarios, including ultra-fast mobile internet, time-sensitive and mission-critical connectivity, ultra-dense IoT, global coverage, and hyper-mobility communications. Beyond-communication wireless technologies, such as high-resolution sensing, accurate positioning, and wireless energy, expand the capabilities of mobile networks beyond the current horizon. For example, mobile networks can provide novel services, like object recognition/identification, location-based services, and wireless powering of lightweight IoT devices. 

\item \textbf{Artificial Intelligence:}
The extensive use of AI forms the basis for the 6G network intelligence. AI will be exploited at all levels of the network, including the core, edge, and air interface. While AI will upgrade the core network from cloud to cloud intelligence, the network edge will evolve from edge computing to edge intelligence. AI has the potential to profoundly change the way how mobile networks are designed, operated, and managed. Ultimately, pervasive AI/ML can make 6G more intelligent, efficient, flexible, scalable, autonomous, automated, proactive, economical, ecological, trustworthy, and secure. 
 
\item \textbf{Internet of Everything:} 
IoE refers to the massive number of diverse types of connected objects/entities, such as sensors, devices, machines, vehicles, drones, robots, systems, networks, processes, and applications. IoE can be seen as a major extension of IoT. Whereas IoT is often seen as connected sensors, IoE is a much broader concept. Strong support for IoE enables novel functionalities and applications for different objects and entities. IoE is a paradigm shift from human- to machine- and application-centric communications, where the connectivity of objects/entities is at the center of attention. This will significantly extend the applicability of mobile networks to versatile vertical industries. 
\end{itemize}

\begin{figure}[htbp]
\center{\includegraphics[width=\columnwidth]
{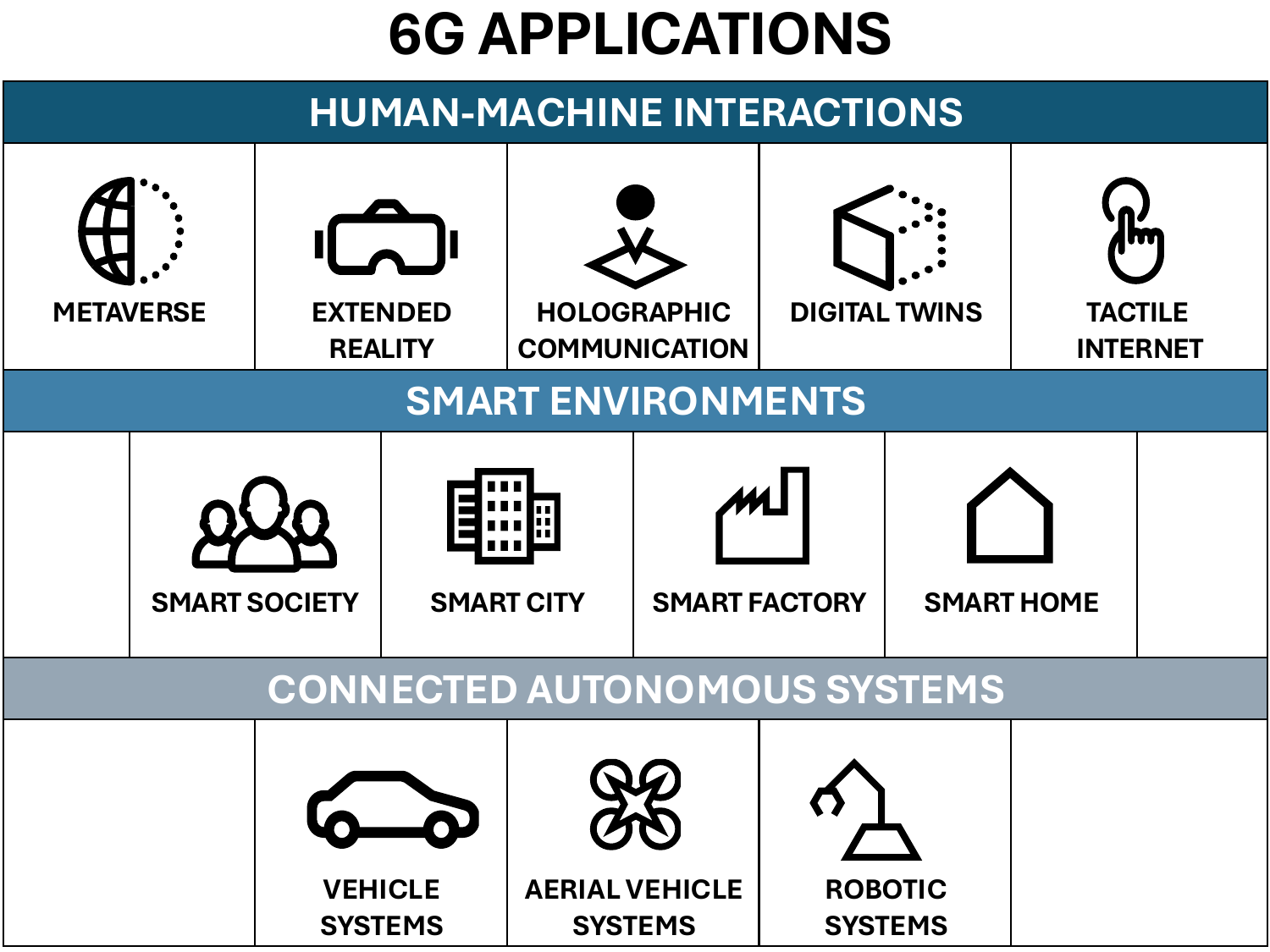}}
\caption{\label{Fig_Applications}Disruptive applications for 6G.}
\end{figure}

\section{DISRUPTIVE APPLICATIONS FOR 6G}
\label{Applications}
In this section, we introduce 12 disruptive 6G applications in the categories of human-machine interactions, smart environments, and connected autonomous systems. These applications are summarized in Figure \ref{Fig_Applications}. 

\subsection{HUMAN-MACHINE INTERACTIONS}
Human-machine interactions refer to the ways in which humans interact with machines, devices, and different types of smart entities. In the following, we review five main interactions for 6G, including metaverse, XR, holographic-type communications, digital twins, and Tactile Internet. 

\subsubsection{METAVERSE}
Metaverse is a computer-generated virtual shared space \cite{Wang-23}. The grand vision of the metaverse is a fully immersive, interoperable, and hyper-spatio-temporal virtual ecosystem merging the physical, human, and digital worlds \cite{Wang-23, Xu-23}. This vision is sometimes referred to as the next-generation Internet \cite{Wang-23} or successor of the mobile internet \cite{Xu-23}. The evolution toward this emerging paradigm has been driven by advances in key technological areas, such as XR, AI/ML, and 5G/B5G/6G \cite{Wang-23, Xu-23}. It has been proposed that a generic evolution road to a fully immersive metaverse consists of three main phases: digital twins, digital natives, and surreality \cite{Lee-21b, Wang-23}. The goal of the first phase is to form digital mirror worlds of real-life objects, systems, and entities using digital twins. The second phase aims to create native digital content inside the virtual world. The third phase completes the evolution of the metaverse by producing surrealistic virtual worlds that seamlessly merge reality with digitality. The current evolutionary state of the metaverse is still in its infancy and far from its ultimate vision. In this picture, 6G is expected to provide a fruitful platform for the metaverse to grow toward its vision \cite{Tang-22, Aslam-23, Zawish-24, Alsamh-24, Khan-24c}. 

The term "metaverse" was first used in science fiction in 1992 \cite{Xu-23}. In the 2010s, the term metaverse re-emerged and evolved toward the current vision. Simultaneously, metaverse began to gain more attention in science, becoming a popular topic on the verge of the 2020s. The COVID-19 pandemic has further increased interest in the metaverse since many physical activities have moved to the virtual space \cite{Xu-23}. The metaverse is widely recognized in the technology industry. Major technology companies are investing in the development of practical metaverse applications. For example, Facebook was renamed Meta in 2021 to better describe the vision of the company \cite{Xu-23}. In practice, there exist some applications that are considered rudimentary versions of the metaverse, mainly in gaming. Second Life, Roblox, and Fortnite are examples of such gaming platforms \cite{Wang-23}. In the standardization domain, there exist two metaverse-related standards \cite{Wang-23}, i.e., ISO/IEC 23005 and IEEE 2888. ISO/IEC 23005 focuses on standardizing the interfaces between the real-virtual and virtual-virtual worlds. IEEE 2888 complements ISO/IEC 23005 by standardizing the synchronization between the real and virtual worlds. Further details on the fundamentals, latest advances, and future challenges of the metaverse can be found in recent survey papers: generic \cite{Lee-21b, Park-22, Wang-23, Xu-23, Khan-24b, Zhu-24} and 6G-related \cite{Tang-22, Aslam-23, Zawish-24, Alsamh-24, Khan-24c}. 

\subsubsection{EXTENDED REALITY}
XR is a broad term that covers all types of combined scenarios between real and virtual environments. Virtual reality (VR), augmented reality (AR), and mixed reality (MR) fall under the umbrella of XR \cite{Akyildiz-22b}. VR refers to all virtual scenarios, whereas AR adds virtual content to the real environment. MR is a blend of the real and virtual environments. Currently, XR is most commonly used for entertainment purposes, such as gaming. Other application scenarios are work, education, business, healthcare, marketing, and training, to mention a few. XR is expected to merge deeply into society and profoundly change how people live, experience, and interact \cite{Akyildiz-22b}. 
Traditionally, XR systems have mainly focused on two out of the five human senses, i.e., sight and hearing, through audiovisual data. The sense of touch through haptic technologies is also popular nowadays. Haptic technologies exploit vibrations, motions, and forces to mimic the experience of touch. Haptic information in combination with audiovisual data makes the XR experience more immersive. The ultimate goal is to also add the senses of smell and taste to achieve a fully immersive experience. This 5-sense experience will be a major step and will make XR a particularly attractive technology.  

Although VR/AR has been developed for decades \cite{Akyildiz-22b}, XR is still far from its potential. Existing XR devices (i.e., head-mounted displays (HMDs)) are rather expensive, relatively large/heavy, and often wired, limiting their popularity. Today, the leading high-end XR HMDs are Apple Vision Pro and Microsoft HoloLens 2. To become mainstream, XR devices need to be affordable, lightweight, and wireless. Current wireless XR devices rely primarily on wireless fidelity (WiFi) connections. However, the limited latency and reliability properties of WiFi links may limit the quality of the XR experience. 5G has a better service quality, making it a promising technology for wireless XR. 5G is currently adopting support for wireless XR. However, future immersive XR applications will be extremely data-hungry and delay-sensitive, thereby setting stringent requirements for wireless connections. Since 5G cannot meet these demands, 6G is expected to take wireless multi-sensory XR to the next level, potentially making it truly immersive, mainstream, and ubiquitous. For example, 6G-level XR is expected to form the core for advanced metaverse applications. For more information on the 5G XR standardization, wireless challenges for XR, and ISAC for XR, see \cite{Petrov-22, Huang-23, Hande-23}, \cite{Akyildiz-22b}, and \cite{Huang-24}, respectively. 

\subsubsection{HOLOGRAPHIC-TYPE COMMUNICATION}
Holographic-type communication refers to the transfer of holograms from a source to a (remote) destination via wireless/wired networks \cite{Akyildiz-22c}. A hologram is a realistic 3D representation of an object based on the depth and parallax information \cite{Akyildiz-22c}. Unlike a traditional 3D image, a hologram changes according to the position of the viewer \cite{Clemm-20}. Holography refers to a technology creating holograms \cite{Akyildiz-22c}. Specifically, a camera array is placed around a 3D object to collect image information from multiple angles and views. This information is then processed to generate a 3D hologram. 
To summarize an end-to-end holographic system, holograms are generated using holography (or stored) at the source, transferred via holographic-type communications to the destination, and finally presented by a holographic display \cite{Akyildiz-22c}. There are three main types of displays: XR HMDs, volumetric displays, and light-field displays \cite{Akyildiz-22c}. XR HMDs are close to the eyes of the user, with two view angles and looser requirements for data transfer. Volumetric displays are suitable for mobile devices due to their small size and limited viewing angles. Light-field displays have the highest quality with thousands of view angles, multi-user support, and extremely high requirements for holographic-type communications. 

Since the late 2000s, an increasing number of demonstrations on end-to-end holographic systems have been carried out \cite{Akyildiz-22c}. However, the quality of holographic experience is still rather limited due to the limited (rate/latency) performance of existing communication networks. Immersive high-quality holograms (e.g., using light-field displays) require extremely high data rates (even Tbps-level) and ultra-low latency (even sub-millisecond-level) \cite{Clemm-20, Akyildiz-22c}. It is expected that 6G and future wired networks will satisfy these extreme requirements and enable novel holographic applications and business opportunities. 
Potential holographic application scenarios include education, training, healthcare, gaming, sports, and marketing \cite{Akyildiz-22c}. Holographic telepresence enables remote participation in meetings and events, as realistic holograms. HMD-based holograms can be used as a part of education and training to make them more interactive, educational, and efficient. In healthcare, holograms can assist in surgeries and remote diagnostics. Holographic technologies enable more immersive ways to experience sports and gaming. For example, a game can be projected onto a 3D hologram. In marketing, holographic advertisements with a 5-sense technology can provide a real-like experience and may revolutionize the whole industry. More information on holographic-type communications can be found in \cite{Clemm-20, Akyildiz-22c, Petkova-24}. 

\subsubsection{DIGITAL TWINS}
A simple definition of a digital twin is a virtual digital copy of a physical object or entity. This real-world object/entity can be, for example, a sensor, device, machine, vehicle, UAV, robot, process, system, network, or even human. Conceptually, a digital twin can be divided into three parts, i.e., the physical object, digital object, and connections between them \cite{Mihai-22}. With these connections, a comprehensive amount of data can be collected from a physical object to form an accurate digital copy. With the aid of digital twins, one can monitor, analyze, evaluate, control, and manage a physical object/entity through the collected information \cite{Mihai-22}. Digital twins can provide many advantages \cite{Mihai-22}, for example, facilitate product development and testing, improve the efficiency and reliability of processes, and reduce the operational and maintenance costs of systems. There are numerous application scenarios for digital twins \cite{Mihai-22}, including manufacturing, product development, process design, system maintenance, network management, robotics, construction, healthcare, automotive industry, smart factory, smart city, and metaverse. Generally, the digital twin technology is seen as a key element in the realization of the fourth industrial revolution (i.e., industry 4.0) \cite{Mihai-22}. 

The first precursor of the digital twin concept dates back to the 1960s, when NASA started to model and analyze space systems on the ground \cite{Khan-22b}. The basic architecture of a digital twin, consisting of real space, virtual space, and the communication link between them, was introduced in 2002, under the name "mirrored spaces model" \cite{Mihai-22}. This landmark is seen as the beginning of modern digital twin research. Due to advances in key technologies, such as AI/ML, IoT, 5G/6G, distributed computing, and XR \cite{Mihai-22}, the attention toward digital twins (in industry and academia) has been growing since the 2010s. Today, the use of digital twins is spreading widely across industries. The digital twin concept is one of the major research areas in the field of technology \cite{Mihai-22}. However, the digital twin technology is still at a rather early stage in its evolution path and far from its envisioned potential. It is expected that digital twins will shape the future by connecting the physical and digital worlds in a revolutionary way, thus becoming an integral part of a 6G-enabled smart society. In the context of 6G and wireless networks, the latest information on digital twins can be found in \cite{Mihai-22, Kuruvatti-22, Khan-22b, Guo-23, Xu-23c, Alkhateeb-23, Lin-23b, Masaracchia-23, Gao-23b, Bariah-23, Mao-24, Sheraz-24}. 

\subsubsection{TACTILE INTERNET}
In a broad sense, Tactile Internet is considered as a communication system that can provide real-time control and interactivity between humans and machines, with the aid of haptic information (e.g., touch, vibration, and motion) in addition to traditional audiovisual data \cite{Promwongsa-21}. In other words, Tactile Internet can be seen as a paradigm shift from conventional content-based communications to control-based communications \cite{Sharma-20}. Tactile Internet is considered the next evolution step of the Internet, after the mobile internet and IoT \cite{Promwongsa-21}. Tactile Internet can be exploited for a wide range of applications \cite{Promwongsa-21}, such as industrial automation, robotics, gaming, healthcare (e.g., remote surgery), intelligent transportation/vehicle systems, and education/training. To realize Tactile Internet, it requires ultra-low latency (up to 1 ms) with high reliability (up to 1-10$^{\text{-}7}$), availability, and security \cite{Fettweis-14, Promwongsa-21}. These requirements are highly dependent on the application scenario. 

The concept of Tactile Internet was first introduced in 2014 \cite{Fettweis-14}. In 2016, a working group was established for the first standards (i.e., IEEE P1918.1) related to Tactile Internet \cite{Promwongsa-21}. The main objective of IEEE P1918.1 was to specify the architectural basis for Tactile Internet. In the research domain, 5G was first seen as an enabler for Tactile Internet due to its support for URLLC with a latency of up to 1 ms \cite{Promwongsa-21}. However, 5G was not designed to support Tactile Internet and its demanding applications. Hence, 5G is not able to serve as an enabling platform for Tactile Internet in practice. Despite the lack of broad support, 5G still has the potential to enable light versions of tactile applications and pave the way for 6G networks as a true enabler. 6G is expected to provide a fruitful platform for Tactile Internet and enable the broad use of novel real-time interactive applications \cite{Hou-21}. The capabilities of 6G go way beyond to those of 5G in every aspect, especially in terms of performance, intelligence, flexibility, and security, all of which are essential for realizing Tactile Internet in practice. Further details on Tactile Internet can be found in \cite{Promwongsa-21, Sharma-20, Hou-21}. 

\subsection{SMART ENVIRONMENTS} 
A smart environment refers to an entity in which advanced digital technologies, such as 5G/6G, AI/ML, IoT/IoE, cloud/edge computing, big/small data, and data security/privacy, are exploited to produce greater value. A greater value can be, for example, a better quality of life for an individual, more efficient manufacturing for a company, or better services for the residents of a city. 6G is seen as a common enabler for mainstreaming many types of smart environments. In the following, we discuss four smart environments: smart society, smart city, smart factory, and smart home.  

\subsubsection{SMART SOCIETY}
At a fundamental level, a smart society can be defined as a concept that exploits a vast variety of advanced digital technologies to provide benefits to all levels of the society. By utilizing these technologies, a smart society is expected to offer diverse benefits for individuals, communities, and private/public sectors. Possible benefits for people include a better quality of life in terms of improved services related to living conditions, healthcare, work, education, social benefits, free-time activities, and public safety. The potential benefits for the private sector include new business opportunities, more efficient processes (e.g., manufacturing, testing, and design), reduced costs (e.g., production, operation, and maintenance), and reduced environmental impact. A smart society comprises numerous vertical segments, such as smart governance, smart economy, smart healthcare, smart energy, smart grid, smart industry, smart infrastructure, smart education, smart cities, smart organizations, smart mobility, smart agriculture, smart culture, smart security, etc. 

\subsubsection{SMART CITY}
By relying on advanced information and communication technologies, the concept of a smart city aims to improve the management, operation, maintenance, and services to offer diverse advantages and a better quality of life for its residents. The synergy among advanced digital technologies provides a fruitful platform to enable smart cities worldwide. Smart cities will be more efficient in numerous ways and will offer a broad range of improved and novel services for their residents. A smart city consists of diverse elements \cite{Kirimtat-20, Alwis-23}, such as smart governance, smart economy, smart environment (nature), smart transportation, smart traffic, smart water, smart waste, smart sanitation, smart schools, smart buildings, smart homes, smart lighting, smart culture, smart sports, smart activities, smart parking, smart shopping, and so on. The opportunities, challenges, priorities, and realizations of smart cities depend on many different factors, such as the population, location, climate, environment, and area of the city \cite{Kirimtat-20, Alwis-23}. Thus, each city is different, and smart elements need to be designed accordingly. Currently, numerous smart city plans and projects are ongoing around the world \cite{Kirimtat-20, Alwis-23}. It is envisioned that smart cities will prosper in the 6G era since key technologies have become sufficiently mature to enable such complex entities \cite{Alwis-23, Murroni-23, Kim-24}. Recent information on smart cities is available in \cite{Kirimtat-20, Alwis-23, Murroni-23, Mishra-23, Yaqoob-23, Kim-24}. 

\subsubsection{SMART FACTORY}
A smart factory is a concept that merges advanced information, communication, networking, computing, and control technologies and processes to increase the intelligence, automation, efficiency, productivity, quality, flexibility, adaptability, and predictability of a factory \cite{Chen-18b, Soori-23}. A smart factory is at the core of industry 4.0 \cite{Chen-18b, Soori-23}. Industry 4.0 refers to the new era of smart manufacturing, where the integration of advanced digital technologies (e.g., IoT, AI, big data, and cloud computing) become mainstream, revolutionizing the entire industrial sector. Smart factories provide diverse benefits, such as improved monitoring, management, maintenance, and repair of processes; increased manufacturing efficiency; reduced manufacturing costs; higher quality products; faster production cycles; more robust processes; enhanced security; better prevention of threats/hazards; and novel production solutions. Although a smart factory is currently a relatively immature concept, it is expected to reach an adequate level of maturity in the 6G era. Detailed discussions on smart factories and industry 4.0 can be found in \cite{Chen-18b, Soori-23, Khang-23, Hu-24, Hazra-24}. 

\subsubsection{SMART HOME}
A smart home refers to a home that is equipped with an intelligent, connected, automated, and integrated control, computing, and communication system that provides comprehensive management and monitoring of the entire home ecosystem, including heating, air conditioning, water, electricity, energy consumption, lighting, entertainment, appliances, household robots, air quality, smoke/gas/leak detection, home access, security, and privacy. AI, IoT, and wireless connectivity are vital elements of a smart-home ecosystem. The concept of a smart home (also known as an automated home) has been studied for over two decades. However, to date, there have been no real breakthroughs. Currently, there is a wide range of smart home products in the market. Typical ones include lighting, security cameras, wireless network elements, voice assistants, TVs, speakers, etc. There are also numerous service providers with (mainly) WiFi-based cloud services. Although there are many products and services, the markets and solutions are somewhat fragmented. Moreover, there is still a lack of comprehensive smart and automated home solutions that include all vital elements. Common policies and standardization are also inadequate, especially in terms of data security and privacy. A real breakthrough in smart homes is still awaiting. 

Smart homes are anticipated to become popular in the 6G era due to the maturation of key technologies, such as intelligent cloud with AI-empowered computing and control, high-accuracy sensing and indoor localization, seamless wireless connectivity and IoT support, as well as data security and privacy. This will open new business opportunities for technology companies (such as mobile network operators, vendors, and service providers) to provide customized and comprehensive smart home solutions with plug-and-play styles. This requires cooperation between technology companies, constructors, and building/homeowners. Further discussions on the smart home concept can be found in \cite{Marikyan-19, Sovacool-20, Hammi-22, Rodriguez-23, Magara-24}. 

\subsection{CONNECTED AUTONOMOUS SYSTEMS}
Connected autonomous systems refer to intelligent entities that are capable of operating independently without human influence by relying on a wide range of technologies, ranging from computation and control to sensing, navigation, and communication. In this section, three types of connected autonomous systems are discussed, including connected autonomous vehicle systems, connected autonomous aerial vehicle systems, and connected autonomous robotic systems. 

\subsubsection{CONNECTED AUTONOMOUS VEHICLE SYSTEMS} 
Connected autonomous vehicle systems (CAVSs) refer to intelligent vehicle entities that consist of connected autonomous vehicles (CAVs) capable of independently driving on the roads with no (or little) human involvement to transport people and goods. Autonomous vehicles, commonly known as self-driving cars, exploit AI-empowered computing and control by relying on comprehensive real-time navigation, tracking, and sensing information about the vehicle itself and its surrounding environment. The term CAV is commonly used to define an autonomous vehicle that is capable of connecting to other vehicles, traffic infrastructure, cellular networks, and other road users. These different types of connectivity in vehicular systems fall under the umbrella of V2X communications. The level of autonomy in vehicles is divided into six categories by the Society of Automotive Engineers \cite{Qayyum-20}, i.e., Level 0: no automation, Level 1: driver assistance (hands on), Level 2: partial automation (hands off), Level 3: conditional automation (eyes off), Level 4: high automation (mind off), and Level 5: full automation. 

The development of autonomous vehicles has gradually evolved from small individual experiments of partially automated cars to thorough testing of truly self-driving cars around the world and launching of numerous pre-commercialized and commercialized local intelligent transportation services. The future trends in autonomous vehicles can be divided into two main branches. First, the automotive industry is evolving toward fully electric cars, with increasing levels of intelligence, automation, and autonomy. It is expected that the level of automation of different driver-assistance features will gradually move toward level 5 (full automation). Second, private companies are launching different types of pre-commercialized and commercialized local intelligent transportation services around the world. This trend will continue in three main ways, i.e., more services, locations, and coverage. 

Since 5G is not optimized for intelligent vehicles, it will reach its limits in providing ubiquitous and reliable connectivity for CAVSs. 6G, instead, will be developed to provide better support for CAVs and intelligent transportation systems. It is envisioned that 6G will serve as a fruitful platform to enable a large-scale usage of CAVs as a part of intelligent vehicular and transportation systems \cite{He-21b, Nguyen-22b}. In the 6G era, there will be many application scenarios for CAVs, such as individual consumer products, public transportation, private transportation, freight traffic, and product delivery. 
In the literature, autonomous vehicles have been extensively studied. The earliest considerable works were published in the 1990s. However, it was until the 2010s when the research really boomed. Since then, numerous aspects of autonomous vehicles have been covered, such as 6G/B5G/5G, AI/ML, control, navigation, sensing, communication, safety, traffic efficiency, security, privacy, liability issues, regulations, social perspectives, intelligent transportation systems, and smart cities. In the context of mobile networks, it is vital to develop advanced V2X communications for CAVs and the Internet of Vehicles (IoV) scenarios. Detailed discussions on 6G communications for CAVs are available in \cite{Yang-21, He-21b, Adhikari-21, Osorio-22, Nguyen-22b, Noorarahim-22, Deng-23, Liu-24b, Kakkavas-24}. 

\subsubsection{CONNECTED AUTONOMOUS AERIAL VEHICLE SYSTEMS}
Connected autonomous aerial vehicle systems (CAAVSs) refer to intelligent, cooperative, and independent entities, which manage their individual units of intelligent aerial vehicles (typically UAVs) capable of flying independently with no (or little) human influence, relying on AI-driven control systems with advanced computation, navigation, sensing, and communication features. Note that a special case of CAAVS is a single connected autonomous aerial vehicle (e.g., a connected autonomous UAV). A typical example of CAAVS is an autonomous UAV swarm. An autonomous UAV swarm comprises a set of UAVs that cooperatively perform diverse tasks as independent intelligent entities. Potential application scenarios include surveillance, environmental and industrial monitoring, malfunction detection, disaster aid, search operations, precision agriculture, and entertainment. In addition, autonomous UAVs for delivering items (i.e., air cargo) or carrying people (i.e., air taxis) can be considered CAAVSs. Currently, there exist many pre-commercialized and commercialized air cargo and taxi services worldwide. These services are expected to continue to spread both horizontally and vertically. 6G is seen as a promising platform to enable CAAVSs since it can potentially provide ubiquitous, high-performance, and reliable mobile connectivity for UAVs. Further information on UAVs and UAV swarms can be found in \cite{Mozaffari-21, Mu-23b, Mohsan-23c, Qazzaz-23} and \cite{Tahir-19, Javaid-23, Javed-24, Cao-24}, respectively. 

\subsubsection{CONNECTED AUTONOMOUS ROBOTIC SYSTEMS} 
Connected autonomous robotic systems (CARSs) refer to intelligent robotic entities that can independently and cooperatively perform complex tasks assigned to them without external human control. The core of CARSs is an AI/ML-based control system that requires advanced computation, data analytics, sensing, positioning, communications, and security capabilities. CARSs are connected to a wireless network, allowing communication and information exchange between a cloud server and other nodes in the network. Network connectivity enables novel capabilities, features, and applications of robotic systems, while advancing collaboration and interaction among them. CARSs have the potential to significantly widen the applicability of robotics to new areas and tasks. CARSs can adapt to dynamic environments and changing conditions. CARSs are applicable to numerous verticals, including manufacturing, warehousing, logistics, transportation, aviation, healthcare, and services. CARSs are expected to play a key role in future smart factories by expanding their capabilities, increasing flexibility, improving efficiency, and reducing costs. From a wider perspective, CARSs are seen as a means of taking robotics to the next level by freeing its potential. In this regard, 6G is considered a promising platform for providing customized network solutions to ensure secure high-performance connectivity with minimal delays, extreme levels of reliability, and advanced IoT capabilities. Further details on connected robotics can be found in \cite{Alsamhi-19, Chen-21e, Khang-23, Groshev-23, Soori-23b, Lessi-24}. 

\begin{figure}[!htb]
\center{\includegraphics[width=\columnwidth]
{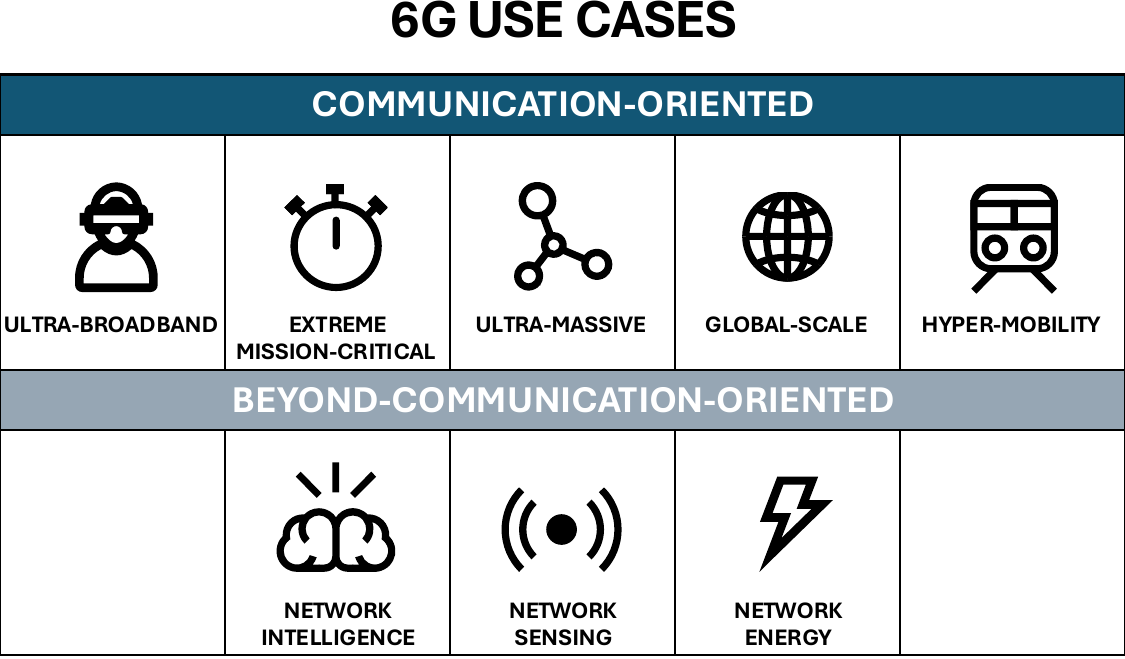}}
\caption{\label{Fig_Usecases}Key use cases for 6G.}
\end{figure}

\begin{table*}[htb!]
\begin{center}
\caption{Summary of key use cases for 6G}
\label{Table_Usecases}
\centering
\begin{tabularx}{\textwidth}{| >{\centering\arraybackslash}X | >{\centering\arraybackslash}X |
>{\centering\arraybackslash}X | 
>{\centering\arraybackslash}X |
>{\centering\arraybackslash}X |
>{\centering\arraybackslash}X |
>{\centering\arraybackslash}X |
>{\centering\arraybackslash}X |}
\hline
\centering
\vspace{3mm} \textbf{Use Cases} \vspace{3mm} & \centering \textbf{Target} & \centering \textbf{Environment} & \centering \textbf{Requirements} & \centering \textbf{Challenges} & \centering \textbf{Enablers} & \vspace{1.5mm} \begin{center} \textbf{Applications} \end{center} \\
\hline
\vspace{3mm} Ultra-Broadband Multimedia Communications \vspace{3mm}  & Immersive multimedia & Hotspot to rural & Extreme rates $\&$ ultra-low latency & Extreme requirements & THz $\&$ UDNs & XR $\&$ HTC \\
\hline
\vspace{3mm} Extreme Time Sensitive Mission Critical Commun. \vspace{3mm}  & Applications with tight constraints & Industrial $\&$ vehicular & Extreme reliability $\&$ ultra-low latency & Conflicting requirements & Fast/robust network processes & Smart factory $\&$ CAVs $\&$ tele-surgery \\
\hline
\vspace{3mm} Ultra-Massive Communications \vspace{3mm}  & Ultra-massive IoT/IoE & Industrial $\&$ urban $\&$ wide-area & Ultra-dense connections & Network flexibility & Flexible network mechanisms & Smart factory $\&$ smart city \\
\hline
\vspace{3mm} Global-Scale Communications \vspace{3mm}  & Global coverage & Land $\&$ air $\&$ sea & Super-coverage & Cost efficiency $\&$ 3D network & NTNs & Global IoT $\&$ remote internet \\
\hline
\vspace{3mm} Hyper-Mobility Communications \vspace{3mm}  & Connectivity at high device velocities & Urban to rural & Extreme mobility & Mobility management & Lower spectrum $\&$ fast handovers & Ultra-high-speed trains \\
\hline
\vspace{3mm} Network Intelligence \vspace{3mm}  & Intelligence as a service & Industrial $\&$ organizational & High level of intelligence & Powerful computing infra & Pervasive AI/ML & Smart environments \\
\hline
\vspace{3mm} Network Sensing \vspace{3mm}  & Ubiquitous sensing & Industrial $\&$ healthcare & Sensing capabilities & Accurate sensing & ISAC & Detection $\&$ tracking \\
\hline
\vspace{3mm} Network Energy \vspace{3mm}  & Wireless energy services & Industrial $\&$ private & WET capabilities & Efficiency $\&$ distances & Energy beamforming & Powering IoT devices \\
\hline
\end{tabularx}
\end{center}
\end{table*}

\section{KEY USE CASES FOR 6G}
\label{Usecases}
6G is expected to greatly expand the capabilities of mobile networks by enhancing "traditional" communication features and introducing new beyond-communication ones. Following this evolution, we introduce five communication-oriented and three beyond-communication-oriented use cases, supporting a wide range of new 6G-level application scenarios. These use cases are summarized in Figure \ref{Fig_Usecases} and Table \ref{Table_Usecases}. 

\subsection{COMMUNICATION-ORIENTED USE CASES}
6G is expected to push the communication performance into its limits. The corresponding performance dimensions are capacity, latency, reliability, density, coverage, and mobility. We identify five communication-oriented use cases to cover all of these dimensions. 

\begin{itemize}

\item \textbf{Ultra-Broadband Multimedia Communications:}
The goal of this use case is to provide wide support for immersive multimedia content in diverse environments, from dense hotspots to sparse rural areas. Examples of data-hungry application scenarios are wireless multi-sensory XR, holographic-type communications, information showers, and ultra-broadband mobile internet. Due to the nature of demanding applications, this use case needs to support many different performance requirements, such as very high link- and network-level capacities, low latency, and high reliability. The key enablers include massive radio resources for capacity (more spectrum, cells, and antennas), fast network mechanisms for low latency, and robust communication methods for reliability. Achieving these requirements simultaneously is an application-specific trade-off since each one of them is (more or less) conflicting to each other. Note that this use case can be considered as an extension of 5G eMBB. 

\item \textbf{Extreme Time-Sensitive and Mission-Critical Communications:}
This use case is intended to support time-sensitive and mission-critical applications in versatile environments, such as smart factories, smart hospitals, autonomous vehicle systems, and emergency operations. Extreme reliability and ultra-low latency are required since there is no room for severe errors. The main enablers for extreme reliability include network- and link-level diversity methods, high channel coding rates, robust modulation orders, and advanced retransmission mechanisms. Extremely low latency can be attained through flat higher-layer management procedures, bringing content close to end-devices, minimized control signaling, fast network access mechanisms, and short transmission frames. Since reliability and low latency are conflicting requirements, there is always a trade-off between them. Note that this use case can be seen as an extension of 5G URLLC. 

\item \textbf{Ultra-Massive Communications:}
This use case aims to support an ultra-massive number of different types of devices in diverse application scenarios. The range of devices is wide, including IoT sensors, devices, wearables, machines, robots, vehicles, drones, etc. Application scenarios vary from smart environments (e.g., cities, factories, hospitals, buildings, warehouses, etc.) to wide-area IoT (e.g., environmental/industrial monitoring) and low-power sensor networks. Due to versatile devices and applications, many performance dimensions need to be considered, such as connection density, coverage, mobility, capacity, and reliability. The key enablers for massive connectivity include flexible spectrum operation, fine-grained frame structure, massive grant-free network access, robust transmission techniques, and efficient resource management. The challenge is to build such a flexible network. This use case is an extension of 5G mMTC. 

\item \textbf{Global-Scale Communications:}
The goal of the use case is to provide worldwide super-coverage, including remote areas on the land, at the sea, and in the air. This use case supports diverse global-level application areas, including remote internet, global IoT, environmental monitoring, industrial tracking, and maritime/aerial communications, to mention a few. Aligned with the United Nations Sustainable Development Goals, digital divide may be alleviated by providing internet access to remote and developing areas around the world. Moreover, we can fight climate change with worldwide environmental monitoring capabilities. Global coverage is enabled by integrated terrestrial and non-terrestrial networks that exploit satellite, aerial, and mobile communications. 

\item \textbf{Hyper-Mobility Communications:} 
This use case is meant to support communications at high velocities, even up to 500--1000 km/h. Example application scenarios are high-velocity transportation systems, such as ultra-high-speed trains. The main enablers include flexible operation within a suitable spectrum range, fast handovers, and enhanced mobility management techniques. In general, lower frequencies are better for higher-mobility scenarios due to their wider coverage and less frequent handovers. 

\end{itemize}

\begin{table*}[t!]
\begin{center}
\caption{Main performance requirements for 6G}
\label{Table_Requirements}
\centering
\begin{tabularx}{\textwidth}{| >{\centering\arraybackslash}X | >{\centering\arraybackslash}X |
>{\centering\arraybackslash}X | 
>{\centering\arraybackslash}X |
>{\centering\arraybackslash}X |}
\hline
\centering
\vspace{3mm} \textbf{Performance Requirements} \vspace{3mm} & \centering \textbf{IMT-2020 (ITU-R)} & \centering \textbf{IMT-2030 (ITU-R)} & \vspace{1.5mm} \begin{center} \textbf{6G (Literature)} \end{center} \\
\hline
\vspace{3mm} Peak Data Rate \vspace{3mm}  & 20 Gbit/s & 50/100/200 Gbit/s & 1 Tbit/s \\ 
\hline
\vspace{3mm} User Experienced Data Rate \vspace{3mm} & 100 Mbit/s & 300/500 Mbit/s & 1/10 Gbit/s \\
\hline
\vspace{3mm} Peak Spectral Efficiency \vspace{3mm} & 1X & 1.5X/3X & 2X--3X \\
\hline
\vspace{3mm} Area Traffic Capacity \vspace{3mm} & 10 Mbit/s/m$^2$ & 30/50 Mbit/s/m$^2$ & 1/10 Gbit/s/m$^2$ \\
\hline
\vspace{3mm} Latency \vspace{3mm} 
Latency & 1 ms & 0.1--1 ms & 0.1 ms \\
\hline
\vspace{3mm} Reliability \vspace{3mm} & 1-10$^{\text{-}5}$ & 1-10$^{\text{-}5}$--1-10$^{\text{-}7}$ & 1-10$^{\text{-}7}$--1-10$^{\text{-}9}$ \\
\hline
\vspace{3mm} Connection Density \vspace{3mm} & 10$^{6}$/km$^2$ & 10$^{6}$--10$^{8}$/km$^2$ & 10$^{7}$--10$^{8}$/km$^2$ \\
\hline
\vspace{3mm} Mobility \vspace{3mm} & 500 km/h & 500--1000 km/h & 1000 km/h \\
\hline
\vspace{3mm} Positioning Accuracy \vspace{3mm} & ND & 1--10 cm & 0.1--10 cm \\
\hline
\end{tabularx}
\end{center}
\end{table*}

\subsection{BEYOND-COMMUNICATION-ORIENTED USE CASES}
Since 6G is expected to extend its capabilities beyond communication, we introduce three corresponding use cases to cover the network-enabled intelligence, sensing, and energy dimensions. 

\begin{itemize}
\item \textbf{Network Intelligence:}
The target of this use case is to provide AI/ML-assisted intelligent network services and applications, such as intelligence as a service, network-assisted intelligent systems (industry, transportation, healthcare, etc), computation offloading, and edge caching. This use case requires powerful computation and storage abilities, advanced AI/ML methods, abundant data acquisition, and efficient AI/ML model training/inference. A major challenge is to provide a ubiquitous communication, computation, and caching infrastructure. 

\item \textbf{Network Sensing:}  
This use case aims to expand the capabilities of cellular networks to sensing and positioning. Example applications include object identification, shape recognition, activity detection, range estimation, velocity evaluation, and movement tracking. Typical application environments include industrial and healthcare. For example, network sensing can be used to detect a gas leakage in a factory or detect a person falling down in an elderly care with an automatic call for help. This use case is enabled by the integration of communication and sensing. The main challenge is to obtain a high accuracy in sensing. 

\item \textbf{Network Energy:}
The aim of this use case is to support wireless energy transfer (WET) services to power network nodes, such as IoT sensors and low-power devices. In particular, wireless energy services aim to advance the long life-time, low-complexity, autonomy, mobility, and novel usages of low-energy networks. This use case can provide customized energy services to diverse vertical industries. Energy beamforming is a key technology to provide wireless energy capabilities. In energy beamforming, typical challenges are to achieve high energy transfer efficiency and sufficiently long transfer ranges.  

\end{itemize}

\section{MAIN PERFORMANCE REQUIREMENTS FOR 6G}
\label{Requirements}
In this section, the main performance requirements are discussed for 6G. In November 2023, ITU-R introduced 15 performance capabilities for IMT-2030 in Recommendation ITU-R M.2160-0, i.e., "IMT-2030 Framework" \cite{ITU-M2160}. The numerical target values were defined for nine of them. In the following, we review these target requirements and compare them to those of IMT-2020 described in Recommendation ITU-R M.2083-0 \cite{ITU-M.2083-0}. In addition, we present some typical values proposed in the 6G literature (see \cite{Wang-23c}, and the references therein). The target requirements are summarized in Table \ref{Table_Requirements}. At the end of this section, we briefly list the rest of the IMT-2030 capabilities identified by ITU-R. Note that ITU-R has not defined any target values for these capabilities since most of them are difficult to measure numerically. 

\begin{itemize}
\item \textbf{Peak Data Rate:} 
The peak data rate is defined as the maximum achievable rate when allocating all available radio resources to a single link and assuming ideal conditions. In IMT-2020, the target peak rate is 20 Gbit/s for downlink transmission. The target values of ITU-R for IMT-2030 are 50, 100, and 200 Gbit/s, depending on the usage scenario. In the literature, the most common target rate for 6G is 1 Tbit/s. 

\item \textbf{User Experienced Data Rate:}
The user experienced data rate refers to the achievable rate which can be obtained by 95 $\%$ of the users in the coverage area (i.e., the 5 $\%$ point of the cumulative distribution function of the user throughput). The target value of IMT-2020 is 100 Mbit/s, whereas ITU-R suggests 300 and 500 Mbit/s for IMT-2030. Greater values can be explored as well. Typically proposed values in the 6G literature are 1 and 10 Gbit/s. 

\item \textbf{Peak Spectral Efficiency:} 
The peak spectral efficiency represents the maximum achievable user throughput per unit bandwidth. For IMT-2030, ITU-R proposes 1.5/3 times higher spectral efficiency than that of IMT-2020. Greater values may also be studied. In the 6G literature, typical targets are 2--3 times higher (i.e., 60--90 bit/s/Hz) than that of 5G (i.e., 30 bit/s/Hz). 

\item \textbf{Area Traffic Capacity:}
The area traffic capacity is defined as the total aggregated throughput per unit area. Compared to 10 Mbit/s/m$^2$ of IMT-2020, ITU-R recommends the values of 30 and 50 Mbit/s/m$^2$ for IMT-2030, while greater values could also be examined. The values typically presented in the 6G literature are much higher than those of ITU-R, i.e., 1 and 10 Gbit/s/m$^2$. 

\item \textbf{Latency:}
The user plane latency is the time spent by the network from sending a packet to receiving it, assuming unloaded conditions. The latency requirement of IMT-2020 is 1 ms. According to ITU-R, the target values could range from 0.1 to 1 ms for IMT-2030. The target of 0.1 ms is typical in the 6G literature. 

\item \textbf{Reliability:}
Reliability is defined as the success probability of a packet transmission within the required time. The reliability requirement of IMT-2020 is 1-10$^{\text{-}5}$. While ITU-R suggests that the success probabilities could be in the range of 1-10$^{\text{-}5}$--1-10$^{\text{-}7}$ for IMT-2030, the values from 1-10$^{\text{-}7}$ up to 1-10$^{\text{-}9}$ have been proposed in the 6G literature. 

\item \textbf{Connection Density:}
The connection density is defined as the total number of devices that can be served with a certain quality per unit area. For IMT-2030, ITU-R proposes 10$^{6}$--10$^{8}$/km$^2$ connection densities which are 1--100 times higher than that of IMT-2020. In the 6G literature, commonly suggested densities are 10$^{7}$/km$^2$--10$^{8}$/km$^2$. A volumetric-based measure has also been proposed in the literature, with a typical value of 100/m$^3$.  

\item \textbf{Mobility:} 
Mobility is the highest device velocity supported by the network for a communication link of a certain quality. The maximum mobility in IMT-2020 is 500 km/h, whereas ITU-R recommends 500--1000 km/h for IMT-2030. The most common target value in the 6G literature is 1000 km/h. 

\item \textbf{Positioning Accuracy:}
The positioning accuracy is a quantifiable value that represents the difference between the network-estimated location of the device and its real location. There is no position accuracy target defined for IMT-2020. ITU-R suggests that the positioning accuracy could range from 1 to 10 cm. In the 6G literature, the proposed targets range from millimeter to centimeter levels. 
\end{itemize}

The remaining six IMT-2030 capabilities include coverage, sustainability, interoperability, security and resilience, sensing-related capabilities, and applicable AI-related capabilities. Further details on these capabilities can be found in \cite{ITU-M2160}. 

\section{AI/ML FOR 6G}
\label{AI}
In this section, we provide a brief introduction to the fundamentals of AI/ML, review three promising ML methods for 6G, and finally have a high-level discussion on the opportunities and challenges of AI/ML in 6G networks. 

\subsection{INTRODUCTION TO AI AND ML}
In the following, the concepts of AI and ML are introduced, focusing on their past, present, and future. Furthermore, the main types of learning and training are described. 

\subsubsection{ARTIFICIAL INTELLIGENCE}
AI can be seen as a process in which a computer program performs functions that are considered intelligent from a human perspective, such as planning, learning, reasoning, and decision making, in order to perform a specific task. Consequently, AI can make different types of objects and entities intelligent, such as software applications, devices, and systems.  
The research field of AI is considered to be born in 1956, when the term AI was first introduced by the organizers of the first AI event at the Dartmouth College in the US \cite{Haenlein-19}. This AI workshop was based on the idea that any feature of intelligence can, in principle, be simulated by a machine. 
Since the late 1950s, AI has been actively researched with constant progress, except for a few skeptical periods in the late 1970s and between the late 1980s and the early 1990s \cite{Haenlein-19}. Over the years, typical research areas were computer board games, natural-language processing, and robotics. 

The potential of AI was recognized early on, but with unrealistic optimism in practical implementations. After decades of great expectations, the use of AI finally boomed in the 2010s. This was due to advances in key areas, such as computer technology, ML, and big data. Since the 2010s, the top technology companies have invested heavily in AI. A detailed history of AI is presented in \cite{Haenlein-19}. 
Today, AI is used everywhere. Through a vast variety of applications, AI has penetrated deeper into the different levels of modern society. Typical applications include Internet search engines, recommendation systems, online advertising, social media, entertainment, gaming, voice assistants, smart phones, smart home devices/appliances, online food delivery, electric scooter services, assisted/self-driving cars, drones, robots, etc. In the future, the use of AI will significantly increase in all areas of life and society. 

\subsubsection{MACHINE LEARNING}
ML can be seen as a process in which a computer program learns from training data to perform a specific task to which the program is not explicitly programmed. In other words, ML transforms experience into expertise. ML is a key tool to realize AI functions \cite{Simeone-18}, currently dominating in AI research and applications. ML algorithms learn through training. More training results in better performance. In general, (data-based) ML is a preferable solution over conventional (model-based) mathematical methods to problems where there exist a model or algorithm deficit \cite{Simeone-18}. A model deficit refers to the non-existence of a mathematical model for the problem at hand. In the case of an algorithm deficit, a mathematical model is available, but the problem is too complicated to be solved using conventional mathematical algorithms. It is worth mentioning that for some multi-phase problems, it may be preferable to use a combination of model- and data-based approaches \cite{Zappone-19}. 

As a technological term, ML was first introduced in 1959 in the context of a computer trained to play the game of checkers \cite{Samuel-59}. However, it was until the 1990s when ML started to bloom. Until then, the dominant methods in AI research were based on reasoning (from the 1950s to the early 1970s) and knowledge (from the mid-1970s to the early 1990s) concepts, such as inductive logical programming and expert systems, respectively \cite{Zhou-21b}. In the mid-1990s, statistical learning, such as kernel methods, became mainstream \cite{Zhou-21b}. After decades of research on neural networks, connectionism learning began to dominate in the 2000s in the form of DL, which is typically realized through multi-layered neural networks \cite{Zhou-21b}. DL showed excellent performance in many problem settings when the training datasets were sufficiently large. DL became popular due to advances in computational power and the rise of big data technology \cite{Zhou-21b}. Currently, DL is the dominant AI technology used in research and applications. ML is widely used in the modern society, benefiting diverse verticals, such as technology industry, healthcare, manufacturing, economics, marketing, security, and agriculture. ML is constantly expanding its frontiers and integrating deeper into devices, systems, and applications. Further details on the past, present, and future of ML can be found in \cite{Zhou-21b}. 

\subsubsection{MAIN TYPES OF ML}
There are three main types of ML algorithms: supervised, unsupervised, and reinforcement learning \cite{Simeone-18}. 

\begin{itemize}
\item {\textbf{Supervised Learning}}:
Supervised learning aims at finding a mapping between inputs and outputs by analyzing labelled training data containing input-output pairs \cite{Simeone-18}. Typical tasks for supervised learning are classification and regression \cite{Simeone-18}. In communication, supervised learning is promising for detection and prediction. 

\item {\textbf{Unsupervised Learning}}:
In unsupervised learning, the training data are unlabelled and contain only inputs, without any indication of the desired outputs \cite{Simeone-18}. The goal of unsupervised learning is to recognize patterns behind data generation. Unsupervised learning is typically applied to density estimation, clustering, feature extraction, and dimensionality reduction \cite{Simeone-18}. In communication, unsupervised learning is particularly suitable for problems related to compressing and clustering.

\item {\textbf{Reinforcement Learning}}:
Unlike the other two methods, reinforcement learning dynamically interacts with the operating environment by receiving feedback (i.e., a reward) after choosing an output (i.e., an action) for a given input (i.e., an observation) \cite{Simeone-18}. Feedback provides information on how well the chosen action satisfies the predefined targets of the learning algorithm. Reinforcement learning is particularly suitable for problems involving sequential decision making \cite{Simeone-18}. In communications, reinforcement learning naturally lends itself to resource allocation problems. 
\end{itemize}

\subsubsection{MAIN TYPES OF TRAINING}
In ML, there are two main types of training: offline and online \cite{Hoi-21}. 

\begin{itemize}
\item {\textbf{Offline Learning}}: 
Offline learning refers to a learning process executed over the entire training dataset before the corresponding ML algorithm is taken into action to solve the problems at hand \cite{Hoi-21}. Offline learning is cost-inefficient and poorly scalable for large-scale real-world applications that change (rapidly) over time and space due to the massive and expensive (re-)training processes \cite{Hoi-21}. 

\item {\textbf{Online Learning}}: 
Online learning refers to an incremental learning process, in which the training data instances appear in a sequential order, and the learner is updated one by one \cite{Hoi-21}. Online learning is particularly suitable for real-world applications under rapidly changing conditions \cite{Hoi-21}. Compared to offline learning, online learning requires fewer computation and storage resources, is faster and cheaper to implement, and can instantly adapt to varying conditions. However, online learning is challenging and may face efficiency and robustness issues in practice due to the poor quality of the training data \cite{Hoi-21}.   
\end{itemize}

\begin{figure}[!htb]
\center{\includegraphics[width=0.7\columnwidth]
{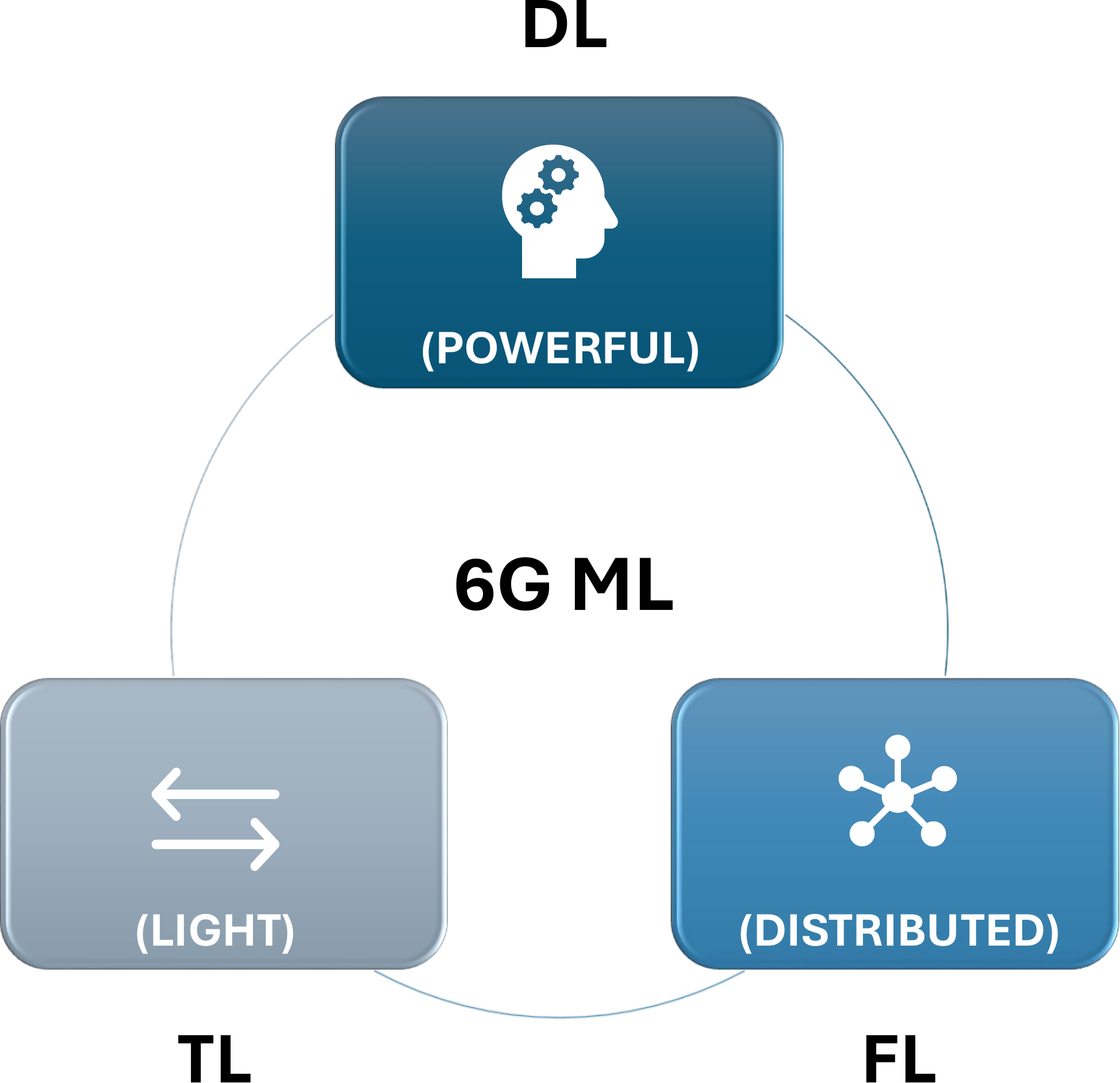}}
\caption{\label{Fig_ML}Key ML methods for 6G.}
\end{figure}

\begin{table*}[htb!]
\begin{center}
\caption{Summary of ML methods for 6G}
\label{Table_ML}
\centering
\begin{tabularx}{\textwidth}{| >{\centering\arraybackslash}X | >{\centering\arraybackslash}X |
>{\centering\arraybackslash}X | 
>{\centering\arraybackslash}X |
>{\centering\arraybackslash}X |
>{\centering\arraybackslash}X |}
\hline
\centering
\vspace{3mm} \textbf{ML methods} \vspace{3mm} & \centering \textbf{Vision} & \centering \textbf{Description} & \centering \textbf{Opportunities} & \vspace{1.5mm} \begin{center} \textbf{Challenges} \end{center} \\
\hline
\vspace{3mm} Deep Learning \vspace{3mm}  & Key ML method for 6G & Multi-layer artificial neural networks & Excellent performance & Lots of training \\
\hline
\vspace{3mm} Federated Learning \vspace{3mm}  & Distributed ML for 6G & Local learning at devices & Reduced latency $\&$ enhanced privacy & Efficiency $\&$ scalability $\&$ asynchronous nature \\
\hline
\vspace{3mm} Transfer Learning \vspace{3mm}  & Enhanced ML training for 6G & knowledge transfer from source to target learner & Reduced training and computation& Negative transfer \\
\hline
\end{tabularx}
\end{center}
\end{table*}

\subsection{KEY ML METHODS FOR 6G}
In this section, we discuss three ML methods that can provide diverse benefits to 6G networks. These methods include DL, federated learning (FL), and transfer learning (TL), as illustrated in Figure \ref{Fig_ML}. The opportunities and challenges of these methods are summarized in Table \ref{Table_ML}. 

\subsubsection{DEEP LEARNING}
DL is a popular sub-class of ML, exploiting multi-layer artificial neural networks that loosely mimic the function of a biological brain \cite{Zhang-19b}. Each layer consists of connected nodes (i.e., artificial neurons) with non-linear processing capabilities \cite{Zhang-19b}. DL can solve complex problems by transforming, progressively layer-by-layer, the raw input data into a higher-level representation, according to the objective of the given problem, through the non-linear processing of the nodes \cite{Zhang-19b}. Numerous artificial neural network architectures have been developed to solve different types of problems. A few classical networks include convolutional neural networks, recurrent neural networks, generative adversarial networks, deep neural networks, and deep belief networks \cite{Zhang-19b}. DL can be applied according to the supervised, unsupervised, or reinforcement learning principles. DL has proven to be an efficient and accurate method if the available training datasets are sufficiently large \cite{Zhang-19b}. 

Generally, DL is applicable to diverse engineering problems. In the context of wireless networks, DL is well-suited to many types of communication problems \cite{Mao-18, Zhang-19b, Dai-20b, Cheng-21, Joshi-23}, such as network resource management, network access, data traffic prediction, routing, user scheduling, resource allocation, channel estimation, and signal detection. DL is a particularly promising technology for 6G networks since it fits well for diverse 6G-specific challenges at different levels of the network \cite{Zhang-21l, Zheng-21, Jagannath-21, Ozpoyraz-22}. Such challenges are related to space-air-ground integration, cell-free network design, network slicing, network security, edge caching, joint communication and sensing, massive MIMO, RISs, etc. Although DL algorithms have been extensively studied, there are still many challenges to be solved to satisfy the stringent 6G requirements. DL has been reviewed for wireless networks and 6G in \cite{Mao-18, Zhang-19b, Dai-20b, Cheng-21, Joshi-23, Wang-24b} and \cite{Zhang-21l, Zheng-21, Jagannath-21, Ozpoyraz-22, Nguyen-23, Abd-24}, respectively. 

\subsubsection{FEDERATED LEARNING} 
FL is a distributed ML method, in which each involved edge device trains its local learning model based on local data at the device itself and sends only the resulting model parameters, instead of raw data samples, to a centralized server/cloud, updating the aggregated global learning model and broadcasting the updated parameters back to the edge devices \cite{Niknam-20, Liu-20k, Yang-22}. This process is repeated until the desired level of convergence is achieved. A distributed FL process can also be performed without a central server by sharing the local parameters among the involved edge devices \cite{Yang-22}. 
Due to its distributed nature, FL provides potential benefits compared with centralized ML methods, including saved wireless resources, reduced latency, and enhanced privacy \cite{Yang-22}. Since only model parameters (not extensive training data) are communicated between the edge devices and the centralized server, the amount of exchanged data is significantly reduced, leading to the possible savings of communication resources. The FL process with local training at each device and global updating at the parameter server may reduce latency since the computational complexity of such algorithms is much lower than that of the centralized algorithm, resulting in faster computation time and shorter computation delays. However, the distributed process is iterative, and thus, the possible latency reductions also depend on the speed of convergence. Protected privacy is inherent in FL since the device-specific training data remain in the device itself. 

FL can be applied to many common problems in wireless networks, such as resource/spectrum management, resource allocation, user behavior prediction, channel estimation, signal detection, and security/privacy issues \cite{Lim-20, Niknam-20, Liu-20k, Yang-22}. FL has also been proposed to assist in diverse emerging technologies, such as RISs, non-orthogonal multiple access (NOMA), integrated terrestrial and non-terrestrial networks, edge caching, vehicular networks, etc \cite{Lim-20, Boyzinis-22, Yang-22}. Although FL is applicable to numerous functions in future mobile networks, there are still many fundamental challenges to be addressed before practical realizations. Typical challenges are related to communication overhead, asynchronous communication, learning efficiency, joint communication and computation design, resource allocation, scalability, network heterogeneity, and security/privacy \cite{Lim-20, Li-20e, Niknam-20, Liu-20k, Boyzinis-22, Yang-22}. In the context of 6G, FL is a promising ML method to enable distributed and privacy-preserving learning. However, the research and development work of mobile FL is still at a rather early stage, requiring extensive efforts in academia and industry in the near future. Recent survey papers have reviewed FL in terms of 6G \cite{Liu-20k, Boyzinis-22, Yang-22, Alquraan-23, Duan-23, Hafi-24}, wireless/mobile networks \cite{Li-20e, Niknam-20, Lim-20, Qin-21, Beitollahi-23, Lee-24}, and IoT \cite{Nguyen-21d, Khan-21d, Zhang-22}. 

\subsubsection{TRANSFER LEARNING}
TL is a branch of ML, which objective is to transfer knowledge from source domains to target learning processes at target domains to enhance the quantity and quality of training data, reduce the computational demands of training processes, and improve the speed, accuracy, and robustness of target learners \cite{Zhuang-21, Nguyen-22c}. In other words, less training data needs to be collected to implement efficient learning algorithms with improved performance and reduced demands \cite{Zhuang-21}. Since the collection of adequate training data is often time-consuming, costly, or sometimes infeasible, TL saves time and expenses, or even enables the use of ML in otherwise impractical situations \cite{Zhuang-21}. 
To attain the aforementioned benefits, the challenge is to find proper source domains to transfer useful knowledge instead of harmful one (i.e., negative transfer \cite{Zhuang-21}), which may lead to poor performance of the TL algorithms. More precisely, source domains should be sufficiently related/similar to target domains to avoid the pitfall of transferring knowledge that has a negative impact on the target learners. In general, there are three main challenges in TL: what, when, and how to transfer \cite{Nguyen-22c}. Based on these challenges, TL can be categorized in four groups, including feature-, parameter-, relational-, and instance-based \cite{Zhuang-21}. TL can also be classified into three categories based on the problem perspective: inductive, transductive, and unsupervised \cite{Nguyen-22c}. TL can be combined with other ML methods, such as DL (i.e., deep transfer learning (DTL)) and FL (i.e., federated transfer learning (FTL)) \cite{Wang-21, Nguyen-22c}. 

Due to its benefits, TL has gained popularity in the field of ML. There are diverse application scenarios for TL, such as recommendation systems, bioinformatics, healthcare, transportation, computer vision, computing, and communications \cite{Zhuang-21}. Recently, TL has attracted increasing interest in the wireless research community \cite{Nguyen-22c}. In this context, TL algorithms have been proposed to address specific issues in resource/spectrum management, energy-efficient design, caching, sensing, localization, and security \cite{Nguyen-22c}. 
In particular, TL has been recognized as a promising ML method for 6G to address emerging challenges in data collection by sharing knowledge between different learning algorithms, potentially leading to more efficient training processes, improved learning performance, mitigated signaling overhead, and enhanced data privacy \cite{Wang-21, Parsaeefard-22, Nguyen-22c}. In the context of 6G, advanced types of TL are of special interest, including DTL, FTL, and online TL. Further research is required to realize practical TL algorithms that address the special features, requirements, and application scenarios of 6G. TL has been comprehensively reviewed for generic purposes, wireless networks, and 6G in \cite{Zhuang-21}, \cite{Nguyen-22c}, and \cite{Wang-21}, respectively. 

\subsection{DISCUSSION}
AI/ML is expected to play a revolutionary role in 6G networks. The extensive use of AI/ML can fundamentally change the way how mobile networks are designed, operated, and managed. While AI/ML holds great potential when integrated into 6G networks, it also brings significant challenges across technological, societal, and ethical domains. To ensure an efficient integration of AI/ML into 6G networks, it requires overcoming substantial technological obstacles in developing a ubiquitous and powerful computing infrastructure from the core to the edge, advanced data management processes, and practical data acquisition methods. In particular, real-time processing in AI/ML-enhanced network environments is a major challenge. Since the extensive use of AI/ML requires substantial amounts of energy, developing sustainable and energy-efficient AI/ML solutions is crucial. These technological challenges are discussed in detail in Section \ref{NI}. 

As pervasive AI/ML requires vast amounts of data, possibly personal or sensitive, to be collected and analyzed, it raises concerns on the privacy, data ownership, and consent of the users. In data ownership, the question is who owns the data, is it the users who generate it, the companies who manage it, or the governments who regulate it. This is a critical question since the ownership affects how data is being used. Due to the lack of transparency in AI/ML systems, it may be difficult for users to fully understand how their data is being used. Thus, getting a genuine consent from the users to use their personal information becomes challenging. The ethical considerations of AI/ML in decision making revolve around transparency and accountability. If there is a lack of transparency, it can be challenging to pinpoint who is accountable when errors occur. Clear regulations are needed to protect the privacy of the users, prevent the misuse of the data, and ensure the responsible use of AI/ML in the 6G era. Further details on responsible AI for mobile networks can be found in \cite{Nokia-AI}. 

\begin{figure}[!htb]
\center{\includegraphics[width=\columnwidth]
{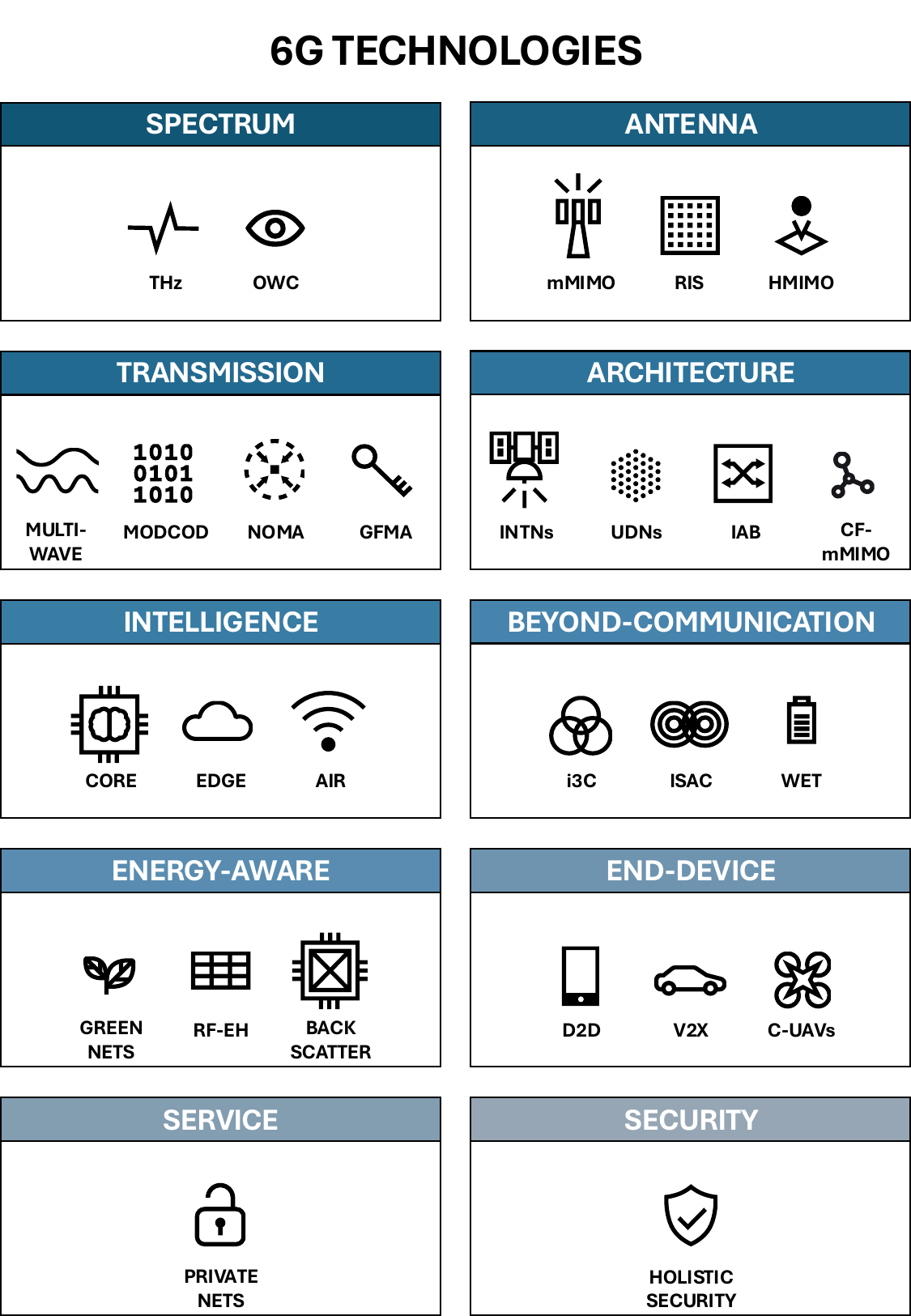}}
\caption{\label{Fig_Technologies}Potential technologies for 6G.}
\end{figure}

\begin{table*}[htb!]
\begin{center}
\caption{Summary of spectrum-level technologies for 6G}
\label{Table_Spectrum}
\centering
\begin{tabularx}{\textwidth}{| >{\centering\arraybackslash}X | >{\centering\arraybackslash}X |
>{\centering\arraybackslash}X |
>{\centering\arraybackslash}X |
>{\centering\arraybackslash}X |
>{\centering\arraybackslash}X |
>{\centering\arraybackslash}X |
>{\centering\arraybackslash}X |}
\hline
\centering
\vspace{3mm} \textbf{Spectrum-Level Technologies} \vspace{3mm} & \centering \textbf{Vision} & \centering \textbf{Description} & \centering \textbf{Opportunities} & \centering \textbf{Challenges} & \centering \textbf{Past} & \vspace{1.5mm} \begin{center} \textbf{Present} \end{center} \\
\hline
THz Communications & Extreme capacity for 6G & Operation at 0.1-10 THz band & Massive spectrum & \vspace{3mm} Propagation losses $\&$ HW impairm \vspace{3mm} & Research since early 2010s & 5G mmWave \\
\hline
Optical Wireless Communications & Complementary access and backhaul for 6G & \vspace{3mm} IR 0.3-400 THz $\&$ VLC 400-750 THz $\&$ UV 0.75-30 PHz \vspace{3mm} & Enormous spectrum $\&$ low-cost & Blocked easily $\&$ reliability issues & Research since 2000s (VLC) $\&$ 1960s (FSO) & Commercial FSO Expermental VLC \\
\hline
\end{tabularx}
\end{center}
\end{table*}

\section{POTENTIAL TECHNOLOGIES FOR 6G} 
\label{Technologies}
Since 6G will be a largely complex and heterogeneous entity, there needs to be a comprehensive set of technological elements upon which such a network can be built. In this section, we identify and review 27 potential 6G technologies in 10 different technological categories. These technologies and categories are introduced in Figure \ref{Fig_Technologies}. For clarity, each technology is introduced using the same template: {vision}, {introduction}, {past and present}, {opportunities and challenges}, {literature and future directions}. In the vision part, we envision briefly what is the expected role of that particular technology in 6G. In the introduction, the description and basic principles of the technology are discussed, along with the main benefits, shortcomings, and application scenarios. In the past and present part, we provide a brief introduction to the background and current status of the technology. Then, we discuss the opportunities that the technology is expected to offer and the main challenges on the road. In the final part, we review recent survey papers in the literature and discuss future research directions. 

\subsection{SPECTRUM-LEVEL TECHNOLOGIES FOR 6G}
To fulfill the demanding performance, service, and application requirements of 6G, it is vital to support a massive frequency range from sub-6 GHz and centimeter wave (cmWave) to mmWave and THz bands. At the cost of smaller system bandwidths, lower carrier frequencies with favorable propagation characteristics are preferable for wider coverage and higher mobility scenarios, such as rural areas, wide-area IoT, and high-mobility systems. At the cost of harsher propagation conditions, higher frequencies with larger bandwidths enable higher capacity and throughput in denser network scenarios, such as hotspot, urban, and sub-urban areas. In this picture, THz communication is seen as a promising technology to achieve extremely high data rates. In addition, optical wireless communication (OWC) is a potential candidate to offer complementary frequency resources for special scenarios, such as short-range indoor access and long-range outdoor backhaul links. These two technologies are summarized in Table \ref{Table_Spectrum} and discussed in detail below. 

\begin{figure}[!htb]
\center{\includegraphics[width=\columnwidth]
{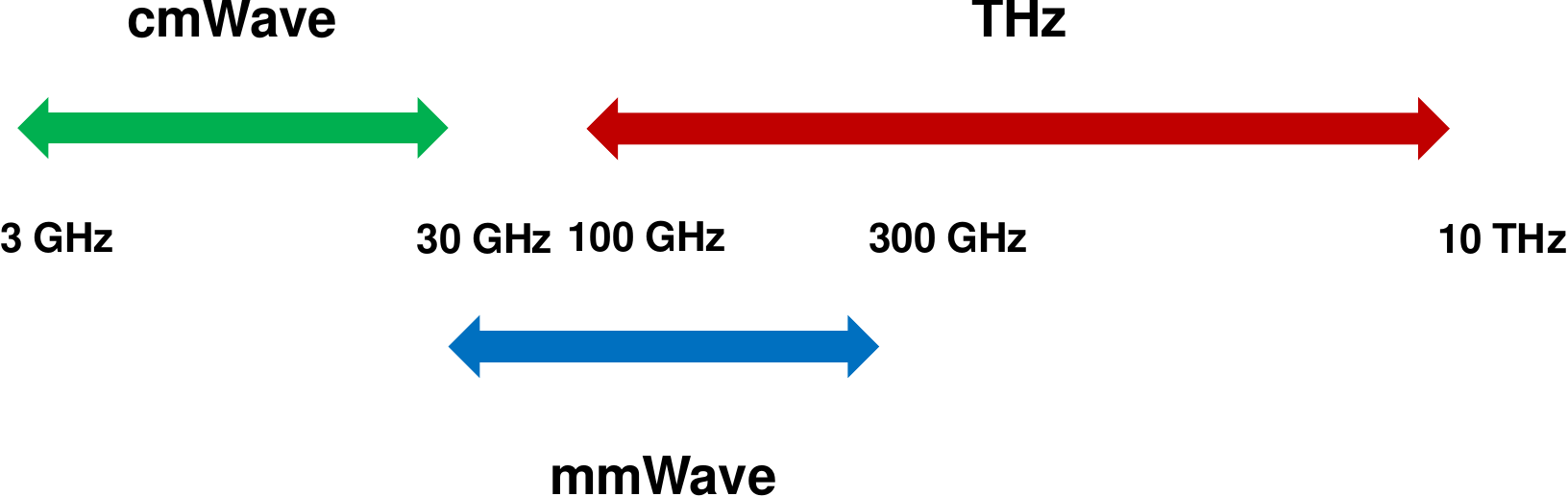}}
\caption{\label{Fig_THz}cmWave, mmWave, and THz spectrum bands.}
\end{figure}

\subsubsection{THZ COMMUNICATIONS}

\begin{itemize}
\item {\textbf{Vision}}:
Due to massive amount of spectrum available at THz frequencies, THz communication has been identified as one of the key enabling technologies to fulfill the extreme capacity requirements of 6G. The THz spectrum is also seen as a key enabler for beyond-communication technologies, such as high-accuracy sensing and localization. 

\item {\textbf{Introduction}}:
THz communication refers to the operation at THz frequencies, ranging from 0.1 to 10 THz \cite{Akyildiz-22}. THz spectrum, in relation to the cm- and mmWave bands, is illustrated in Figure \ref{Fig_THz}. THz communication can achieve extremely high data rates because of the massive spectrum available in the THz band \cite{Akyildiz-22}. However, the communication ranges are short since propagation losses are high at the THz spectrum \cite{Faisal-20}. Due to the aforementioned properties, THz communication is applicable to data-hungry and close-proximity network scenarios. Other potential application scenarios range from nano- to space-scale communications and from sensing to localization \cite{Akyildiz-22}. 
 
\item {\textbf{Past and Present}}:
In the literature, THz research on mobile networks started to gain increasing interest around the mid-2010s \cite{Akyildiz-14}. However, back then, the main focus of higher frequency research was on mmWave communication for 5G. Since 5G, with mmWave operation as a key technology, was launched in the late 2010s, the worldwide research focus was shifted to THz communications for 6G. Currently, academia is extensively studying THz communication from a wide range of perspectives, from theoretical aspects and propagation characteristics to transceiver designs, assisting technologies, and performance evaluations \cite{Akyildiz-22}. In addition, cooperation between academia and industry is narrowing the gap between theory and practice, producing fruitful ideas and leading to novel designs and demonstrations. 

In the industry, the focus is on practical implementations, prototyping, and trialing. For example, Samsung, LG, Nokia, and Keysight Technologies have showcased successful THz communications at frequencies above 100 GHz \cite{RCRWireless}. In 2021, Samsung demonstrated THz communication at a 140 GHz carrier frequency with 2 GHz bandwidth, achieving a data rate of 6.2 Gbit/s over a 50-feet distance. In 2022, LG Electronics performed a trial with a successful THz outdoor link over a distance of 320 meters, operating at a frequency range of 155-175 GHz. In 2023, Nokia, with NTT and NTT Docomo, showcased a data rate of 25 Gbit/s via THz communication at a 144 GHz operating frequency. Also in 2023, Keysight Technologies, along with its partners, obtained over a 100 Gbit/s data rate at a 300 GHz carrier frequency.  

In the spectrum regulatory domain, WRC 2019 defined the conditions for the use of a spectrum ranging from 275 to 450 GHz by land mobile and fixed services \cite{Kurner-20}. As a result, there is currently a spectrum of 160 GHz for THz communications, with no specific conditions to protect Earth exploration satellite services. As discussed in Section \ref{ITU}, at WRC 2023, ITU-R identified five frequency bands above 100 GHz to be further studied for IMT-2030, with potential approval at WRC 2031 \cite{ITU-WRC}. These bands were 102--109.5 GHz, 151.5--164 GHz, 167--174.8 GHz, 209--226 GHz and 252--275 GHz. ITU-R has also examined the feasibility of mobile communications at THz frequencies, concluding that it is feasible to operate at the spectrum above 100 GHz \cite{ITU-M2541}. 
Further spectrum policies for THz communications will be defined in the future WRCs. In the standardization domain, IEEE specified the first standard (IEEE Std. 802.15.3d-2017) for fixed wireless point-to-point THz communications at 252-321 GHz frequencies, with target applications of wireless backhaul/fronthaul, wireless links in data centers, kiosk downloading, and intra-device communications \cite{Petrov-20}. 

Although there are no standards for mobile THz communications, IEEE Std. 802.15.3d-2017 will pave the way toward 6G THz standardization. In 2021, the IEEE Communications Society Radio Communications Committee established a special interest group on THz communications to advance research, development, and standardization activities toward 6G and beyond \cite{Akyildiz-22}. 
In the existing mobile networks, 5G is the first generation that operates at frequencies higher than 6 GHz. Specifically, 5G supports mmWave communications at two frequency ranges, i.e., FR2-1: 24.5-52.6 GHz (Release 15 \cite{Rel-15}) and FR2-2: 52.7-71 GHz (Release 17 \cite{Rel-17}). 6G will continue this trend by going beyond 100 GHz, becoming the first generation to enter THz frequencies. 

\item {\textbf{Opportunities and Challenges}}:
The main goal of THz communication is to provide extremely high data rates and capacities at the link and system levels by exploiting the massive spectrum available at THz frequencies. Unlocking the potential of the THz spectrum opens many opportunities for 6G. In particular, a massive spectrum is the key in achieving one of the main 6G targets of 1 Tbit/s peak rate, which in turn enables new revolutionary applications, such as holographic-type communication and wireless XR \cite{Akyildiz-22}. To provide extreme system-level capacity in hotspot areas, ultra-dense THz cell deployments come into play. In addition to ultra-high data rates, operation at the THz spectrum can provide many other benefits, such as decreased latency, improved PHY layer security, close-proximity connectivity, and beyond-communication services \cite{Akyildiz-22}. 

Due to the larger bandwidths, the frame duration is shorter, leading to decreased latency. Since THz signals propagate relatively short distances, they are more difficult to eavesdrop or attack. THz communication naturally lends itself to close-proximity connectivity due to its short coverage. Due to the very narrow beams, THz ultra-massive MIMO can be used for high-accuracy sensing and localization, enabling services beyond communication. THz communication can be applied in numerous scenarios, such as data-hungry applications (holographic telepresence, wireless XR, information showers), close-proximity communications (D2D, intra-device connectivity), integrated communication, sensing, localization, and imaging (object/gesture recognition, health monitoring, air quality detection, object scanning, dark vision), wireless backhaul/fronthaul (IAB), satellite communications, and nano-scale communications (nano-networks) \cite{Akyildiz-22}. 

There are many challenges in integrating THz communication as a key part of 6G networks and freeing its great potential. The main challenges include high propagation losses (spreading and molecular absorption losses), easy blockages, severe hardware impairments (oscillator phase noise and RF non-linearity), complex hardware design (signal generation and detection), and high power consumption of THz devices \cite{Rappaport-19}. Due to the high molecular absorption loss (especially above 800 GHz), there are large chunks of spectrum in the THz band that may be unsuitable for mobile communications \cite{Rappaport-19}. Other parts of the spectrum, where attenuation is tolerable, are called transmission windows and considered more preferable for 6G purposes \cite{Rappaport-19}. There are massive amounts of spectrum in these windows, ranging from tens to even hundreds of GHz. 

In general, high propagation losses at THz frequencies, especially along with low transmission powers, lead to short communication distances, which is a major drawback in the context of mobile networks \cite{Faisal-20}. This calls for ultra-dense cell deployments. In addition, a potential solution to this so-called distance problem is to use ultra-massive MIMO beamforming to combat harsh propagation conditions by high array gains \cite{Faisal-20}. In theory, ideal beamforming gains can compensate or even overcome free-space pathloss at higher frequencies \cite{Rappaport-15}. In practice, however, accurate beamforming is very difficult at THz frequencies. For very narrow THz beams, even a slight misalignment can significantly degrade link performance. Thus, beam misalignment is a major problem in THz ultra-massive MIMO communications. Cell and device discovery is also challenging. To some extent, the lessons learned from mmWave research can be used in THz studies. For example, lower frequencies with favorable propagation characteristics can be used for initial access. 

A major challenge is how to handle hardware impairments, such as phase noise and power amplifier non-linearity, as they become more severe at higher frequencies. To a certain extent, a wider SCS can alleviate hardware impairments in multi-carrier-based systems. Another solution is to employ single-carrier waveforms for the operation at very high frequencies since they are more robust against phase noise and other hardware impairments \cite{Tervo-22}. In terms of transceiver hardware, signal generation and detection are particularly challenging for {THz communications} \cite{Faisal-20}, and novel solutions are required for cost-, power-, and energy-efficient communications. As THz communication is prone to blockages, a possible solution is to utilize RIS technology, which can reflect the transmitted signals to bypass blockages \cite{Chen-21, Chen-21c}. With controlled reflections, RISs can also be used to alleviate the distance problem by extending the coverage of THz communications \cite{Chen-21, Chen-21c}. 
High power consumption of THz devices can become a limiting factor, highlighting the need for energy-efficient transmission techniques, low-power components, and optimized circuit design. Hybrid beamforming, for instance, offers an excellent balance between power efficiency and performance, making it a potential solution. THz devices also generate a substantial amount of heat, possibly degrading performance and shortening lifespan. This calls for efficient thermal management mechanisms in terms of innovative materials, heat-aware circuit design, and advanced cooling systems. 

\item {\textbf{Literature and Future Directions}}:
Mobile THz communication has been one of the major 6G topics since the late 2010s. The THz literature has covered a broad range of topics, such as device technologies, channel modeling, communication designs, networking aspects, application scenarios, regulation/standardization, and experiments \cite{Akyildiz-22}. The latest THz research has been comprehensively reviewed in numerous studies \cite{Elayan-20, Sarieddeen-20, Faisal-20, Petrov-20, Polese-20, Chen-21c, Sarieddeen-21, Wang-21d, Chen-21d, Chen-21, Ning-23, Chaccour-22, Akyildiz-22, Tan-22, Han-22, Shafie-22, Serghiou-22, Wang-22c, Elbir-22, Hao-22, Farhad-23, Petrov-23, Jiang-23, Cai-24, Han-24, Xue-24}. 

THz communication systems were reviewed in \cite{Elayan-20} from the perspective of signal generation, channel modeling, contender technologies, applications, standardization, and future directions. In \cite{Sarieddeen-20}, the THz band was considered for sensing, imaging, and localization applications, especially in the context of 6G use cases. THz ultra-massive MIMO was comprehensively examined in \cite{Faisal-20} and \cite{Ning-23}. Standardization for sub-THz communications in IEEE 802.15.3d was discussed in \cite{Petrov-20}. 6G THz networks were studied from the full-communication stack perspective in \cite{Polese-20}, addressing link- and system-level challenges. RIS-assisted THz communication was explored for 6G networks in \cite{Chen-21c, Chen-21, Hao-22}. In \cite{Sarieddeen-21}, a tutorial was provided on signal processing techniques for THz communications, with an emphasis on ultra-massive MIMO and RIS systems. 

In \cite{Wang-21d}, key technologies for 6G THz communications were reviewed, considering  channel modeling, multi-beam antennas, front-end chip design, baseband signal processing, and resource management. The authors of \cite{Chen-21d} discussed THz communications for the 6G era in terms of ISAC, ultra-massive MIMO, RISs, and ML. In \cite{Chaccour-22}, seven defining features of THz communications were examined, i.e., quasi-opticality of the THz band, THz wireless architectures, synergy with lower frequency bands, ISAC, PHY layer designs, spectrum access, and real-time network optimization. THz channel characteristics, modeling, and measurements were surveyed for 6G in \cite{Han-22, Serghiou-22, Wang-22c}. 6G THz precoding was discussed in \cite{Tan-22}. In \cite{Shafie-22}, THz communication was explored for 6G and beyond from the aspects of physical, link, and network layers. 

The article in \cite{Akyildiz-22} examined THz research advancements over the last decade and presented future directions for the next decade, covering topics from devices, channels, and communications to networking and experiments. In \cite{Elbir-22}, THz-band ISAC was reviewed. The focus was on the antenna array design, hybrid beamforming, RISs, and ML. The work \cite{Farhad-23} considered AI/ML-assisted THz communications. In \cite{Petrov-23}, the authors focused on near-field communications at THz frequencies for 6G. In \cite{Jiang-23}, a thorough survey was provided on THz communication and sensing toward 6G and beyond. The paper \cite{Cai-24} discussed THz channel propagation, measurements, and modeling for 6G. The authors in \cite{Han-24} reviewed the PHY layer aspects of THz ISAC. THz/mmWave beam management was investigated in \cite{Xue-24}. 
 
There are still many open problems to be addressed in future research toward efficient THz communications and networking in the 6G era. At a high level, the main future topics can be divided into different categories, such as fundamentals, device technology, channel modeling, communication designs, networking, and experiments. As discussed in the aforementioned surveys, the key areas that need further research include overcoming the distance problem (ultra-massive MIMO, RISs, UDNs), close proximity scenarios and applications (D2D, machine-to-machine (M2M), vehicle-to-vehicle (V2V)), suitable THz spectrum bands for 6G, alleviating hardware impairments (multi-waveform design), efficient transceiver design (electronic/photonic transceivers, signal generation/detection, RF design), integrated THz communication and sensing, and practical channel and simulation models (realistic link- and system-level performance evaluation). 

For THz ultra-massive MIMO, the essential research direction is to develop accurate beamforming and practical initial access methods. In the RIS domain, it is important to study coverage extension and blockage avoidance to alleviate the distance and blockage problems. To achieve extreme capacities at the system level, an ultra-dense network design needs to be further studied with extra-small THz cell deployments. More research is required to provide support for a wide range of close-proximity services and applications. In particular, THz-enabled D2D-type connections are essential for the success of IoT, IoV, and the Internet of UAVs (IoU). To meet 6G performance expectations, it is crucial to find and study the most suitable THz frequency bands for mobile networks. 

Flexible multi-waveform design is a promising approach to alleviate hardware impairments by utilizing single-carrier waveform at higher THz frequencies and providing flexible spectrum use at lower frequencies via multi-carrier technology. Major efforts are required for practical THz transceiver design, particularly in terms of electronic vs. photonic components, signal generation and detection, and RF/antenna designs. An important future research avenue is to use THz communication to aid accurate localization and sensing. For realistic link- and system-level performance evaluations, more practical channel and simulation models need to be designed. Further information on the required THz research for the next decade can be found in \cite{Akyildiz-22}. 
\end{itemize}

\begin{figure}[!htb]
\center{\includegraphics[width=\columnwidth]
{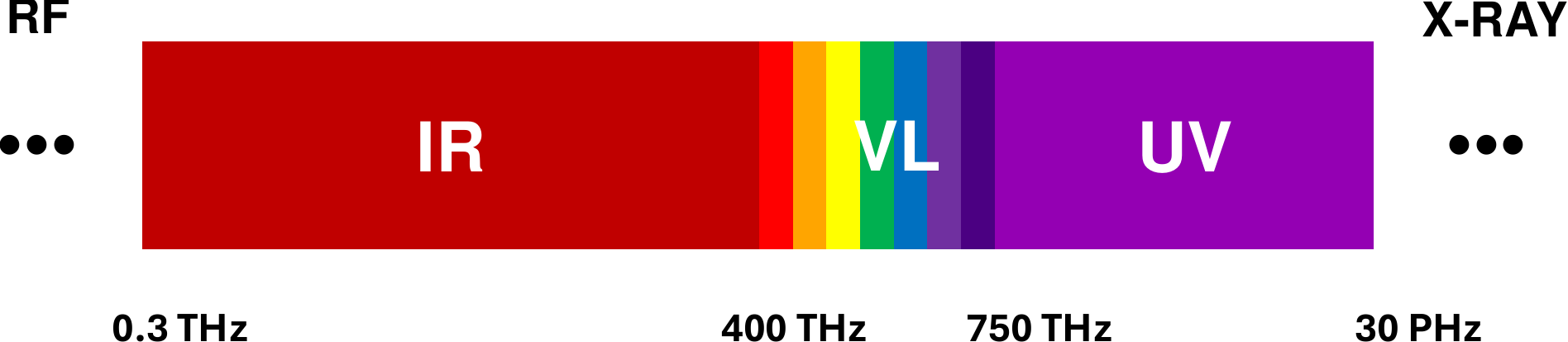}}
\caption{\label{Fig_OWC}Optical wireless spectrum.}
\end{figure}

\subsubsection{OPTICAL WIRELESS COMMUNICATIONS}

\begin{itemize}
\item {\textbf{Vision}}:
OWC is considered as a potential technology to complement 6G networks in special communication scenarios, such as short-range indoor access and long-range outdoor backhauling. 

\item {\textbf{Introduction}}:
OWC refers to the operation at the optical frequency bands, i.e., infrared (IR) (0.3-400 THz), visible light (VL) (400-750 THz), and ultraviolet (UV) (0.75-30 petahertz (PHz)) \cite{Chowdhury-18}. A schematic illustration of optical wireless spectrum is shown in Figure \ref{Fig_OWC}. The main types of OWC are VLC, utilizing the visible light spectrum \cite{Matheus-19}, and {{free-space optical (FSO)}}, usually operating at infrared frequencies \cite{Hamza-19}. VLC is mostly applicable to short-range indoor scenarios, exploiting light-emitting diodes (LEDs) and laser diodes (LDs), whereas FSO is commonly used for short- and long-range point-to-point communications, relying on laser links. Other OWC technologies include light fidelity, light detection and ranging, and optical camera communications \cite{Chowdhury-18}. In the following, the main focus is on VLC and FSO, as they are the most promising candidates to complement 6G networks. 

\item {\textbf{Past and Present}}: 
While fire, smoke, and other ancient types of visible signaling methods can be seen as the ancestors of OWC, the first experiment was conducted in 1880, when Alexander Graham Bell transmitted voice over 213 meters by modulating sunlight using his patented photophone device \cite{Khalighi-14}. In the modern era, the earliest demonstrations of terrestrial FSO date back to the 1960s and 1970s \cite{Khalighi-14}. For example, FSO LED and laser technologies were employed to transmit data over 48 and 14 km long distances in 1962 and 1970, respectively \cite{Khalighi-14}. Multi-Gbit/s data rates were demonstrated in the late 1990s, whereas a record-breaking aggregated rate of 40 Gbit/s was achieved over a distance of 4.4 km in 2000 \cite{Nykolak-00}. 
In 2008, the first Tbit/s-level FSO transmission was demonstrated for a link range of 212 meters \cite{Ciaramella-09}. In 2018, 13.16 Tbit/s data rate was reported for a 10.45 km long FSO link \cite{Dochhan-19}. 

For short-range FSO experiments, i.e., in the order of meters, multi-Tbit/s rates were shown in 2012. A speed of 100 Tbit/s was achieved over a 1 meter link range in a laboratory experiment in 2014. State-of-the-art demonstrations can support multi-Tbit/s rates over multiple kilometers for long-range FSO communication and Petabit/s-level rates over multiple meters for short-range links \cite{Trichili-20}. In practice, commercial FSO systems are capable of transmitting data at tens of Gbit/s rates over link distances of several kilometers \cite{Son-17, Trichili-20}. 
Despite the promising experiments and decades of research, the commercial use of terrestrial FSO systems is still rather limited due to their poor link reliability in practice. State-of-the-art research has been focusing on finding practical solutions to the major challenges, ranging from atmospheric turbulence and attenuation to beam divergence and pointing errors. Further details on the latest FSO technologies and solutions can be found in recent surveys \cite{Trichili-20, Algailani-21, Jahid-22}. 

Early research of VLC dates back to the early 2000s \cite{Komine-04}. However, it was until the 2010s that the real interest in VLC started to rise \cite{Jovicic-13}. Experiments have shown increasing data rates and communication distances over the years, e.g., 1.1 Gbit/s/0.23 m (2012), 3.22 Gbit/s/0.25 m (2013), 4 Gbit/s/0.2 m (2015), 5 Gbit/s/0.75 m (2016), 15.73 Gbit/s/1.6 m (2019), 16.6 Gbit/s/2 m (2021), 24.25 Gbit/s/1.2 m (2021) for LED-based setups \cite{Bian-19, Hu-21c} and 4 Gbit/s/0.15 m (2015), 6 Gbit/s/0.15 m (2017), 20 Gbit/s/1 m (2018), 40 Gbit/s/2 m (2019), and 46.4 Gbit/s/0.3 m (2021) for laser diode (LD)-based setups \cite{Hu-22}. By 2022, the longest transmission distances for the multi-Gbit/s LED and LD VLC systems were 20 and 100 meters, achieving 2.12 and 6 Gbit/s data rates, respectively. The latest details on LED and LD technologies and experiments can be found in \cite{Hu-21c} and \cite{Hu-22}, respectively. 

In 2011, the OWC standard IEEE 802.15.7 was published, including short-range VLC with PHY/MAC layer functionalities. The standard was revised in 2018, expanding the operating frequency range and including new OWC technologies. In 2023, the standard IEEE 802.15.13 was released for multi-Gbit/s OWC, with stationary and mobile devices and link distances up to 200 m. 

\item {\textbf{Opportunities and Challenges}}: 
OWC has great potential due to its enormous unregulated spectrum, large bandwidths, low implementation cost and complexity, high energy efficiency, ultra-low latency, robustness to electromagnetic interference, high spatial reusability, and secure communication channels \cite{Jahid-22}. Due to massive bandwidths, OWC can provide ultra-high-speed communication for indoor and outdoor scenarios, with link distances ranging from nanometers to thousands of kilometers \cite{Chowdhury-18}. License-free spectrum and low implementation costs result in economic benefits. Low energy consumption facilitates the design of green communication systems. Immunity to RF interference allows ubiquitous usage and facilitates the network design. Since OWC does not penetrate walls, its spatial reusability is high. The PHY layer security of OWC links is inherently high due to their locality and/or high directivity. 

Given the aforementioned benefits, OWC is applicable to a wide range of scenarios, such as short-range cellular access, cellular backhaul links, hybrid radio-optical networks, D2D/M2M/V2V communications, optical IoT, positioning and sensing, intra-device connections, on-board communications, short-range indoor communications, long-range outdoor point-to-point communications, space communications, underwater communications, and underground communications \cite{Chowdhury-18, Chi-20, Jahid-22}. 
Different OWC technologies have their own characteristics and the corresponding advantages which differ from one to another. While VLC supports fast, low-cost, and energy-efficient short-range communication in indoor environments, FSO provides ultra-high-speed, highly directive, and low-cost long-range point-to-point outdoor connectivity. Consequently, VLC and FSO are promising technologies to complement 6G networks, with the main application scenarios being short-range indoor access and long-range outdoor backhaul links, as illustrated in Figure \ref{Fig_OWC2}. 

\begin{figure}[!tb]
\center{\includegraphics[width=\columnwidth]
{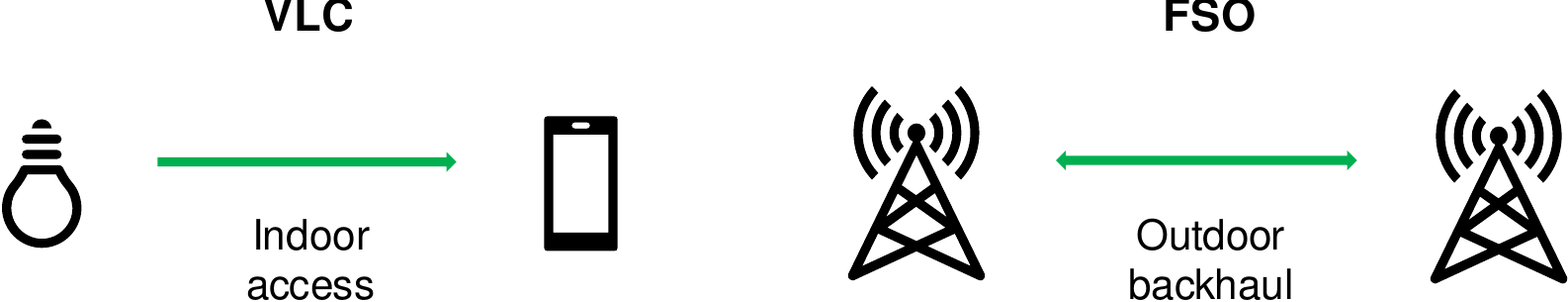}}
\caption{\label{Fig_OWC2}Main applications of visible light communication and free-space optical.}
\end{figure}

There are many challenges to overcome before freeing the potential of OWC and integrating it into future mobile networks as a complementary technology. Each OWC technology has its own set of unique challenges. For VLC, the main fundamental challenges include limited communication range, easy blockage, ambient light interference, line-of-sight (LOS) requirement, random receiver orientation, uplink communication, mobility, and the non-linearity of devices \cite{Jovicic-13, Matheus-19, Chi-20}. Hybrid RF-VLC networks have been recognized as a promising solution to tackle many of these challenges \cite{Chowdhury-20b, Abuella-21, Wang-22d}. In general, hybrid systems consist of two or more networks that can be accessed in different ways based on the network and application types \cite{Chowdhury-20b}. There are many possible access types, such as accessing the strongest network, accessing both networks simultaneously, accessing one network for downlink and the other for uplink, access based on the quality of service (QoS) requirements, and access the highest priority network and the other is backup \cite{Chowdhury-20b}. The design of hybrid networks is highly complicated, with many challenging tasks, such as network selection, network access, handover, load balancing, and resource allocation \cite{Chowdhury-20b}. AI/ML is a promising tool to assist in many of these complex tasks. For example, AI/ML can be applied to resource management and traffic prediction problems. More details on the hybrid RF-VLC systems can be found in \cite{Chowdhury-20b, Abuella-21, Wang-22d}. 

Even though FSO is already commercially available, its usage is rather limited due to poor link reliability. The corresponding challenges range from atmospheric turbulence and attenuation to beam divergence and pointing errors \cite{Jahid-22}. Various weather conditions, such as rain, fog, smog, haze, dust, and snow, cause atmospheric turbulence/attenuation, degrading the link performance and reliability. Since point-to-point FSO links are based on extremely narrow laser beams, beam divergence and pointing errors become a serious problem. Even a slight misalignment between the transmitter and receiver can significantly deteriorate link performance. To obtain high link reliability, the mitigation of the aforementioned impairments is vital. Potential mitigation solutions include pointing, acquisition, and tracking techniques, diversity schemes, and hybrid RF-FSO systems \cite{Jahid-22}. 

Pointing errors can be mitigated by pointing the transmitter toward the receiver to obtain an LOS condition, acquiring the transmitted signal at the receiving end by aligning the receiver in the direction of the beam, and maintaining accurate pointing and acquisition via adaptive tracking with appropriate measurement and feedback mechanisms. Further details on pointing, acquisition, and tracking techniques can be found in \cite{Jahid-22}. Diversity methods aim to improve the link reliability by transmitting multiple copies of the same information. Three main diversity dimensions are time, frequency, and space. At a high level, the objective of hybrid RF-FSO systems is to eliminate common disadvantages while exploiting the benefits of both technologies. As discussed earlier, hybrid networks introduce many new challenges that must be tackled before successful deployment. Further details regarding the challenges of hybrid RF-FSO systems can be found in \cite{Mohsan-23}. 

\item {\textbf{Literature and Future Directions}}:
Recent survey papers have reviewed the fundamentals, latest advances, open problems, and main literature of OWC \cite{Chowdhury-20b, Mohsan-22, Wei-22b, Celik-22, Zhang-23b, Luo-23, Shehata-23, Chow-24}, VLC \cite{Chi-20, Arfaoui-20, Memedi-21, Yu-21b, Abuella-21, Oyewobi-22, Naser-22, Wang-22d, Sejan-23, Bastiaens-24}, and FSO \cite{Trichili-20, Zafar-21, Algailani-21, Jamali-21, Jahid-22, Jeon-23, Mohsan-23, Saiyyed-24}. In \cite{Chowdhury-20b}, the authors surveyed hybrid RF-optical networks. The paper \cite{Mohsan-22} presented a survey on OWC technologies, with the main focus on classification, enabling technologies, link design, mitigation of impairments, security issues, and future challenges. The authors of \cite{Wei-22b} discussed a roadmap for OWC research toward 6G deployments. The work \cite{Celik-22} provided a thorough survey of OWC for four IoT domains: terrestrial, underwater, biomedical, and underground. In \cite{Zhang-23b}, the security of OWC was considered. The survey \cite{Luo-23} reviewed the latest advancements in OWC in the context of 6G and WiFi. In \cite{Shehata-23}, the paper focused on optical wireless and THz communications for 6G. The authors of \cite{Chow-24} discussed the present and future of OWC and FSO toward the 6G era. 

In \cite{Chi-20}, VLC was studied in the context of 6G. VLC PHY layer security was discussed from the information-theoretic and signal processing perspectives in \cite{Arfaoui-20}. Vehicular VLC was explored in \cite{Memedi-21}, reviewing state-of-the-art and identifying open problems. The article \cite{Yu-21b} surveyed the VLC system technology in terms of devices, architectures, and applications. In \cite{Abuella-21}, hybrid RF-VLC systems were discussed from the perspectives of network topology, performance evaluation, potential applications, and future challenges. VLC was reviewed for IoT in \cite{Oyewobi-22}. The work \cite{Naser-22} examined the design of federated learning-enabled VLC for 6G networks. In \cite{Wang-22d}, the authors focused on the access and applications of hybrid VLC-RF systems. A thorough review of MIMO-VLC was provided in \cite{Sejan-23}. The tutorial in \cite{Bastiaens-24} discussed visible light indoor positioning. 

In \cite{Trichili-20}, the authors discussed a research roadmap toward next-generation FSO from the perspective of system design. The work \cite{Zafar-21} studied FSO communication for indoor and outdoor application scenarios. In \cite{Algailani-21}, FSO was reviewed in terms of single- and multi-beam systems, the effect of certain weather conditions, and the scalability of an FSO network. The paper \cite{Jamali-21} focused on intelligent reflecting surface (IRS)-aided FSO communication systems. In \cite{Jahid-22}, a comprehensive survey was given on FSO communication, with the main emphasis on application scenarios, link reliability, mitigation methods, and multi-user systems. The authors in \cite{Jeon-23} examined a field-programmable gate array-based prototype for 6G long-range FSO communication. The study \cite{Mohsan-23} reviewed hybrid FSO-RF systems in terms of switching methods, routing protocols, modulation schemes, research projects, applications, challenges, and potential solutions. In \cite{Saiyyed-24}, a thorough survey of FSO systems was provided. 

In order to realize the potential of OWC, interdisciplinary research efforts and close cooperation between academia, industry, and standardization bodies are needed. Based on the review of the aforementioned survey papers, the future research directions of OWC can be divided into six high-level categories, including fundamentals, channels, systems, emerging technologies, applications, and experiments. The primary research directions in the fundamentals category are device technologies, transceiver designs, and PHY layer techniques. These are the core elements of OWC systems, thus requiring constant development. A key direction for future research is to develop more efficient devices, transceivers, and PHY layer designs, especially in terms of spectrum, energy, and cost. 

The research of OWC channels covers propagation characteristics, channel measurements, and channel modeling. These are the key elements in understanding OWC and providing proper tools for performance evaluation. In the future, it is essential to develop realistic OWC channel models for 6G-specific application scenarios. Further details on the OWC channel modeling can be found in \cite{Anbarasi-17, Alkinani-18, Yahia-21}. In the systems category, OWC is considered from a system-level perspective. While traditional OWC research has mainly focused on individual links and PHY layer techniques, system-level design has gained increasing interest over the past decade. Since OWC aims to complement future mobile networks, system-level design with network management and higher-layer considerations is highly important. In this domain, a particular future direction is to study hybrid RF-optical networks, especially for VLC and FSO. Hybrid RF-optical networks were reviewed in \cite{Chowdhury-20b, Abuella-21, Wang-22d, Mohsan-23}. 

\begin{table*}[htb!]
\begin{center}
\caption{Summary of antenna system technologies for 6G}
\label{Table_Antenna}
\centering
\begin{tabularx}{\textwidth}{| >{\centering\arraybackslash}X | >{\centering\arraybackslash}X |
>{\centering\arraybackslash}X | 
>{\centering\arraybackslash}X |
>{\centering\arraybackslash}X |
>{\centering\arraybackslash}X |
>{\centering\arraybackslash}X |
>{\centering\arraybackslash}X |}
\hline
\centering
\vspace{3mm} \textbf{Antenna System Technologies} \vspace{3mm} & \centering \textbf{Vision} & \centering \textbf{Description} & \centering \textbf{Opportunities} & \centering \textbf{Challenges} & \centering \textbf{Past} & \vspace{1.5mm} \begin{center} \textbf{Present} \end{center} \\
\hline
\vspace{3mm} Ultra-Massive MIMO \vspace{3mm}  & 6G THz umMIMO & mMIMO extension & High BF gains & Cost efficient design & mMIMO concept invented in 2010 & 5G mmWave mMIMO \\
\hline
\vspace{3mm} Reconfigurable Intelligent Surfaces \vspace{3mm}  & Controllable wireless environment & Electromagnetic metasurface reflectors & Energy efficient $\&$ low-cost & Real-time control $\&$ HW design $\&$ ch estimation & Research since early 2010s & RIS prototypes and field-trials \\
\hline
\vspace{3mm} Holographic MIMO \vspace{3mm}  & Beyond mMIMO technology & \vspace{3mm} Electromagnetic metasurface transceivers \vspace{3mm} & High spatial multiplexing gains & Practical implementations & Research since late 2010s & Early experiments \\
\hline
\end{tabularx}
\end{center}
\end{table*}

In the category of emerging technologies, OWC is studied in combination with other emerging technologies, such as AI/ML and IRSs. Since AI/ML will be a key technology in 6G, it is essential to study it for OWC systems as well, especially in the context of hybrid RF-optical networks. With proper design, AI/ML can provide benefits to all communication layers, from PHY to MAC and higher layers. The IRS technology is also expected to be an integral part of 6G, promising diverse benefits from coverage extensions and obstacle avoidance to energy- and cost-efficient deployments. In the context of OWC, the optical IRS technology has been proposed for FSO communication to relax the LOS requirement \cite{Jamali-21}. This is an interesting topic for future research with practical benefits. 

In the applications category, OWC research ranges from nano- to space-scale communications. In particular, 6G-relevant application scenarios for OWC include short-range indoor access, long-range outdoor backhauling, D2D/V2V communications, hybrid RF-optical networks, optical IoT, localization/sensing, and satellite communications. The experiments category consists of demonstrations, trials, and prototyping. The results of these experiments are essential benchmarks in the evolution path of OWC technologies from theory to practice, showcasing the state-of-the-art performance usually in terms of data rates and communication ranges. The latest experimental results of VLC and FSO were reviewed in the previous sections. In this category of research, an important future direction is to perform more practical experiments and customize them for 6G-specific application scenarios. 

For VLC, a particular aim of future research is to tackle its fundamental problems, which range from short communication distances and easy blockages to uplink communication, ambient light interference, and mobility. A promising future research direction is to study AI/ML-aided hybrid RF-VLC networks, as they can alleviate these problems by efficiently switching between RF and VLC access. In particular, hybrid networks need to be considered for 6G-specific communication and application characteristics and scenarios (e.g., human/machine-centric, home/office/factory, public/private, and IoT/IoV/IoU) since different environments have different opportunities and challenges. For example, blockages are more random and less predictable/controllable for human-centric communication in public networks than for machine-centric communication in private networks. Further discussions on the future research directions of hybrid RF-VLC networks can be found in \cite{Abuella-21, Wang-22d}. 

For FSO, the main aim of future research is to improve link reliability by mitigating beam divergence/pointing errors and fighting against atmospheric turbulence/attenuation. A particular research direction is to study advanced mitigation methods, such as pointing, acquisition, and tracking techniques, diversity schemes, and hybrid RF-FSO systems, in the 6G-relevant backhaul scenarios. The latest literature, advancements, and future directions in these domains were discussed in \cite{Jahid-22, Mohsan-23}.  
\end{itemize}

\subsection{ANTENNA SYSTEM TECHNOLOGIES FOR 6G}
In the 6G era, antenna systems are expected to be pushed to their limits in the forms of {ultra-massive MIMO}, RISs, and {holographic MIMO} (HMIMO). Ultra-massive MIMO is a promising technology to enable THz communication in cellular environments by combating severe propagation losses through high beamforming gains. An RIS is a revolutionary technology to control a wireless environment in a desirable way. HMIMO has the potential to provide very high spatial multiplexing gains due to efficient utilization of spatial dimension. These three technologies are reviewed below and summarized in Table \ref{Table_Antenna}. 

\begin{figure}[!htb]
\center{\includegraphics[width=0.9\columnwidth]
{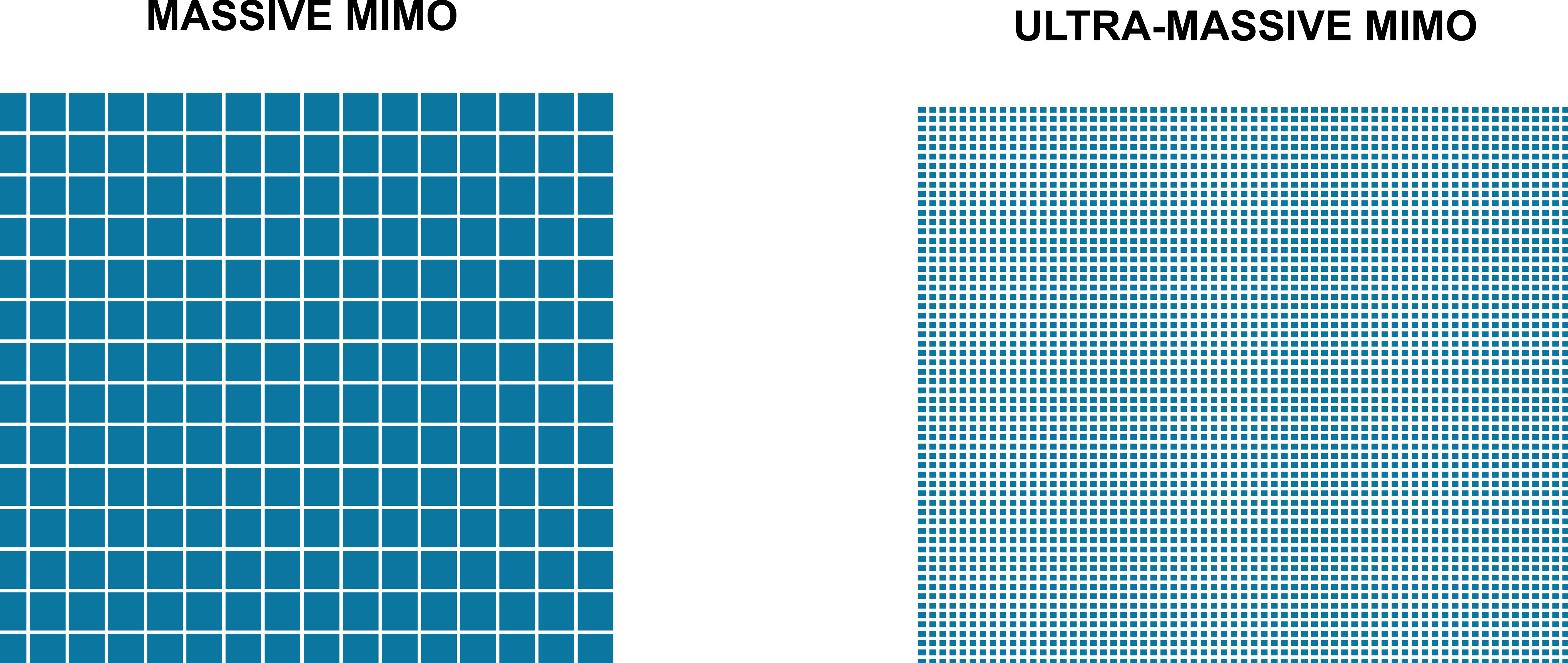}}
\caption{\label{Fig_umMIMO}From massive MIMO to ultra-massive MIMO.}
\end{figure}

\subsubsection{ULTRA-MASSIVE MIMO} 

\begin{itemize}
\item {\textbf{Vision}}: 
Ultra-massive MIMO is expected to become the core antenna system technology in 6G, enabling THz communications with increased capacity and coverage. Moreover, very narrow beams of THz ultra-massive MIMO may be used for high-accuracy sensing and localization, while expanding the capabilities of 6G networks beyond communication. 

\item {\textbf{Introduction}}: 
Ultra-massive MIMO (also known as extremely large-scale MIMO (XL-MIMO)) can be seen as an extension of massive MIMO by employing even a larger number of antennas in practice \cite{Faisal-20}. Whereas practical implementations of massive MIMO employ tens to hundreds of antennas, ultra-massive MIMO will go from hundreds to thousands, as illustrated in Figure \ref{Fig_umMIMO}. Due to the huge number of antennas, very narrow beams can be formed, leading to extreme beamforming gains. Thus, ultra-massive MIMO is considered a promising technology to enable THz communications with extended coverage by combating harsh propagation conditions with high beamforming gains \cite{Faisal-20}. In addition, increased spectral efficiency can be achieved through multi-stream transmissions with high spatial multiplexing gains \cite{Faisal-20}. Due to the miniature size of THz antennas, the physical size of antenna arrays with an ultra-massive number of elements remains feasible in practice \cite{Akyildiz-16}. 

\item {\textbf{Past and Present}}: 
The theoretical concept of massive MIMO was developed in 2010 for a multi-user multiple-input single-output (MISO) system with the number of antennas approaching infinity \cite{Marzetta-10}. Since then, massive MIMO has been widely studied in the literature \cite{Bjornson-19}. The first practical implementations of massive MIMO were witnessed in the early phase of 5G in the late 2010s, relying on antenna arrays with 64 dual-polarized elements and fully digital transceiver chains \cite{Bjornson-19}. Since then, the number of antennas in practical implementations has been steadily increasing, reaching already several hundreds \cite{Wang-23c}. In the literature, the first studies on ultra-massive MIMO were conducted in the mid-2010s \cite{Akyildiz-16}. Currently, ultra-massive MIMO is considered as an extension of 5G mmWave massive MIMO to 6G THz frequencies, with potential applications in diverse scenarios, ranging from ultra-broadband cellular access to wireless backhauling, sensing, and localization \cite{Faisal-20}. 

\item {\textbf{Opportunities and Challenges}}: 
Ultra-massive MIMO has great potential to become an enabling technology for THz communications with increased data rates and extended coverage in 6G networks \cite{Faisal-20}. Furthermore, high spectral efficiency is enabled via spatial multiplexing using multi-stream transmissions, especially for multi-user scenarios. To further extend the coverage and avoid blockages, THz ultra-massive MIMO beamforming can be jointly designed with the RIS technology. Due to very narrow beams, THz ultra-massive MIMO can also be used to obtain high-accuracy localization and sensing in 6G networks \cite{Faisal-20}. 
 
However, there are many fundamental challenges to overcome before THz ultra-massive MIMO systems can be realized in practice. The cost-, energy-, and spectrum-efficient design of practical ultra-massive MIMO technology is challenging, particularly in terms of RF, antenna, and beamforming designs. As the number of antennas increases, hardware design becomes more costly, power consuming, and complex. The choice of beamforming architecture plays a major role in this picture. There are three main beamforming architectures: digital, analog, and hybrid \cite{Heath-16}. Digital beamforming is costly for large antenna arrays since it requires one RF chain per antenna. Analog beamforming is a significantly cheaper option. However, it cannot support multi-stream transmission. Hybrid beamforming provides a compromise between performance and cost, supporting multi-stream communication with a reduced number of RF chains. Hybrid beamforming is a viable solution for such antenna settings where digital beamforming is too expensive. 

As the beams become narrower, the initial access, channel estimation, accurate beamforming, and mobility support become more challenging \cite{Faisal-20}. Cell and device discovery is particularly challenging at the cell-edge region since very narrow beams are required for the distance, while making it more difficult to discover the target and establish a connection. Enlarged channel dimensions lead to the high complexity of channel estimation. For sharp beams, even a slight misalignment may significantly degrade the performance of the communication link. Beam misalignment becomes more severe in mobile scenarios. Further information on the opportunities and challenges can be found in \cite{Faisal-20, Bjornson-24}. 

\item {\textbf{Literature and Future Directions}}: 
While massive MIMO theory has been well studied in the literature, the main research focus of ultra-massive MIMO is on practical designs/implementations at THz frequencies to overcome fundamental challenges, provide extreme performance, and expand the capabilities of 6G networks. Recent survey papers have reviewed the latest literature, advancements, and open problems \cite{Faisal-20, Shlezinger-21, Ning-23, Cui-23, Huo-23, Wang-23e, Wang-24, Bjornson-24, Lu-24b}. 
In \cite{Faisal-20}, THz ultra-massive MIMO was reviewed in terms of transceiver design, channel modeling, research advancements, and future directions. In \cite{Shlezinger-21}, the authors studied active dynamic metasurface antennas for transmission and reception in 6G extreme massive MIMO systems. The focus was on hardware architectures, transceiver design, implementation challenges, and open research problems. Recommended future directions for metasurface-based beamforming included frequency-selective beamforming, channel estimation, hybrid passive/active metasurfaces, use cases, and experiments. 

The article \cite{Ning-23} focused on beamforming technologies for THz ultra-massive MIMO. The covered topics included transceiver architectures, beamforming design principles, wideband beamforming, RIS-assisted beamforming, existing array technologies, and emerging applications. In \cite{Cui-23}, the authors reviewed recent massive MIMO trends toward 6G in terms of localization and sensing, AI, and non-terrestrial communications. The work \cite{Huo-23} explored near-field communications for 6G using extremely large-scale antenna arrays. In \cite{Wang-23e}, extreme MIMO was reviewed, with a focus on the hardware design. The article \cite{Wang-24} provided a comprehensive survey on XL-MIMO for 6G, discussing hardware architectures, signal processing, and channel modeling. In \cite{Bjornson-24}, the authors explored 6G ultra-massive MIMO in terms of spatial multiplexing, degrees of freedom, electromagnetic characteristics, and signal processing. The authors of \cite{Lu-24b} discussed 6G XL-MIMO in the context of near-field communications. 

At a high level, the main focus of future research needs to be on THz ultra-massive MIMO for diverse 6G applications. In this domain, there are many fundamental problems that still need to be tackled and promising application scenarios to be further explored. Critical open challenges toward practical THz ultra-massive MIMO systems include the initial access, beam misalignment, and mobility. Further research and development work is needed to develop more aware, accurate, and adaptive beamforming systems, calling for more efficient beam searching, training, and tracking methods. Channel estimation is also a highly complex task, requiring more practical pilot designs and more efficient estimation methods, potentially based on AI/ML. The essential future directions to further expand the capabilities and applicability of THz ultra-massive MIMO are RIS-assisted communications, integrated wireless backhaul and access, cellular UAV communications, vehicular connectivity, and joint communication, sensing, and localization. 
\end{itemize}

\begin{figure}[!htb]
\center{\includegraphics[width=0.6\columnwidth]
{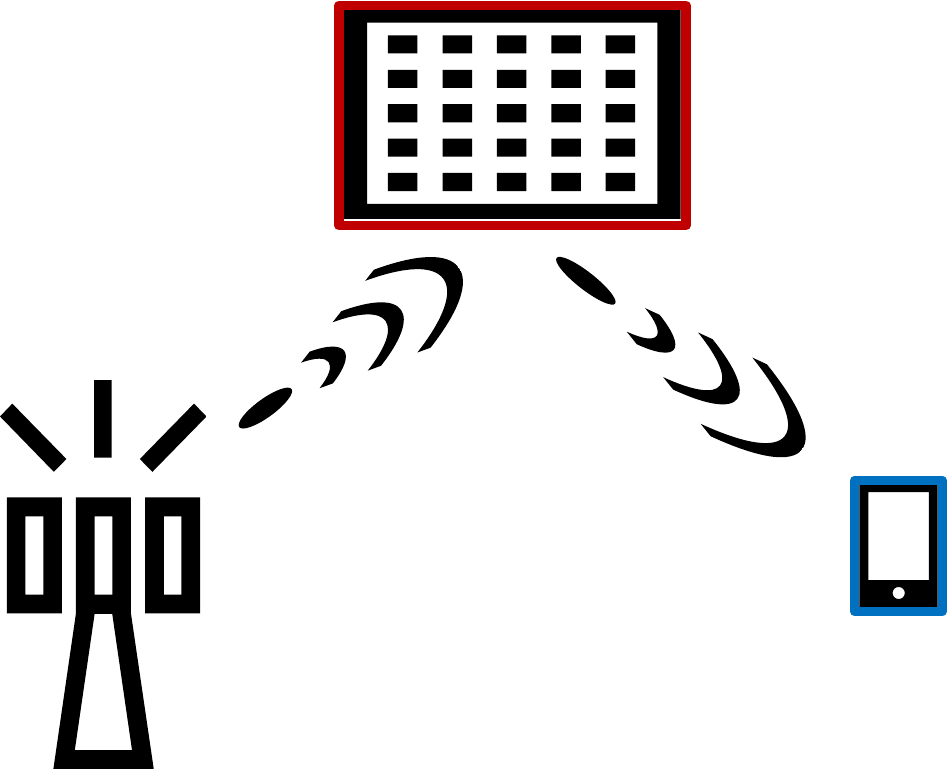}}
\caption{\label{Fig_RIS}RIS-assisted communications.}
\end{figure}

\subsubsection{RECONFIGURABLE INTELLIGENT SURFACES} 

\begin{itemize}
\item {\textbf{Vision}}: 
Energy-efficient, low-cost, and easily deployable RISs are expected to become a mainstream 6G technology, providing diverse benefits from improved signal quality to coverage extensions and blockage avoidance. Ultimately, RISs have the potential to revolutionize mobile communication by turning an unpredictable and destructive wireless propagation environment into a programmable smart entity that can be favorably controlled. 

\item {\textbf{Introduction}}: 
The fundamental idea of RISs is to shape the wireless environment in a favorable manner by reflecting radio signals via passive beamforming toward the desired receivers \cite{Basar-19, Rajatheva-21}. Specifically, an RIS consists of a high number of sub-wavelength size reconfigurable electromagnetic elements that are electronically controlled to adjust the phase of the incoming signal to direct and shape the reflected beam in a way such that the signal quality is improved at the receiver \cite{Basar-19}. RISs can be used to strengthen signal quality, mitigate interference, expand coverage, overcome difficult propagation conditions, and even control signal polarization and channel rank \cite{Basar-19, Rajatheva-21}. A schematic illustration of RIS-assisted communication is shown in Figure \ref{Fig_RIS}. 

Without the need for complex decoding/encoding processes and power-consuming RF operations, the passive type of RIS is a cost- and energy-efficient technology \cite{Basar-19}. Another attractive feature of passive RISs is their natural support for full-duplex operation without typical antenna noise amplification and self-interference effects \cite{Basar-19}. Typical implementations of RISs include traditional reflect-arrays, liquid-crystal surfaces, and software-defined metasurfaces \cite{Basar-19}. Metasurfaces are particularly promising for advanced implementations due to their unique electromagnetic properties. RISs are relatively easy to deploy since they are lightweight and can be installed on flat surfaces (e.g., walls and ceilings) \cite{Pan-21}. 

In addition to typical passive RISs, active and hybrid RIS concepts also exist. The concept of active RIS refers to the capability to amplify reflected signals to overcome the so-called double fading effect of a typical RIS environment \cite{Zhang-23}. Compared with passive RISs, active RISs can offer improved performance at the cost of increased power consumption \cite{Zhang-23}. While active RISs do not necessarily require active RF components, there exist hybrid architectures that rely on the combinations of passive RIS elements and some additional active RF chains \cite{Zhang-23}. Among these three different RIS concepts, there is a performance-energy/complexity/cost trade-off to be tackled to find preferable implementations for different application scenarios. 

While the aforementioned passive, active, and hybrid RIS concepts operate in the relay mode, RISs can operate in the transmitter mode as well \cite{Basar-21}. In the transmitter mode, a passive RIS is used as part of an access point (AP) by adjusting the phase and amplitude of the incoming unmodulated carrier signal from the nearby RF source to transfer information by creating virtual amplitude-phase modulation constellations over-the-air \cite{Basar-19b}. The transmitter-mode RIS allows a rather simple RF chain-free hardware architecture, facilitating cost- and energy-efficient design. While the coverage of reflective-only RISs is limited to the front side of the surface, a concept called transmissive-reflective RIS aims to achieve 360 degrees coverage by allowing transmissive communication through the RIS to cover the backside of the surface as well \cite{Xu-21b}. 

\item {\textbf{Past and Present}}: 
Since the early 2010s, the concept of RIS has been studied under many different names, such as reconfigurable reflectarrays, programmable metasurfaces, software-controlled metasurfaces, and IRS. The term RIS was introduced in the late 2010s. At present, RIS and IRS are the most popular names, with RIS being dominant. The RIS concept was extended from passive to active (without RF chains) and hybrid versions in the early 2020s \cite{Zhang-23}. Generally, many aspects of RIS technology have been explored, ranging from fundamentals and algorithmic designs to prototyping and experiments \cite{Renzo-20, Liu-22}. Recently, a special focus of research has been directed toward RIS-assisted 6G networks, with a wide range of application scenarios \cite{Pan-21, Chen-21c, Zhu-22, Basharat-22, Chen-22}. 

RISs have also evoked enthusiasm in the industry due to their low cost and energy efficiency. Many major technology companies are currently developing various types of RIS solutions. Recently, prototypes and trials have been introduced \cite{Liu-22}. In 2018, NTT DOCOMO demonstrated a metasurface reflect-array-aided 5G communication system in the 28 GHz frequency band, obtaining a data rate of 560 Mbit/s in comparison to 60 Mbit/s with no reflector assistance \cite{Liu-22}. In 2020 and 2021, NTT DOCOMO conducted further trials on transparent dynamic metasurfaces using the same 5G frequency band \cite{Liu-22}. In 2022, LG showcased four different RIS prototypes, i.e., PIN diode RIS, liquid crystal RIS, transparent planar RIS, and transparent flexible RIS \cite{RCRWireless}. Also in 2022, Japanese Kyocera introduced a transmissive metasurface to expand coverage and avoid obstacles in 5G/6G networks \cite{RCRWireless}. 

Although RIS technology has received significant worldwide attention in academia and industry, standards exist only at regional levels \cite{Liu-22}. However, coordinated efforts toward common understanding and guidelines have been taken in terms of large-scale projects and new interest and specification groups organized by international standardization bodies. For example, the European Horizon 2020 program has recently funded many RIS-dedicated projects \cite{Liu-22}. Moreover, the IEEE Communications Society has established two interest groups and an emerging technology initiative focusing on RISs to promote multidisciplinary research and international collaborations, including academic and industrial players \cite{Liu-22}. In 2021, ETSI founded an industry specification group on RISs, with a wide range of partners from vendors and operators to research institutes and universities \cite{Liu-22}. The expected outcome consists of white papers, technical reports, and proof-of-concepts. Further details on the industrial and standardization aspects of RISs can be found in \cite{Liu-22}. 

\item {\textbf{Opportunities and Challenges}}: 
The main promise of RISs is that they can turn an uncontrollable and hostile wireless propagation environment into a programmable entity that can be shaped in a favorable way. This can be achieved with an energy-efficient, low-cost, and easily deployable RIS technology operating in a (nearly) passive relay-type mode. RISs can provide diverse benefits, such as high beamforming gain, improved received signal quality, extended coverage, and obstacle avoidance. There are many opportunities to exploit RISs. In particular, RISs can be used to enhance a broad range of emerging 6G technologies \cite{Pan-21, Liu-21b}. Ultimately, RISs have the potential to become a mainstream technology and revolutionize the manner in which mobile networks are designed in the 6G era. 

There are many challenges to overcome before large-scale deployment in mobile networks is possible. First of all, there is a lack of comprehensive theoretical models and performance limits, as well as realistic channel models \cite{Basar-19}. Furthermore, the main practical problems exist in hardware design, real-time control, channel estimation, channel state information (CSI) acquisition, mobility management, and resource allocation \cite{Rajatheva-21, Pan-21}. The current hardware solutions have limited controllability. Since passive RISs cannot receive or send pilot signals, channel estimation, CSI acquisition, real-time control, and mobility management become particularly problematic \cite{Bjornson-20b, Pan-21}. Adding an RIS to a wireless environment further complicates the resource allocation problems. 

\item {\textbf{Literature and Future Directions}}: 
Since the late 2010s, RIS technology has been one of the major 6G topics in the literature. Over the years, a vast variety of research and development work has been conducted, from theory to practice. However, there still exist unresolved issues remain to be addressed. Further efforts are needed to find practical solutions for hardware design, real-time control, channel estimation, CSI acquisition, and mobility management. Another vital future direction is to study RISs in the context of 6G, considering its special characteristics and potential application scenarios. In particular, RISs can be used to enhance emerging 6G technologies, such as THz communications, massive MIMO, cell-free massive MIMO, UAV communications, NOMA, D2D, backscattering, wireless power transfer (WPT), PHY layer security, sensing, and localization. 

Recently, numerous survey articles on RISs have been published \cite{Basar-19, Elmossallamy-20, Wu-20, Renzo-20, Dai-20, Gong-20, Alghamdi-20, Bjornson-20b, Yuan-21, Pan-21, Liu-21b, Chen-21c, Zhu-22, Liu-22, Faisal-22, Jian-22, Basharat-22, Zhang-22b, Chen-22, Rana-23, Naeem-23}. These can be divided into three high-level categories: generic \cite{Basar-19, Elmossallamy-20, Wu-20, Renzo-20, Gong-20, Yuan-21, Liu-21b, Jian-22}, specific \cite{Dai-20, Bjornson-20b, Liu-22, Faisal-22, Zhang-22b, Rana-23, Naeem-23}, and 6G-oriented \cite{Alghamdi-20, Pan-21, Chen-21c, Zhu-22, Basharat-22, Chen-22}. While the generic surveys reviewed RISs from a wide perspective, the specific surveys focused on a particular narrow aspect of RISs. The generic papers covered operating principles, state-of-the-art technologies, recent research advances, opportunities, applications, and open challenges \cite{Basar-19, Elmossallamy-20, Wu-20, Renzo-20, Gong-20, Yuan-21, Liu-21b, Jian-22}. The specific surveys explored prototyping and experiments \cite{Dai-20}, myths and critical questions \cite{Bjornson-20b}, industrial perspective \cite{Liu-22}, ML approaches \cite{Faisal-22}, sensing/localization \cite{Zhang-22b}, hardware \cite{Rana-23}, air-to-ground communications \cite{Gu-24}, and security/privacy \cite{Naeem-23}. The 6G-oriented survey papers examined RISs for 6G from the perspective of generic overviews \cite{Pan-21, Basharat-22}, performance optimization \cite{Alghamdi-20}, THz communications \cite{Chen-21c, Alexandropoulos-24}, vehicular communications \cite{Zhu-22}, positioning in IoT \cite{Chen-22}, near-field communications \cite{Mu-24}, PHY layer security \cite{Kaur-24}, and NTNs \cite{Khan-24}. 
\end{itemize}

\begin{figure}[!htb]
\center{\includegraphics[width=0.4\columnwidth]
{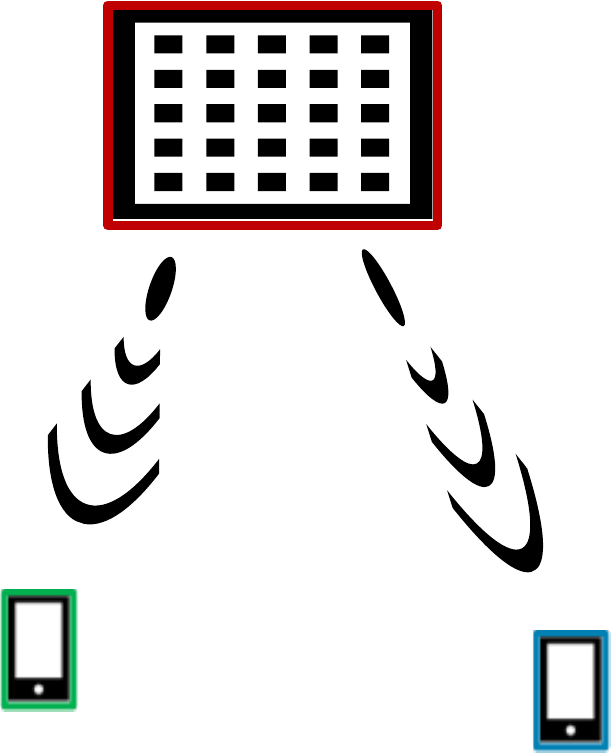}}
\caption{\label{Fig_HMIMO}An active holographic MIMO surface.}
\end{figure}

\subsubsection{HOLOGRAPHIC MIMO}
\begin{itemize}
\item {\textbf{Vision}}: 
HMIMO is seen as a promising beyond massive MIMO technology for 6G networks to obtain very high spatial multiplexing gains via transceiver-based active HMIMO or controllable wireless environments via low-power/cost passive HMIMO reflectors (i.e., RISs). 

\item {\textbf{Introduction}}: 
HMIMO aims to go beyond massive MIMO by transmitting, receiving, or reflecting communication signals using electromagnetically-driven surfaces \cite{Bjornson-19, Huang-20}, known as HMIMO surfaces (HMIMOS) \cite{Huang-20}. HMIMOS can be classified into active/passive and contiguous/discrete categories \cite{Huang-20}. The difference between active and passive is that an active HMIMOS is used as a transceiver with sophisticated RF and signal processing capabilities, whereas a passive HMIMOS acts as a reflector with low-cost and low-power passive elements. An active HMIMOS is also known as a large intelligent surface (LIS), whereas a passive HMIMOS is widely known as an RIS or IRS \cite{Huang-20}. A schematic illustration of an active HMIMOS is presented in Figure \ref{Fig_HMIMO}. 

The idea behind a contiguous HMIMOS is to create a contiguous aperture by incorporating an (virtually) uncountable number of elements into a two-dimensional surface of a finite size \cite{Huang-20}. A discrete HMIMOS relies on a discrete aperture that typically consists of a large number of software-controllable unit cells made of metamaterials \cite{Huang-20}. Based on the aforementioned categorization, the four operation modes of HMIMOS are defined as contiguous active transceiver HMIMOS, discrete active transceiver HMIMOS, discrete passive reflector HMIMOS, and contiguous passive reflector HMIMOS \cite{Huang-20}. 

Among these modes, the typical ones are contiguous active transceiver HMIMOS and discrete passive reflector HMIMOS \cite{Huang-20}. Since the former mode approximates a virtually infinite number of elements, it implies that it could potentially achieve the fundamental limits of massive MIMO and enable extremely high spatial multiplexing gains. Furthermore, the implementations of active HMIMOS are more energy- and cost-efficient than massive MIMO technology. The most attractive features of the latter mode are software-controlled reflections, energy/cost-efficient implementation, and flexible deployment. In this section, the main focus is on the active HMIMOS since the passive ones (i.e., RISs) were already reviewed in the previous section.  

\item {\textbf{Past and Present}}: 
In the literature, the first works on active HMIMOS were published in 2017, under the names of holographic RF systems, holographic beamforming, and LISs \cite{Huang-20}. Active HMIMOS began to gain more interest at the beginning of the 2020s. The main focus of HMIMO research is on passive HMIMOS (i.e., RISs) due to its energy/cost-efficient implementation and relatively easy deployment. 
In current mobile networks, 5G supports codebook-based massive MIMO with limited resolution in the azimuth and elevation angles \cite{Rel-15, Li-20}. This beam-space approach is highly sub-optimal, obtaining a performance far from the theoretical limits of massive MIMO \cite{Rajatheva-21}. Active HMIMOS is gaining an increasing amount of interest toward 6G, as it is seen as a potential solution to this need. 

\begin{table*}[htb!]
\begin{center}
\caption{Summary of transmission scheme technologies for 6G}
\label{Table_Transmission}
\centering
\begin{tabularx}{\textwidth}{| >{\centering\arraybackslash}X | >{\centering\arraybackslash}X |
>{\centering\arraybackslash}X | 
>{\centering\arraybackslash}X |
>{\centering\arraybackslash}X |
>{\centering\arraybackslash}X |
>{\centering\arraybackslash}X |
>{\centering\arraybackslash}X |}
\hline
\centering
\vspace{3mm} \textbf{Transmission Scheme Technologies} \vspace{3mm} & \centering \textbf{Vision} & \centering \textbf{Description} & \centering \textbf{Opportunities} & \centering \textbf{Challenges} & \centering \textbf{Past} & \vspace{1.5mm} \begin{center} \textbf{Present} \end{center} \\
\hline
\vspace{3mm} Multi-Waveform Scheme \vspace{3mm}  & Ultra-flexible waveform scheme & More than one waveform & Supports massive spectrum range & Finding suitable waveforms & 4G CP-OFDM and DFT-s-OFDM & 5G CP-OFDM and DFT-s-OFDM \\
\hline
\vspace{3mm} Advanced Modulation and Coding Methods \vspace{3mm}  & High-performance modulaton/coding & Bits into symbols $\&$ error correction & High throughput $\&$ high reliability & High-order mod $\&$ coding for URLLC & 4G Turbo/conv $\&$ up to 256-QAM & 5G LDPC/polar $\&$ up to 1024-QAM \\
\hline
\vspace{3mm} Non-Orthogonal Multiple Access \vspace{3mm}  & Next-generation multiple access & User separation in power or code & More efficient than orthogonal & Receiver complexity & 5G study item & Under study for 6G \\
\hline
\vspace{3mm} Grant-Free Medium Access \vspace{3mm}  & Massive access & Network access wo grant from BS & Fast/efficient network access & Preamble collisions & 4G four-step random access & 5G two-step random access \\
\hline
\end{tabularx}
\end{center}
\end{table*}

\item {\textbf{Opportunities and Challenges}}: 
The main opportunity that active HMIMOS offers is to potentially obtain extremely high spatial multiplexing gain and performance close to the theoretical limits of massive MIMO in practice, while being more energy- and cost-efficient than massive MIMO technology. On this road, diverse issues must be addressed. These open challenges are primarily related to theoretical limits, channel modeling, algorithm design, practical implementations, and experiments \cite{Gong-22}. 

A comprehensive characterization of the theoretical limits with a proper electromagnetic domain analysis is still to be defined, providing performance upper bounds for practical algorithms. Channel models must be updated by considering the special properties of HMIMOS, such as LOS and non-line-of-sight (NLOS) near-field propagation. Channel estimation becomes difficult due to the extremely high number of antenna elements. HMIMOS also sets major challenges for beamforming and beam focusing designs, calling for efficient low-complexity algorithms to enable practical systems. In practical implementations, many challenges arise, e.g., identification and compensation of hardware impairments, handling of mutual coupling between densely located antenna elements, and the support of numerous beams for high spatial multiplexing gains. In order to verify the performance of practical implementations, experiments need to be executed in realistic mobile network scenarios and setups. Further details of the main challenges can be found in \cite{Gong-22}. 

\item {\textbf{Literature and Future Directions}}: 
Since the late 2010s, many aspects of active HMIMOS have been researched. The main topics included fundamental theory, channel modeling, algorithm design, practical implementation, experiments, and sensing/localization. Since HMIMOS is a new research area, all the aforementioned directions need major research and development efforts in the near future to attain an adequate level of maturity for practical implementation. 

The fundamental theory of active HMIMOS is largely lacking, requiring a comprehensive characterization of the theoretical performance limits. Efficient algorithms need to be designed to approach these limits in practice. Since contiguous HMIMOS is considered as a whole, conventional far-field channel models may not be valid in certain cases \cite{Dardari-21}. Specifically, the size of the surface is much larger than the wavelength and is somewhat comparable to the link distances. Thus, near-field models must be carefully studied. Radio waves behave very differently in the near-field than in the far-field. Although this brings new challenges for the design of HMIMOS systems, it also opens up novel opportunities. For example, the directive nature of the near-field propagation environment may lead to improved spatial multiplexing gains through efficient beam focusing methods. Near-field HMIMOS operation may also open new possibilities for WET. 

In particular, important future directions include practical implementations, prototyping, experiments, and field-trials. This work is ongoing in academia and industry. However, further efforts are needed to prove the feasibility of HMIMOS in practice. Moreover, the special characteristics of 6G networks must be considered, such as mmWave/THz operation, stringent performance requirements, and a heterogeneous network architecture. 
Recently, a handful of survey papers have been published on HMIMOS \cite{Huang-20, Dardari-21, Gong-22, An-23, An-23b, An-23c, Gong-24, Wei-24}. In \cite{Huang-20}, HMIMOS was reviewed from the 6G perspective. A broad range of topics was covered, such as hardware architectures, fabrication methodologies, operation modes, functionality types, characteristics, communication applications, design challenges/opportunities, and case studies. 

The survey \cite{Dardari-21} explored LIS technology in terms of information-theoretical limits, communication modes, power-scaling law, and future research directions. 
In \cite{Gong-22}, a comprehensive survey was given on HMIMOS, reviewing physical aspects, theoretical foundations, enabling technologies, extensions, and open problems. In \cite{An-23, An-23b, An-23c}, the authors provided a three-part tutorial on HMIMO, covering channel modeling and estimation in part 1, performance analysis and beamforming aspects in part 2, and opportunities and challenges in part 3. The paper \cite{Gong-24} discussed channel modeling for HMIMO communications in the near-field domain. In \cite{Wei-24}, the authors introduced electromagnetic information theory for HMIMO communications.  
\end{itemize}

\subsection{TRANSMISSION SCHEME TECHNOLOGIES FOR 6G}
The role of a well-designed transmission scheme is crucial for 6G since it determines how efficiently the available spectral resources can be used. In other words, the 6G transmission scheme plays a key role in turning a massive spectrum range into extreme performance to support stringent requirements and a wide range of demanding usage scenarios. To this end, 6G requires an ultra-flexible transmission scheme with novel technological elements. Some potential technologies include highly tunable {multi-waveform scheme}, {advanced modulation and coding methods}, efficient NOMA, and fast {grant-free medium access}, as summarized in Table \ref{Table_Transmission}. In the following, we provide a detailed discussion on each of these technologies. 

\subsubsection{MULTI-WAVEFORM SCHEME}

\begin{itemize}
\item {\textbf{Vision}}:
A flexible multi-waveform scheme is expected to play a crucial role in the 6G air interface by supporting efficient operation in a massive spectrum range from sub-6 GHz to THz frequencies. 

\item {\textbf{Introduction}}:
At a high level, the purpose of a waveform is to define a transmission framework to carry modulated information symbols from the transmitter to the receiver over a wireless communication channel. A flexible multi-waveform scheme refers to an approach in which more than one waveform, with flexible features, is employed in a communication system. Multi-numerology is a key element to achieve flexibility for a given waveform by using multiple sets of parameters to choose from, depending on the prevailing conditions. Specifically, different sets of parameters are used for different transmission settings and channel conditions, leading to improved performance and enabling efficient communication in diverse wireless environments. 

\item {\textbf{Past and Present}}:
The earliest realization of a multi-waveform scheme is from 4G LTE, where different waveforms are used for downlink (cyclic prefix (CP)-OFDM) and uplink (discrete Fourier transform-spread-OFDM (DFT-s-OFDM)), and two CP lengths exist (i.e., normal and extended) \cite{Ankarali-17}. Spectral efficiency is preferred in downlink, whereas power efficiency in uplink. An extended CP is used for worse channel conditions, with the aim of eliminating inter-symbol interference. CP-OFDM was adopted in 4G due to its robust nature toward multipath fading as well as straightforward support for low-complexity receivers and MIMO communications. While CP-OFDM has many advantages, there are also apparent disadvantages, such as reduced spectral efficiency due to CP and orthogonality, poor power efficiency induced by a high peak-to-average-power ratio (PAPR), and sensitivity to hardware impairments and mobility. 

Due to the shortcomings of CP-OFDM, many other waveforms were studied for 5G in the literature \cite{Schaich-14, Guan-17}. The main ones included filter bank multi-carrier (FBMC), universal filtered multi-carrier (UFMC), and generalized frequency-division multiplexing (GFDM) \cite{Schaich-14}. In addition, different types of OFDM variants were proposed, such as filtered (F)-OFDM and windowing (W)-OFDM \cite{Guan-17}. Nevertheless, a tunable version of CP-OFDM technology was chosen for 5G, with flexible multi-numerology as a key feature \cite{Dahlman-20}. The main elements of multi-numerology are scalable SCSs and CP lengths. The former provides robustness against hardware impairments and lower latency at higher frequencies, whereas the latter aims to eliminate the effects of multipath propagation and inter-symbol interference in all channel conditions. While CP-OFDM is used for both downlink and uplink directions in 5G, there are certain uplink setups, i.e., single-layer transmissions, where a single-carrier-based DFT-s-OFDM technology is an alternative option, enabling extended coverage. Flexible CP-OFDM plays a key role in 5G for supporting diverse use cases, such as eMBB, URLLC, and mMTC. Currently, different types of waveforms are being studied in academia and industry to support a wide range of demanding 6G application scenarios. 4G and 5G waveforms are summarized in Figure \ref{Fig_Wave}. 

\begin{figure}[!tb]
\center{\includegraphics[width=0.7\columnwidth]
{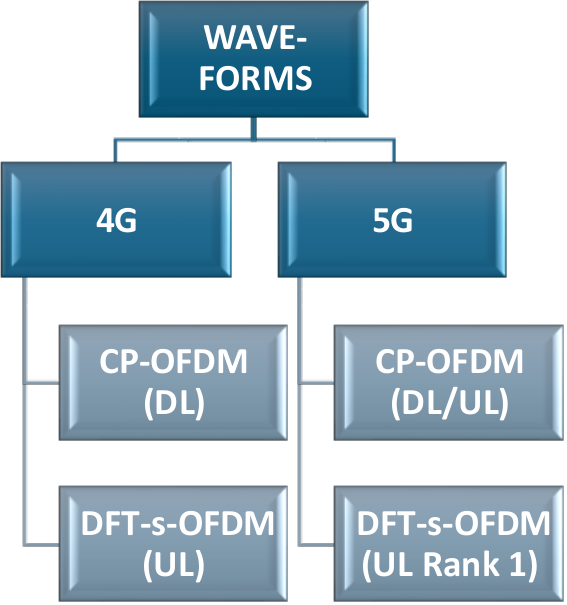}}
\caption{\label{Fig_Wave}4G and 5G waveforms.}
\end{figure}

\item {\textbf{Opportunities and Challenges}}: 
The potential of a multi-waveform design, with highly tunable numerology, is to provide an efficient core to an ultra-flexible transmission structure to support a vast range of possible 6G frequencies, requirements, use cases, and application scenarios. Due to the massive spectrum range of 6G, relying solely on a single waveform may not be a feasible solution. A preferable approach is to select a set of waveforms to support all types of 6G scenarios. To this end, the main challenge is to develop a proper set of waveforms to support a massive spectrum range from sub-6 GHz to THz frequencies, and also to find a comprehensive set of parameters for those waveforms and the associated frame structure. 

At a high level, a possible solution for the multi-waveform problem is to choose one primary waveform technology to handle most of the scenarios and one or many complementary technologies for special cases. For example, an ultra-tunable OFDM-based waveform strategy could be chosen as the main technology due to its proven performance and flexibility in 5G. The flexible OFDM scheme can then be complemented by a single carrier-type waveform technology, such as DFT-s-OFDM or its variant, for higher-frequency operation in the THz band since this type of waveform has a low PAPR and is more robust against severe hardware impairments at very high frequencies \cite{Tervo-22, Sarajlic-23}. It is worth noting that if an OFDM-based multi-waveform scheme will be selected, the first version of 6G could be designed to interact closely with 5G, enabling a faster and smoother launch. Such a design could exploit the lessons learned from the transition from 4G to 5G. 

\item {\textbf{Literature and Future Directions}}: 
In the literature, a vast variety of single-carrier, multi-carrier, and OFDM-based waveforms has been studied for 5G and beyond \cite{Conceicao-21, Adoum-23, Sarajlic-23, Solaija-24}. Typical single-carrier waveform candidates include different variations of DFT-s-OFDM \cite{Sahin-16, Tervo-22, Sarajlic-23}. In the multi-carrier domain, popular waveforms include FBMC, UFMC, GFDM, and orthogonal time frequency space modulation \cite{Wei-21, Kebede-22, Adoum-23, Shtaiwi-24}. Numerous OFDM variants have been developed, such as F-OFDM, block-filtered OFDM, W-OFDM, cyclic postfix windowing-OFDM, time-interleaved block-windowed burst OFDM, and weighted overlap and add OFDM \cite{Conceicao-21}. Since CP-OFDM has established its status in mobile networks, competing technologies aim to overcome its shortcomings. However, considering individual waveforms, each of these technologies has its own pros and cons, but there is no single method that fits well to all 6G scenarios. Consequently, multiple waveforms must be utilized in ultra-heterogeneous 6G networks. 

In the context of multi-waveform schemes, there are two main future research directions, i.e., developing a comprehensive set of waveforms to support different types of 6G scenarios and finding a flexible set of parameters for each of these waveforms. As mentioned earlier, a logical starting point for this category of research is to study an enhanced OFDM-based scheme as a primary technology, complemented by one or more waveform technologies. Single-carrier waveforms, such as DFT-s-OFDM and its variants, seem potential candidates for complementary technologies, as they fit well for THz frequencies due to their low PAPR and robustness toward hardware impairments \cite{Tervo-22, Sarajlic-23, He-23}. The paper \cite{Sarajlic-23} provided a survey on the waveform design for 6G THz communications. As 6G is expected to expand its capabilities beyond communication, an important future topic is the study of waveform design for a joint communication and sensing paradigm \cite{Liyanaarachchi-21, Mao-22b, Zhou-22d, Wang-23d, Shtaiwi-24}. 

\end{itemize}

\begin{figure}[!htb]
\center{\includegraphics[width=0.9\columnwidth]
{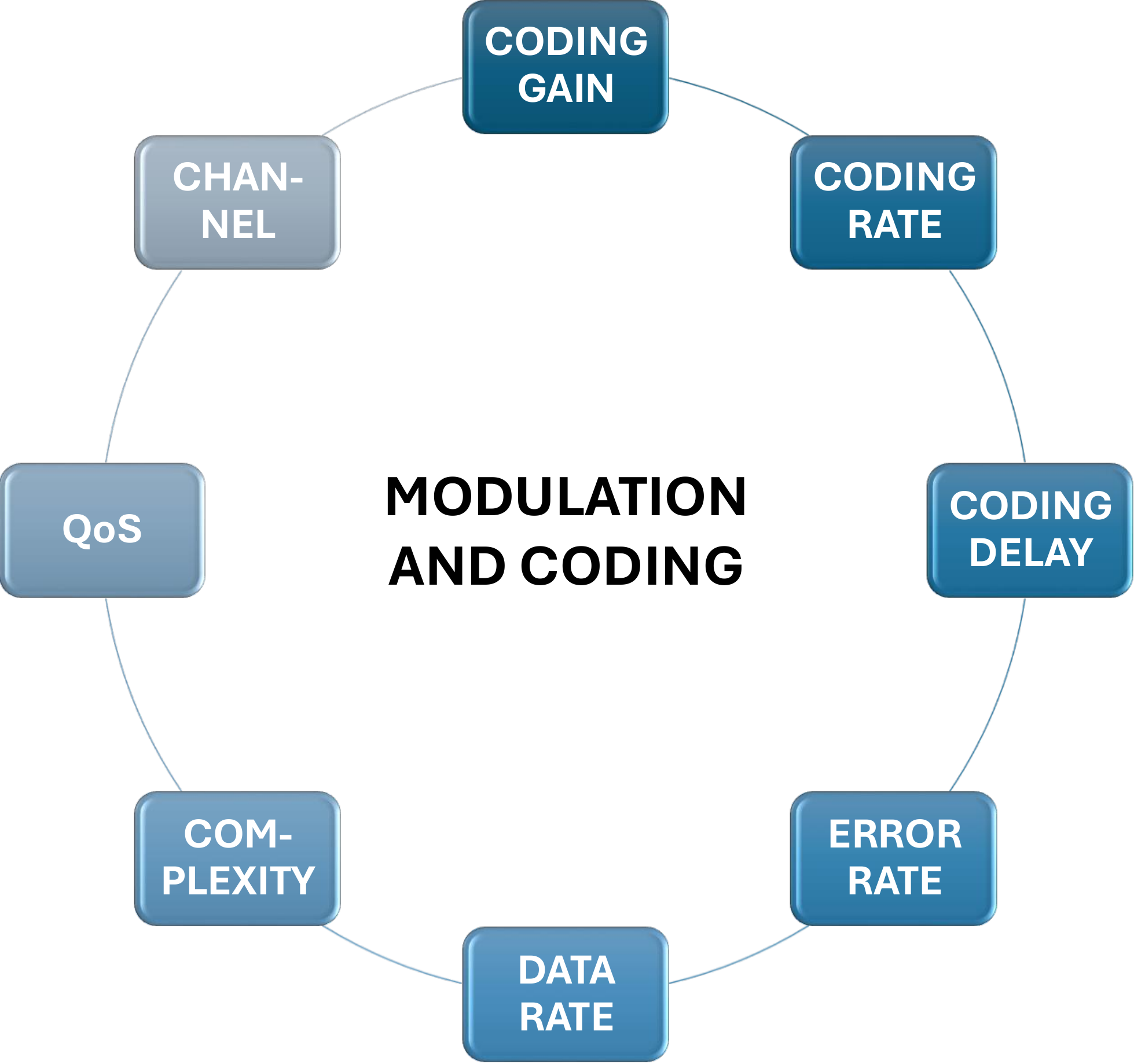}}
\caption{\label{Fig_ModCod}Factors to consider in the design of modulation and coding schemes \cite{Hanzo-02}.}
\end{figure}

\subsubsection{ADVANCED MODULATION AND CODING METHODS}

\begin{itemize}
\item {\textbf{Vision}}:
Advanced modulation and channel coding methods will play key roles in the 6G PHY layer, especially in terms of high-throughput and extremely reliable communications. 

\item {\textbf{Introduction}}:
A modulation method is in charge of mapping the bits into symbols. The order of the modulation is defined by the number of bits per symbol. Higher-order modulations are used to achieve higher data rates, whereas lower-order modulations are meant for better reliability. Channel coding fights against errors occurring in a wireless medium by adding redundancy to the transmitted data sequence. Higher coding rates are employed for reliability, whereas lower code rates are used for better throughput.

Modulation and channel coding are tight together in a concept known as adaptive modulation and coding, in which modulation orders and coding rates are bundled into modulation and coding schemes (MCSs), each of which has different spectral efficiency and reliability properties \cite{Torres-20}. Adaptive modulation and coding is used to adapt MCSs to the prevailing channel conditions by relying on the signal quality information fed back from the receiver \cite{Torres-20}. Adaptive modulation and coding plays an important role in the link-level performance of current mobile networks, particularly in terms of data rate and reliability \cite{Torres-20}. Enhanced modulation and coding methods are needed for every new mobile generation to meet the ever-growing performance demands. Figure \ref{Fig_ModCod} summarizes the main factors that need to be considered in the design of modulation and coding schemes. 

\item {\textbf{Past and Present}}:
As summarized in Figure \ref{Fig_ModCod2}, each mobile generation has adopted a novel or enhanced channel coding concept, i.e., 2G: convolutional coding (data and control), 3G: Turbo coding (data) and convolutional coding (control), 4G: Turbo coding (data) and convolutional coding (control), 5G: low-density parity check (LDPC) coding (data), and polar coding (control) \cite{Geiselhart-23}. Currently, LDPC codes are used for data channels and polar codes for control channels in 5G \cite{Hui-18}. Turbo coding was the first channel coding method approaching theoretical channel capacity. Linear LDPC coding also belongs to the class of capacity-approaching codes while outperforming Turbo coding, particularly in the higher coding rate region and in terms of the error floor \cite{Bae-19}. Polar coding is a linear block-error correction code with low-complexity encoding and decoding designs, and the first channel code that achieves theoretical channel capacity \cite{Bae-19}. LDPC and polar codes have been reviewed for 5G NR in \cite{Hui-18, Bae-19}. 

\begin{figure}[!tb]
\center{\includegraphics[width=\columnwidth]
{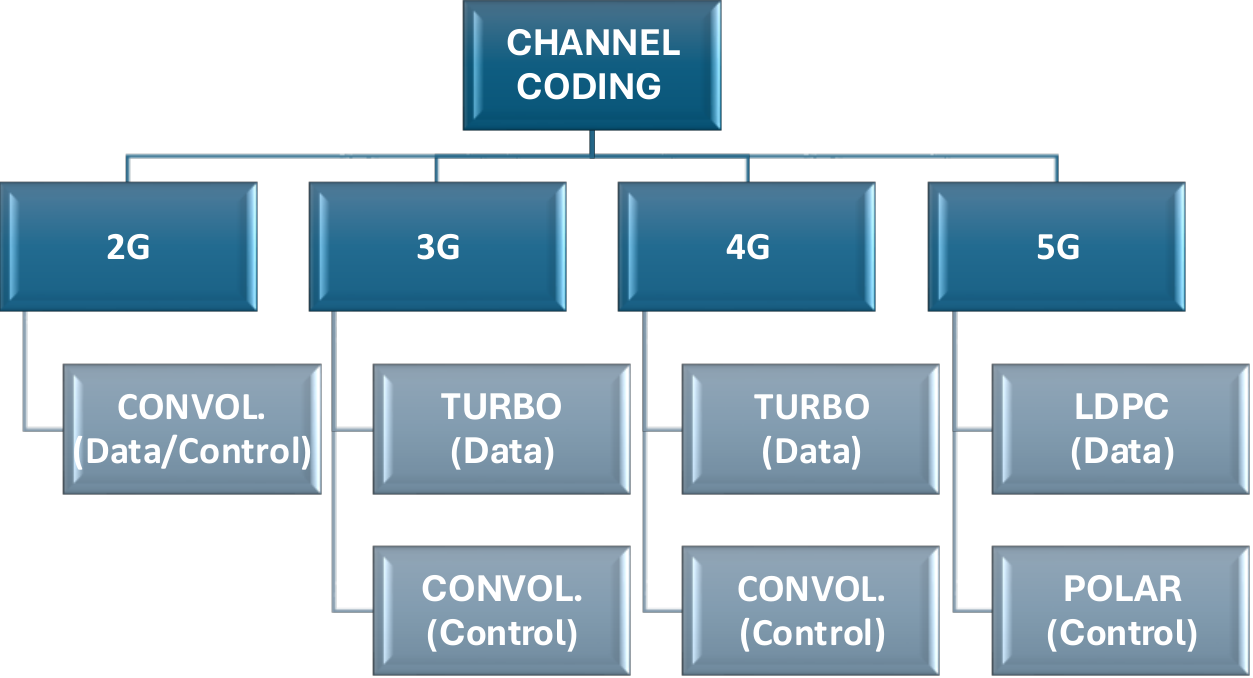}}
\caption{\label{Fig_ModCod2}Evolution of channel coding from 2G to 5G.}
\end{figure}

The most commonly used modulation methods in mobile communication are phase shift keying (PSK) and quadrature amplitude modulation (QAM)\cite{Singya-21}. PSK is a digital modulation technique in which the phase of the carrier wave is altered. Digital QAM adjusts the amplitudes of the two orthogonal carriers according to a digital modulation process. 
The used modulation methods and their constellation sizes for different mobile generations are as follows \cite{Singya-21}: Gaussian minimum shift keying (GMSK) in 2G; binary PSK (BPSK), quadrature PSK (QPSK), 16-QAM, and 64-QAM in 3G; BPSK, QPSK, 16-QAM, 64-QAM, and 256-QAM in 4G; $\pi/2$-BPSK, QPSK, 16-QAM, 64-QAM, 256-QAM, and 1024-QAM in 5G. Each new generation has raised the highest constellation size a step further. This trend suggests that 6G could adopt 4096-QAM, provided that it will be proven beneficial in practice. 

\item {\textbf{Opportunities and Challenges}}: 
The main advantage of advanced modulation and coding methods is improved link-level performance in terms of reliability and data rates. Higher modulation orders and lower coding rates provide better throughput, whereas lower orders and higher code rates provide reliability. In particular, advanced channel coding will play a key role in providing extremely reliable communication for many 6G application scenarios, such as smart factory and smart healthcare environments. Enhanced polar codes seem promising for 6G due to their favorable properties for highly reliable communication, particularly with stringent latency requirements \cite{Shirvanimoghaddam-19, Rajatheva-21, Yue-22}. 

At a high level, there are two main challenges in modulation and coding designs toward 6G, i.e., developing efficient methods for ultra-high throughput and extremely reliable communication links. The former design problem calls for efficient high-order modulation and low coding rate methods with decent error-correcting properties. A straightforward solution is to increase the modulation order beyond 1024-QAM used in 5G, requiring a thorough performance comparison between different variations of QAM. However, increasing the order of QAM may not provide the needed performance gains at higher 6G frequencies due to the inefficiency of power amplifiers \cite{Rajatheva-21}. To alleviate this problem, coded modulation with signal shaping is a promising concept \cite{Pikus-17}, especially in combination with polar codes \cite{Iscan-18, Iscan-19, Chiu-22, Runge-22}. Polar codes have favorable features to efficiently support lower coding rates in combination with higher-order modulations \cite{Iscan-18, Tong-21b, Wehn-21}. Coded modulation with probabilistic shaping is already in use in optical fiber communications \cite{Rajatheva-21}. 

In the latter design problem, a major challenge is to develop channel codes that fit well with 6G-level reliability and latency requirements. A challenging trade-off exists between reliability and latency in the design of channel coding in the URLLC-type scenarios \cite{Yue-22}. In other words, long block-length codes are required for reliability, whereas short ones are required for latency. Polar coding has been shown to strike a good balance between reliability and latency, making it a promising candidate for the 6G-level URLLC scenarios \cite{Shirvanimoghaddam-19, Yue-22}. In general, polar codes have favorable properties for highly reliable communications, such as the lack of error-floor and efficient error-correction capabilities \cite{Shirvanimoghaddam-19, Rajatheva-21}. Further discussions of the latest enhancements for polar codes can be found in \cite{Rajatheva-21, Wehn-21, Yue-22}. 

\item {\textbf{Literature and Future Directions}}: 
In the 2010s, LDPC and polar codes were studied for 5G \cite{Hui-18}. Research beyond 5G is ongoing, with a special emphasis on enhanced polar codes \cite{Wehn-21, Yue-22, Rowshan-24}. Many different variations of polar codes have been proposed in the literature \cite{Oliveira-21, Liao-22, Choi-23, Zunker-23}. In particular, polar codes have been examined for 6G-level scenarios, such as ultra-high throughput and extremely reliable low-latency communications \cite{Shirvanimoghaddam-19, Tong-21b, Wehn-21, Yue-22, Jiang-23b, Niu-23}. Shaped polar-coded modulation has been another active and particularly interesting topic for 6G \cite{Iscan-19, Rajatheva-21, Chiu-22, Runge-22, Wu-23, Runge-23}. The past, present, and future of channel coding were discussed in recent surveys \cite{Zhang-23c, Geiselhart-23, Rowshan-24}. Further research is needed in these directions to achieve a required level of maturity for practical implementation in the 6G era. 

In the modulation domain, QAM is the most widely used and studied modulation method in mobile networks \cite{Singya-21}. Currently, 5G supports QAM with modulation orders of 16, 64, 256, and 1024 \cite{Singya-21}. 6G calls for even higher-order modulations. Recent studies have shown special interest in hexagonal QAM due to its promising performance properties \cite{Singya-21}. A comprehensive survey of QAM and its diverse variants was provided in \cite{Singya-21}. 
\end{itemize}

\begin{figure}[!htb]
\center{\includegraphics[width=\columnwidth]
{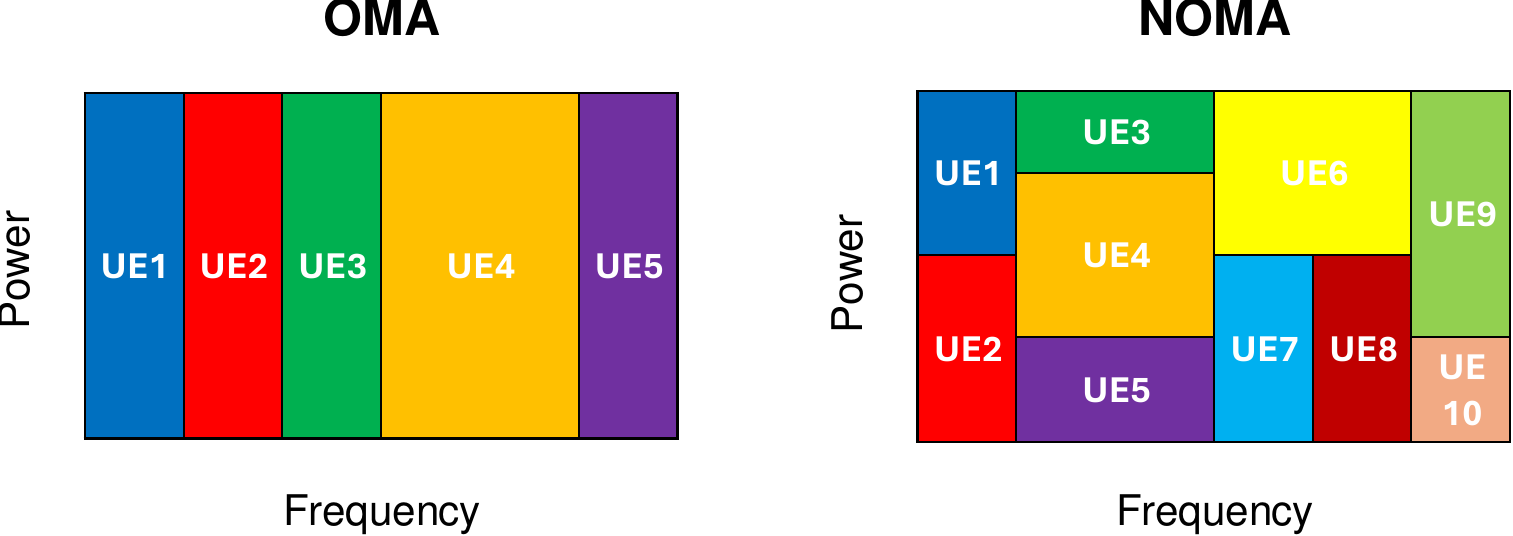}}
\caption{\label{Fig_NOMA}Orthogonal versus non-orthogonal multiple access \cite{Cheon-17}.}
\end{figure}

\subsubsection{NON-ORTHOGONAL MULTIPLE ACCESS}

\begin{itemize}
\item {\textbf{Vision}}:
NOMA is considered a potential technology candidate for improving the spectral efficiency and system capacity of 6G. A particularly promising 6G usage scenario for NOMA is to facilitate massive IoT communications in combination with fast grant-free access. 

\item {\textbf{Introduction}}:
NOMA is based on the principle that multiple users/devices can be served simultaneously by utilizing the same transmission resource element (e.g., a time-frequency resource block) \cite{Makki-20}. At the cost of increased interference levels and receiver complexity, NOMA can improve spectral efficiency, system throughput, and user-fairness compared to orthogonal multiple access (OMA) \cite{Makki-20}. The difference between OMA and NOMA is illustrated in Figure \ref{Fig_NOMA}. Furthermore, latency properties can be improved by combining NOMA with a grant-free access scheme \cite{Shahab-20}. 

Typically, NOMA schemes are categorized into two main classes \cite{Budhiraja-21}: power-domain NOMA (PD-NOMA) and code-domain NOMA (CD-NOMA). In the former strategy, the users are separated in the power domain. For example, using superposition coding at the transmitter and successive interference cancellation (SIC) at the receiver. The latter approach exploits code-level separation by relying on, for example, spreading codes, interleaving, scrambling, or other types of user-specific re-modification of data sequences. CD-NOMA can be further categorized into dense and sparse coding classes \cite{Budhiraja-21}. There are also hybrid NOMA schemes that combine the PD-NOMA and CD-NOMA strategies. A comprehensive categorization of the numerous NOMA variants can be found in \cite{Budhiraja-21}. 

\item {\textbf{Past and Present}}:
In the past decade, there has been great interest in studying the NOMA technology for 5G and beyond \cite{Budhiraja-21}. Consequently, NOMA was added as a study item to the 3GPP 5G NR standardization process \cite{Makki-20}. Many different types of NOMA schemes were proposed and evaluated using link- and system-level simulations. However, it was decided in 3GPP that NOMA studies do not proceed to the work item phase, but are left for possible future usage in beyond 5G networks. According to \cite{Makki-20}, the main reason behind this decision was that there were no clear gains shown in the evaluation process of different NOMA schemes compared to the existing Release 15 technologies (e.g., multi-user MIMO), and given that the implementation complexity of NOMA receivers is rather high. A detailed discussion of the proposed NOMA candidates and the corresponding evaluation and decision process are provided in \cite{Makki-20}. Nevertheless, the research focus of NOMA has shifted from 5G to beyond 5G, as it is still considered a promising technology candidate for many communication scenarios in the future.  

\item {\textbf{Opportunities and Challenges}}: 
The main benefit of NOMA technology is that it provides improved spectral efficiency and fairness by serving multiple devices per each time-frequency resource unit. The most promising 6G application scenario for NOMA is ultra-massive IoT communications with grant-free access \cite{Shahab-20}. For a typical overloaded massive connectivity scenario, it is assumed that there are two defining features present \cite{Shahab-20}. First, there are much more devices requesting service than the number of available orthogonal resource units. Second, it is assumed that narrowband IoT sensors will send only a small amount of uplink data every now and then. 

In this type of scenario, conventional orthogonal multiple access schemes may be inadequate to provide service for all the devices in need, and waste too much resources for each narrowband IoT sensor, as the smallest orthogonal resource unit may significantly exceed the need of an individual IoT device. NOMA, instead, is seen as a more efficient method to serve all devices in need due to its ability to serve multiple devices in a single resource unit and better match the needs of each individual device, and not waste the scarce spectral resources. Moreover, combining NOMA with fast grant-free medium access will further improve the system performance, especially in terms of decreased latency and reduced signaling \cite{Shahab-20}. 

For the aforementioned reasons, grant-free NOMA is an attractive technology for 6G ultra-massive IoT scenarios. There are also numerous other application scenarios for NOMA in the context of 6G networks \cite{Budhiraja-21}. Combining NOMA with other emerging 6G technologies (e.g., RIS, D2D, V2X, VLC, etc.) may provide further benefits in terms of spectral efficiency, system throughput, service balancing/fairness, and energy efficiency \cite{Vaezi-19}. A comprehensive review of the combination of NOMA and many other emerging technologies has been presented in \cite{Vaezi-19, Maraqa-20}. 

As mentioned earlier, NOMA was studied for 5G NR as a study item, but 3GPP decided not to include it in the official 5G standards \cite{Makki-20}. This 3GPP process showed that a critical challenge of NOMA is the development of schemes that provide a sufficient balance between performance and complexity. In particular, a typical problem of PD-NOMA is the implementation complexity of the required multi-user detection-based receiver (e.g., the SIC-receiver). Consequently, reducing the receiver complexity is an essential challenge, as discussed in \cite{Makki-20}. Another key challenge is to find suitable 6G usage scenarios in which the NOMA technology is beneficial, and to develop customized NOMA methods for each of these usage scenarios. 

\item {\textbf{Literature and Future Directions}}: 
As discussed earlier, NOMA was not included in the 5G standards but left to be studied further for beyond 5G networks. Since then, a plethora of NOMA schemes have been proposed for promising beyond 5G usage scenarios. Comprehensive surveys on a wide range of future NOMA methods were presented in \cite{Budhiraja-21, Liu-21, Liu-22e, Liu-22d}. In particular, combinations of NOMA with many other emerging 6G technologies were studied in \cite{Vaezi-19}. 
 
In the literature, numerous survey papers have reviewed different aspects of NOMA and captured its evolutionary path from the preliminary 5G studies to promising 6G candidate technology, with numerous potential application scenarios. At a high level, NOMA surveys can be classified into several categories, i.e., 5G and beyond \cite{Wu-18, Dai-18, Budhiraja-21, Elnaby-23}, generic \cite{Maraqa-20, Makki-20, Mohsan-23d, Liu-24c}, specific \cite{Vaezi-19b, Vaezi-19, Maraqa-20, Shahab-20, Ding-22b, Mohsan-23f, Mohsan-23e, Sadia-23, Mu-23, Bepari-23, Sarkar-24, Nasser-24}, and 6G \cite{Yu-21, Liu-21, Liu-22e, Liu-22d, Clerckx-24}. In the 5G and beyond domain, the main focus was on the applicability of NOMA to 5G and expanding the possible NOMA application scenarios to better fit the needs of beyond 5G networks. The generic surveys reviewed NOMA from a wider perspective, covering current status, recent advances, state-of-the-art, application scenarios, and open problems. The specific surveys examined various narrow topics, including the combination of NOMA and other emerging technologies \cite{Vaezi-19, Maraqa-20}, NOMA myths and critical questions \cite{Vaezi-19b}, grant-free NOMA for IoT \cite{Shahab-20}, RIS-assisted NOMA \cite{Ding-22b, Sadia-23, Sarkar-24}, DL-assisted NOMA \cite{Mohsan-23f}, NOMA-based VLC \cite{Mohsan-23e}, NOMA for ISAC \cite{Mu-23, Nasser-24}, and cache-aided NOMA \cite{Bepari-23}. The 6G-oriented surveys explored NOMA as the next-generation multiple access technology. 

To include the NOMA technology in the 6G standards, there are many open issues to be addressed. For instance, the implementation complexity of NOMA schemes, especially at the receiver side, must be reduced for practical usage. Hence, complexity reduction is an essential target for future research. Another research avenue is to study what are the most suitable 6G scenarios for different types of NOMA schemes. For each potential scenario, tailored NOMA solutions need to be developed and thoroughly evaluated. This category of research is widely conducted worldwide. Recently, different types of NOMA solutions have been proposed for various future application scenarios. In particular, the NOMA concept has been combined with many other emerging technologies. Currently, the most promising research direction is grant-free NOMA for massive IoT communications \cite{Shahab-20}. Due to its promising nature, it is advisable to direct an extra focus to that research area. Other potential topics, with disruptive nature, include NOMA-D2D, NOMA-V2X, NOMA-UAV, NOMA-RIS, NOMA-massive MIMO, and NOMA-VLC \cite{Vaezi-19, Maraqa-20}. Applying AI/ML methods to assist in NOMA solutions is also an important research direction. 
\end{itemize}

\begin{figure}[!htb]
\center{\includegraphics[width=\columnwidth]
{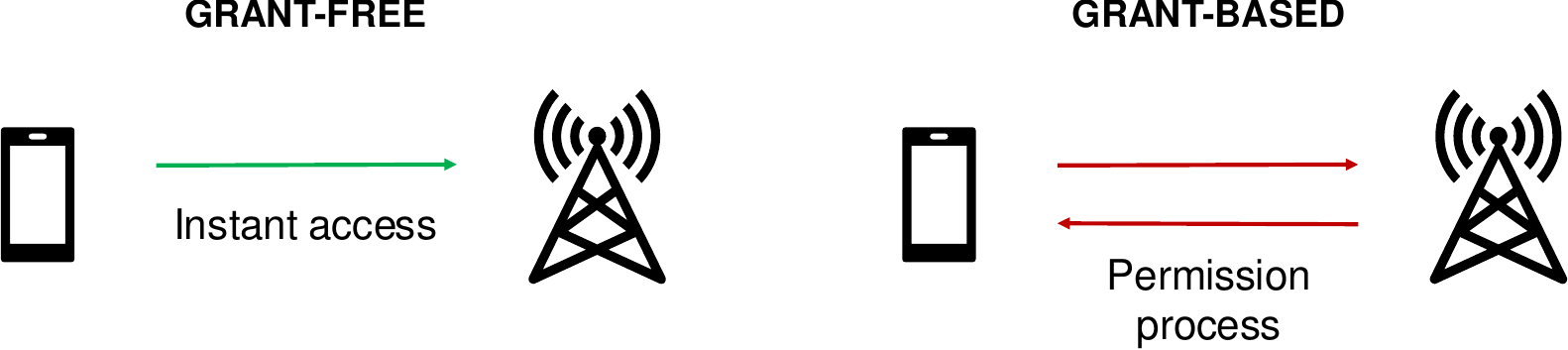}}
\caption{\label{Fig_GFMA}Grant-free versus grand-based medium access.}
\end{figure}

\subsubsection{GRANT-FREE MEDIUM ACCESS}

\begin{itemize}
\item {\textbf{Vision}}:
Fast grant-free medium access is considered a key technology to enable efficient network access in the 6G era, especially for ultra-massive IoT scenarios. 

\item {\textbf{Introduction}}:
Grant-free medium access refers to an access scheme where devices can instantly access the network without a time-consuming permission process, as illustrated in Figure \ref{Fig_GFMA}. Grant-free access aims to offer a fast and efficient network access procedure. Its main benefits are decreased latency, reduced signaling overhead, and improved energy efficiency \cite{Choi-22}. Typical drawbacks include the collisions of transmissions and a lack of priority among service-needing devices \cite{Choi-22}. All these features make grant-free access applicable to IoT scenarios, with many devices sending small amounts of data every now and then. 

\item {\textbf{Past and Present}}:
In the literature, grant-free access for mobile communications started to gain considerable interest in the mid-2010s due to the recognized relevance of diverse IoT scenarios in future mobile networks. A particular focus of the corresponding studies was on mMTC for 5G. Recent research on grant-free access looks beyond 5G to enable ultra-massive connectivity in 6G \cite{Choi-22, Gao-23}. In standardization, 4G LTE supports a four-step random access procedure, while 5G NR updated its random access to a grant-free two-step procedure in Release 16 \cite{Kim-21b}. 

Whereas the four-step process in 4G requires a response from the base station (BS) before sending its payload data, the two-step access of 5G allows devices to transmit data without a particular response \cite{Choi-22}. This grant-free random access provides benefits in lowering latency and reducing control signaling overhead, particularly for MTC scenarios with many burst transmissions of small data packets from a massive number of devices \cite{Choi-22}. However, the performance of the two-step access is limited by the preamble collision problem when multiple devices send the same preamble to the BS simultaneously, causing a collision of the transmissions. In \cite{Kim-21b}, a survey was provided on the two-step random access procedure in 5G NR, reviewing also the latest literature and potential solutions to the preamble collision problem. 

\item {\textbf{Opportunities and Challenges}}: 
The potential of advanced grant-free access is to serve as a key enabler for ultra-massive IoT and its numerous application scenarios in the 6G era by providing a fast and efficient network access procedure. As grant-free access offers many benefits by reducing latency, releasing communication resources from excess signaling, and facilitating energy-efficient designs, it is naturally applicable to IoT scenarios with massive numbers of devices occasionally sending small data packets. On the road to this vision, there are still many challenges to overcome. The main ones are related to the special characteristics of 6G, such as massive connectivity, stringent performance requirements, and network heterogeneity \cite{Choi-22}. 

First, the greater the number of access-needing devices is, the higher is the probability of preamble collisions. This requires advanced collision avoidance strategies \cite{Kim-21b, Choi-22}. Grant-free access becomes more challenging when massive connectivity is combined with the stringent latency requirements of devices \cite{Choi-22}. Since transmission collisions increase latency, novel solutions must be adopted. For example, a contention-free access strategy with reserved preambles can be used to achieve collision-free transmissions within the target delay time \cite{Choi-22}. However, when the number of devices increases, reserved preambles may run short. Adding stringent reliability constraints to this scenario, that is, massive URLLC, further complicates the problem \cite{Choi-22, Ding-22}. In this case, there is a compromise between connection density, latency, and reliability. To alleviate the problem of finding a proper balance between these conflicting requirements, AI/ML-aided traffic prediction methods can be exploited \cite{Choi-22}. Accurate traffic prediction facilitates preamble management and collision avoidance. 

Integrated space-air-ground networks also bring new challenges to grant-free access due to unconventional 3D channel and interference characteristics with non-terrestrial APs (UAVs and satellites) and end-devices (UAVs). Cell-free network design with distributed lightweight APs, another architectural evolution expected to be employed in 6G, will also change the traditional network topology, creating new opportunities and challenges. While spatial sparsity can be exploited to support massive access, synchronization becomes trickier. In general, AI/ML is a promising tool to tackle diverse issues in grant-free access by utilizing device activity and traffic patterns for efficient design and management. More information regarding the aforementioned challenges of grant-free access in 6G networks can be found in \cite{Choi-22}. 

\begin{table*}[htb!]
\begin{center}
\caption{Summary of network architectural technologies for 6G}
\label{Table_Architecture}
\centering
\begin{tabularx}{\textwidth}{| >{\centering\arraybackslash}X | >{\centering\arraybackslash}X |
>{\centering\arraybackslash}X | 
>{\centering\arraybackslash}X |
>{\centering\arraybackslash}X |
>{\centering\arraybackslash}X |
>{\centering\arraybackslash}X |
>{\centering\arraybackslash}X |}
\hline
\centering
\vspace{3mm} \textbf{Network Architectural Technologies} \vspace{3mm} & \centering \textbf{Vision} & \centering \textbf{Description} & \centering \textbf{Opportunities} & \centering \textbf{Challenges} & \centering \textbf{Past} & \vspace{1.5mm} \begin{center} \textbf{Present} \end{center} \\
\hline
\vspace{3mm} Integrated Non-Terrestrial and Terrestrial Networks \vspace{3mm}  & Space-air-ground network access & Space-air-ground network layers & Global coverage & Cost efficiency $\&$ industry incentive & Satellite commun since 1960s & 5G NTNs \\
\hline
\vspace{3mm} Ultra-Dense Networks \vspace{3mm}  & THz UDNs & Highly densified networks & Extreme capacity & Interference management & Denser networks from 1G to 5G & 5G mmWave small cells \\
\hline
\vspace{3mm} Cell-Free Massive MIMO \vspace{3mm}  & Complementary 6G architecture & Lots of distributed low-cost APs & Stable QoS over network coverage & Scalability $\&$ clustering & Concept invented in 2015 & Under study for 6G \\
\hline
\vspace{3mm} Integrated Access and Backhaul \vspace{3mm}  & THz IAB & Same resources for A$\&$B & Faster/cheaper to install than fiber & Resource alloc between A$\&$B & Separated A$\&$B & 5G IAB \\
\hline
\end{tabularx}
\end{center}
\end{table*}

\item {\textbf{Literature and Future Directions}}: 
In the literature, there are a handful of survey papers on grant-free access that cover different aspects and review the latest advancements \cite{Shahab-20, Ding-22, Choi-22, Ye-22, Gao-23}. A comprehensive survey was provided on grant-free NOMA for IoT in \cite{Shahab-20}. In \cite{Ding-22}, grant-free URLLC was studied, with a special focus on the potential enhancements provided by massive and cell-free massive MIMO technologies. The paper \cite{Choi-22} discussed grant-free random access for MTC. In this context, massive MIMO and NOMA concepts were reviewed, as well as future challenges toward 6G. In \cite{Ye-22}, the authors explored grant-free access for massive satellite-based IoT in the 6G era. The work \cite{Gao-23} discussed grant-free network access based on compressive sensing for 6G massive communications. 

Although 5G has already adopted a grant-free access process, it needs to be upgraded to match the unique characteristics of 6G, such as massive access, extremely tight performance requirements, novel network architectures, and versatile application scenarios. In this respect, it is vital to study and design grant-free access mechanisms that consider emerging 6G technologies, such as AI/ML, NOMA, RIS, cell-free massive MIMO, and space-air-ground network architecture. AI/ML is a versatile tool for making grant-free access more efficient and robust \cite{Kim-21b, Choi-22}. NOMA allows more devices to be served simultaneously by allocating multiple devices to a single resource unit and separating them into power or code domains. Thus, grant-free NOMA is a potential candidate for ultra-massive IoT connectivity in 6G \cite{Shahab-20}. 

Recently, grant-free NOMA has been extended to RIS-assisted scenarios to further improve the network access performance by providing potential benefits in terms of spectral efficiency or reliability \cite{Tasci-22}. Cell-free massive MIMO with grant-free access is a promising concept to provide massive connectivity with stable coverage and decreased latency for low-power devices in IoT scenarios \cite{Ding-22}. Grant-free access can exploit the distributed nature of cell-free networks, but needs to address the corresponding synchronization issues \cite{Choi-22, Ding-22}. Although 3D networks offer great opportunities for 6G, they also set new challenges for grant-free access \cite{Choi-22, Ye-22}. 
\end{itemize}

\subsection{NETWORK ARCHITECTURAL TECHNOLOGIES FOR 6G}
The design of mobile networks is becoming increasingly complex for each new generation. From the network design perspective, 6G is expected to support 3D network architectures, UDNs, flexible cell deployments, and cell-free operations. A 3D network architecture will be enabled by integrating terrestrial and non-terrestrial access, including space, air, and ground layers. Satellite communication is a key technology for global coverage on the land, in the air, and at the sea. Due to THz communications, 6G will rely on an ultra-dense cell design to provide sufficient coverage for extreme-capacity scenarios. Flexible cell deployments will be facilitated by integrating wireless backhaul and access. In addition to the traditional cell-based design, 6G is expected to be complemented by a cell-free network architecture to provide stable QoS in the coverage area. This section provides a detailed discussion on the aforementioned network architectural technologies, i.e., INTNs, UDNs, IAB, and {cell-free massive MIMO}. These technologies are summarized in Table \ref{Table_Architecture}. 

\subsubsection{INTEGRATED NON-TERRESTRIAL AND TERRESTRIAL NETWORKS} 

\begin{itemize}
\item {\textbf{Vision}}:
INTNs are expected to play a key role in 6G, potentially providing global coverage and enabling a myriad of novel applications, ranging from high-speed remote area connectivity to global IoT and worldwide environmental/industrial monitoring. 

\item {\textbf{Introduction}}:
INTNs are communication systems that typically consist of ground, air, and space layers. This architecture is also known as a space-air-ground integrated network (SAGIN) \cite{Liu-18}, as illustrated in Figure \ref{Fig_INTN} \cite{Rajatheva-20, Rajatheva-21}. For this network architecture, the corresponding layers are called ground-based, airborne, and spaceborne \cite{Rajatheva-21}. The main elements of the ground-based layer include mobile networks, satellite ground stations, and mobile/satellite devices \cite{Rajatheva-21}. 

The airborne layer consists of high- and low-altitude aerial platforms. HAPSs are aerial network nodes, such as aircrafts, airships, and balloons, operating as relay-type entities at the stratospheric altitudes of up to 20 km \cite{Rajatheva-21}. Due to their relatively high altitudes, HAPSs can provide wide area coverage from urban to rural environments. In the lower airborne layer, UAVs can be used as aerial BSs or relays to improve the performance of mobile networks, especially in terms of capacity and coverage \cite{Saad-20b}. UAVs are considered as the low-altitude platform stations (LAPSs), serving as agile and flexible network nodes \cite{Rajatheva-21}. UAVs can provide temporal or long-term performance enhancements, depending on the application scenario. 

The spaceborne layer is based on satellite communication. Satellite communication refers to communication between widely separated locations on the globe by sending radio signals from the ground-based transmitters by relaying and amplifying satellites to the ground-based receivers. Since satellites are high in space, each one of them can cover large geographical areas on Earth. Consequently, satellites can provide global coverage, including land, sea, and air. There are three primary orbits for communication satellites: geostationary orbit (GEO) (35786 km), medium Earth orbit (MEO) (7000-25000 km), and low Earth orbit (LEO) (300-1500 km) \cite{Azari-21}. Each orbit has its advantages and disadvantages. In general, wider coverage is obtained using satellites in higher orbits, while the propagation characteristics are better and satellites cheaper in lower orbits. 

\begin{figure*}[!htb]
\center{\includegraphics[width=\textwidth]
{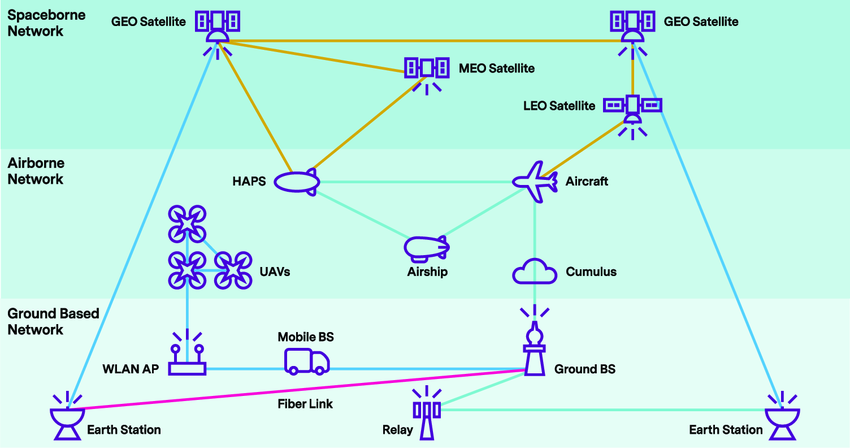}}
\caption{\label{Fig_INTN}Integrated space-air-ground network \cite{Rajatheva-20, Rajatheva-21}.}
\end{figure*}

\item {\textbf{Past and Present}}: 
Different layers of the INTNs have evolved over different timescales. Although the history of satellite communications dates back to the 1960s and the first terrestrial cellular networks were opened in the 1980s, UAV-assisted communication is still in its infancy. The era of satellite communication began in the early 1960s when the first communication satellite was launched. Since then, satellites have been an important part of worldwide communications, currently providing a vast range of services from television broadcasts and satellite phones to high-speed Internet access and global navigation systems. The next major step is to integrate satellite communication into mobile networks, thereby opening up a broad range of novel application scenarios. The earliest research on HAPSs was conducted in the 1990s. However, HAPSs have gained more interest only recently, as they are seen as part of the next generation space-air-ground network architecture. UAV-assisted communications have been extensively studied over the past decade, with the main focus on 5G networks. In the late 2010s, the focus of research started to shift toward 6G networks \cite{Saad-20b}. 

5G is already taking the first steps toward the integration of terrestrial and non-terrestrial networks. Currently, NTNs, especially satellite access (with implicit support for HAPS and air-ground networks), are under standardization in 3GPP 5G NR \cite{Lin-21b, Geraci-23}. The work toward supporting NTNs started in 2017 as a Release 15 study item on deployment scenarios and channel models, followed by Release 16 studies identifying necessary physical/higher-layer features and key use cases with the associated requirements \cite{Lin-21b}. NTNs were formally incorporated into the 5G NR standards in Release 17. The corresponding work items specified the necessary enhancements to support LEO/GEO satellites (also implicitly HAPS and air-ground scenarios), stage 1 service requirements, and produced normative specifications based on the previous studies \cite{Lin-21b, Saad-24}. Release 17 also studied non-terrestrial IoT, satellite access/backhaul, business roles, service management, and public land mobile network (PLMN) selection (in international areas) \cite{Lin-21b, Saad-24}. Release 18 studied NR NTN enhancements in terms of coverage, access beyond 10 GHz, mobility, and service continuity \cite{Geraci-23, Saad-24}. Detailed reviews on the standardization efforts of 5G NR NTNs are presented in \cite{Lin-21b, Geraci-23, Saad-24}. It is worth noting that in 3GPP, UAVs are studied in a separate track than other non-terrestrial components, focusing only on cellular-enabled UAV communications \cite{Lin-21b}. 

\item {\textbf{Opportunities and Challenges}}: 
Incorporating non-terrestrial components as part of terrestrial networks provides many opportunities for expanding capabilities and introducing novel application scenarios. In a ground-air-space network, each layer offers its own set of benefits and challenges. The main advantage of the spaceborne layer is global-scale coverage, vastly expanding the capabilities of traditional mobile networks. Other benefits include robustness against security attacks and natural disasters. For mobile networks, satellite-assisted communication enables a wide range of new application scenarios, such as high-speed remote area connectivity, global IoT, worldwide environmental/industrial monitoring, high-quality maritime and aeronautical communications, and remote area emergency/safety/disaster communications. 

In the airborne layer, HAPSs can provide wide area coverage and high-speed connectivity for diverse types of environments, from urban and rural to remote and disaster areas. Compared to satellites, HAPSs offer much smaller coverage. However, they are cheaper and faster to deploy, which makes them more flexible network nodes. In the lower airborne layer, UAVs can provide dynamic improvements to system-level performance, particularly in terms of capacity and coverage. Potential application scenarios include flexible support for temporal hotspots and mass events, communication for disaster areas, and coverage for outage/remote areas. Due to their numerous benefits and potential applications, NTNs are expected to become an integral part of 6G networks. 

Integrating non-terrestrial communications into mobile networks brings many challenges, such as satellite integration at a reasonable cost and QoS, network management in a heterogeneous 3D network architecture, realistic channel modeling, and regulatory aspects. In terms of satellite integration, LEO and mini-satellites are promising candidates for 6G due to their reasonable balance between cost and performance. Satellites in the lower orbits have better propagation features, shorter delays, and cheaper prices than those in the higher orbits. A heterogeneous 3D network architecture with space-air-ground layers makes network and resource management challenging, calling for advanced cooperation mechanisms between different layers. AI/ML is a promising tool for addressing diverse network management issues. Realistic channel models are needed for appropriate performance evaluation in the 6G-specific application scenarios. Regulatory challenges arise when aiming at global coverage. Regulations in different countries and international waters must be considered when designing INTNs. The main challenges of INTNs were discussed in \cite{Liu-18, Geraci-23}. 

\item {\textbf{Literature and Future Directions}}: 
In the literature, INTNs have received a considerable amount of attention since the mid-2010s, becoming one of the key topics in 6G research. Numerous survey articles have recently been published \cite{Lin-21b, Azari-21, Araniti-21, Giordani-21, Ray-22, Cheng-22b, Geraci-23, Fontanesi-23, Iqbal-23, Ozger-23, Fu-23, Saad-24, Majamaa-24, Mahboob-24, Nguyen-24, Shahid-24}. They explored a broad variety of INTN aspects. The work in \cite{Lin-21b} reviewed 5G NTNs in 3GPP, covering radio access, systems, services, protocols, and IoT. In \cite{Azari-21}, the authors surveyed the evolution of NTNs from 5G to 6G networks. NTNs were discussed from the perspectives of 5G, mmWave, IoT, multi-access edge computing (MEC), AL/ML, higher layers, field trials, industry progress, and 6G. NTNs were explored toward 6G in \cite{Araniti-21}. In \cite{Giordani-21}, a survey was conducted on NTNs in the 6G era. The focus was on enabling technologies, open issues, and a case study. 6G SAGIN was studied in \cite{Ray-22}. The main topics included space and mobile networking, key enablers, UAV-as-a-service, design aspects, applications, challenges, and future research avenues. 

In \cite{Cheng-22b}, 6G service-oriented SAGIN was reviewed in terms of applications, requirements, resource management, cloud-edge synergy, and future research. The study in \cite{Geraci-23} focused on INTNs, discussing standardization, architecture, use cases, opportunities, challenges, and future directions. In \cite{Fontanesi-23, Iqbal-23, Mahboob-24}, AI for NTNs was reviewed. The authors in \cite{Ozger-23} provided a 6G connected sky vision on the integration of terrestrial and non-terrestrial components, discussing use cases, architecture, and network design. The work \cite{Fu-23} examined satellite-terrestrial convergence from 5G to 6G. In \cite{Saad-24}, the authors reviewed recent advances of NTNs in 5G, focusing on 3GPP Release 17 and 18. Multi-connectivity was explored for beyond 5G NTNs in \cite{Majamaa-24}. The paper \cite{Nguyen-24} studied AI/ML, network slicing, and O-RAN technologies to facilitate the design of 6G NTNs from the academia and industry perspectives. In \cite{Shahid-24}, the authors discussed the challenges of radio access technologies for 6G NTNs, focusing on the waveform design, spectrum coexistence, and radio resource management. 

To free the potential of INTNs, major efforts are still required in the interdisciplinary research and development work. Important future research directions include 3D network management, cooperation between layers, performance-cost trade-offs, network selection, traffic offloading, resource allocation, multi-connectivity, mobility management, distribution of computation/caching resources across layers, realistic channel modeling, and practical experiments \cite{Liu-18, Geraci-23}. Further details on these research aspects can be found in \cite{Liu-18, Geraci-23}. 
\end{itemize}

\begin{figure}[!htb]
\center{\includegraphics[width=0.8\columnwidth]
{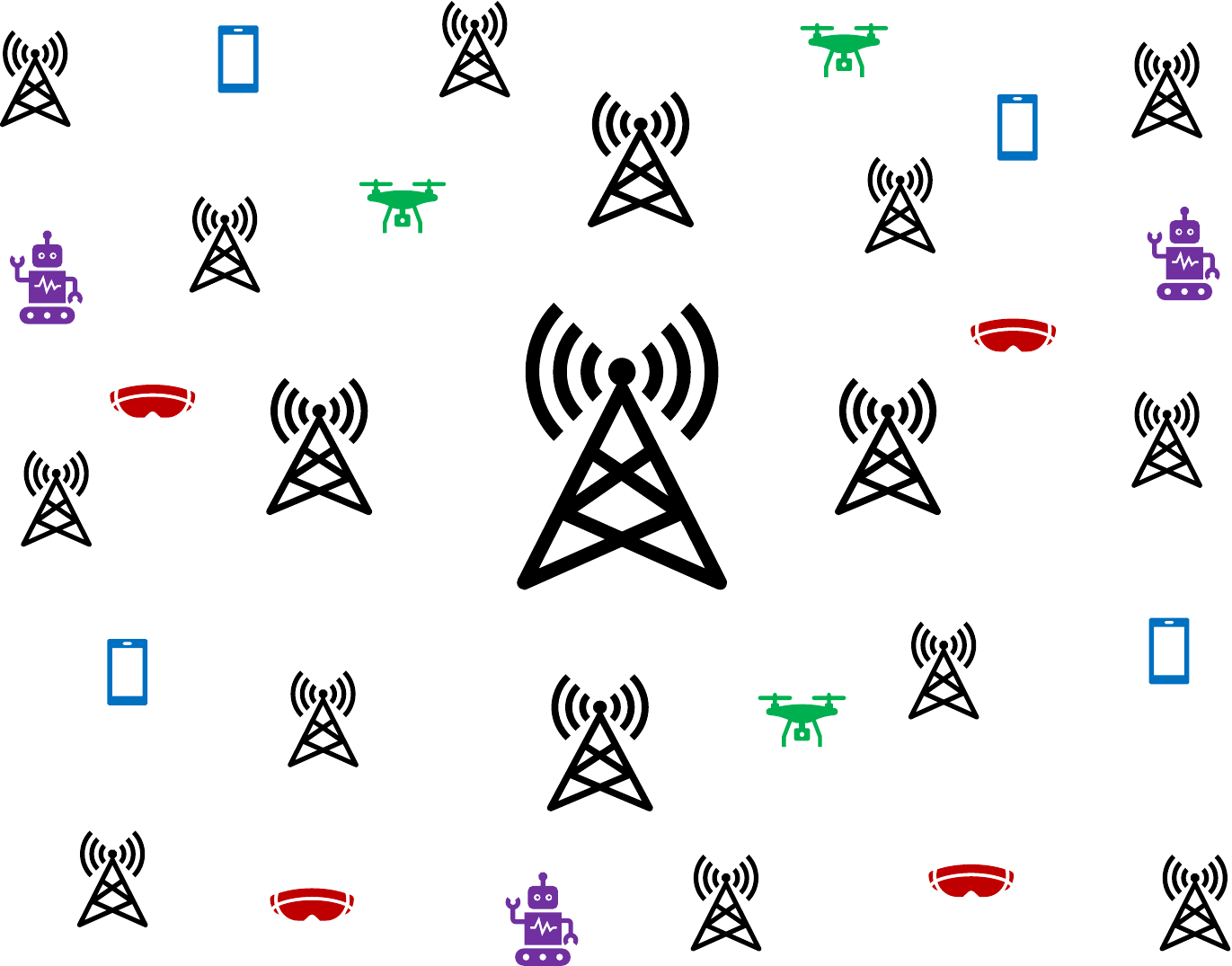}}
\caption{\label{Fig_UDN}A heterogeneous ultra-dense network.}
\end{figure}

\subsubsection{ULTRA-DENSE NETWORKS}

\begin{itemize}
\item {\textbf{Vision}}: 
UDNs are considered a key element in meeting the extreme capacity needs of 6G networks, especially via dense THz cell deployments. 

\item {\textbf{Introduction}}: 
UDNs refer to extremely densified wireless networks, where there are a massive number of heterogeneous access nodes and end-devices. The typical access nodes of next-generation heterogeneous UDNs range from macro and small/tiny BSs to roadside units and aerial/satellite relay nodes \cite{Stoynov-23}. End-devices can vary from mobile devices and IoT sensors to vehicles, drones, and robots \cite{Stoynov-23}. A schematic illustration of a heterogeneous UDN is shown in Figure \ref{Fig_UDN}. Network densification is a means to correspond to the exponential growth of mobile data traffic and the constant need for more capacity. Higher-frequency communication, with larger bandwidths and shorter link distances, naturally lends itself to the network densification paradigm, providing extreme capacities and data rates in ultra-dense deployments. Network densification brings access nodes and end-devices closer to each other, leading to lower propagation losses, reduced transmission powers, and improved received signal qualities. In general, network densification aims to improve capacity, data rates, spectral efficiency, coverage, and energy efficiency \cite{Mughees-21, Stoynov-23}. On the other hand, densification makes the management of the network more challenging, especially in terms of resource, interference, and mobility management \cite{Mughees-21}. 

\item {\textbf{Past and Present}}: 
In the history of mobile communication, network densification has always been one of the key ways to satisfy the ever-increasing demand for higher network capacity \cite{Stoynov-23}. Each generation, from 1G to 5G, has developed denser and denser networks with a more heterogeneous nature. In 5G networks, the main types of access nodes include macro and small cell BSs \cite{Adedoyin-20}. High-power macro BSs typically operate at sub-6 GHz frequencies, and provide a relatively large coverage, with efficient mobility support. Lower-power small BSs, with limited coverage and mobility support, offer a higher capacity, primarily relying on mmWave communications. 5G also supports a heterogeneous network architecture, with the emerging rise of IoT, vehicular, and non-terrestrial communications. 6G will continue the trend of developing denser and more heterogeneous networks by expanding to higher operating frequencies with denser deployments and aiming to fully integrate space, air, mobile, vehicular, and IoT components into the network architecture. 

\item {\textbf{Opportunities and Challenges}}: 
The potential of UDNs is to fulfill the extreme network capacity demands of the 6G era. In particular, ultra-dense THz cell deployments enable novel data-hungry multimedia applications in densely populated areas. Although UDNs hold great potential, they also pose major challenges. A major limiting factor of the network capacity in UDNs is interference \cite{Tinh-22}. Hence, it is of the utmost importance to develop efficient interference management methods to free the potential of UDNs and achieve the desired capacity gains. There are three main types of interference in UDNs: inter-tier, inter-cell (inter-tier), and intra-cell interference (intra-tier) \cite{Tinh-22}. The key to managing interference is cooperation among neighboring/interfering cells. Cooperation requires computation power, information sharing, and additional signaling, increasing the demands of computation hardware and backhaul/fronthaul links. 

Another challenge is the mobility management \cite{Tinh-22}. Due to smaller cells, handovers are more frequent for mobile end-devices. Handovers require network control, end-device assistance, and communication resources from the source and target cells. Thus, more frequent handovers consume more computational and communication resources, leading to a decrease in capacity and an increase in latency, energy consumption, and computational complexity. Resource management is also more challenging for UDNs \cite{Mughees-21}. For example, efficient load-balancing mechanisms are required to prevent service imbalances and user fairness issues. Backhauling may also become a bottleneck in an ultra-dense deployment \cite{Adedoyin-20}. Because conventional fiber links are costly and time-consuming to install, the deployment of a large number of access nodes is very expensive and slow. In this respect, the ultra-dense integrated wireless access and backhaul concept is a promising solution due to its flexibility, low cost, and fast installation. 

\item {\textbf{Literature and Future Directions}}: 
In the literature, UDNs have been extensively explored during the past decade \cite{Kamel-16, Mughees-21, Stoynov-23}. The original driver behind the increased research interest was 5G mmWave communications with dense small-cell deployments \cite{Chen-16}. Popular research directions for UDNs have been resource allocation, interference management, user association, energy efficiency, mobility/handover management, and backhauling \cite{Kamel-16, Adedoyin-20, Mughees-21, Stoynov-23}. Further research is still needed to address these fundamental challenges. Recently, the research focus has shifted to beyond 5G UDNs \cite{Andreev-19, Tinh-22}. Consequently, the variety of studied topics has widened, while taking the special characteristics of 6G networks into account. Accordingly, interesting future topics include the exploration of UDNs from the perspective of AI/ML assistance, THz operation, optical wireless, ultra-massive IoT, vehicular networks, ultra-dense IAB, and cell-free network design. In particular, AI/ML is applicable to solve many complex network management problems in the design of UDNs \cite{Mughees-21, Tinh-22, Stoynov-23}. Further details on UDNs can be found in recent surveys that have reviewed the latest literature, research achievements, open issues, and future directions \cite{Adedoyin-20, Mughees-21, Tinh-22, Stoynov-23, Ngo-24}. 
\end{itemize}

\begin{figure}[!htb]
\center{\includegraphics[width=\columnwidth]
{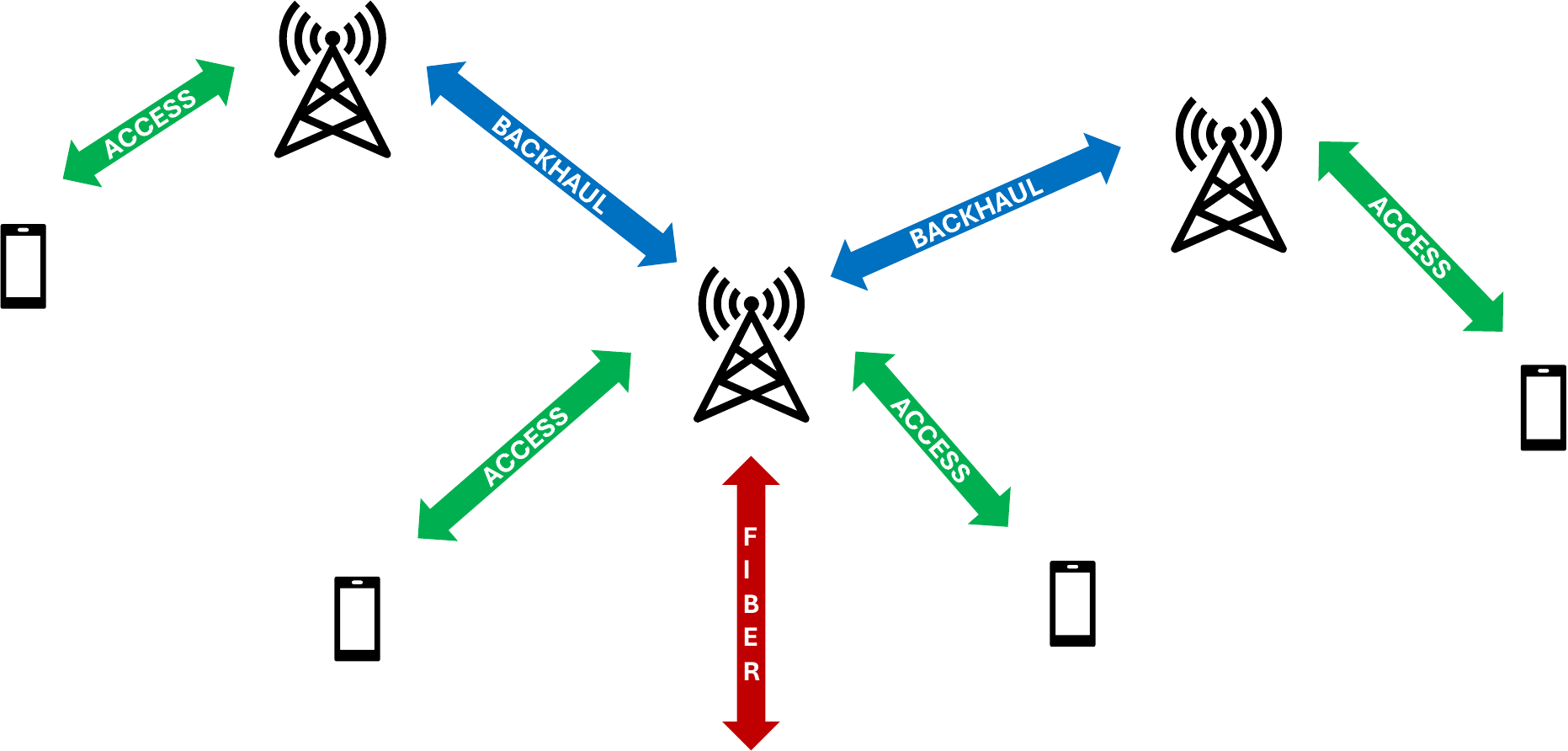}}
\caption{\label{Fig_IAB}Integrated access and backhaul.}
\end{figure}

\subsubsection{INTEGRATED ACCESS AND BACKHAUL}

\begin{itemize}
\item {\textbf{Vision}}: 
While IAB is already part of 5G standards, its role in 6G will grow due to continuing cell densification. Since IAB is a cheaper and faster alternative to fiber links, it is expected to become an integral element in ultra-dense 6G networks.

\item {\textbf{Introduction}}: 
IAB is a concept in which the same radio resources of a BS are used for both wireless access and backhaul \cite{Zhang-21k}, as illustrated in Figure \ref{Fig_IAB}. IAB is a flexible, agile, and low-cost method to provide backhauling, being a promising alternative to fiber links, especially in dense networks due to faster and cheaper installation \cite{Zhang-21k}. 

\item {\textbf{Past and Present}}:
IAB was first standardized in 4G under the name LTE relaying \cite{Madapatha-20}. However, this approach has not achieved much success owing to scarce spectral resources \cite{Madapatha-20}. IAB started to gain considerable interest in the mid-2010s, with a special focus on 5G mmWave networks. Compared to 4G, 5G is more suitable for the widespread use of IAB due to dense small-cell deployments, mmWave spectrum, and massive MIMO technology \cite{Rajatheva-21}. IAB was first standardized in 5G NR Release 16 and further enhanced in Release 17 \cite{Zhang-21k}. Overviews of NR IAB can be found in \cite{Polese-20c, Madapatha-20, Zhang-21k}. 

\item {\textbf{Opportunities and Challenges}}: 
Commonly used optical fiber backhaul links are very fast and reliable, but they are also costly and time-consuming to install \cite{Rajatheva-21}. In this context, IAB is a promising alternative to fiber links due to its lower cost, higher flexibility, and faster deployment \cite{Zhang-21k}. The role of IAB will become more important in the future since mobile networks will become denser in the 6G era. Thus, IAB is expected to play a key role in ultra-dense 6G networks operating at mmWave/THz frequencies \cite{Zhang-21k, Rajatheva-21}. Other possible application scenarios for IAB include UAV-assisted networks, RIS-empowered networks, optical wireless networks, cache-enabled networks, and NTNs \cite{Zhang-21k}. 
 
The main challenges in IAB are related to resource and interference management \cite{Zhang-21k}. For example, resource allocation between access and backhaul is not a trivial task depending on the traffic loads at the source and destination BSs. If too much resources are allocated to either side, system performance degrades. AI/ML is a promising tool to facilitate such resource allocation problems by predicting traffic loads and capacity demands. Another fundamental challenge is the mitigation of interference. In IAB, two types of interference exist \cite{Zhang-21k}: interference between access and backhaul (i.e., inter-technology interference) and interference within access and backhaul (i.e., intra-technology interference). The former is a (rather) new type of interference in mobile networks, which makes interference management more challenging. This calls for novel interference mitigation mechanisms, possibly AI/ML-assisted. Integrating IAB with other emerging 6G technologies, such as NTNs, RISs, OWC, and caching, opens up new opportunities for IAB, but also introduces a new set of challenges \cite{Zhang-21k}. 

\item {\textbf{Literature and Future Directions}}: 
In recent years, IAB has been extensively studied \cite{Zhang-21k}, considering diverse topics from resource and interference management to network topology optimization and end-to-end performance evaluations. A comprehensive overview of the IAB literature was provided in \cite{Zhang-21k}. As the standardization work on NR IAB started in the late 2010s, a special research focus has been directed toward 5G mmWave networks \cite{Cudak-21}. Further information on NR IAB and the related literature can be found in \cite{Polese-20, Madapatha-20}. Recent IAB topics on 5G evolution and beyond include UAV-aided IAB \cite{Tafintsev-20}, AI/ML-aided IAB \cite{Lei-20}, cache-aided IAB \cite{Zhang-22e}, RIS-aided IAB \cite{Fiore-21}, mmWave IAB \cite{Sadovaya-22}, and cell-free IAB \cite{Jazi-23}. Further discussion of the latest research areas has been provided in \cite{Zhang-21k}. 

Future IAB research can be divided into two main areas: practical enhancements to 5G NR IAB and IAB solutions beyond 5G. In the former category, essential research directions include enhancements to beamforming, interference management, resource allocation, routing, and topology optimization. For instance, massive MIMO is a potential technology to enhance NR IAB such that the same frequency-time units can be used for access and backhaul links, leading to improved spectral efficiency. Hybrid beamforming is a particularly promising method due to its good balance between performance and implementation complexity. In the latter category, it is important to study the synergy between IAB and other emerging 6G technologies, such as THz IAB, AI/ML-aided IAB, RIS-aided IAB, UAV IAB, cache-enabled IAB, and optical IAB \cite{Zhang-21k}. Since 6G is expected to enter the THz spectrum, a special research focus needs to be directed toward ultra-dense 6G IAB networks operating at THz frequencies. 
\end{itemize}

\begin{figure}[!htb]
\center{\includegraphics[width=0.85\columnwidth]
{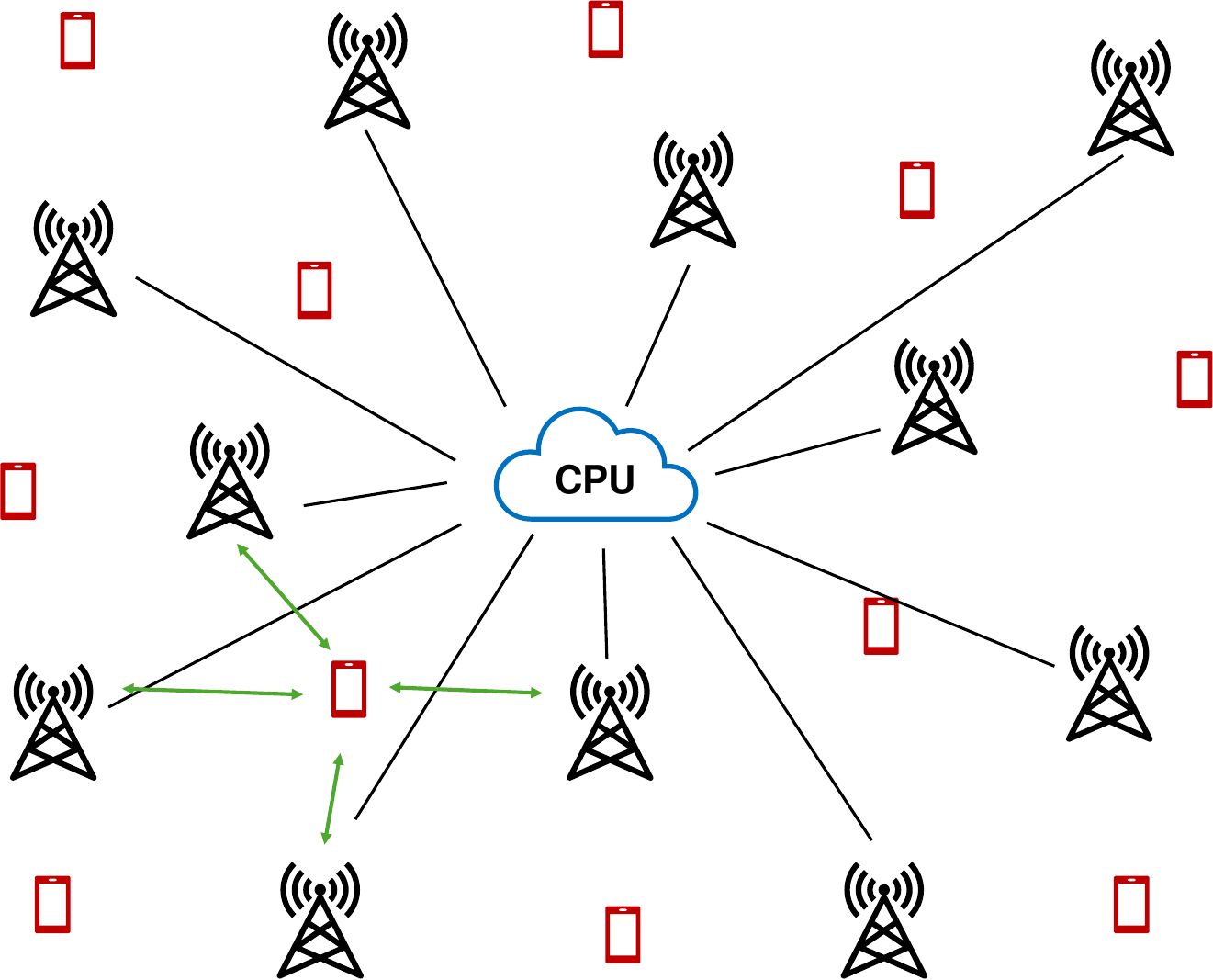}}
\caption{\label{Fig_CFmMIMO}Cell-free massive MIMO \cite{Maryopi-19}.}
\end{figure}

\subsubsection{CELL-FREE MASSIVE MIMO}

\begin{itemize}
\item {\textbf{Vision}}:
Cell-free massive MIMO is a promising technology to complement the traditional cell-based network design in the 6G era. It is applicable to scenarios that require stable QoS levels over the entire network coverage. 

\item {\textbf{Introduction}}: 
Cell-free massive MIMO aims at providing stable performance across the network coverage area, thus alleviating the traditional problem of large performance discrepancies between the cell-center and cell-edge regions \cite{Demir-21, Rajatheva-21}. The main idea is to use lots of distributed low-cost APs with a few antennas each instead of having a lower number of traditional APs with a large number of co-located antennas \cite{Demir-21, Rajatheva-21}. Each device in the network area is jointly served by its nearby APs. Cooperation between APs is enabled by the central processing units (CPUs) and fronthaul links. 

Compared with the traditional cell-based network design, cell-free massive MIMO provides shorter link distances, lower propagation losses, reduced transmission powers, alleviated interference profiles, and improved signal qualities \cite{Demir-21, Rajatheva-21}. Consequently, cell-free operation leads to improved stability of the user-experienced performance across the network coverage area and increased system-level energy efficiency. Cell-free design is applicable to scenarios in which there is a need for steady QoS within the entire network. Scalability is a typical problem in cell-free network architectures \cite{Bjornson-20c}. A schematic example of a cell-free massive MIMO network is illustrated in Figure \ref{Fig_CFmMIMO}. 

\item {\textbf{Past and Present}}:
The concept of cell-free massive MIMO was introduced in 2015 \cite{Ngo-15, Nayebi-15}. The proposed idea, that is, a large number of distributed APs jointly serving users in the network area \cite{Ngo-17, Nayebi-17}, has flavors from three different concepts, namely massive MIMO, network MIMO, and UDNs \cite{Demir-21}, as shown in Figure \ref{Fig_CFmMIMO2}. Since the total number of antennas serving each user is large, the PHY layer technologies of massive MIMO are applicable. The foundation of a cell-free operation that jointly serves each user via many APs is based on the principles of network MIMO (i.e., joint processing CoMP). A large number of geographically distributed APs in a relatively small area reflects the nature of UDNs. The challenge lies in the co-design of these three elements to obtain a cell-free operation with sufficient scalability for practical implementations \cite{Demir-21}. 

\begin{figure}[!tb]
\center{\includegraphics[width=0.6\columnwidth]
{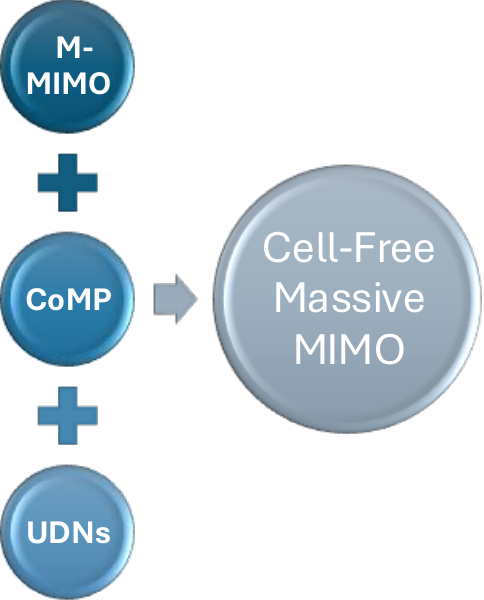}}
\caption{\label{Fig_CFmMIMO2}Three technologies behind cell-free massive MIMO \cite{Demir-21}.}
\end{figure}

In the original concept, all distributed APs are assumed to jointly serve each user in the network area, leading to scalability issues due to the increasing computational complexity and signaling overhead as the number of APs and users grows, making the concept infeasible for practical large-scale implementations. To solve scalability issues, a refined version of the cell-free operation was developed, known as user-centric cell-free massive MIMO \cite{Buzzi-17}, in which each user is served by a limited set of nearby APs. Clustering APs for individual users makes the network more scalable, leading to better suitability for large-scale deployments. In \cite{Bjornson-20c}, the user-centric approach was further enhanced by applying a clustering framework from the network MIMO literature, known as dynamic cooperation clustering. This clustering concept has been proven scalable for any network size \cite{Bjornson-20c}. The latest details of cell-free massive MIMO research have been discussed in \cite{Demir-21, Elhoushy-22, Ammar-22, Zheng-24, Ngo-24}. Cell-free massive MIMO is expected to be utilized in 6G to complement the traditional cell-based network architecture in scenarios that require stable QoS in the network coverage area \cite{He-21, Zheng-24, Ngo-24}. 

\begin{table*}[htb!]
\begin{center}
\caption{Summary of network intelligence technologies for 6G}
\label{Table_NI}
\centering
\begin{tabularx}{\textwidth}{| >{\centering\arraybackslash}X | >{\centering\arraybackslash}X |
>{\centering\arraybackslash}X | 
>{\centering\arraybackslash}X |
>{\centering\arraybackslash}X |
>{\centering\arraybackslash}X |
>{\centering\arraybackslash}X |
>{\centering\arraybackslash}X |}
\hline
\centering
\vspace{3mm} \textbf{Network Intelligence Technologies} \vspace{3mm} & \centering \textbf{Vision} & \centering \textbf{Description} & \centering \textbf{Opportunities} & \centering \textbf{Challenges} & \centering \textbf{Past} & \vspace{1.5mm} \begin{center} \textbf{Present} \end{center} \\
\hline
\vspace{3mm} Intelligent Core \vspace{3mm}  & AI-native core & AI-enhanced core & \vspace{3mm} Enhanced network management \vspace{3mm} & E2E optim $\&$ cloud-edge coop  & Cloud-native 5G & AI/ML in 5G-Advanced \\
\hline
\vspace{3mm} Intelligent Edge \vspace{3mm}  & AI-native edge & AI-enhanced edge & \vspace{3mm} Enhanced edge management \vspace{3mm} & Infra $\&$ AI algs $\&$ data acquisition & Edge computing in 5G & AI/ML in 5G-Advanced \\
\hline
\vspace{3mm} Intelligent Air Interface \vspace{3mm}  & AI-native air interface & AI-enhanced air interface & Enhanced PHY/MAC & Fast changing ch conditions & Flexible air interface in 5G  & AI/ML in 5G-Advanced \\
\hline
\end{tabularx}
\end{center}
\end{table*}

\item {\textbf{Opportunities and Challenges}}: 
The key advantages of cell-free massive MIMO are stable QoS across the entire network coverage area, energy- and cost-efficient deployment, macro diversity, and favorable propagation. Cell-free network design is particularly applicable to scenarios where there is a need for robust and steady performance over the network, such as industrial environments. Cell-free massive MIMO is expected to serve as a complementary technology to conventional cellular network topology in the 6G era \cite{Zheng-24, Ngo-24}. However, this vision is still far from reality, as the research and development work of cell-free massive MIMO has mainly been theoretically-oriented. Hence, there is a lack of practical performance validations, experiments, and field-trials. Before practical large-scale realizations are possible, the fundamental problems must be resolved. 

The basic issue in cell-free operation is scalability \cite{Bjornson-20c}, as discussed earlier. This problem can be solved by dynamic user-centric clustering, which limits the computational burden in the processing units and signaling overhead in the fronthaul/backhaul links \cite{Bjornson-20c}. Therefore, clustering plays a critical role. The number of APs in a cooperative cluster cannot be too large since the complexity increases significantly. Cluster sizes cannot be too small either due to the increased interference levels, degrading network performance. Another fundamental issue in cell-free massive MIMO is synchronization. A synchronized network is required because each user is jointly served by multiple APs via coherent transmission. Inaccurate synchronization leads to degraded performance. Therefore, proper synchronization mechanisms are pivotal. 

Since cell-free architecture is a new way to construct a network, resource allocation strategies must be redesigned accordingly. Mobility is also problematic since it leads to rapid chances in clustering. Theoretical channel models are commonly used in the performance evaluation of the proposed cell-free solutions, leading to unrealistic results. Practical channel models based on the dedicated channel measurements are required to achieve more realistic results and more useful performance analyses. Prototyping and field trials are required to shift cell-free massive MIMO from a theoretical concept to a practical network architecture. More detailed discussions on the fundamental challenges of cell-free networks can be found in \cite{Demir-21, Elhoushy-22, Ammar-22, Zheng-24, Ngo-24}. 

\item {\textbf{Literature and Future Directions}}: 
In the literature, cell-free massive MIMO has been actively researched since the mid-2010s. The concept has been thoroughly reviewed in recent surveys \cite{He-21, Demir-21, Elhoushy-22, Ammar-22, Chen-22b, Demir-24, Shi-24, Zheng-24, Ngo-24}. A broad range of aspects was discussed, including theoretical foundations, network scalability, dynamic user-centric clustering, synchronization, resource allocation, uplink/downlink operations, AP selection, receive combining, transmit precoding, power optimization, channel hardening, channel estimation, pilot assignment, hardware impairments, mobility issues, mmWave/THz operation, fronthauling, deployment challenges, and future research toward 6G. Although substantial research efforts have been made to date, further work is still needed to overcome the fundamental challenges discussed earlier. Moreover, promising 6G-related future topics include AI/ML for cell-free, cell-free with RISs, cell-free for IoT, cell-free in O-RAN, practical implementations, and experiments \cite{He-21, Demir-24, Shi-24, Zheng-24, Ngo-24}. 
\end{itemize}

\begin{figure}[!htb]
\center{\includegraphics[width=0.7\columnwidth]
{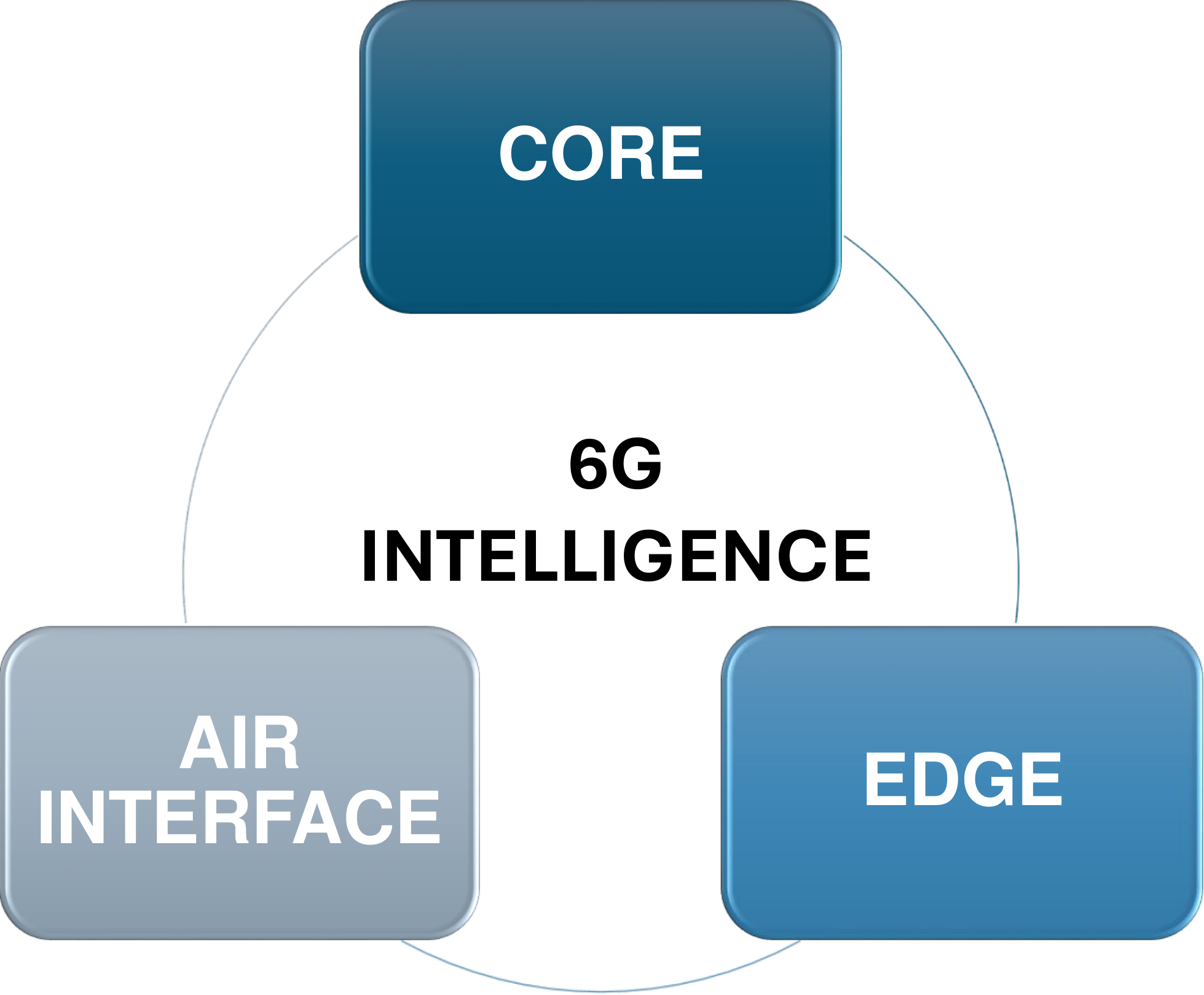}}
\caption{\label{Fig_NI}The cornerstones of 6G network intelligence.}
\end{figure}

\subsection{NETWORK INTELLIGENCE TECHNOLOGIES FOR 6G}
\label{NI}
The ultimate goal of AI/ML is to revolutionize the design, operation, and management of mobile networks by making them more intelligent, efficient, flexible, scalable, automated, proactive, economical, ecological, and secure. While this vision is currently far from reality, the first concrete steps toward it have been taken. AI/ML has been adopted in the 5G standards from Release 18 onward (i.e., 5G-Advanced). In 6G, pervasive AI/ML is expected to be a major leap toward the grand vision. This section discusses how AI/ML can be exploited at different levels of the network, i.e., the core, edge, and air interface. These are the cornerstone of 6G network intelligence, as illustrated in Figure \ref{Fig_NI} and summarized in Table \ref{Table_NI}. 

\subsubsection{INTELLIGENT CORE}

\begin{itemize}
\item {\textbf{Vision}}:
In the core network, 6G will go beyond the softwarization of 5G by adopting intelligence via pervasive AI/ML. This can be seen as an evolution from cloud to cloud intelligence. 

\item {\textbf{Introduction}}: 
Mobile networks consist of three main components: core network, RAN, and mobile devices, as illustrated in Figure \ref{Fig_Core}. The core and RAN operate cooperatively between the Internet and mobile devices. The core network is responsible for the overall network operation, management, and security, whereas the RAN handles wireless access to serve mobile devices. For example, one of the key tasks of the core network is to guarantee secure and reliable end-to-end communication between the source and destination nodes. The current core networks are software-based and cloud-native. The next major step in the evolution of core networks is the adoption of intelligence via AI/ML in the 6G era. The intelligent network core is based on the pervasive utilization of AI/ML in order to improve the design, operation, management, maintenance, and security of the network. DL is an attractive technology for the core since it achieves excellent performance when the training datasets are large. Data acquisition is a crucial component of the intelligent core. 

\begin{figure}[!tb]
\center{\includegraphics[width=\columnwidth]
{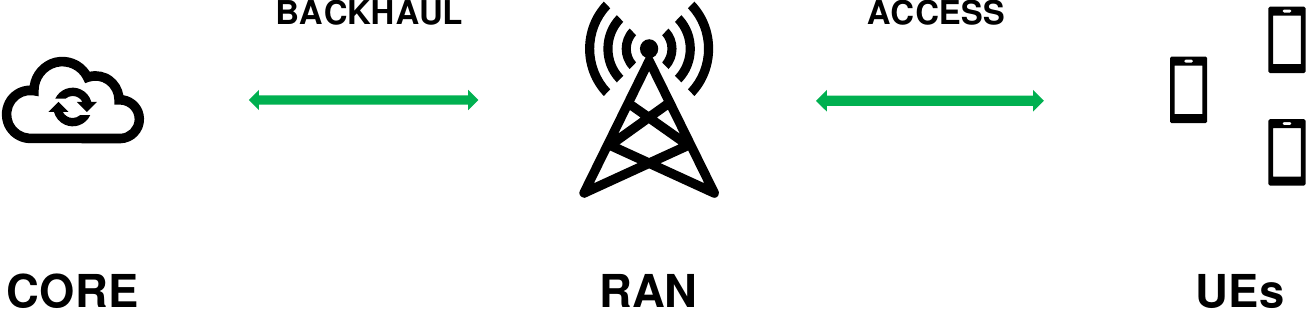}}
\caption{\label{Fig_Core}Three main elements of mobile networks.}
\end{figure}

\item {\textbf{Past and Present}}: 
Each mobile network generation has its own core network that responds to the demands and expectations of that particular generation. Since the requirements of mobile networks are constantly growing, the capabilities of the core networks evolve generation by generation. In short, the core networks have evolved from the circuit-switched network of 2G through the packet cores of 3G and 4G to the latest software/service-based architecture of 5G. Specifically, a new cloud-native service-oriented core network was developed for 5G, as defined by 3GPP in Release 15 \cite{Rel-15}. The main features of the 5G Core include SBA, SDN, NFV, and network slicing. SBA relies on a vast variety of interconnected network functions, each of which provides one or more services that are accessible by other network functions. This architecture allows the agile addition of novel functionalities. SDN enables cloud-native and flexible network management. NFV and network slicing enable the division of the network resources into independent virtual slices, each of which can be dedicated to a specific service. Virtual slicing facilitates the introduction of new services and tailored customer solutions. 

\item {\textbf{Opportunities and Challenges}}: 
While softwarization made 5G core network cloud native, AI/ML is expected to make 6G core intelligent and take mobile networks to the era of intelligence. AI/ML can potentially improve the operation, management, and security of the network. Although much research has been conducted, there are many issues to be resolved in harnessing the synergy of AI/ML and mobile networks. In \cite{Mao-18, Sun-19, Zhang-19b, Ahmad-20, Tang-21, Noman-23}, many challenges and future directions were discussed. These can be divided into several categories: infrastructure, AI/ML algorithms, network optimization, core-edge cooperation, and security. First of all, network infrastructure needs to be highly capable from the core to the edge (in terms of communication, computation, and information exchange) to enable the pervasive and efficient use of AI/ML throughout the network. There is a challenging trade-off between the cost and performance in the update of the infrastructure. 

In mobile networks, the used AI/ML algorithms need to be efficient, fast, reliable, and data-protected. This makes the design of algorithms challenging and calls for advanced AI/ML methods, inference/training approaches, and data acquisition mechanisms. There are many challenging network optimization problems that can be enhanced by AI/ML, e.g., routing, traffic control, and network slicing, to mention a few. A close core-edge cooperation and cross-layer design are required to obtain the proper end-to-end performance of the network. AI/ML-based security solutions are vital to cope with a changing threat landscape and to tackle outside attacks. Further discussion on open problems and future research opportunities can be found in \cite{Mao-18, Sun-19, Zhang-19b, Ahmad-20, Tang-21, Noman-23}. 

\item {\textbf{Literature and Future Directions}}: 
In the literature, mobile network intelligence has been extensively studied since the late 2010s, with a focus on 6G networks. Recent research progress and open issues regarding intelligent networking have been reviewed in numerous survey articles \cite{Xie-19, Sun-19, Zhang-19b, Chen-19, Ahmad-20, Tang-21, Banchs-21, Cheng-21, Wu-22c, Shen-22, Zhou-22c, Noman-23, Alhammadi-24, Celik-24}. In \cite{Zhang-19b, Chen-19, Cheng-21}, the works focused on DL, discussing its applications in wireless networks to achieve network intelligence. In \cite{Zhang-19b}, a thorough overview was provided of DL in mobile and wireless networking, with the main focus on the technological enablers of DL in networking, DL-driven network aspects, and customizing DL for networks. 

The authors of \cite{Chen-19} provided a tutorial on the applications of artificial neural networks to solve diverse issues in wireless networking. The work \cite{Cheng-21} reviewed DL for wireless network optimization, discussing universal modeling, complexity mitigation, algorithm design, and latent knowledge exploration. The surveys \cite{Xie-19, Sun-19, Ahmad-20} explored ML from a wider perspective in wireless networks. In \cite{Xie-19}, ML was discussed in the context of SDN, surveying traffic classification, routing optimization, QoS prediction, resource management, and network security. The article \cite{Sun-19} focused on ML applications for resource management, networking, mobility management, and localization. In \cite{Ahmad-20}, ML was reviewed for the PHY, MAC, and network layers, edge computing, SDN, and network security. 

In \cite{Tang-21, Banchs-21, Wu-22c, Shen-22, Zhou-22c, Noman-23, Alhammadi-24, Celik-24}, AI/ML was reviewed in the context of 6G networks. In \cite{Tang-21}, ML was considered for intelligent end-to-end network optimization in various layers toward 6G. The focus was on MAC layer network access, network layer routing, network layer traffic control, and application layer streaming adaptation. The paper \cite{Banchs-21} studied network intelligence in 6G, discussing its role, challenges, architecture, and orchestration possibilities. In \cite{Wu-22c}, the authors explored intelligent network slicing for 6G networks, introducing an AI-native slicing architecture as well as discussing AI for slicing and slicing for AI. 

The tutorial in \cite{Shen-22} presented an architectural framework for 6G holistic network virtualization and pervasive network intelligence, reviewing also topics like the interplay between network slicing and digital twins, AI for networking, and networking for AI. In \cite{Zhou-22c}, the work introduced a concept of intelligence-endogenous networks for 6G, utilizing AI and knowledge graph technologies. The article \cite{Noman-23} surveyed ML-based resource management for 6G networks. The covered aspects included resource allocation, task offloading, mobility management, energy efficiency maximization, and latency minimization. In \cite{Alhammadi-24}, the authors provided a comprehensive survey on AI for 6G networks, focusing on AI-aided technologies and applications. The work \cite{Celik-24} discussed generative AI for 6G wireless intelligence. Future research directions were discussed earlier in the opportunities and challenges part of this section. 
\end{itemize}

\begin{figure}[!htb]
\center{\includegraphics[width=0.65\columnwidth]
{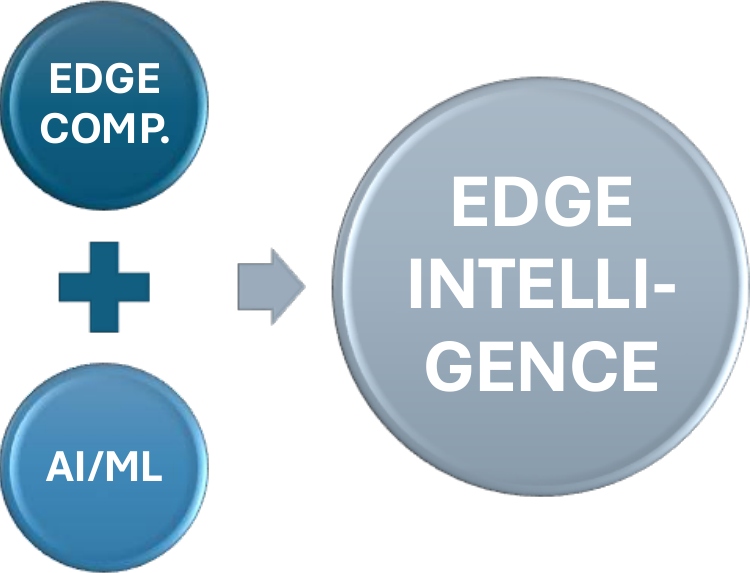}}
\caption{\label{Fig_EI2}Two technologies behind edge intelligence.}
\end{figure}

\subsubsection{INTELLIGENT EDGE}

\begin{itemize}
\item {\textbf{Vision}}: 
Intelligent edge is anticipated to become an integral part of 6G networks, enabling novel services and applications with reduced latency and communication overhead. 

\item {\textbf{Introduction}}: 
Intelligent edge, commonly known as edge intelligence (EI), refers to a combination of edge computing and AI \cite{Zhou-19}, as shown in Figure \ref{Fig_EI2}. EI benefits from the synergy of both technologies. Edge computing drives computational capabilities, tasks, and applications from the core to the network edge \cite{Zhou-19}. The network edge consists of different types of edge nodes, typical ones being edge servers and end-devices (e.g., mobile and IoT devices) \cite{Zhou-19}. The edge computing paradigm shift, which moves computation resources closer to the edge information sources, naturally leads to reduced latency and communication overhead. 

AI, the other part of EI, is typically based on ML methods, which consist of two main phases, including model training and inference \cite{Zhou-19}. In the training phase, the AI/ML model is optimized for the given objective using training datasets that aim to represent a comprehensive set of practical data realizations. In the inference phase, the trained AI/ML model is executed by feeding it the real-world input data, acquired from the information sources, and obtaining an output, optimized according to the target objective through the knowledge learned in the training process. DL, the most popular AI/ML method, has been recognized as a natural fit for edge computing, achieving excellent performance with large datasets. FL is another promising AI/ML method for the network edge, particularly targeted for distributed optimization at the end-devices. 

As shown in Figure \ref{Fig_EI}, there are six levels of EI based on where AI/ML training and inference are performed in the cloud-edge-device hierarchy \cite{Zhou-19}: cloud-edge co-inference and cloud training, in-edge co-inference and cloud training, on-device inference and cloud training, cloud-edge co-training and inference, all in-edge, and all on-device. As the levels increase, less computation offloading and information exchange are required. The protection of data privacy is also facilitated. However, the network is disposed to increased computation latency and energy consumption issues due to the limited computation and energy resources of the edge devices. It is dependent on the application which level is the best fit for it. 

\begin{figure}[!tb]
\center{\includegraphics[width=\columnwidth]
{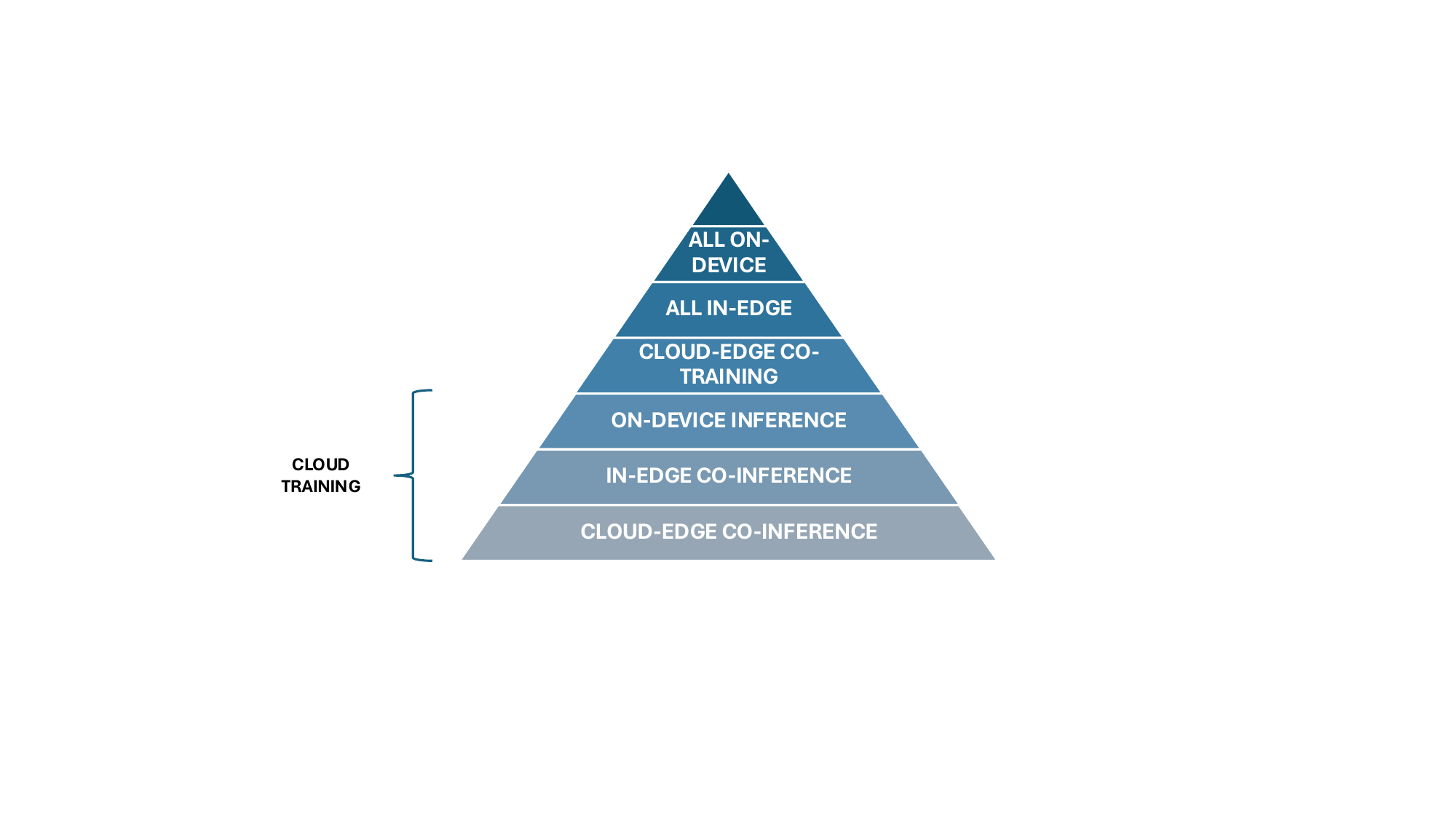}}
\caption{\label{Fig_EI}Six levels of edge intelligence \cite{Zhou-19}.}
\end{figure}

\item {\textbf{Past and Present}}: 
While EI is a new concept, its components, AI and edge computing, have been studied separately in the past. AI research started already in the 1950s, with ups and downs on the road, until it reached a level of constant progress through the introduction of DL in the 2000s. Due to advances in other key technologies, AI boomed in the 2010s, reaching also mobile networks. AI was introduced for 5G in 3GPP Release 18 \cite{Rel-18}. The history of AI was discussed in more detail in Section \ref{AI}. Although it took decades for AI to mature, edge computing has evolved rapidly. Specifically, edge computing became a popular topic in the 2010s, attracting wide interest in academia and industry. Consequently, edge computing has been supported by 5G since Release 15 \cite{Rel-15}, with constant enhancements in Release 16 \cite{Rel-16}, 17 \cite{Rel-17}, and 18 \cite{Rel-18}. 3GPP is continuing its work on edge computing toward EI. The term EI was first introduced in the white paper of the Gartner Hype Cycle in 2018 \cite{Zhou-19}. Since then, EI has been one of the key 6G topics in academia and industry, spurring a broad range of publications, projects, collaborations, and development efforts. 

\item {\textbf{Opportunities and Challenges}}: 
Combining edge computing and AI/ML is a natural step in the evolution of mobile networks toward pervasive network intelligence. Edge computing offers a computation- and data-intensive platform, with the beneficial features of low latency, reduced communication overhead, high energy efficiency, protected privacy, and context awareness. AI has the ability to free the potential offered by edge computing by optimizing the network edge (e.g., performance, efficiency, adaptability, automation, and security/privacy) and enabling a vast range of disruptive applications. Due to its versatile benefits, EI is expected to play a major role in the 6G network intelligence. To realize large-scale EI in practice, fundamental challenges must be addressed and resolved. At a high level, the main challenges are related to the computation infrastructure, AI/ML models/algorithms, data availability, network management, real-time processing, network security, and privacy protection. 

First, large-scale EI requires a ubiquitous computation infrastructure, with powerful computation and storage capabilities at the edge servers, which is a highly non-trivial trade-off between cost and performance. Given the stringent performance requirements of 6G, particularly in time-sensitive and mission-critical applications, real-time processing in AI/ML-driven network management is crucial to ensure fast and adaptable decision-making and network operations. This emphasizes the need for powerful computing, fast AI/ML algorithms, and efficient data acquisition mechanisms at the network edge. For each EI purpose/task, a dedicated AI/ML approach must be developed. Since EI will change the threat landscape, novel security approaches are required. Due to massive amounts of generated data, EI will be vulnerable to privacy threats, calling for advanced privacy protection techniques. For the success of EI, all the aforementioned issues must be properly addressed. Comprehensive discussions of open challenges and future directions can be found in \cite{Zhou-19, Deng-20, Wang-20, Xu-21, Joshi-23}. 

\item {\textbf{Literature and Future Directions}}: 
While edge computing has been a well-studied subject, its combination with AI introduced a new paradigm. In the literature, the earliest studies on EI are from the late 2010s. The research boomed at the beginning of the 2020s. Over the years, EI has been explored from diverse perspectives, such as AI models and algorithms, end-to-end architectures and performance, security and privacy approaches, resource management techniques, and application requirements with technical enablers. The latest progress in these directions and the corresponding literature have been reviewed in numerous surveys \cite{Zhou-19, Deng-20, Wang-20, Xu-21, Joshi-23, Xiao-20, Letaief-22, Zhang-20, Amin-21, Qi-21, Yang-21, Zhou-21, McEnroe-22, Lim-22, Mao-23, Alquraan-23, Lin-24, Yang-24}. These surveys can be divided into generic, 6G-oriented, and specialized categories. The generic and specialized surveys considered EI from broad and narrow perspectives, respectively, whereas the 6G-oriented surveys focused on EI in 6G. 

The generic EI surveys discussed the basics, key enablers, potential applications, open challenges, and future research opportunities. In the earliest and most popular survey on EI \cite{Zhou-19}, the authors focused on DL model training and inference at the network edge. In \cite{Deng-20}, AI was discussed for edge and on edge. The paper \cite{Wang-20} provided a thorough overview of the combination of edge computing and DL, covering a vast variety of topics. In \cite{Xu-21}, an intelligent edge was rigorously surveyed in terms of caching, training, inference, and offloading. The authors of \cite{Joshi-23} reviewed all in-edge DL, with the main focus on computation architectures, technological enablers, model adaption, and key performance metrics. 

The 6G-oriented surveys envisioned EI as a key element of 6G networks. In \cite{Xiao-20}, self-learning EI was discussed for 6G, also introducing a self-learning architecture, with a case study to confirm its effectiveness. The paper \cite{Letaief-22} surveyed trustworthy and scalable edge AI for 6G in terms of communication efficient training and inference, resource allocation techniques, end-to-end architecture, standardization (computing and learning), hardware and software platforms, and applications (IoT, healthcare, and vehicles). In \cite{Lin-24}, the authors explored split learning for 6G edge networks. Each specialized survey focused on exploring EI from a narrow perspective, including IoV \cite{Zhang-20}, IoT in healthcare \cite{Amin-21}, vehicular systems in 6G \cite{Qi-21}, autonomous driving in 6G \cite{Yang-21}, on-device learning systems \cite{Zhou-21}, UAVs \cite{McEnroe-22}, metaverse \cite{Lim-22}, security/privacy \cite{Mao-23}, federated learning \cite{Alquraan-23}, and reinforcement learning \cite{Yang-24}. 

Further studies and development efforts are needed to reach the required level of maturity for the standardization and commercialization of EI in mobile networks. The previously discussed challenges need to be properly solved in the near future. Future research topics have been discussed in \cite{Zhou-19, Deng-20, Wang-20, Xu-21, Joshi-23}. 
\end{itemize}

\begin{figure}[!htb]
\center{\includegraphics[width=0.85\columnwidth]
{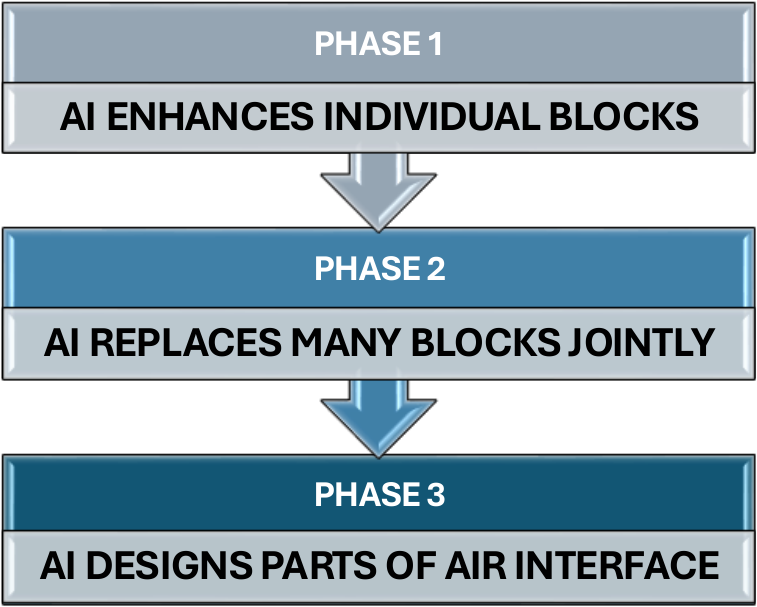}}
\caption{\label{Fig_IA}Three evolution phases toward AI-native air interface \cite{Hoydis-20}.}
\end{figure}

\subsubsection{INTELLIGENT AIR INTERFACE}

\begin{itemize}
\item {\textbf{Vision}}:
The air interface of 6G networks is expected to be enhanced by AI/ML, potentially revolutionizing the design, standardization, and implementation of mobile networks. 

\item {\textbf{Introduction}}:
An intelligent air interface refers to an advanced radio interface where the PHY and MAC layers are enhanced by AI/ML \cite{Hoydis-20}. In this evolution path, there exist three main phases, as defined in \cite{Hoydis-20} and summarized in Figure \ref{Fig_IA}. In the first phase, AI/ML is used to enhance or replace individual air interface processing blocks, such as channel estimation or symbol demapping in the PHY layer. In the second phase, AI/ML replaces many blocks by jointly designing them. An example of this is the joint design of channel estimation, equalization, and symbol demapping. In the third phase, AI/ML is expected to design parts of the air interface in the PHY and MAC layers, relying on end-to-end learning processes. The third phase would be a revolutionizing paradigm shift in the design of mobile networks, leading to substantial performance benefits and significantly reduced standardization \cite{Hoydis-20}.  

Traditionally, the PHY layer consists of individual processing blocks, which are designed separately. Conventional model-based mathematical approaches are generally efficient for individually designed PHY layer modules. However, the overall performance tends to be suboptimal when the PHY layer is considered as a whole \cite{Rajatheva-21}. The joint design of different PHY layer blocks is mathematically extremely complex \cite{Rajatheva-21}. Conventional mathematical tools are incapable of optimally solving such problems. Due to this algorithm deficit, the use of AI/ML is justified in the joint design problems \cite{Rajatheva-21}. 

The MAC layer is responsible for the network access process of the service-needing devices and the corresponding radio resource allocations. Efficient allocation of resources in diverse dimensions, such as frequency, time, space, power, and/or code, is one of the most essential and challenging tasks of the air interface. Due to their mathematical complexity, resource allocation problems are well suited for AI/ML. Generally, resource optimizations are inherently non-convex combinatorial problems, being too complex to be solved using conventional mathematical methods \cite{Rajatheva-21}. Due to this algorithm deficit, AI/ML solutions may be beneficial for many types of resource allocation problems. In this respect, DRL has been recognized as a promising tool to efficiently solve different types of MAC layer problems, particularly those related to resource allocation and network access \cite{Abbasi-21}. The reason for this is that DRL naturally lends itself to solving sequential decision making and optimization problems \cite{Abbasi-21}. 

\item {\textbf{Past and Present}}:
In the literature, AI/ML research on the air interface began around the mid-2010s, with a special focus on the PHY layer design. AI/ML-aided MAC layer design also became an interesting topic. In the past few years, an AI/ML-enhanced 6G air interface has gained increasing interest in the wireless community and is currently a widely studied topic in academia and industry. In practice, 5G was the first mobile generation to adopt AI/ML. Specifically, AI/ML was introduced in Release 18, which is the first realization of 5G-Advanced \cite{Rel-18, Lin-22}. However, the use of AI/ML is somewhat limited since 5G was not originally optimized for pervasive AI/ML. 6G will be the first generation designed for pervasive AI/ML, including intelligent core, edge, and air interface.   

\item {\textbf{Opportunities and Challenges}}: 
Exploiting AI/ML in the design of the air interface provides many potential benefits, ranging from improved performance to easier design and reduced standardization \cite{Hoydis-20}. Possible performance enhancements include reduced complexity, decreased latency, and increased spectral/energy/cost efficiency. The design of the PHY/MAC layers also becomes easier since there is less algorithm/protocol design needed. Moreover, less standardization is required mainly due to the reduction in the PHY/MAC layer options and parameters. Ultimately, the pervasive use of AI/ML for the air interface would be a major paradigm shift and revolutionize the way how mobile networks are designed, standardized, operated, and deployed \cite{Hoydis-20}. To this end, there are major challenges ahead. 

\begin{table*}[htb!]
\begin{center}
\caption{Summary of beyond-communication technologies for 6G}
\label{Table_Beyond}
\centering
\begin{tabularx}{\textwidth}{| >{\centering\arraybackslash}X | >{\centering\arraybackslash}X |
>{\centering\arraybackslash}X | 
>{\centering\arraybackslash}X |
>{\centering\arraybackslash}X |
>{\centering\arraybackslash}X |
>{\centering\arraybackslash}X |
>{\centering\arraybackslash}X |}
\hline
\centering
\vspace{3mm} \textbf{Beyond-Communication Technologies} \vspace{3mm} & \centering \textbf{Vision} & \centering \textbf{Description} & \centering \textbf{Opportunities} & \centering \textbf{Challenges} & \centering \textbf{Past} & \vspace{1.5mm} \begin{center} \textbf{Present} \end{center} \\
\hline
\vspace{3mm} Integrated Communication, Computation, and Caching \vspace{3mm}  & Freeing synergy between 3C & Joint design of 3C & Expanded capabilities & 3C resource management & Research since mid-2010s & 5G edge computing \\
\hline
\vspace{3mm} Integrated Sensing and Communication \vspace{3mm}  & Perceptive 6G networks & Joint radio sensing and comm & Ubiquitous network sensing & Trade-off between S$\&$C performance & Research since late 2010s & Under study for 6G \\
\hline
\vspace{3mm} Wireless Energy Transfer \vspace{3mm}  & Powering IoT sensors/devices & Electrical energy over-the-air & Sustainable IoT & Efficiency $\&$ distance & Decades of research & Under study for mobile networks \\
\hline
\end{tabularx}
\end{center}
\end{table*}

In general, 6G presents many challenges for the design of the air interface due to the significantly increased network complexity and tightened requirements. Specifically, there will be a massive number of diverse types of devices with numerous service classes, extremely high QoS requirements, integrated terrestrial and non-terrestrial networks, converged communication and beyond-communication technologies, and a wide range of application scenarios. Even though AI/ML is seen as one of the key enablers corresponding to these 6G challenges, it will be highly challenging to practically implement pervasive AI/ML for every level of the network, including the air interface. 

The main challenges of implementing an AI/ML-enhanced air interface are related to rapidly varying channel conditions, a vast variety of wireless environments, stringent and diverse QoS requirements, data collection, heterogeneous network architecture, multiple resource dimensions, and massive network access. Hybrid offline-online learning is a vital element to tackle the challenge of providing accurate training that matches well with the real-world channel conditions and environments. Extreme latency and reliability targets also set high requirements for the AI/ML-based solutions, calling for parallel computing and accurate training. 

Adequate data collection is a critical element for the practical implementation of AI/ML-based approaches, particularly for the resource allocation and network access problems of the MAC layer. However, this requires plenty of (over-the-air) signaling between different network nodes. Integrated satellite-air-ground access and communication-computation-sensing-energy resources further complicate the design of AI/ML-based MAC and higher layer functions and interactions between them. Overall, efficient AI/ML-aided air interface solutions are needed, which take the special characteristics of 6G into account, in cooperation with higher layers. Further discussions on the specific PHY and MAC layer challenges and future guidelines can be found in \cite{Hoydis-20, Han-20b}. 

\item {\textbf{Literature and Future Directions}}: 
In the literature, the main research focus of AI/ML-enhanced air interface has been on DL-aided PHY layer designs, ranging from enhancing/replacing single or multiple processing blocks (such as channel coding, symbol mapping, channel estimation, equalization, detection, decoding, and symbol demapping) to replacing the entire PHY layer with an end-to-end learning process \cite{Hoydis-20}. Furthermore, AI/ML-assisted MAC layer design has been widely examined, often related to resource allocation problems with DRL approaches \cite{Abbasi-21}. Due to the immaturity of this field of research, major efforts are needed in all of these directions in the future, including AI/ML assistance from smaller to larger entities. 

The AI/ML-assisted air interface has been reviewed in the recent literature \cite{Qin-19, He-19, Gunduz-19, Huang-20e, Kim-20c, Restuccia-20, Han-20b, Hoydis-20, Zhang-21l, Elbir-21, Ozpoyraz-22, Mao-23c, Islam-24}. In general, these surveys covered the main literature, recent advances, and open challenges. Most studies considered PHY layer design using DL \cite{Qin-19, He-19, Gunduz-19, Huang-20e, Kim-20c, Restuccia-20, Zhang-21l, Ozpoyraz-22, Mao-23c, Islam-24}. These DL works can be further divided into the 6G-specific \cite{Zhang-21l, Ozpoyraz-22, Mao-23c}, B5G/5G-specific \cite{Huang-20e, Restuccia-20}, and generic surveys \cite{Qin-19, He-19, Gunduz-19, Kim-20c, Islam-24}. Additionally, federated learning was reviewed for the PHY layer in \cite{Elbir-21}. The design of the MAC layer functions was discussed in \cite{Gunduz-19, Han-20b, Hoydis-20}. The work \cite{Hoydis-20} provided a vision toward an AI/ML-native 6G air interface, ideally enabling optimized communication schemes for any hardware, wireless environment, and application. The paper described the main transition phases on the road to achieving AI/ML-nativeness and discussed the needed PHY/MAC layer learning procedures. In \cite{Han-20b}, the authors introduced a framework for an AI/ML-enabled 6G air interface, covering PHY/MAC layer designs and interactions with the higher layers. 
Major research efforts are required in the near future to realize the vision of an intelligent 6G air interface. The corresponding challenges were discussed earlier. In particular, the 6G-specific characteristics and requirements must be properly addressed in the design of AI/ML-enhanced air interface. 
\end{itemize}

\subsection{BEYOND-COMMUNICATION TECHNOLOGIES FOR 6G}
\begin{figure}[!tb]
\center{\includegraphics[width=0.7\columnwidth]
{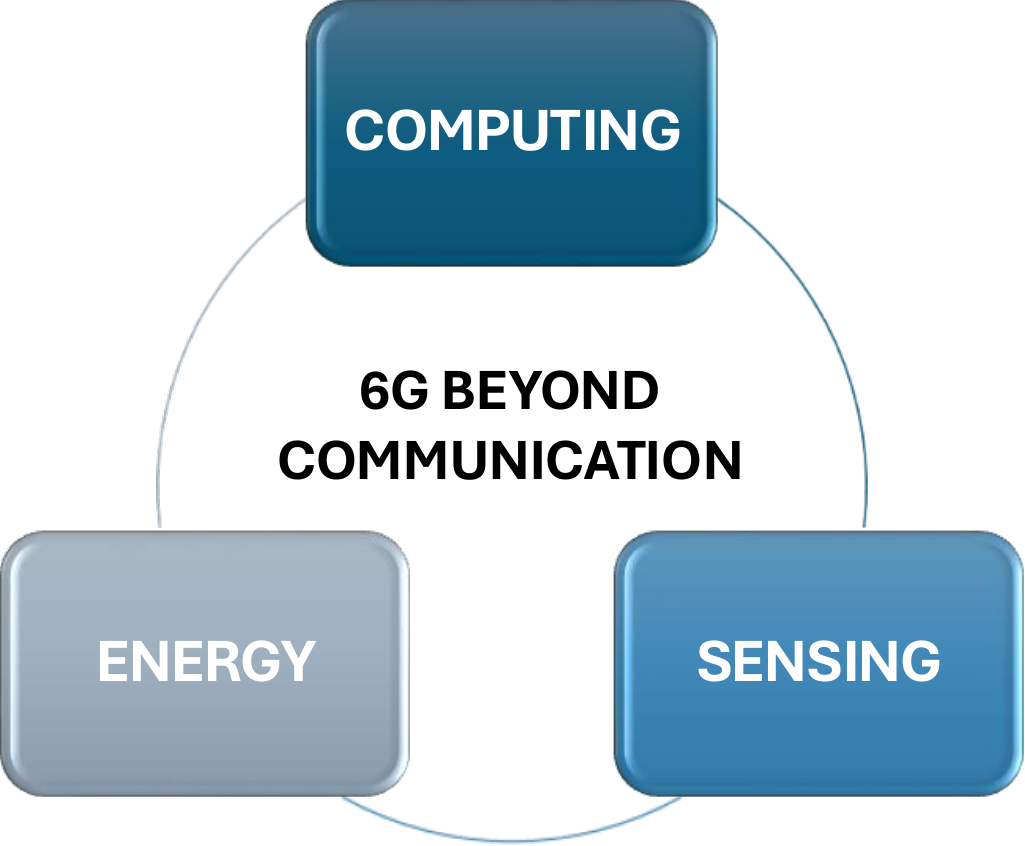}}
\caption{\label{Fig_Beyond}Beyond-communication technologies for 6G.}
\end{figure}

The integration of beyond-communication technologies will be a major paradigm shift in the evolution of mobile networks, significantly extending the way how mobile networks can be exploited. The main beyond-communication technologies include computation, sensing, and energy, as shown in Figure \ref{Fig_Beyond}. Currently, 5G networks have adopted edge computing capabilities and enhanced positioning in its palette of technologies. This paves the way for the broader usage of beyond-communication technologies in the 6G era. Beyond-communication technologies are expected to enable a broad range of novel capabilities and applications for 6G. In the following, we discuss {integrated communication, computation, and caching (i3C)}, ISAC, and WET. These technologies are summarized in Table \ref{Table_Beyond}. 

\subsubsection{INTEGRATED COMMUNICATION, COMPUTATION, AND CACHING}

\begin{itemize}
\item {\textbf{Vision}}: 
Communication, computation, and caching (3C) integration is considered as one of the key technologies for 6G networks, exploiting the synergy between 3C capabilities at the network edge. 

\item {\textbf{Introduction}}:
i3C refers to the joint design of 3C technologies at the edge of the network \cite{Wang-17}, as shown in Figure \ref{Fig_3C}. Through joint design, 3C technologies can be used to benefit each other, enhancing the performance and extending the capabilities of mobile networks. Consequently, i3C enables novel services and applications. i3C is considered a major paradigm shift in the design of mobile networks \cite{Adam-23}. 

\item {\textbf{Past and Present}}:
In the literature, i3C started to gain increasing interest after the mid-2010s. Early studies focused on applying i3C to 5G networks. Back then, the novel concepts of softwarization, virtualization, and edge computing played key roles in i3C research \cite{Wang-17}. At the end of the 2010s, research started to shift toward 6G, with a focus on AI/ML-assisted i3C. Although 5G networks support SDN, NFV, and edge computing technologies, i3C is still too immature for large-scale usage in practice. 

\begin{figure}[!tb]
\center{\includegraphics[width=0.7\columnwidth]
{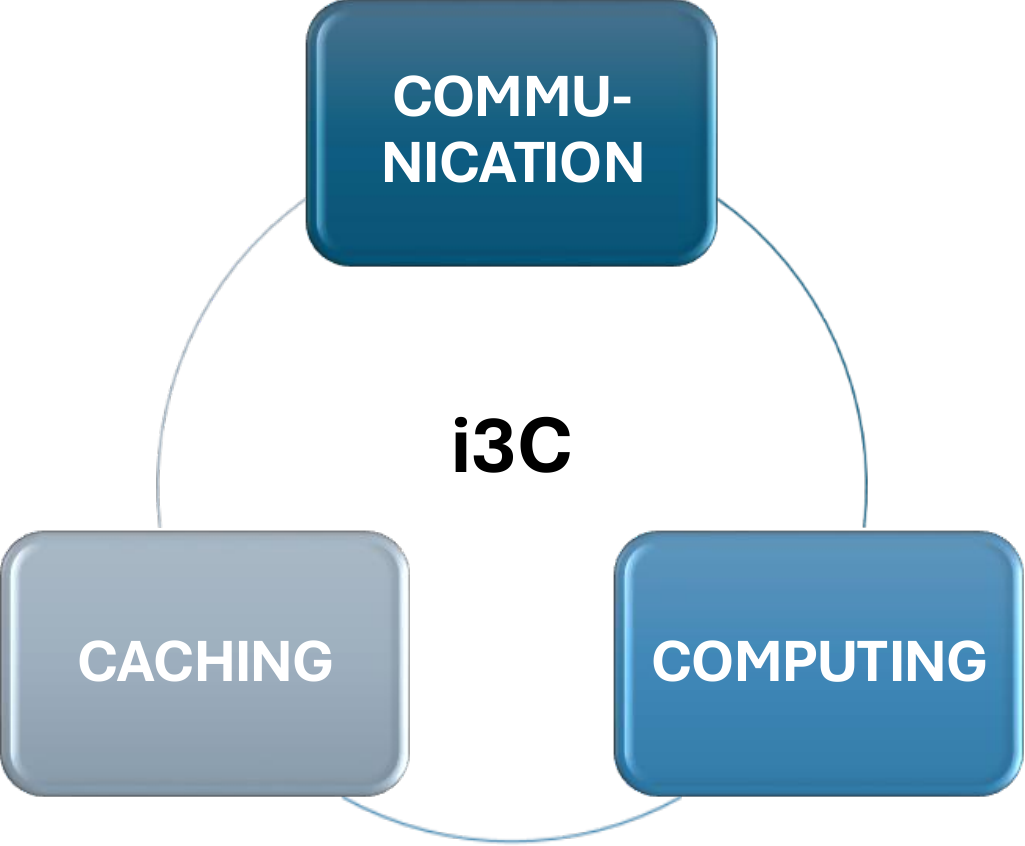}}
\caption{\label{Fig_3C}Integrated 3C exploits the synergy between communication, computing, and caching at the network edge.}
\end{figure}

\item {\textbf{Opportunities and Challenges}}: 
The potential benefits which i3C offers to future mobile networks include extended capabilities, enhanced performance, and novel applications \cite{Wang-17, Adam-23}. Caching enables bringing popular content to the network edge and closer to end-devices, reducing latency and communication overhead. Advanced computing and processing enable bringing more intelligence to the edge, enhancing network management. The main performance enhancements of i3C are lower latency, higher data rates, and better energy efficiency. i3C can play a key role in diverse future applications, such as immersive XR, smart city/factory, and intelligent vehicle systems. Due to its potential, i3C is expected to become an integral part of 6G networks \cite{Adam-23}. 

Before the practical implementation of i3C is possible, diverse challenges must be resolved. Key challenges are related to heterogeneity, resource management, latency requirements, real-time analytics, mobility, security, and privacy \cite{Wang-17, Adam-23}. In the 6G era, the heterogeneity of networks, devices, and applications makes the joint design of 3C challenging, calling for novel solutions. In particular, resource management and allocation become more complicated due to multiple resource types and their diverse performance requirements. A major challenge is to satisfy the stringent performance requirements of future services and applications, especially minimal latency. Due to the complexity of i3C networks, real-time analytics and processing are difficult in practice. Mobility management also becomes trickier since neighboring BSs may have different 3C capabilities, affecting service quality. Since the introduction of i3C exposes mobile networks to the new types of threats and attacks, novel security and privacy solutions need to be developed. AI/ML has been recognized as a promising tool for addressing many of the aforementioned issues \cite{Adam-23}. The fundamental challenges and future topics of i3C have been discussed in \cite{Wang-17, Adam-23, Maia-24}.  

\item {\textbf{Literature and Future Directions}}: 
In the literature, i3C has been reviewed from many different perspectives in a handful of survey papers \cite{Wang-17, Wang-19e, Tang-20d, Bouras-20, Adam-23, Maia-24}. In \cite{Wang-17}, a survey was provided on mobile edge networks from the perspective of the 3C convergence. Computing and caching at the edge were first reviewed individually, and then the synergy between the 3C technologies was discussed. The work \cite{Wang-19e} discussed FL-assisted MEC, communication, and caching to provide more intelligence to the network edge. In \cite{Tang-20d}, 3C resource sharing was explored for D2D IoT scenarios. The authors in \cite{Bouras-20} presented a review of the 3C convergence in IoT. In \cite{Adam-23}, a comprehensive survey was provided on the integration of 3C and control (i4C) for beyond 5G networks. The study \cite{Maia-24} gave a thorough discussion on i3C in terms of cloud-edge cooperation, resource management, and intelligence.  

Further research is needed in the near future before large-scale usage is possible. In addition to solving the challenges discussed earlier, the main future directions are related to AI/ML-aided i3C/i4C, for example, DL/FL assistance, big/small data analytics, and real-time decision-making. Further details on the future research directions can be found in \cite{Adam-23}. 
\end{itemize}

\begin{figure}[!htb]
\center{\includegraphics[width=\columnwidth]
{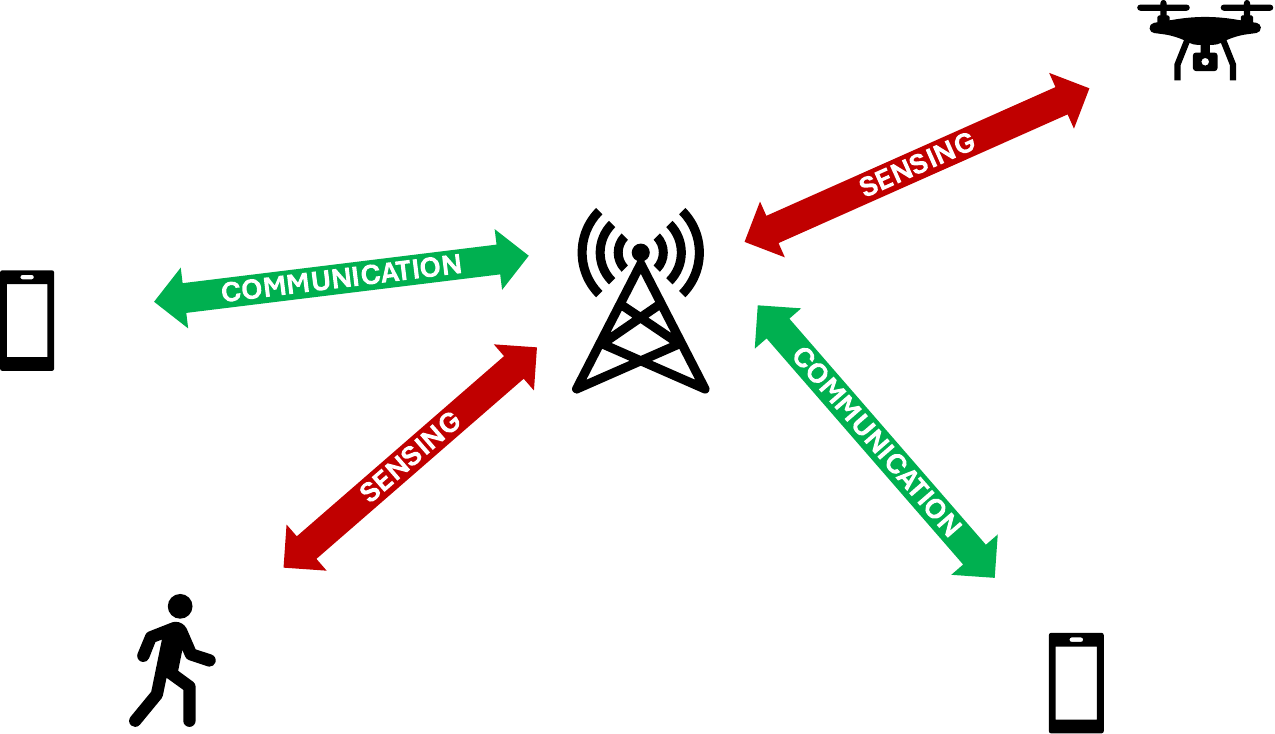}}
\caption{\label{Fig_ISAC}Joint sensing and communication in a cellular environment.}
\end{figure}

\subsubsection{INTEGRATED SENSING AND COMMUNICATION}

\begin{itemize}
\item {\textbf{Vision}}: 
ISAC is considered a revolutionary technology for 6G, expanding the services and applications of mobile networks beyond traditional communication toward ubiquitous sensing. 

\item {\textbf{Introduction}}:
ISAC refers to the integration of radio sensing and communication capabilities into the same wireless system \cite{Liu-22c}, as illustrated in Figure \ref{Fig_ISAC}. Although radio sensing and communications are different technologies with distinct objectives, they have similarities in hardware and signal processing designs, making it possible to integrate them into one system in a cost-, energy-, and spectrum-efficient manner \cite{Zhang-22g, Liu-22c}. ISAC is also known by different names, such as radar-communication, joint radar and communication, joint communication and radar, dual-functional radar communication, joint (radar/radio) sensing and communication, and joint communication and (radar/radio) sensing \cite{Zhang-22g, Liu-22c}. There are three types of ISAC design: communication-centric, radar-centric, and joint optimization \cite{Zhang-22g}, as shown in Figure \ref{Fig_ISAC2}. The communication-centric design refers to merging radio sensing functions into wireless communication systems, whereas the radar-centric design merges communication functions into radar systems. In the joint optimization design, there is no bias toward the underlying systems, but the system can be jointly optimized to meet the needs of the desired applications. 

In the context of mobile networks, ISAC refers to joint radio sensing and communications using the same cellular spectrum and network infrastructure, sharing the majority of hardware and signal processing modules \cite{Zhang-21m}. This concept is also known as the perceptive mobile networks \cite{Zhang-21m}. In the perceptive networks, radio sensing is profoundly fused into mobile networks, greatly expanding network capabilities and enabling ubiquitous sensing services and applications. 
Radio sensing refers to the retrieval of information from the surrounding environment through the received radio signals and the measurements of sensing parameters (such as angle of arrival, angle of departure, time delay, and Doppler frequency) and feature parameters (such as the pattern signals of objects, activities, and events) \cite{Zhang-21m, Zhang-22g}. There are three types of cellular radio sensing, i.e., downlink active sensing, downlink passive sensing, and uplink sensing \cite{Zhang-21m}. While downlink sensing signals are from the BSs, uplink sensing signals originate from the UEs. Active and passive sensing refer to the signals transmitted from the BS and other nearby BSs, respectively. 

Radio sensing enables a broad range of novel capabilities for mobile networks, including the detection (size, shape, material, flaw), recognition (gesture, activity, event), localization (indoor/outdoor), mapping (2D/3D), tracking (industrial, environmental, consumer), imaging (biomedical, security), and monitoring (health, medical, security, environmental, industrial, agricultural) of objects and entities \cite{Lima-21, Zhang-22g, Liu-22c}. 
As there are many similarities between sensing and communication, they can be integrated into mobile networks, sharing the same wireless resources \cite{Zhang-22g, Liu-22c}. The integration is becoming more natural since radars and communications have evolved in the same direction, i.e., toward higher frequencies with larger bandwidths and larger-scale antenna arrays with higher array gains. This development path has provided more resources in the spectral and spatial dimensions, leading to increased capacity and connection density in communication systems and improved range and accuracy in radar sensing systems. Moreover, the development of mobile networks toward denser cell deployment and pervasive AI/ML will further enhance the integration of sensing and communication. 

\begin{figure}[!tb]
\center{\includegraphics[width=\columnwidth]
{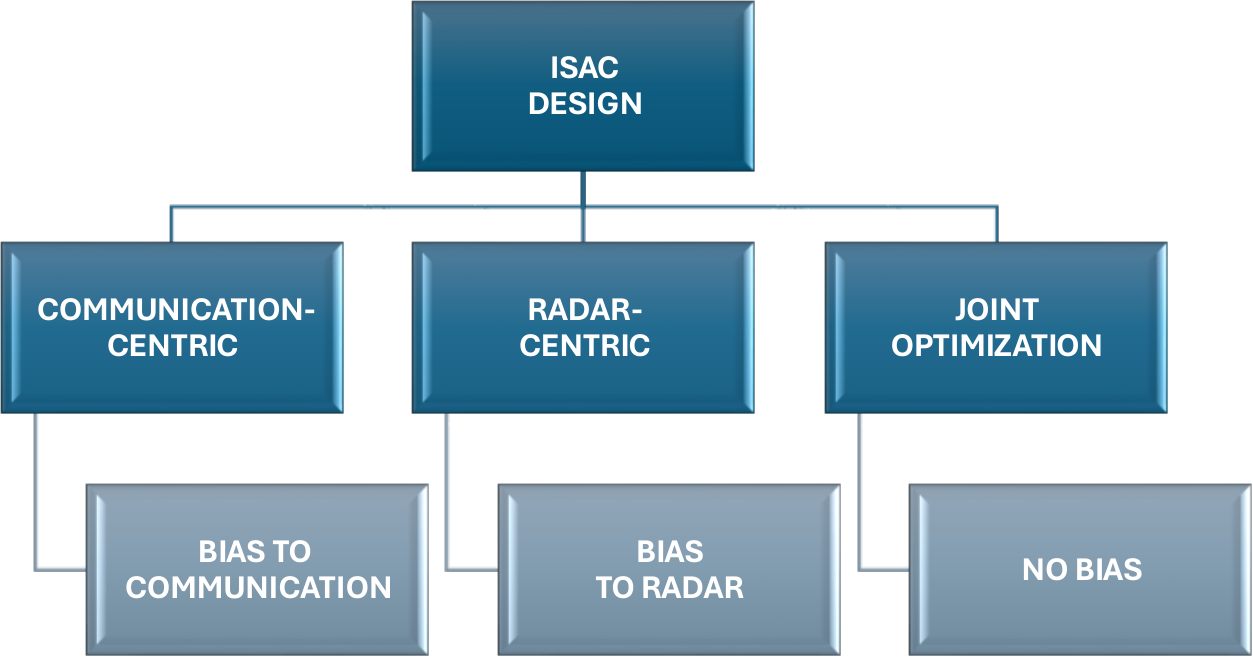}}
\caption{\label{Fig_ISAC2}Main types of ISAC design.}
\end{figure}

Due to the sharing of wireless resources and mutual assistance between sensing and communication, cellular ISAC provides diverse benefits compared to the separated sensing and communication systems. In other words, these benefits originate from the integration and coordination gains, defined in \cite{Liu-22c}. The integration gain originates from the sharing of the same wireless network resources between sensing and communication, providing improved spectrum, energy, size, and cost efficiency. The coordination gain is achievable through mutual assistance between sensing and communication, that is, communication-assisted sensing and sensing-assisted communication. For example, the coordination gain may appear in terms of improved beamforming and sensing efficiency/accuracy. 

Due to the different nature of sensing and communication, cellular ISAC has certain shortcomings in practice \cite{Zhang-21m}. First, ISAC requires a full-duplex operation or equivalent. Second, the sensing distance may be limited due to the limited transmission power of the cellular BS. Third, performance trade-offs exist between sensing and communication due to their conflicting targets and requirements. Typical trade-offs are defined as the PHY layer, spatial degrees of freedom, and cross-layer trade-offs, as discussed in \cite{Liu-22c}. 

\item {\textbf{Past and Present}}:
Although ISAC is a new topic in the context of mobile networks, integrating communication into radar sensing systems has been studied since the 1960s \cite{Liu-22c}. Since then, ISAC research was dominated by the radar community for a long time. In the 2010s, the integration of radar/radio sensing into wireless communication systems started to gain considerable interest in the wireless community. 
It was until the introduction of the perceptive mobile network concept in the late 2010s \cite{Zhang-21m} that the wireless community noticed the game-changing potential of ISAC for future mobile networks. Recent advances in mobile networks and radar systems have made the vision of a perceptive network possible. In particular, the development toward larger-scale antenna arrays, higher frequencies with wider bandwidths, denser cell deployments, and pervasive AI/ML are the key elements to enable the beneficial integration of radar/radio sensing into mobile infrastructure. 

Currently, ISAC is one of the key 6G topics in academia, industry, and standardization. For example, ITU-R considers ISAC as one of the emerging technology trends/enablers for IMT-2030 \cite{ITU-M2516, ITU-M2160}. Recent survey papers provide further details on the history and present of ISAC, from the perspectives of academia, industry, and standardization bodies \cite{Wild-21, Liu-22c}. 

\item {\textbf{Opportunities and Challenges}}: 
Ultimately, the potential of ISAC is to realize the concept of perceptive network by turning mobile networks into a ubiquitous sensing entity (i.e., "network as a sensor" \cite{Zhang-21m}), with a vast variety of innovative services and applications. As this grand vision significantly extends the capabilities of mobile networks, it opens new business opportunities for mobile network operators and vertical industries. Due to the synergy between sensing and communication, cellular ISAC is seen as a key technology to be exploited in many future application scenarios, ranging from immersive context-aware human-machine interactions and smart factory/city/home environments to intelligent vehicle/transportation systems and e-health/energy/agriculture \cite{Zhang-22g, Liu-22c}. 

In addition to expanding the capabilities of mobile networks, ISAC also offers other benefits. Compared to two separated systems, integrating communication and sensing into the same system provides direct benefits in terms of spectrum, energy, and cost efficiency \cite{Liu-22c}. Additional benefits are achievable through mutual assistance between communication and sensing, potentially leading to performance improvements in beamforming, resource allocation, PHY layer security, and sensing efficiency/accuracy \cite{Liu-22c}. Due to its potential, ISAC is expected to become a revolutionary element for 6G, providing sensing capabilities in mobile networks \cite{Liu-22c, Dong-24}. 

Due to the infancy of the cellular ISAC concept, diverse unresolved issues and design challenges exist on the way toward the commercialization of perceptive networks with ubiquitous sensing capabilities \cite{Lima-21, Zhang-22g, Liu-22c}. First, the previously mentioned drawbacks of ISAC need to be properly addressed, i.e., developing practical solutions for full-duplex or equivalent operation, compensating for limited sensing ranges when needed (e.g., by cooperating among neighboring BSs), and thoroughly studying fundamental performance trade-offs between sensing and communication to fully understand their nature and find a suitable balance between the performances in potential cellular application scenarios. As communication is the primary function, it has the highest priority in the design of cellular ISAC networks. In this regard, a critical challenge is to integrate sensing without compromising communication performance. 

Other essential challenges are related to the fundamental performance bounds, joint waveform/array optimization, clutter suppression, and sensing parameter estimation \cite{Zhang-21m, Zhang-22g, Liu-22b, Liu-22c}. The characterization of information-theoretical limits on cellular ISAC is largely unknown, which limits the understanding of the theoretical foundations of perceptive networks. Deriving performance bounds is far from trivial due to the special characteristics of mobile ISAC. For example, significant differences between sensing and communication signals lead to fundamental performance trade-offs, which have a critical impact on the system design. While the research work is in progress with some existing studies \cite{Zhang-21m, Zhang-22g, Liu-22b, Liu-22c}, it is still a widely open challenge to develop practical signaling methods that adequately satisfy the (more or less) conflicting requirements of both functions with given priority weights (usually higher on communication). The center of this development work is the joint waveform optimization and antenna array design since they have quite different requirements for sensing and communication, notably affecting the performance of both functions. 

In cellular ISAC networks, rich multipath environments are challenging for sensing due to the presence of harmful clutter \cite{Zhang-22g}. Clutter refers to the unwanted multipath signals that contain only a small amount of new information. It is vital to remove clutter signals at the receiver since they can significantly degrade the performance of sensing algorithms by increasing the number of estimated sensing parameters. Although clutter suppression has been well studied for radar networks, it is still mostly an unresolved issue in the context of cellular ISAC \cite{Zhang-22g}. In general, sensing parameter estimation is a challenging task due to the complex signal structures in mobile networks. Since most of the existing techniques from radar systems cannot be directly applied, many novel methods have recently been studied. For example, compressed-sensing-based parameter estimation methods have shown emerging promise, but they still have major limitations in practice \cite{Zhang-22g}. 
To summarize, all of the aforementioned topics require major advances to find proper solutions for their corresponding problems. Detailed reviews on these topics and the related literature can be found in \cite{Lima-21, Zhang-21m, Zhang-22g, Liu-22b, Liu-22c}. 

\item {\textbf{Literature and Future Directions}}:
Since the late 2010s, ISAC has been studied from a vast range of aspects, such as information theoretical limits, performance trade-offs, communication-assisted sensing, sensing-assisted communication (beamforming, resource allocation, PHY layer security), signal processing (waveform optimization, MIMO design, receiver processing), network management (resource allocation, higher layer designs), integration with other emerging technologies (AI/ML, THz, RISs, V2X, UAVs, satellites), and potential applications (smart home/factory/city/healthcare, digital twins, XR, vehicular systems, remote sensing, environmental monitoring, etc.). 

The aforementioned topics and the related literature have been thoroughly reviewed in numerous surveys \cite{Zhang-21m, Lima-21, Wild-21, Cui-21, Zhang-22g, Chaccour-22, Liu-22b, Liu-22c, Wang-22b, Zhou-22b, Cheng-22, Demirhan-22, Zhang-22h, Elbir-22b, Elbir-22, Chepuri-22, Liu-23b, Meng-23, Xie-23, Wei-23, Zhu-23, Dong-24, Jiang-23, Han-24, Lu-24, Liu-24, Guo-24, Chen-24, Wymeersch-24, Strinati-24, Kaushik-24, Gonzales-24}. At a high level, these survey papers can be classified into three categories: generic, specific, and 6G-oriented. In \cite{Zhang-21m, Zhang-22g, Liu-22b, Wang-22b, Xie-23, Lu-24}, the generic surveys discussed ISAC from a wide perspective, covering topics such as fundamentals, state-of-the-art designs, recent advancements, potential applications, open problems, and future research guidelines. The specific surveys reviewed narrower aspects of ISAC, including fundamental limits \cite{Liu-22b}, channel modeling \cite{Liu-24}, waveform design \cite{Zhou-22b}, ISAC signals \cite{Wei-23}, THz ISAC \cite{Chaccour-22, Elbir-22, Jiang-23, Han-24}, ISAC for IoT \cite{Cui-21}, RIS-assisted ISAC \cite{Elbir-22b, Chepuri-22, Liu-23b}, holographic ISAC \cite{Zhang-22h}, UAV-assisted ISAC \cite{Meng-23}, ISAC for vehicular communication networks \cite{Cheng-22}, and AI/ML-assisted ISAC \cite{Demirhan-22, Zhu-23}. The 6G-oriented surveys provided comprehensive explorations of ISAC in the context of 6G \cite{Lima-21, Wild-21, Liu-22c, Dong-24, Guo-24, Chen-24, Wymeersch-24, Strinati-24, Kaushik-24, Gonzales-24}. 

Since the development of cellular ISAC networks is in a rather early phase, further research is needed in the coming years. Some important future topics toward 6G include a thorough characterization of the fundamental performance limits/trade-offs, sensing-integrated channel models, joint waveform/signaling/array optimization, compressed sensing-based clutter suppression, networked sensing, cooperative distributed sensing, sensing-assisted mmWave/THz beamforming, pervasive AI/ML assistance, and sensor fusion \cite{Lima-21, Wild-21, Liu-22c, Dong-24, Lu-24}. 
\end{itemize}

\begin{figure}[!htb]
\center{\includegraphics[width=0.7\columnwidth]
{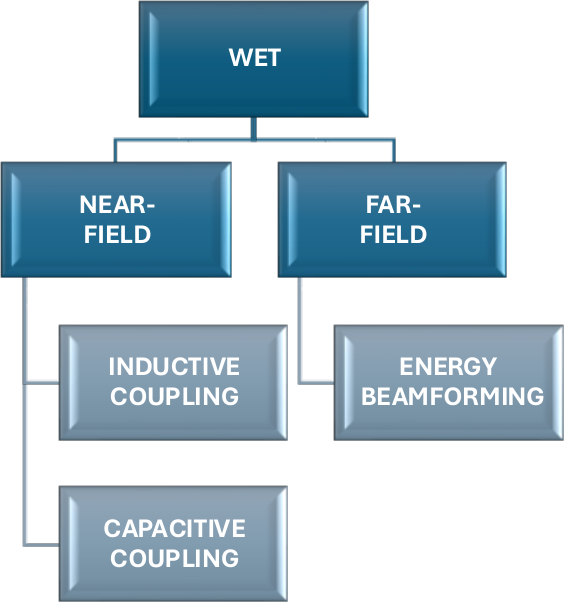}}
\caption{\label{Fig_WET2}Main categories of WET.}
\end{figure}

\subsubsection{WIRELESS ENERGY TRANSFER}

\begin{itemize}
\item {\textbf{Vision}}: 
WET is considered a revolutionary technology to potentially energize lightweight IoT networks in the 6G era. 

\item {\textbf{Introduction}}: 
WET (also known as WPT) is a technology that transmits electrical energy over the air through a wireless medium \cite{Clerckx-21, Lopez-21}. WET aims to power and charge wireless devices to promote autonomy, mobility, long life-time, and novel applications \cite{Clerckx-21, Lopez-21}. There are two main categories of WET: near-field and far-field \cite{Zhang-19f}, as summarized in Figure \ref{Fig_WET2}. Near and far fields refer to the different regions of the electromagnetic field around a radiating source. In the near-field region, the behavior of the electromagnetic field is far different from that in the far-field region. While the near-field behavior dominates in the close proximity of the source, far-field characteristics, i.e., typical electromagnetic radiation, dominate at longer ranges. Due to the different radiation characteristics, different types of technologies are required for WET in the near and far fields. 

In the near-field category, energy is transferred by inductive coupling via magnetic fields or capacitive coupling via electric fields \cite{Zhang-19f}. In the near-field technologies, the distances over which the energy can be efficiently transferred are short. Typical commercial applications of the near-field WET include wireless charging of mobile phones, tablets, smart watches, electric toothbrushes, electronic medical implants, and electric vehicles, to mention a few. 

In the far-field category, energy is typically transferred by focusing electromagnetic radiation, such as microwaves, toward the dedicated receivers via directive transmissions \cite{Shinohara-21, Lopez-21}. This technology is known as energy beamforming (power beamforming) \cite{Lopez-21}. Energy beamforming aims to provide efficient energy transfer over relatively long ranges, much greater than that of the near-field technologies. However, the operating efficiency, i.e., the portion of the transmitted energy received, decreases with the radiation distance. Thus, achieving efficient energy transfer over long ranges is a fundamental challenge in the far-field technologies. A typical application proposed for the far-field WET is the powering of lightweight IoT devices \cite{Lopez-21}. More details on the energy beamforming can be found in \cite{Lopez-21}. 

\begin{figure}[!tb]
\center{\includegraphics[width=0.6\columnwidth]
{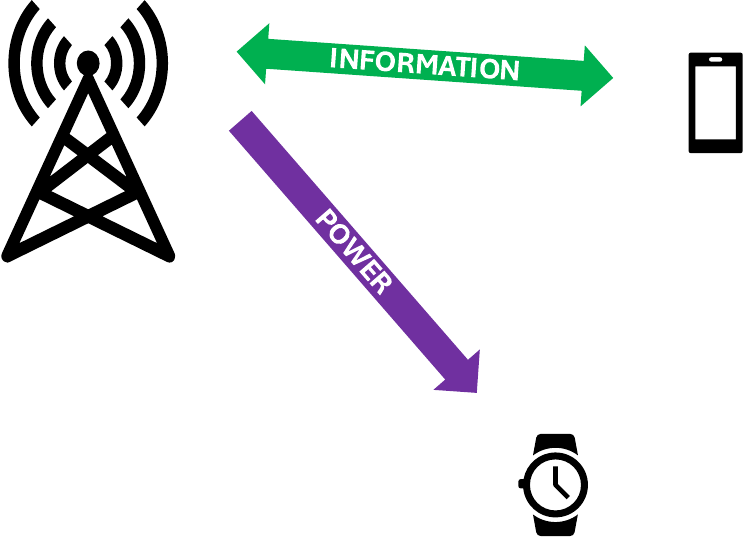}}
\caption{\label{Fig_WET}Simultaneous wireless information and power transfer.}
\end{figure}

While the traditional concept of WET is considered separately from wireless communication systems, a relatively new idea is to transfer both energy and information simultaneously using the same network \cite{Clerckx-21}, as illustrated in Figure \ref{Fig_WET}. This concept is known as wireless information and power transfer (WIPT), which can be divided into three categories: simultaneous WIPT (SWIPT), wireless powered communication networks, and wireless powered backscatter communication \cite{Clerckx-21}. In the first category, information and energy are simultaneously transferred from one or many transmitters to one or many receivers. There are two types of receivers: information receivers and energy receivers \cite{Clerckx-21}. Information and energy receivers can be co-located in the same device or separated into different devices. In the second category, downlink transmissions are used for energizing devices, whereas devices use this energy to transmit information in the uplink. In the third category, the downlink is for energy and uplink for information, while the devices are low-power, low-complexity tags that directly modulate the downlink signals by their own information without the need to generate carrier signals by themselves.

\item {\textbf{Past and Present}}:
The earliest studies on WET date back to the end of the 19th century, when Nicola Tesla conducted a wireless high-power experiment using a massive device called the Tesla Tower \cite{Shinohara-21}. However, the experiment failed. After Tesla's experiment, it took a long time before anything significant occurred in the field of WET. In the late 1940s, the concept of backscatter communication was introduced \cite{Stockman-48}. Backscatter communication is currently considered a form of the WIPT concept \cite{Clerckx-21}. The most popular commercial application of backscatter communications is radio frequency identification (RFID) \cite{Huynh-18}. Active research on WET via microwaves began in the 1960s, when William Brown conducted his first experiments \cite{ITU-SM.2392-1}. Brown achieved a remarkable 54 $\%$ overall efficiency in one of his laboratory experiments in 1975 \cite{Shinohara-21}. In the 1990s, phased arrays began to gain popularity, leading to the rise of beamforming-based WET \cite{Shinohara-21}. 

Over the years, WET with microwave energy beamforming has been extensively studied. Traditional research has mainly focused on the RF aspects, while recent studies have expanded to signal and system designs as well, with a special focus on integrating WET into future wireless networks \cite{Clerckx-21}. In this respect, the concept of SWIPT started to gain interest in the 2010s \cite{Ponnimbaduge-18}. Currently, WET is being studied for 6G networks, with a special emphasis on energizing low-power IoT devices \cite{Lopez-21}. While there exist no large-scale commercial applications of the far-field WET, there are some in the near-field domain, such as wireless chargers for wireless/mobile devices, electronic medical implants, and electric vehicles. Although WET has been extensively studied over the decades, with some commercialized products, it is far from maturity. From a regulatory perspective, ITU-R published its first report on radio-frequency WET in 2016, with an updated version in 2021 \cite{ITU-SM.2392-1}. This report focused on the possible applications of WET via radio waves, corresponding technologies, and candidate spectrum bands. 

\item {\textbf{Opportunities and Challenges}}: 
WET has great potential to revolutionize wireless ecosystems and expand the capabilities of wireless networks beyond communications by wireless charging and powering electric devices, particularly low-power IoT sensors. In particular, WET is considered a key enabler for sustainable IoT in the 6G era by promoting the long life-time, mobility, and autonomy of devices \cite{Lopez-21}. There are many design challenges in the path toward efficient and ubiquitous WET as an integral part of future wireless networks. The fundamental challenges are mainly related to efficiency, distance, availability, mobility, and safety \cite{Clerckx-21}. These elements are crucial for integrating WET into mobile networks. 
The seamless integration of wireless energy with communication, computation, sensing, and positioning is a major challenge that requires multidisciplinary research efforts \cite{Clerckx-21}. Further challenges arise from the trends toward 6G, such as higher frequencies, larger number of antennas, denser networks, greater number of devices, and a higher level of intelligence. On the other hand, by overcoming the aforementioned design challenges, integrated WET may enable many novel applications, such as the wireless powering of low-power IoT, lightweight autonomous systems, crowd sensing, and distributed EI \cite{Clerckx-21}. 

\item {\textbf{Literature and Future Directions}}:
In the literature, there are lots of survey articles which have reviewed a wide range of WET aspects \cite{Clerckx-19, Zhang-19f, Lopez-21, Clerckx-21, Xie-21c, Wu-22b, Clerckx-22, Zhou-22, Zhang-22f, Papanikolaou-23, Chen-24, Psomas-24}. The work \cite{Clerckx-19} reviewed WIPT from the perspectives of RF, signal, and system designs. In \cite{Zhang-19f}, a survey was provided on non-radiative near-field WET, covering fundamentals, challenges, metamaterials, and applications. In \cite{Lopez-21}, the authors studied massive WET for sustainable IoT in the 6G era. A detailed discussion was given on the system architecture, applications, technological enablers, energy beamforming schemes, and future directions. 

In \cite{Clerckx-21}, WPT was studied for future networks from the perspectives of signal processing, ML, computation, and sensing. This comprehensive review covered challenges, key technologies, rate-energy trade-off, system design methodologies, and wireless-powered IoT. In \cite{Xie-21c}, a tutorial overview was provided on UAV-enabled WET. The main topics included single- and multi-UAV scenarios as well as UAV-enabled wireless powered communication networks and mobile edge computing. In \cite{Wu-22b}, the authors presented an overview of the RIS-aided WIPT. In \cite{Clerckx-22}, the work reviewed the theory, prototypes, and experiments on WPT, with and without simultaneous information transfer. Metamaterials and -surfaces were explored for WPT and energy harvesting in \cite{Zhou-22}. In \cite{Zhang-22f}, the near-field WPT was considered in the context of IoE for 6G. The paper \cite{Papanikolaou-23} discussed simultaneous lightwave information and power transfer for 6G in terms of transceiver architectures, optical beam propagation, design trade-offs, synergy with other emerging technologies, and application scenarios. In \cite{Chen-24}, the authors discussed a multi-functional 6G, with the integration of sensing, communication, and energy. The study \cite{Psomas-24} reviewed the design of WIPT and its fusion into 6G networks. 

Although WET has been studied for decades, it is far from its true potential. Further research is needed in all areas from the fundamentals and experiments to system design and emerging technologies. In particular, 6G with its novel technologies and application scenarios brings new opportunities and challenges that need to be widely addressed. In the fundamentals category, more studies are needed in the RF design to further improve energy transfer efficiency and increase distances to be practically reasonable for 6G purposes, i.e., from meters to tens of meters. More experiments are needed in the context of 6G technological trends, such as higher frequencies, a larger number of antennas, and a greater number of low-power devices. 

In the system design, special research efforts need to focus on a paradigm shift in which wireless networks integrate energy, communication, computation, sensing, and positioning. There are some emerging technologies which can be used to aid WET and expand its capabilities, such as RISs \cite{Wu-22b} and UAVs \cite{Xie-21c}. RISs can be used to improve efficiency and increase range, whereas UAVs can provide flexibility and availability. Further research is needed in these promising directions. Other interesting future directions include wireless powered 6G systems, such as massive lightweight IoT networks \cite{Lopez-21}. 

\begin{table*}[htb!]
\begin{center}
\caption{Summary of energy-aware technologies for 6G}
\label{Table_EE}
\centering
\begin{tabularx}{\textwidth}{| >{\centering\arraybackslash}X | >{\centering\arraybackslash}X |
>{\centering\arraybackslash}X | 
>{\centering\arraybackslash}X |
>{\centering\arraybackslash}X |
>{\centering\arraybackslash}X |
>{\centering\arraybackslash}X |
>{\centering\arraybackslash}X |}
\hline
\centering
\vspace{3mm} \textbf{Energy-Aware Technologies} \vspace{3mm} & \centering \textbf{Vision} & \centering \textbf{Description} & \centering \textbf{Opportunities} & \centering \textbf{Challenges} & \centering \textbf{Past} & \vspace{1.5mm} \begin{center} \textbf{Present} \end{center} \\
\hline
\vspace{3mm} Green Networks \vspace{3mm}  & \vspace{3mm} Sustainable 6G networks \vspace{3mm} & Energy-efficient design at all levels & Ecological and economic benefits & High energy consumption & Research since early 2010s & Energy-efficient design in 5G \\
\hline
\vspace{3mm} Energy Harvesting \vspace{3mm}  & Sustainable lightweight IoT & External energy collection & Lightweight IoT devices & Efficiency $\&$ consumption & RF-EH research since 2000s & Under study for mobile networks \\
\hline
\vspace{3mm} Backscatter Communications \vspace{3mm}  & Autonomous low-power IoT & External energy to transmit data & Battery-free low power IoT devices & Low rates $\&$ short distances & Concept invented in 1940s & Under study for mobile networks \\
\hline
\end{tabularx}
\end{center}
\end{table*}

While most of the future research directions are focused on the far-field studies, also the radiating near-field WET is an active research area for 6G purposes \cite{Zhang-22f}. As the far-field energy transfer suffers from relatively low efficiency, radiative near-field technologies may provide some benefits in certain scenarios, especially with higher frequencies and a larger number of antennas. In this direction, far-field studies may no longer be valid, thus many aspects of energy beamforming need further research, such as channel estimation, beam design, and waveform design \cite{Zhang-22f}. In addition, a rethinking of SWIPT is needed, as radiating near-field characteristics may provide benefits in some scenarios \cite{Zhang-22f}. Additionally, the use of metasurfaces to aid near-field energy beamforming is an interesting topic for future research \cite{Zhou-22}.  
\end{itemize}

\subsection{ENERGY-AWARE TECHNOLOGIES FOR 6G}
As the absolute energy consumption is constantly growing, it is of utmost importance to further improve the energy efficiency of mobile networks. In this respect, developing more efficient energy management technologies is one of the main goals and challenges of 6G. In this section, we focus on three energy-aware technologies: {green networks}, {energy harvesting (EH)}, and {backscatter communications}. These technologies are summarized in Table \ref{Table_EE}. 

\begin{figure}[!htb]
\center{\includegraphics[width=\columnwidth]
{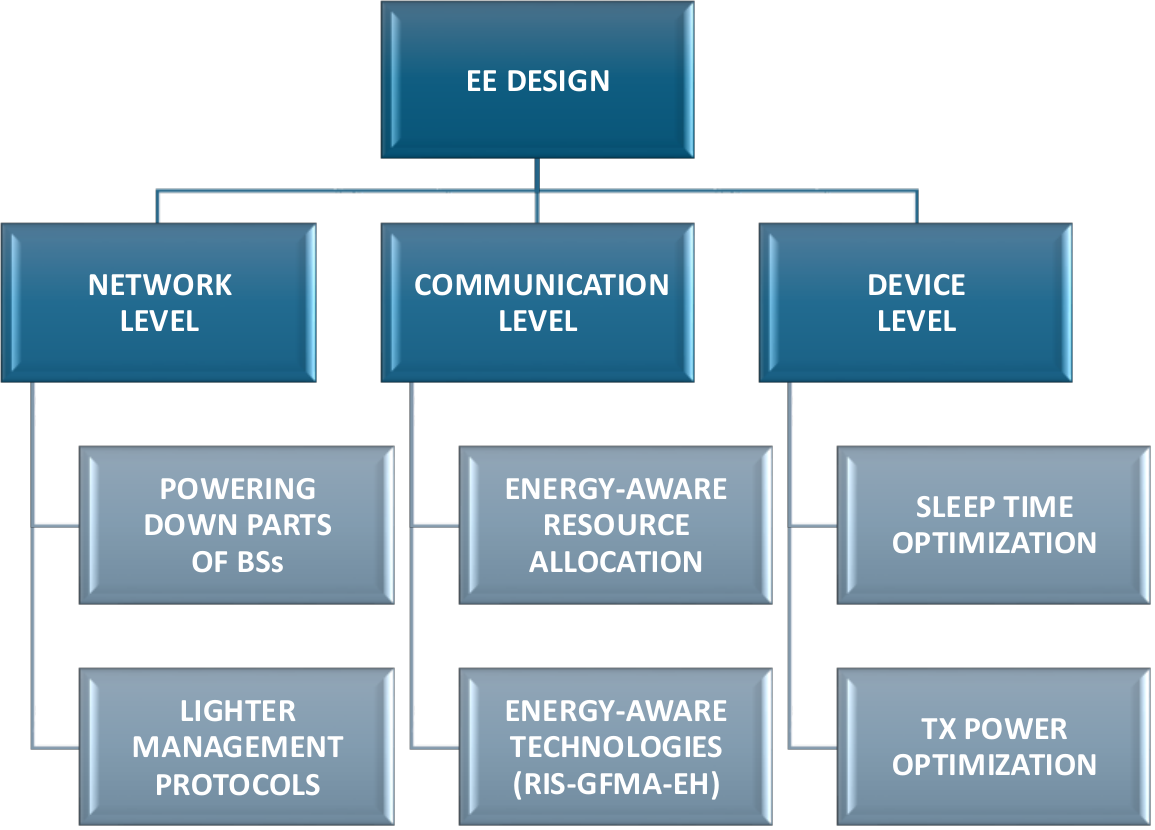}}
\caption{\label{Fig_Green}Energy-efficient design at different levels of green mobile networks.}
\end{figure}

\subsubsection{GREEN NETWORKS}

\begin{itemize}
\item {\textbf{Vision}}: 
6G networks are anticipated to be highly energy efficient at all levels of the network, providing ecological and economic benefits. 

\item {\textbf{Introduction}}: 
Green mobile networks aim to alleviate energy overhead and reduce the negative environmental impact caused by mobile networks \cite{Mao-22}. Green communication and networking focus on reducing energy consumption through energy-efficient design at different levels of the system, including the network, communication, and device levels, as summarized in Figure \ref{Fig_Green}. At the network level, the main trends include temporally powering down underutilized parts of the BSs (e.g., parts of the transmission equipment), virtual resource sharing among operators, and energy-aware network management approaches. 
At the communication level, the focus is on enhancing resource allocation in frequency, time, space, power, and code dimensions, utilizing energy-efficient technologies (e.g., RISs, grant-free access, EH, and backscatter communications), and designing energy-aware PHY layer techniques. 

At the device level, greener approaches are particularly needed for transmission power and sleep time optimization. The aforementioned green design aspects have been discussed in detail in \cite{Han-21b, Mao-22, Lopez-22, Larsen-23, Lopez-24}. In addition to the environmental impact, there is also an economic incentive since decreasing the energy overhead reduces energy costs as well. As the overall energy consumption of mobile networks is constantly growing, it is of utmost importance to develop novel energy-aware approaches for 6G networks. AI/ML is expected to play a key role in green 6G networks \cite{Mao-22}. 

\item {\textbf{Past and Present}}: 
Due to increasing energy costs and environmental impact, energy efficiency has been one of the key design principles in modern mobile networks. The concept of green networks dates back to the early 2010s \cite{Hasan-11}. This research boomed during the 2010s, with the main focus on green 5G networks \cite{Wu-17, Masoudi-19}. Green 5G networks have been intensively studied in theory and practice. Consequently, 5G supports many energy-efficient approaches, such as massive MIMO, ultra-lean carrier design, and sleep modes \cite{Lopez-22}. More information on the energy-efficient technologies used in 5G RAN is given in \cite{Lopez-22}. In the late 2010s, the focus of research started to shift toward green 6G networks \cite{Huang-19}. 

\item {\textbf{Opportunities and Challenges}}: 
In the big picture, the information and communication technology (ICT) industry is a major electricity consumer, causing massive usage of fossil fuels and high energy costs \cite{Mao-22, Larsen-23}. The total energy consumption of the ICT industry is exponentially growing since the volume of network infrastructure, number of devices, and amount of data and computing are constantly increasing \cite{Mao-22, Larsen-23}. Since mobile networks comprise a major part of the ICT industry, green communication and networking technologies are vital for alleviating this problem \cite{Mao-22, Larsen-23}. Diverse opportunities and challenges exist in the design of green mobile networks. The potential of green design in the 6G era is to significantly reduce the annual growth rate of the total energy consumption of mobile networks. Thus, green 6G networks are expected to provide ecological and economic benefits by reducing fossil fuel usage and energy costs \cite{Mao-22}. 

Diverse challenges must be resolved to obtain the potential benefits of green networks in practice. Mobile networks are becoming more complex and heterogeneous, with more resources, leading to more energy consuming networks. 6G is expected to utilize macro, small, and tiny BSs, as well as aerial/space APs. Moreover, the main 6G resources are expected to be communication, computation, caching, sensing, and energy. This heterogeneity makes the design and management of green networks challenging. AI/ML is seen as a key tool to alleviate the design of green 6G networks \cite{Mao-22}. Possible energy savings can be attained via AI/ML-optimized energy-aware network/resource management, network node cooperation/signaling, relaying, traffic/routing control, user scheduling, resource allocation, and mobility management \cite{Mao-22}. Pervasive AI/ML itself is energy-aggressive since it requires lots of computation power. Thus, it is another challenge to design energy-efficient and lightweight AI/ML approaches \cite{Mao-22}. Further details on the opportunities and challenges of green mobile networks can be found in \cite{Mao-22, Larsen-23}. 

\item {\textbf{Literature and Future Directions}}:
Recently, several surveys have been published on green communication and networking for 6G, reviewing the latest research progress and related literature \cite{Huang-19, Han-21b, Jiang-22, Mao-22, Larsen-23, Kumar-23, Naser-23, Lopez-24}. In \cite{Huang-19}, a survey was provided on green 6G networks. The focus was on network architectures (space-air-ground-sea, intelligence network, new network protocol stack) and promising technologies (THz/VLC, EH, molecular/quantum communication, blockchain-based security, intelligent materials). In \cite{Han-21b}, the authors discussed greener PHY layer technologies for 6G, analyzing the joint energy-spectral efficiency design in NOMA and waveform overlapping multiple access frameworks. The work \cite{Jiang-22} reviewed green-UAV communications for 6G in terms of power consumption models, trends, enabling techniques, applications, and open research problems. 

In \cite{Mao-22}, a comprehensive review was presented of AI models for green communication toward 6G networks. In this framework, the work considered the existing literature, 6G paradigms, AI models, mobile network communications, MTC, computation-oriented communications, and open research challenges. The authors of \cite{Larsen-23} explored greener 5G/B5G access networks, focusing on energy consumption modeling, energy-efficient network architectures, technological evolution, and network sharing. The paper \cite{Kumar-23} provided a thorough discussion on sustainable 6G toward greener networks. In \cite{Naser-23, Lopez-24}, the surveys discussed 6G green communication in terms of zero-energy devices, focusing on technological enablers. 
Although AI/ML has been widely studied for green 6G networks since the late 2010s, major efforts are still needed for practical realizations in the near future. As discussed earlier, at the center of future research is AI/ML-optimized green designs at all levels of the 6G networks \cite{Mao-22}. 
\end{itemize}

\begin{figure}[!htb]
\center{\includegraphics[width=0.9\columnwidth]
{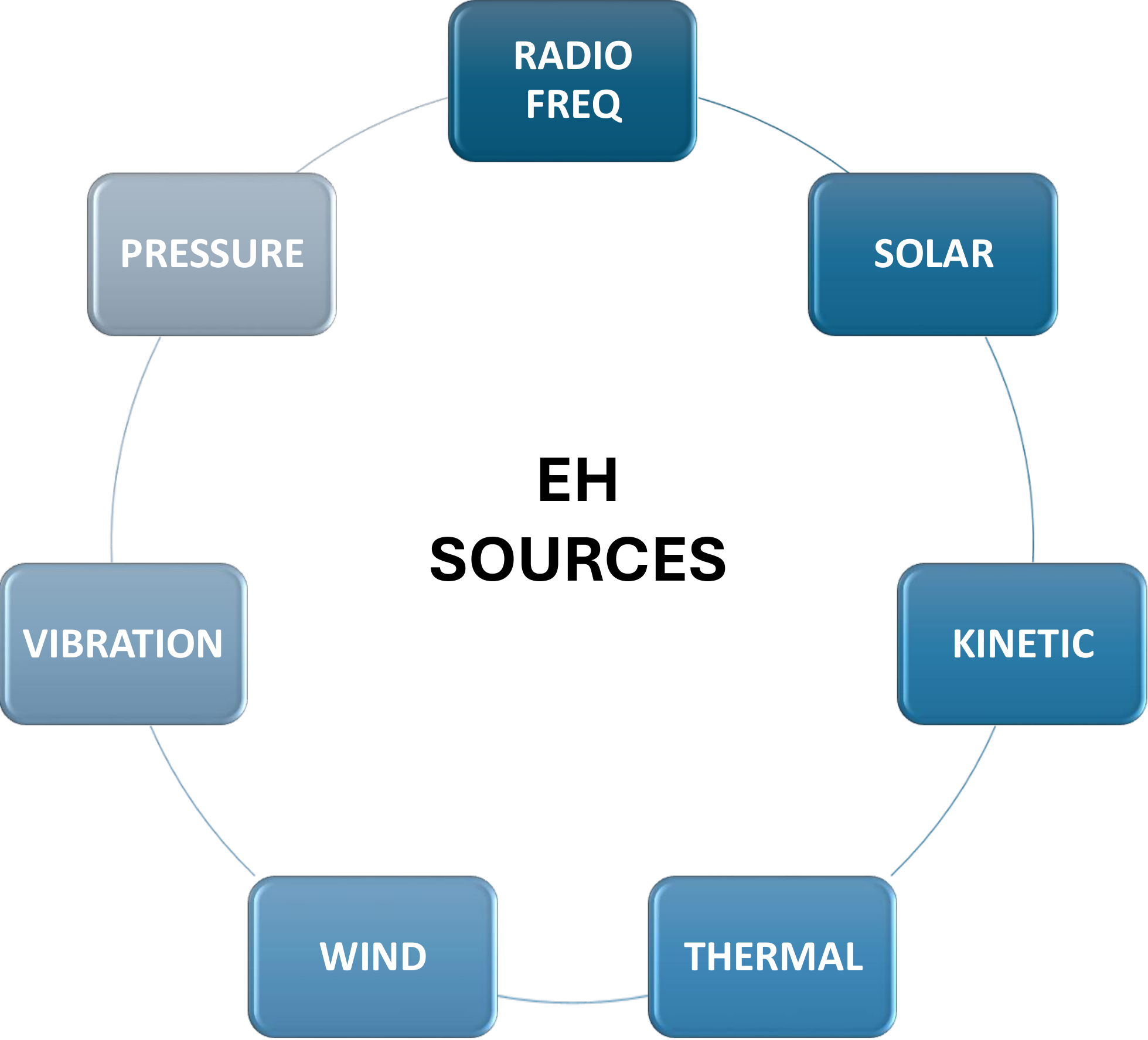}}
\caption{\label{Fig_EH}Typical energy sources for energy harvesting.}
\end{figure}

\subsubsection{ENERGY HARVESTING}

\begin{itemize}
\item {\textbf{Vision}}: 
EH has been recognized as a promising technology for promoting energy-constrained, low-power, and sustainable IoT networks in the 6G era. 

\item {\textbf{Introduction}}: 
EH is defined as a process in which energy is collected from external sources in order to promote the autonomy, mobility, and sustainability of wireless devices and networks \cite{Mao-22}. Typical external energy sources are RF, solar, kinetic, thermal, wind, vibration, and pressure \cite{Mao-22}, as illustrated in Figure \ref{Fig_EH}. Due to its properties, RF energy is well-suited for mobile network applications, such as low-power IoT. Thus, the focus is on RF-EH from now on. 
RF-EH can be divided into two types: ambient and dedicated \cite{Lu-15}. In the ambient RF-EH, energy is harvested from existing radio signals, such as cellular communications, television broadcasts, and WiFi connections. The benefit of ambient RF-EH is that no dedicated infrastructure is needed for energy sources, while the drawback is the possible availability and energy level issues. In the dedicated RF-EH, dedicated energy transmitters are used to transfer energy to the EH devices. This enables guaranteed availability and energy levels. Typical application scenarios of RF-EH include low-power IoT networks, wireless sensor networks, and wireless-powered communication networks. 

RF-EH networks can be divided into centralized (infrastructure-based) and decentralized (infrastructure-less) architectures, as introduced in \cite{Lu-15}. A typical centralized architecture consists of three main elements: information gateways (BSs), energy sources (dedicated or ambient), and network nodes (devices) \cite{Lu-15}, as shown in Figure \ref{Fig_EH2}. The key components of an RF-EH device include an RF energy harvester, power management module, energy storage, application, lightweight microcontroller, and lightweight RF transceiver \cite{Lu-15}. Typically, an RF energy harvester comprises an antenna, impedance matching module, voltage multiplier, and capacitor \cite{Lu-15}. The aforementioned EH architecture is known as harvest-store-use. Another main class is a harvest-use architecture that lacks the capability of storage. This leads to a reduced cost of EH devices. 

\begin{figure}[!tb]
\center{\includegraphics[width=\columnwidth]
{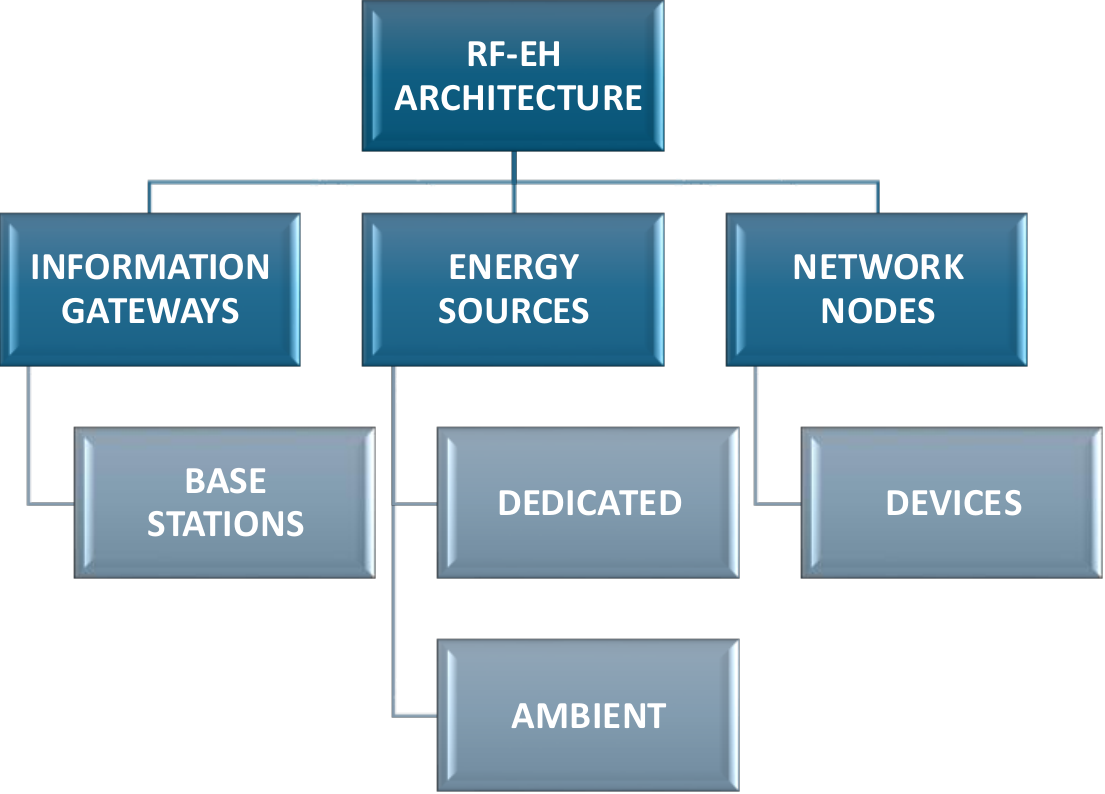}}
\caption{\label{Fig_EH2}Three main elements of a centralized RF-EH architecture.}
\end{figure}

\item {\textbf{Past and Present}}:
RF-EH has been actively researched since the 2000s, with the main focus on wireless sensor networks. In the 2010s, RF-EH for cellular network applications (e.g., IoT) began to receive more attention in the wireless community. Currently, the focus of cellular RF-EH research is on 6G, with special interest in energy-limited IoT networks. Although commercialized RF-EH solutions exist, RF-EH is still too immature for large-scale use in cellular networks. 

\item {\textbf{Opportunities and Challenges}}: 
RF-EH offers many new opportunities for mobile networks to advance sustainability, autonomy, and mobility. Ultimately, RF-EH acts as a key enabler for the large-scale usage of lightweight communication networks in the 6G era, such as different types of energy-limited IoT and sensor networks. There are still many technological obstacles to overcome before RF-EH can become a mainstream technology in mobile networks. The main challenges of RF-EH are related to the optimization of energy conversion efficiency, energy consumption, distance between energy sources and harvesters, operating frequency, locations of the dedicated energy sources, and operational/capital expenses \cite{Lu-15, Sherazi-22}. 

\item {\textbf{Literature and Future Directions}}:
In recent years, a handful of survey papers have been published, discussing different aspects of RF-EH and reviewing recent advances \cite{Amer-20, Tedeschi-20, Sherazi-22, Ibrahim-22, Rahmani-23, Moloudian-24, Khan-24d}. RF-EH-based metasurface structures were reviewed in \cite{Amer-20}. RF-EH and metasurfaces were first considered separately, and then RF-EH antenna and rectenna designs based on metasurfaces were discussed. The work \cite{Tedeschi-20} reviewed security in EH networks from the perspective of threats/attacks, PHY layer secrecy, lightweight cryptography, and additional PHY layer countermeasures. In \cite{Sherazi-22}, the authors provided a comprehensive review of RF-EH, discussing applications, evaluation metrics, energy propagation models, rectenna architectures, MAC layer protocols, open challenges, and future directions. Another comprehensive survey of RF-EH was presented in \cite{Ibrahim-22}. This study reviewed many topics, including RF-EH systems, techniques, principles, evaluation metrics, environments, circuits, and applications. In \cite{Rahmani-23}, RF-EH was briefly discussed in the context of next-generation IoT devices. The paper \cite{Moloudian-24} studied RF-EH techniques for low-energy devices in the context of IoT, industry 4.0, and wireless sensing. The authors in \cite{Khan-24d} examined multi-directional rectennas in RF-EH. In \cite{Elahi-20, Williams-21, Calautit-21, Sanislav-21}, the generic surveys explored EH in terms of different energy sources. 

Since RF-EH is still far from its potential, significant research efforts are needed in the future. Therefore, it is vital to address the main challenges mentioned earlier. For instance, improvements in the conversion efficiency and energy consumption of EH devices call for advanced antenna and circuit designs. Further studies are required to obtain and maintain an appropriate level of harvested power density in the devices, especially in mobile environments. Moreover, the locations of dedicated energy transmitters need to be properly determined to ensure fairness among EH devices. More experiments are also required in realistic environments to validate the performance of RF-EH systems. Further details of the aforementioned issues and future directions can be found in \cite{Sherazi-22}. 
\end{itemize}

\subsubsection{BACKSCATTER COMMUNICATIONS}

\begin{itemize}
\item {\textbf{Vision}}: 
Backscatter communication is seen as a promising technology to provide low-power, low-complexity, and low-cost communication systems for 6G, enabling lightweight IoT networks. 

\begin{figure}[!tb]
\center{\includegraphics[width=\columnwidth]
{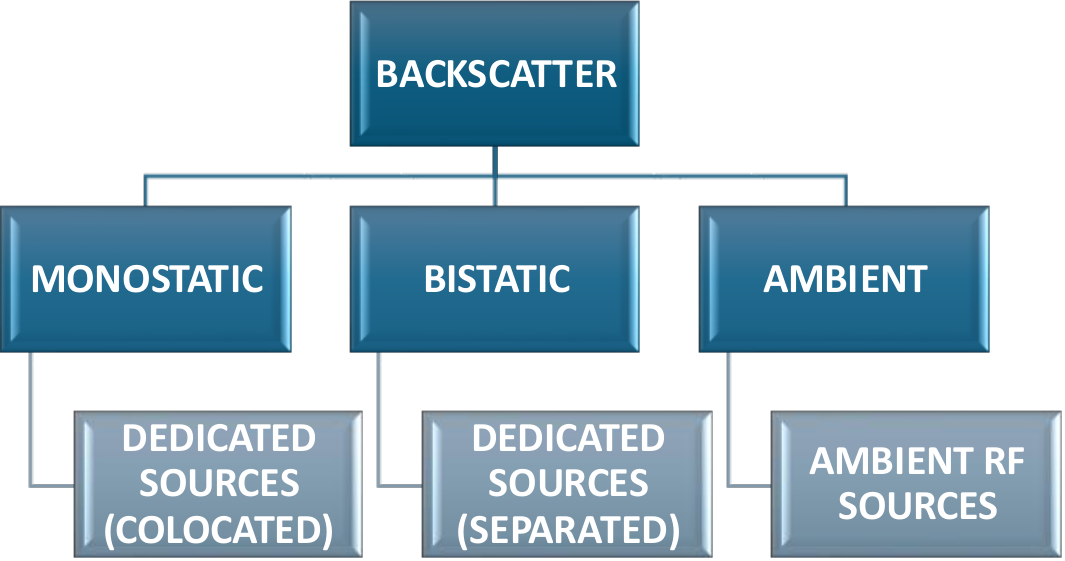}}
\caption{\label{Fig_BC}Three main types of backscatter communication systems.}
\end{figure}

\item {\textbf{Introduction}}: 
A typical backscatter system consists of three main elements, including an RF signal source (i.e., an emitter carrier), a backscatter transmitter, and a backscatter receiver \cite{Huynh-18}. The basic principle of backscatter communications is that the backscatter transmitter absorbs RF signals from an external source, modulates the signals with its own encoded data, and reflects the modified signals toward the backscatter receiver, which decodes and extracts the desired information from the received signals \cite{Huynh-18}. A key feature of backscatter communication is that the backscatter transmitter does not need to generate RF signals by itself \cite{Huynh-18}. As a result, the low-complexity backscatter transmitters can operate with minimal power and be implemented at low cost. The goal of backscatter communications is to provide low-power, low-complexity, and low-cost communication systems for different types of application scenarios, typical ones ranging from RFID systems to sensor and IoT networks. 

There are three main types of backscatter communication systems: monostatic, bistatic, and ambient \cite{Huynh-18}, as summarized in Figure \ref{Fig_BC}. The monostatic system consists of two main components, i.e., a backscatter transmitter and a reader device that contains an RF source and a backscatter receiver \cite{Huynh-18}. Since the receiver is in the same device as the RF source, monostatic systems suffer from a round-trip pathloss and the doubly near-far problem, limiting communication ranges and data rates. Typical usage scenarios of the monostatic systems are short-range RFIDs applications. In the bistatic system, the dedicated RF sources are separated from the backscatter receivers, avoiding the round-trip pathloss and doubly near-far problems. This leads to improved system performance, coverage, and flexibility. 

The ambient backscatter systems exploit existing wireless systems, such as cellular networks, WiFi, and television broadcasts as the RF sources \cite{Huynh-18}. In contrast to the bistatic approach, ambient systems do not require any dedicated spectrum or RF sources for communication, leading to improved spectral, power, and cost efficiency. However, ambient RF signals are uncontrollable and unpredictable, leading to more complicated system design and unstable performance. Nevertheless, the ambient backscatter communication concept is the most evolved and promising one to be utilized in 6G for low-power, low-complexity, and low-cost communication scenarios, enabling applications such as energy-constrained battery-free IoT networks \cite{Duan-20, Nawaz-21}. A comprehensive overview of the backscatter communications systems was presented in \cite{Huynh-18}. 

\item {\textbf{Past and Present}}:
The concept of backscatter communication was first introduced in the late 1940s \cite{Stockman-48}. Later on, this innovation has contributed to many practical applications, such as RFID, remote switches, tracking devices, and medical telemetry \cite{Huynh-18}. Currently, RFID is the most popular commercial application of backscatter communications, typically used for commodity identification in retailing and logistics. However, the applicability of traditional (monostatic) backscattering is rather limited due to its inherent constraints \cite{Huynh-18}. First, the receivers and RF sources are in the same device. Second, the transmitters must be located near their dedicated RF sources. Third, the backscatter transmitters can communicate only when requested by their dedicated receivers. To overcome these limitations, enhanced backscatter systems, i.e., bistatic \cite{Kimionis-12} and ambient \cite{Liu-13}, were developed in the early 2010s, receiving a considerable amount of research interest. 

The bistatic approach separates the RF sources and receivers, enabling a more flexible system design, improved performance, and extended coverage \cite{Kimionis-14}. Ambient backscatter systems go a step further by exploiting existing RF sources, such as cellular and WiFi signals. Consequently, their applicability is much wider than that of the monostatic systems. In particular, advanced backscatter communication is seen as an enabler for future low-power, low-complexity, and low-cost communication systems. Consequently, practical bistatic and ambient backscatter systems are currently under active study, particularly for 6G IoT scenarios \cite{Nawaz-21}. Recently, different types of modified backscatter concepts, with their own pros and cons, have been introduced, such as multi-antenna, full-duplex, NOMA-aided, UAV-aided, and RIS-aided backscatter systems. Detailed reviews of the state-of-the-art backscatter approaches can be found in \cite{Huynh-18, Xu-18, Yao-20, Duan-20, Nawaz-21, Wu-22}. For example, performance comparisons were provided among a set of backscatter systems in \cite{Xu-18, Wu-22}. In particular, the transmission powers, data rates, and communication ranges were compared. 

\item {\textbf{Opportunities and Challenges}}: 
There are many benefits in backscatter communications, offering lots of opportunities for future wireless systems. However, backscatter communication has some shortcomings, placing many obstacles on the road toward large-scale practical deployments. Due to its inherent nature, backscatter communications support energy-constrained lightweight wireless systems \cite{Huynh-18}. As a result, the backscatter concept is seen as a promising approach to realize sustainable IoT applications in the 6G era \cite{Nawaz-21}. Potential application scenarios include smart environments, healthcare, logistics, retail, health monitoring, and sport innovations, to mention a few. For example, backscatter-based sensor networks can be used to detect toxic gases, carbon dioxide, smoke, and movement in smart homes, offices, and buildings. Backscatter communication is also applicable to logistics, warehousing, and retailing for tracking, identification, and monitoring. In health and sport monitoring, backscatter approaches can enable in-body, on-body, and wearable sensor networks. Further details on the aforementioned application scenarios can be found in \cite{Huynh-18, Liu-19e, Nawaz-21, Wu-22}. 

The main shortcomings of backscatter communications include low data rates, short communication distances, lack of strict QoS guarantees, and security issues \cite{Huynh-18}. Since the backscatter transmitters rely on the external RF sources, their transmission powers are low, leading to low data rates and short link ranges. Due to the unpredictability of ambient RF signals, QoS cannot be guaranteed. Since the backscatter transmitters and receivers are low-complex devices, they are prone to security attacks, such as eavesdropping and jamming. The corresponding challenges are to maximize rate and range, increase robustness, and provide secure communication. In addition, other fundamental challenges include minimizing energy consumption and cost. Since most of these objectives conflict with each other, there need to be priorities set and trade-offs made, depending on the purpose of the system. Recent backscatter systems, aiming to tackle these challenges with different types of trade-offs, have been reviewed in \cite{Huynh-18, Xu-18, Duan-20, Nawaz-21, Wu-22}. 

\begin{table*}[htb!]
\begin{center}
\caption{Summary of end-device-oriented technologies for 6G}
\label{Table_Device}
\centering
\begin{tabularx}{\textwidth}{| >{\centering\arraybackslash}X | >{\centering\arraybackslash}X |
>{\centering\arraybackslash}X | 
>{\centering\arraybackslash}X |
>{\centering\arraybackslash}X |
>{\centering\arraybackslash}X |
>{\centering\arraybackslash}X |
>{\centering\arraybackslash}X |}
\hline
\centering
\vspace{3mm} \textbf{End-Device-Oriented Technologies} \vspace{3mm} & \centering \textbf{Vision} & \centering \textbf{Description} & \centering \textbf{Opportunities} & \centering \textbf{Challenges} & \centering \textbf{Past} & \vspace{1.5mm} \begin{center} \textbf{Present} \end{center} \\
\hline
\vspace{3mm} D2D Communications \vspace{3mm}  & Ubiquitous D2D & Direct comm between devices & Data off-loading $\&$ power savings & Interference management & 4G sidelink & 5G sidelink \\
\hline
\vspace{3mm} V2X Communications \vspace{3mm}  & Intelligent vehicular systems & Comm bw vehicle and any entity & Ubiquitous vehicle connectivity & Management $\&$ cooperation & 4G V2X & 5G V2X \\
\hline
\vspace{3mm} Cellular UAV Communications \vspace{3mm}  & Freeing potential of UAVs & Mobile network support for UAVs & New usages for UAVs & 3D coverage $\&$ mobility & Research since early 2010s & 5G UAS \\
\hline
\end{tabularx}
\end{center}
\end{table*}

\item {\textbf{Literature and Future Directions}}:
Modern backscatter research began to gain notable attention after the introduction of the bistatic and ambient backscatter concepts in the early 2010s. Since then, backscatter communication has been studied from various perspectives in the literature. This research can be divided into three main categories: fundamentals, emerging systems, and applications. The main topics in fundamental research included channel coding, modulation, channel modeling, channel estimation, detection, resource allocation, multiple access, performance analysis, and PHY layer security. Literature reviews of these topics can be found in \cite{Huynh-18, Memon-19, Liu-19e, Nawaz-21}. 

The research on emerging systems can be divided into two main categories, i.e., new system concepts and combinations with other emerging technologies. Several novel backscatter systems have recently been proposed \cite{Xu-18, Yao-20, Wu-22}. Typically, they aim to provide (relatively) high throughput or long range with low transmission power and implementation costs. For example, the passive WiFi and Interscatter systems achieved 11 Mbit/s data rates, while the LoRea and LoRa backscatter concepts obtained 3.4 km and 2.8 km communication distances, respectively \cite{Xu-18}. Recently, backscatter communications have been studied in combination with other emerging technologies \cite{Nawaz-21, Xu-23b}, such as MIMO, NOMA, UAVs, RISs, VLC, and AI/ML. These combinations can potentially provide different types of benefits, ranging from improved throughput or range to increased reliability. 

In the application domain, backscatter communications have been studied for various scenarios, such as green IoT, ultra-low power sensor networks, smart homes, healthcare, biomedical applications, environmental monitoring, logistics, transportation, smart cities, agriculture, and body-area networks. In general, backscatter systems can assist in many important functions in those application scenarios, such as sensing, localization, tracking, identification, and monitoring. Backscatter-assisted applications have been reviewed in \cite{Huynh-18, Liu-19e, Nawaz-21, Wu-22}. Since backscatter communication is a relatively immature technology in the context of mobile networks, further research is needed on the aforementioned aspects. A special focus should be directed on the 6G applications. 

Insightful surveys on backscatter communications can be found in \cite{Huynh-18, Xu-18, Memon-19, Liu-19e, Memon-20, Yao-20, Duan-20, Nawaz-21, Wu-22, Xu-23b, Ahmed-24}, from which \cite{Huynh-18, Memon-20, Duan-20, Wu-22} focused on the ambient backscatter systems. These papers covered a broad range of aspects, including fundamentals, state-of-the-art, opportunities, applications, challenges, possible solutions, open problems, and future topics. In \cite{Huynh-18}, a generic survey was presented, with the main focus on the ambient and bistatic backscatter systems. A practical tutorial was provided in \cite{Xu-18}, concentrating on the signal processing perspective and reviewing state-of-the-art backscatter systems and their performance. The authors in \cite{Memon-19} reviewed backscatter communications as a solution to the limited battery life problem. The next-generation backscatter communication systems, techniques, and applications were surveyed in \cite{Liu-19e}. The ambient backscatter concept was explored as an enabler for energizing IoT devices in \cite{Memon-20}. In \cite{Yao-20}, the latest backscatter systems were discussed in the context of IoT. 

The ambient backscatter technologies were examined for ultra-low-energy MTC in \cite{Duan-20}. The work in \cite{Nawaz-21} reviewed backscatter communications to enable ultra-massive connectivity in 6G. A comprehensive survey of the ambient backscatter technologies was presented in \cite{Wu-22}, focusing on the latest concepts proposed in the literature and providing a taxonomy for ambient systems. The paper \cite{Xu-23b} discussed AI/ML-empowered backscatter communications for 6G. In \cite{Ahmed-24}, the authors explored NOMA-based backscatter systems, focusing on technological principles, performance optimization, and potential applications. Moreover, backscatter communication was briefly discussed in terms of next-generation IoT and zero-energy devices in \cite{Rahmani-23} and \cite{Naser-23, Lopez-24}, respectively. 
Further work is needed in all of the aforementioned areas, especially in the context of 6G and its unique characteristics, requirements, and applications. In addition, the fundamental challenges, discussed earlier, must be properly addressed toward the 6G era. 
\end{itemize}

\subsection{END-DEVICE-ORIENTED COMMUNICATION TECHNOLOGIES FOR 6G}
6G is expected to support massive amounts of wireless devices, vehicles, and drones, thereby setting diverse requirements for mobile networks. In this picture, D2D, V2X, and cellular-connected UAV communications will play key roles. Each of these end-device-related technologies is discussed below and summarized in Table \ref{Table_Device}. 

\begin{figure}[!htb]
\center{\includegraphics[width=0.55\columnwidth]
{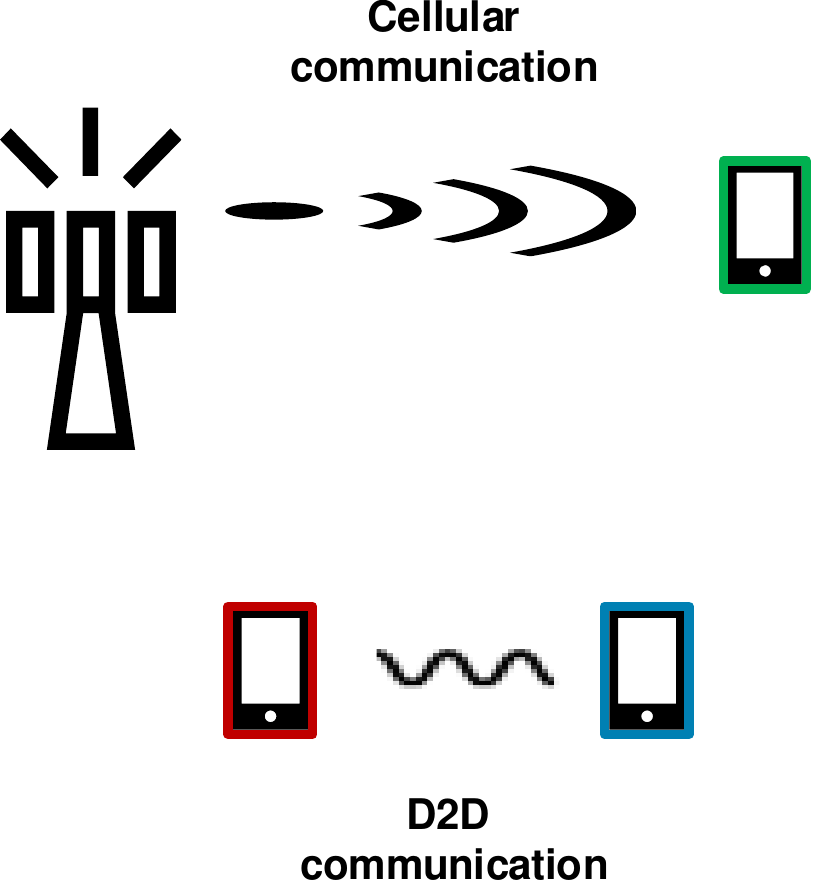}}
\caption{\label{Fig_D2D2}Cellular versus D2D communications.}
\end{figure}

\subsubsection{D2D COMMUNICATIONS}

\begin{itemize}
\item {\textbf{Vision}}:
6G is expected to provide wide support for D2D communications in cellular, industrial, vehicular, and aerial environments. 

\item {\textbf{Introduction}}:
D2D communication refers to direct communication between close-proximity devices without BS involvement \cite{Jameel-18}, as illustrated in Figure \ref{Fig_D2D2}. D2D communications can be classified into two main categories: in-band and out-band \cite{Jameel-18, Gismalla-22}, as summarized in Figure \ref{Fig_D2D}. In the in-band case, there are two modes, i.e., underlay, using the same spectrum as the cellular communication, and overlay, using a dedicated part of the cellular spectrum. Out-band communication is either autonomous between devices or controlled by the BS. D2D communications can provide diverse benefits, such as data traffic off-loading, improved spectral efficiency, decreased latency, increased throughput, better reliability, and improved energy efficiency \cite{Gismalla-22}. Common drawbacks include challenging interference and resource management \cite{Jameel-18, Gismalla-22}. Typical application scenarios include mobile traffic off-loading, relaying, content sharing/distribution, proximity-aware services, and broadcasting road safety messages \cite{Jameel-18}. 

\begin{figure}[!tb]
\center{\includegraphics[width=0.75\columnwidth]
{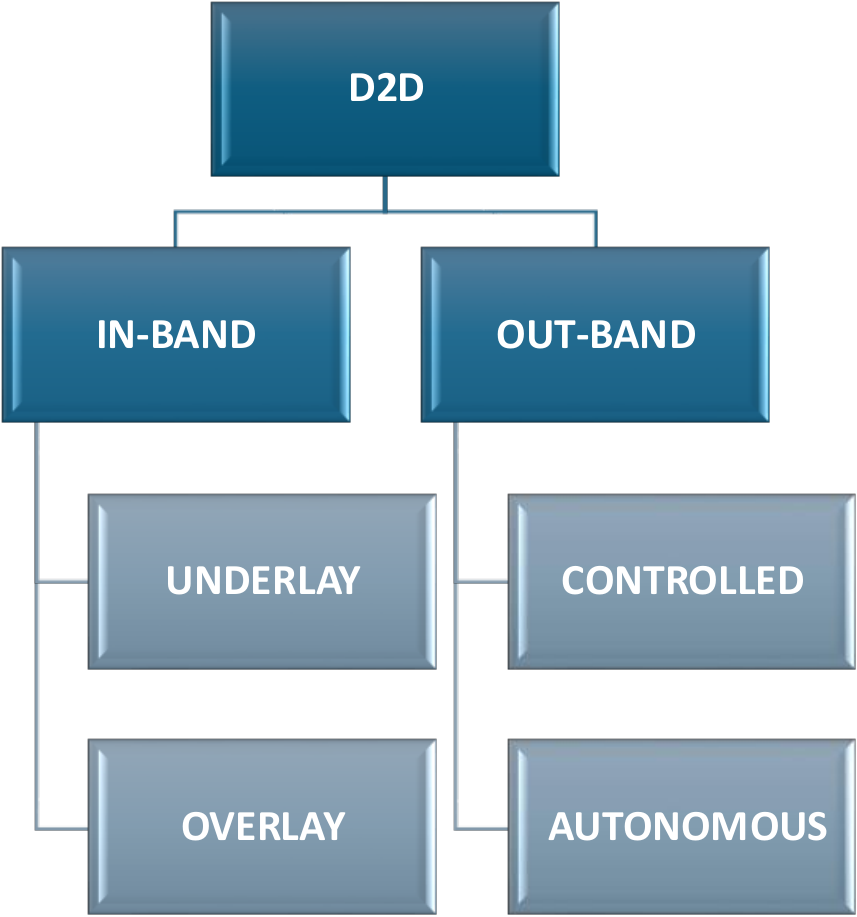}}
\caption{\label{Fig_D2D}Different types of D2D communication \cite{Gismalla-22}.}
\end{figure}

\item {\textbf{Past and Present}}: 
The concept of D2D communications was first proposed in the context of cellular multi-hop relays in 2000 \cite{Lin-00}. However, it was until the late 2000s when D2D research started to boom, with a special focus on LTE scenarios \cite{Liu-15}. Since then, D2D communication has been an active research topic for practical use in cellular networks. In the early 2010s, D2D communication, or sidelink communication, as it is known in the 3GPP specifications, was adopted for 4G LTE-A evolution \cite{Harounabadi-21}. Specifically, sidelink communication was first specified in Release 12, followed by sidelink relaying in Release 13. Releases 14 and 15 provided further enhancements for LTE sidelink, for example, in the context of V2X communications. Recently, sidelink communication has been further enhanced in the 5G NR standards \cite{Harounabadi-21}. NR sidelink was first introduced in Release 16, with the main focus on V2X scenarios. Release 17 defined further enhancements, such as reliability/latency improvements and UE power savings \cite{Rel-17}. Release 18 introduced sidelink positioning, NR sidelink evolution, and NR sidelink relay enhancements \cite{Rel-18, Lin-22}. 

\item {\textbf{Opportunities and Challenges}}: 
The ultimate potential of D2D communications is that any device is a possible D2D node, providing connectivity, content, and assistance to nearby devices, systems, and applications. Depending on their capabilities, D2D nodes can potential offer assistance beyond communication as well, such as computation, caching, positioning, sensing, and energy. Furthermore, D2D nodes can form clusters to be more powerful as a group. Although this vision is hypothetical, it shows the way forward. 

6G provides a major opportunity to take D2D communication to the next level and make it a mainstream cellular technology. In this picture, the synergy with other emerging 6G technologies plays a key role by providing joint benefits and broadening their combined applicability and capabilities \cite{Gismalla-22}. For instance, D2D communications at THz and VL frequencies enable ultra-high data rates. For IoT scenarios, D2D communication can improve connectivity and robustness. Extending D2D communications and relaying to aerial scenarios enables novel applications for UAVs and provides benefits, such as improved coverage, efficiency, and safety. The potential benefits of RIS-aided D2D communications are coverage extensions and obstacle avoidance. AI/ML is a promising technology to enhance resource management and mode selection mechanisms in D2D communications. As the devices and networks become more powerful and intelligent toward 6G, devices could also offer more than connectivity to each other, e.g., computation power, caching, positioning/sensing assistance, and useful interactions as a part of smart applications. The formation of D2D clusters can make these interactions more powerful. These clusters can be autonomous or network-controlled, being also part of the network-driven EI. 

In general, D2D communications can provide performance improvements in terms of spectral efficiency, throughput, latency, energy efficiency, reliability, and coverage \cite{Jameel-18, Gismalla-22}. System-level spectral efficiency can be improved by using underlay in-band D2D communication, which uses the same spectral resources as the cellular network. The potential throughput improvements are due to short communication ranges. Latency reduction is possible due to fast connection establishments and short link and content distances. The reduced energy consumption is due to the low transmission powers and light signaling. D2D relaying can provide coverage extensions and increased reliability for cell-edge and out-of-cell scenarios. Numerous applications can benefit from D2D communications \cite{Jameel-18}, such as location-aware services, online gaming, social networking, interactive advertising, content sharing, public safety broadcasting, road safety/efficiency messages, interactive entertainment, etc. 

In contrast to the numerous opportunities, there are many challenges in D2D communications. Typical ones are related to interference management, device discovery, mobility management, and security/privacy \cite{Jameel-18, Gismalla-22}. The integration of underlay D2D communications into cellular networks makes interference management more challenging since a new tier is added to the network architecture, leading to the new types of interference scenarios. In a such network, there exist two tiers, i.e., a cell-tier and a D2D-tier \cite{Jameel-18}, which cause interference to each other. In addition to inter-tier interference, there exists also intra-tier interference within both tiers. More complicated interference scenarios call for novel interference management solutions without overwhelming signaling overhead and computational complexity. 

Another fundamental challenge in D2D communications is the discovery of other D2D users in the network. Device discovery can be classified into two categories, centralized and distributed \cite{Jameel-18}. In the centralized case, device discovery is handled completely or partially through the BS. The complete involvement of the BS is more controlled, requiring more signaling and time to establish a connection between D2D users. In the case of partial involvement, device discovery is faster with reduced signaling overhead, but less controlled and may cause interference issues in the network. In the distributed case, device discovery is handled by the D2D users, without any involvement of the BS, through periodic beacon signaling. Compared to the centralized discovery, distributed discovery is less time- and resource-consuming at the cost of possible interference issues. Generally, distributed discovery is preferable in delay-constrained application scenarios. Typical challenges in distributed discovery are related to the design (structure/frequency), synchronization, and interference of beacon messages \cite{Jameel-18}. 

D2D communication can be interrupted when D2D users are mobile and moving away from one another. Thus, developing appropriate mobility management mechanisms for D2D communications is highly important. Due to its direct nature, D2D communication exposes the network and users to diverse security and privacy threats. Typical threats include, among others, malicious attacks, eavesdropping, modified messages, and node impersonation \cite{Jameel-18}. Also the anonymity, confidentiality, and integrity of data may be threatened \cite{Jameel-18}. To ensure safe D2D communications, different security and privacy threats need to be properly addressed and efficient solutions developed. More information on the opportunities and challenges of D2D communications can be found in \cite{Jameel-18, Gismalla-22}. 

\item {\textbf{Literature and Future Directions}}: 
In the literature, the early research focus of D2D communications was first on 4G LTE use cases \cite{Asadi-14, Liu-15} but it shifted to 5G \cite{Ansari-18} around the mid-2010s and has recently turned toward 6G \cite{Gismalla-22}. Over the years, a broad range of D2D topics has been examined, such as device discovery, mode selection, interference management, power control, resource allocation, relaying, mobility management, and security/privacy \cite{Jameel-18, Adnan-20}. Recent survey articles have reviewed the latest literature, state-of-the-art solutions, recent advances, and open challenges of D2D communications \cite{Adnan-20, Zhang-20l, Gismalla-22, Mach-22, Iqbal-23b, Alibraheemi-23, Areqi-23, Rathod-24}. These surveys discussed D2D in terms of 5G \cite{Adnan-20}, 6G \cite{Zhang-20l, Gismalla-22, Mach-22}, cognitive D2D \cite{Iqbal-23b}, resource management \cite{Alibraheemi-23}, state-of-the-art solutions \cite{Areqi-23}, and AI-based resource allocation \cite{Rathod-24}. 

Since 6G provides diverse opportunities and challenges for D2D communications \cite{Gismalla-22}, future research needs to be adapted accordingly. Many emerging 6G technologies can benefit D2D communications if combined properly \cite{Gismalla-22}. In addition, the capabilities of D2D can be broadened from communication to computation, caching, sensing, and positioning. In general, 6G aims to provide broad support for D2D communications in diverse scenarios, including cellular, industrial, vehicular, and aerial environments. Future research directions can be divided into two main categories, i.e., extending standards and studying novel topics. 5G supports D2D mainly in V2X, public safety, and specific application scenarios. Release 18 extended the support for higher frequencies (i.e., FR2), unlicensed spectrum, carrier aggregation, and positioning \cite{Rel-18, Lin-22}. For 6G, important future studies include extensions to the THz spectrum, IoT scenarios, UDNs, and UAV environments \cite{Jameel-18, Gismalla-22}. Promising future topics worth studying include AI/ML-aided D2D, RIS-aided D2D, D2D-aided EI, VLC D2D, and D2D sensing. In addition, D2D security and privacy concepts need a major update for the 6G era. 
\end{itemize}

\begin{figure}[!htb]
\center{\includegraphics[width=\columnwidth]
{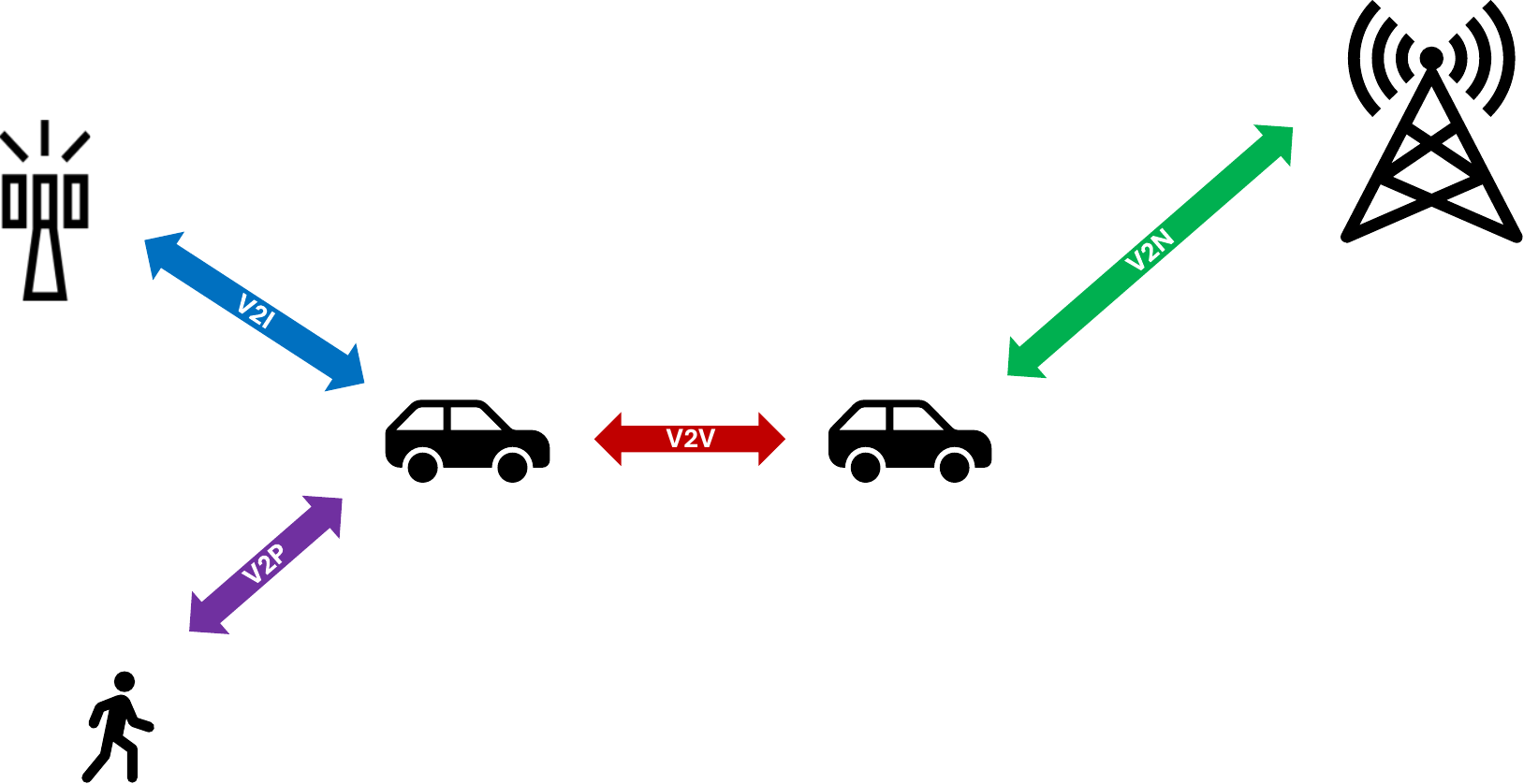}}
\caption{\label{Fig_V2X}Four main types of V2X communications.}
\end{figure}

\subsubsection{V2X COMMUNICATIONS}

\begin{itemize}
\item {\textbf{Vision}}:
V2X communication is expected to act as a key enabler for future intelligent vehicular and transportation systems in the 6G era. 

\item {\textbf{Introduction}}:
V2X communication is generally defined as the wireless connectivity between a vehicle and any type of an object or entity that is under the influence of the vehicle, or vice versa. V2X communication comprises various connectivity types, the main ones being V2V, vehicle-to-infrastructure (V2I), vehicle-to-network (V2N), and vehicle-to-pedestrian (V2P) communications, as defined in 3GPP \cite{Bagheri-21} and illustrated in Figure \ref{Fig_V2X}. V2V refers to the D2D-type direct communication among nearby vehicles. The V2I technology provides communication between vehicles and road infrastructure (e.g., road site units, traffic lights, traffic signs, etc.). V2N connects vehicles and cellular networks, extending the capabilities of V2X. Vehicles and pedestrians are connected via V2P communication. The goal of V2X communications is to promote road safety, traffic efficiency/management, vehicle cooperation, driver assistance, connected autonomous driving, and intelligent vehicle/transportation systems. 

In the standardization-wise, there are two types of V2X communication technologies, i.e., IEEE WiFi-based V2X, known as Dedicated Short-Range Communication (DSRC), and 3GPP cellular-based V2X (C-V2X) \cite{Noorarahim-22}. WiFi-based DSRC, which supports V2V and V2I technologies, does not require any communication network infrastructure to form ad hoc-type connections among vehicles and between vehicles and traffic infrastructure. The shortcoming of WiFi-based V2X communications is that strict service requirements cannot be guaranteed due to unlicensed spectrum operation. In addition to V2V and V2I, C-V2X supports V2N connectivity, which is particularly important for enabling future intelligent vehicular and transportation systems. 

\item {\textbf{Past and Present}}: 
The concept of V2X communications was originally standardized by IEEE in 802.11p in 2010 \cite{Noorarahim-22}. The first generation of this WiFi-based concept was named DSRC, which covers the V2V and V2I technologies. The second-generation DSRC was released in IEEE 802.11bd in 2019 \cite{Noorarahim-22}. In 2017, 3GPP standardized 4G LTE V2X (i.e., 4G-V2X or LTE-V2X) in Release 14 \cite{Garcia-21}. Two communication modes were introduced: LTE-Uu for long-range network connectivity via V2N and LTE sidelink for short-range direct connectivity via V2V and V2I \cite{Noorarahim-22}. Release 15 updated 4G-V2X direct communication parts (named LTE-eV2X) to support transmit diversity for better reliability as well as higher-order modulation and carrier aggregation for improved throughput \cite{Noorarahim-22}. In general, 4G-V2X supports basic road safety and traffic management use cases. 

The first 3GPP studies on 5G NR V2X (i.e., 5G-V2X or NR-V2X) were introduced in Technical Report (TR) 22.886 (Release 15/16), examining the enhancement of 3GPP support for 5G-V2X services \cite{TR22886}. The report identified advanced V2X use cases in four main groups, i.e., vehicle platooning, advanced driving, extended sensors, and remote driving. In Release 16, 5G-V2X was officially included in the 3GPP standards \cite{Garcia-21}. The Technical Specification (TS) 22.186 defined the service requirements for the aforementioned 5G-V2X use cases \cite{TS22186}. Compared to 4G-V2X, 5G-V2X supports more advanced use cases with more stringent requirements, reducing latency and enhancing reliability, throughput, and flexibility \cite{Garcia-21}. 
Releases 17 and 18 provided further enhancements for 5G-V2X. The key aspects of Release 17 included sidelink enhancements for resource allocation, beamforming, and UE relaying \cite{Garcia-21, Saad-23}. Release 18 focused on architectural AI/ML integration and further enhancements for sidelink resource allocation, beamforming management, UE relaying, UE power savings, and the introduction of Uu multicasting services for wide-area V2N communications \cite{Saad-23}. 

\item {\textbf{Opportunities and Challenges}}: 
The potential of next-generation V2X communications is to advance the connectivity, intelligence, autonomy, and automation of large-scale vehicular systems and individual vehicles; traffic tracking, monitoring, management, cooperation, and efficiency; cooperation between different objects and entities in the framework of vehicular systems; and ubiquitous road safety. From a broader perspective, V2X communication is seen as a key technology to enable intelligent vehicular and transportation systems in the 6G era. To realize this vision in practice, V2X must rely on a wide range of emerging technologies. Some potential enablers include a flexible utilization of a broad spectrum range (from sub-6 GHz to THz frequencies), pervasive use of AI/ML (from system architecture to air interface), coordinated utilization of 3D network architecture (ground-air-space), and broad usage of beyond-communication technologies (computing, sensing, localization, and control) \cite{Noorarahim-22}. While these technologies offer novel opportunities for V2X communications, they also introduce diverse challenges for the corresponding network design. The survey \cite{Noorarahim-22} provided a detailed discussion of the key enabling technologies for 6G-V2X, discussing their opportunities and challenges. 

The benefit of a wide spectrum range is that different frequency bands can be used for different V2X purposes, based on their favorable features. For example, THz frequencies, with massive bandwidths, can be used for high-throughput V2V communications and other close-proximity V2X applications. In contrast, lower frequencies, with better coverage and mobility support, can be utilized for long-range V2N communications and other wide-area V2X applications. The utilization of broad spectrum requires dynamic network architecture, with numerous configurations and QoS requirements. Each spectrum band has unique challenges that need to be addressed. Typical ones are related to transceiver design, transmission schemes/parameters, and interference management. For example, propagation losses are high and hardware impairments severe at THz frequencies, calling for highly directional beamforming and careful hardware design, with efficient compensation mechanisms. 

AI/ML is a promising technology to improve V2X network and communication optimization, efficiency, flexibility, and predictability. For example, DL is well-suited for prediction and management problems, DRL for resource allocation, and FL for distributed optimization. However, practical AI/ML implementations are particularly challenging in the V2X framework due to the fast-varying channel conditions (highly mobile scenarios), heterogeneous and dynamic wireless environments (IoV), and stringent latency requirements (real-time processing). This calls for efficient online learning approaches and fast AI/ML algorithms. 

Non-terrestrial communication is recognized as a means to enhance the coverage, reliability, and capacity of V2X scenarios. While satellites are the key for wide and remote area coverage, UAVs enable flexible capacity boosts in the needed areas. The integration of non-terrestrial components requires advanced (AI/ML-based) cooperation mechanisms between different layers. New channel and interference characteristics arise from the 3D network architecture, calling for enhanced resource allocation and interference management approaches, as well as novel 3D channel models for V2X environments. Furthermore, satellite communication is problematic for time-sensitive V2X applications due to its long delays. 

Integrated beyond-communication technologies can potentially advance intelligence, efficiency, awareness, and autonomy in vehicular systems. Powerful computation, storage, and caching capabilities are vital for pervasive AI/ML, whereas network sensing and positioning improve the situational awareness of V2X systems. The integration of beyond-communication technologies into cellular and V2X networks requires significant enhancements to the system architecture, network resource management procedures, cooperation and information exchange mechanisms, and data collection and analytics, among others. 

\item {\textbf{Literature and Future Directions}}:
In the literature, cellular V2X communication has been widely examined for 4G and 5G. 4G-V2X research has mainly focused on developing technological enablers to support road safety and traffic management applications. In the context of 5G, V2X has been extended to support more advanced features related to remote driving, driver assistance, vehicle platooning, and extended sensors. For 6G, V2X research is currently ongoing. Major enhancements are under study, e.g., related to the use of AI/ML, broad spectrum range, non-terrestrial connectivity, beyond-communication technologies, and security. Further research efforts are needed in these directions in the near future. The status, recent advances, and future research areas of V2X communications have been thoroughly discussed in recent survey articles \cite{Garcia-21, Saad-23, Bagheri-21, Alalewi-21, Ficzere-23, Christopoulou-23, Yoshizawa-23, Rammohan-23, Noorarahim-22, Sedar-23, Annu-24, Roger-24, Cai-24b, Gularte-24}. 

The paper \cite{Garcia-21} presented an in-depth overview of 5G-V2X in Release 16. Release 17 and 18 enhancements were surveyed in \cite{Saad-23}. 5G-V2X was also discussed in \cite{Bagheri-21} and \cite{Alalewi-21}. The work \cite{Ficzere-23} studied large-scale V2X deployments, presenting also a vision toward 6G. In \cite{Christopoulou-23}, the authors explored AI/ML technology to enable advanced V2X features. Security and privacy challenges in V2X communications were discussed in \cite{Yoshizawa-23}. The article \cite{Rammohan-23} reviewed V2X technology in the context of intelligent transportation systems. The authors in \cite{Noorarahim-22} provided a detailed discussion of 6G for V2X communications, focusing on the revolutionary and evolutionary technologies. In \cite{Sedar-23}, V2X cyber-security was considered, discussing reactive and proactive security solutions. The work \cite{Annu-24} provided a comprehensive study on resource allocation for 6G-V2X sidelink. Sustainable V2X communication was explored in \cite{Roger-24}, with the focus on relevant use cases and technology trends. In \cite{Cai-24b}, task-oriented V2X was studied from the perspective of digital twins and edge computing. The survey \cite{Gularte-24} provided a thorough discussion of the V2X cyber-security threat landscape. 
\end{itemize}

\begin{figure}[!htb]
\center{\includegraphics[width=0.85\columnwidth]
{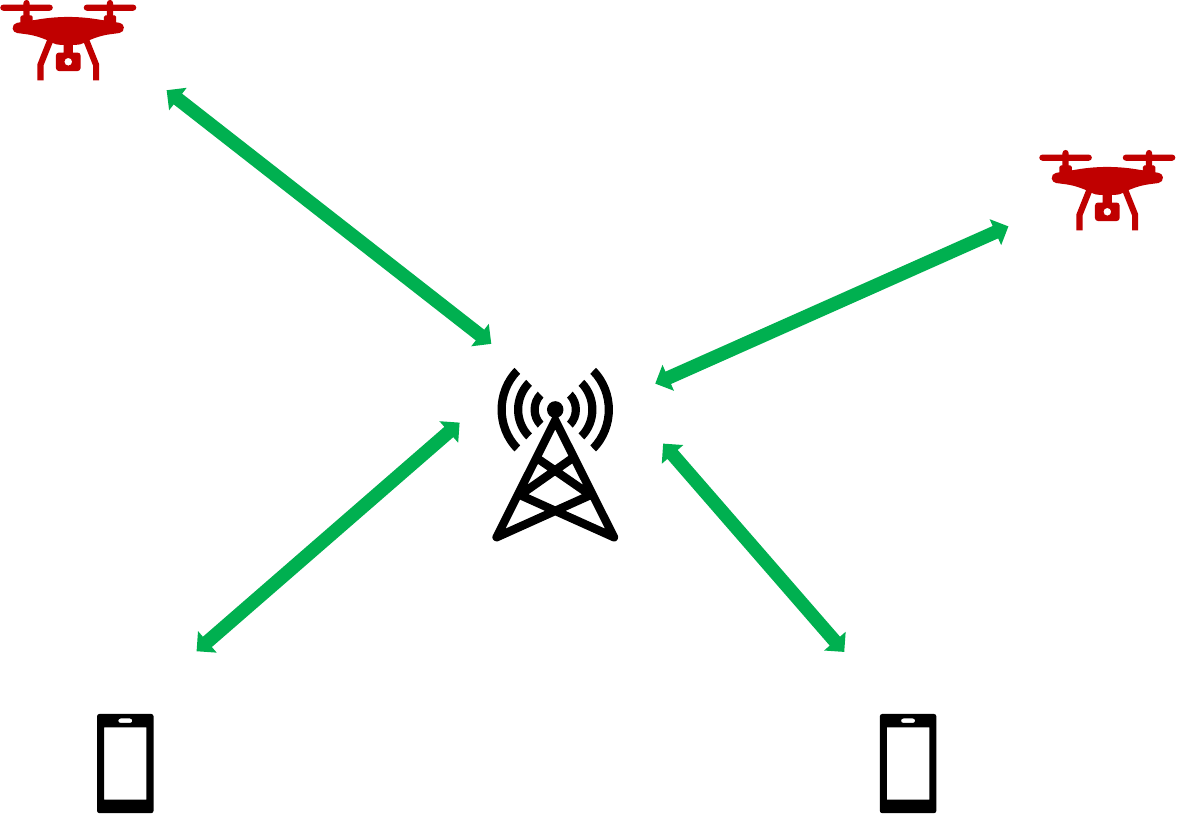}}
\caption{\label{Fig_UAV}Cellular-operated UAV communications.}
\end{figure}

\subsubsection{CELLULAR-CONNECTED UAV COMMUNICATIONS}

\begin{itemize}
\item {\textbf{Vision}}:
Cellular-connected UAV communication is seen as a means to enable novel aerial applications, and ultimately free the potential of UAVs by providing ubiquitous, seamless, and high-quality mobile connectivity for UAVs in the 6G era. 

\item {\textbf{Introduction}}: 
Cellular-connected UAV communication refers to the support of UAVs by mobile networks through cellular connectivity, as illustrated in Figure \ref{Fig_UAV}. The promise of cellular involvement is the networks' potential ability to provide high-quality and reliable mobile connectivity everywhere, expanding the opportunities to exploit UAVs for versatile purposes. However, this requires fundamental architectural changes to the network structure since current mobile networks have been designed to serve ground users. Currently, the cellular support of UAVs is under 5G standardization, while academia and industry are seeking novel ways to enhance it further. 

From a communication perspective, the new concept of UAV-to-everything (U2X) communications, introduced in \cite{Zhang-20t}, is at the center of future cellular UAV connectivity. A generic definition of U2X communications is the connectivity between UAVs and other nodes in a cellular network, such as BSs, ground-devices, and other UAVs \cite{Zhang-20t}. The main connectivity modes of U2X include UAV-to-network (U2N), UAV-to-UAV (U2U), and UAV-to-device (U2D) communications \cite{Zhang-20t}. Other potential modes are UAV-to-satellite (U2S) and UAV-to-vehicle (U2V) communications. Ultimately, U2X communications aim to take advantage of all layers of the network, from ground to space and from cellular to vehicular. In particular, U2X communication is seen as an enabler for IoU \cite{Zhang-20t}. A schematic illustration of potential U2X types is shown in Figure \ref{Fig_U2X}. 

\begin{figure}[!tb]
\center{\includegraphics[width=\columnwidth]
{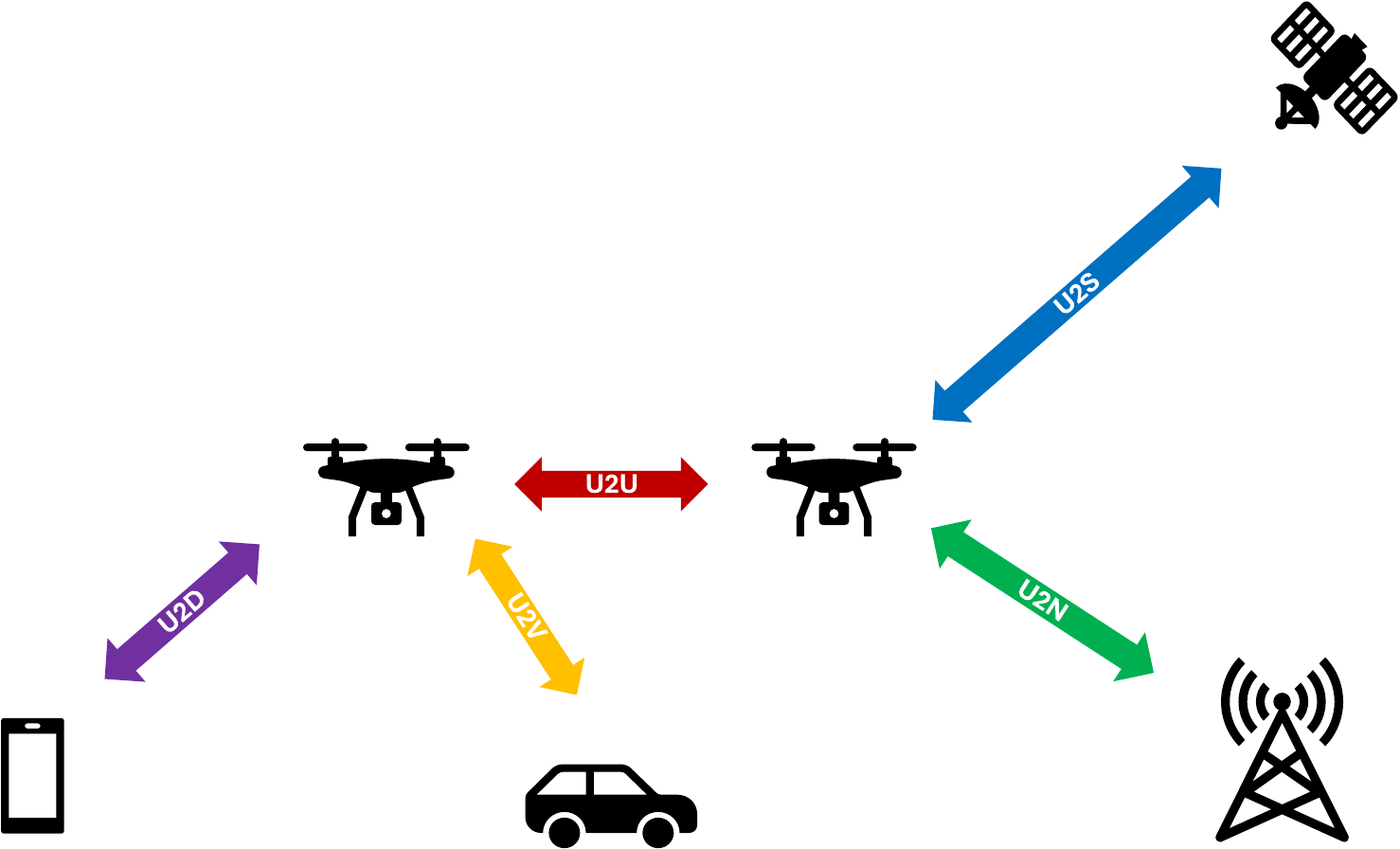}}
\caption{\label{Fig_U2X}Five potential types of U2X communications.}
\end{figure}

\item {\textbf{Past and Present}}: 
In the research-wise, the interest toward cellular UAV connectivity boomed in the 2010s. Due to advances in electronics, AI/ML, wireless communications, and other related fields, UAVs were becoming smaller, cheaper, more capable, and easier to control. Consequently, the use of UAVs for civilian purposes was spreading around the world. Academia and industry saw the potential of combining UAVs and cellular communications, and started to take actions to make the visions real. Lots of research and development work has been conducted, including theoretical studies, channel measurements/modeling, algorithm/protocol/architectural design, technology development, performance evaluations, experiments, and field trials. Strong collaboration between academia and industry has also been involved. In the 2010s, the research on UAVs mainly focused on 5G networks. On the verge of the 2020s, the focus began to gradually shift toward 6G. 

In standardization, 3GPP has been exploring cellular-connected UAV communications from Release 15 onward \cite{Abdalla-21}. In 3GPP, UAVs are studied under the concept of UAS, which consists of two main elements: a UAV and a UAV controller \cite{TS22125}. A UAV can be controlled using a remote controller via a 3GPP cellular network or non-3GPP control mechanisms \cite{TS22125}. The main cellular-based communication services for the UAS ecosystem can be classified into two categories, i.e., command/control data and payload data services \cite{TS22125}. UAS also interacts with UAS traffic management (UTM), which provides many important services for the safe operation of UAS \cite{TS22125}, such as identification, tracking, authorization, regulatory information, and storage of operational data. 

In Release 15, the work on cellular-operated UAVs started with a study item on the required enhancements of LTE networks to support UAVs, resulting in TR 36.777 \cite{Abdalla-21}. In Release 16, the work continued with TR 22.829, which identified UAV use cases and the required enhancements for 3GPP networks, calling for 5G support \cite{Abdalla-21}. Moreover, TS 22.125 defined the requirements for UAS operation over 3GPP networks \cite{Abdalla-21}. In Release 17, TR 23.754 and TR 23.755 studied mechanisms to support connectivity, identification, and tracking in UAS, as well as the architectural requirements and solutions to support UAS applications, respectively \cite{Abdalla-21}. Release 18 provided further enhancements for the NR support of UAVs, especially in terms of identification, reporting, broadcasting, and beamforming \cite{Qazzaz-23}. A comprehensive survey on the 3GPP standardization efforts for UAS was given in \cite{Abdalla-21}. 

\item {\textbf{Opportunities and Challenges}}: 
Diverse opportunities arise from the cellular support of UAVs since mobile networks have unique capabilities to provide reliable, high-performance, and ubiquitous mobile connectivity. Ultimately, cellular-connected UAV communications aim to free the potential of UAVs. Cellular UAVs can potentially enhance or enable many types of applications, including monitoring (e.g., industrial, environmental, and traffic), surveillance (e.g., public safety), logistics (e.g., air cargo and delivery), transportation (e.g., air taxis), entertainment (e.g., immersive VR experiences), search/rescue operations (e.g., difficult terrain and large areas), disaster management (e.g., humanitarian and medical aid), and agriculture (e.g., precision monitoring), among others \cite{Zeng-19, Fotouhi-19, Mishra-20, Abdalla-21, Geraci-22, Qazzaz-23}. It is expected that 6G will provide a fruitful platform for cellular-connected UAV communication to grow toward its potential. 

Many challenges exist on the way toward comprehensive cellular support for UAVs. The UAV communication scenario differs significantly from that of the traditional cellular environment on the ground. The main differences include the high altitude of end-devices, unique 3D channel characteristics, air-ground interference, dynamic 3D network topology, 3D mobility, and emphasized uplink communications \cite{Zeng-19, Fotouhi-19}. To manage these peculiarities, novel solutions are needed. 
As UAVs have a relatively high altitude compared to the ground-users, conventional cellular networks, based on down-tilted BS antennas, lack the ability to provide a broad coverage for UAVs up in the air. For successful UAV communications, cellular networks need to adopt a large-scale 3D network architecture, where the airspace is also ubiquitously covered. This calls for a 3D cell planning/design, up-tilted BS antennas, and 3D beamforming. In particular, massive MIMO is seen as one of the potential transmission technologies to facilitate 3D coverage by forming beams toward UAVs in the sky \cite{Geraci-22}. However, massive MIMO has its own challenges and limitations, such as pilot contamination and CSI acquisition \cite{Geraci-22}. 

Due to the unique nature of a 3D radio channel environment, novel channel models are of great importance to evaluate the performance of the proposed communication algorithms. If UAVs are served using the same spectral resources as the ground-users, a new type of interference is present \cite{Fotouhi-19}, i.e., air-ground interference. This calls for new interference management and power control techniques. For example, the massive MIMO technology has the potential ability to efficiently suppress air-ground interference by accurate 3D beamforming \cite{Geraci-22}. Since UAVs are mobile in 3D space, possibly with relatively high velocity, 3D handovers become frequent, leading to challenging cell design and mobility management \cite{Geraci-22}. AI/ML is seen as an efficient tool to assist in the 3D network design and mobility management problems. Recent surveys provide more details on the challenges and potential solutions of cellular-supported UAV communications \cite{Zeng-19, Fotouhi-19, Mishra-20, Abdalla-21, Geraci-22}.  

\begin{table*}[htb!]
\begin{center}
\caption{Summary of service-oriented technologies for 6G}
\label{Table_PN}
\centering
\begin{tabularx}{\textwidth}{| >{\centering\arraybackslash}X | >{\centering\arraybackslash}X |
>{\centering\arraybackslash}X | 
>{\centering\arraybackslash}X |
>{\centering\arraybackslash}X |
>{\centering\arraybackslash}X |
>{\centering\arraybackslash}X |
>{\centering\arraybackslash}X |}
\hline
\centering
\vspace{3mm} \textbf{Service-Oriented Technologies} \vspace{3mm} & \centering \textbf{Vision} & \centering \textbf{Description} & \centering \textbf{Opportunities} & \centering \textbf{Challenges} & \centering \textbf{Past} & \vspace{1.5mm} \begin{center} \textbf{Present} \end{center} \\
\hline
\vspace{3mm} Private Networks \vspace{3mm}  & \vspace{3mm} Secure PNs for diverse verticals \vspace{3mm} & Customized for vertical customers & New business opportunities & Security $\&$ guaranteed QoS & 4G private networks & 5G private networks \\
\hline
\end{tabularx}
\end{center}
\end{table*}

\item {\textbf{Literature and Future Directions}}: 
In recent years, researchers have conducted numerous surveys on cellular-connected UAVs, reviewing topics that range from fundamentals to standardization, recent advancements, open challenges, and future research guidelines \cite{Zeng-19, Fotouhi-19, Zhang-20t, Mishra-20, Abdalla-21, Sahoo-22, Geraci-22, Qazzaz-23, Amodu-23, Do-24, Hurst-24, Javaid-23, Javed-24, Cao-24}. 
In \cite{Zeng-19}, the article reviewed cellular-connected UAV operations in terms of communication/spectrum requirements, design aspects, potential technologies, and future directions. The authors in \cite{Fotouhi-19} presented a survey on cellular UAV communications, with the focus on UAV types, standardization, UAV-BSs, prototyping/field testing, regulations, security, and future research areas. The work \cite{Zhang-20t} introduced a novel concept of U2X communications from the perspectives of fundamentals, key techniques, reinforcement learning framework, and future extensions. Cellular-operated UAVs were explored in \cite{Mishra-20}, covering applications/use cases, challenges, 5G/B5G innovations, trials/prototyping, standardization, and future research guidelines. 

The paper \cite{Abdalla-21} reviewed 3GPP standardization of cellular UAS support, and introduced the key research drivers. In \cite{Sahoo-22}, the authors discussed enabling technologies, standardization efforts, security, and open issues for advanced cellular UAV operations. The authors envisioned an evolution road for cellular UAV connectivity from 5G to 6G in \cite{Geraci-22}. Many aspects were discussed, including 5G NR and beyond, sub-6 GHz massive MIMO, mmWave/THz, AI/ML, NTNs, RISs, UAV-to-UAV communications, and future challenges. The study \cite{Amodu-23} discussed cellular UAV support using THz frequencies. An overview of UAV clients for beyond 5G was presented in \cite{Qazzaz-23}. The paper \cite{Do-24} reviewed advanced air mobility in cellular networks, with a particular focus on use cases and beamforming techniques. In \cite{Hurst-24}, the authors explored the use of unmanned vehicles in 6G networks. The surveys \cite{Javaid-23, Javed-24, Cao-24} examined cellular-supported cooperative UAV swarms. 

Major research and development efforts are required in the future to provide reliable and ubiquitous cellular support for UAVs. The earlier discussed challenges need to be carefully addressed. Although some studies have been conducted to overcome these issues, extensive research is still needed. In this regard, key future topics on cellular-connected UAV communications include 3D network design (e.g., up-tilted cell design, resource allocation, interference control, mobility management), emerging technologies (e.g., AI/ML, massive MIMO, mmWave, THz, RISs, NTNs), U2X communications, collaborative UAVs (e.g., UAV swarms), channel modeling (e.g., channel measurements, system-level 3D models), and security (e.g., cyber security, physical security) \cite{Zeng-19, Fotouhi-19, Mishra-20, Abdalla-21, Geraci-22, Amodu-23, Qazzaz-23, Javaid-23}. 
\end{itemize}

\subsection{SERVICE-ORIENTED TECHNOLOGIES FOR 6G}
It is anticipated that 6G will significantly expand the services of mobile networks, especially in vertical domains. Private networks will be at the core of this evolution, offering customized services for vertical clients and creating novel business opportunities. Private networks are reviewed in the following and summarized in Table \ref{Table_PN}. 

\begin{figure}[!htb]
\center{\includegraphics[width=0.75\columnwidth]
{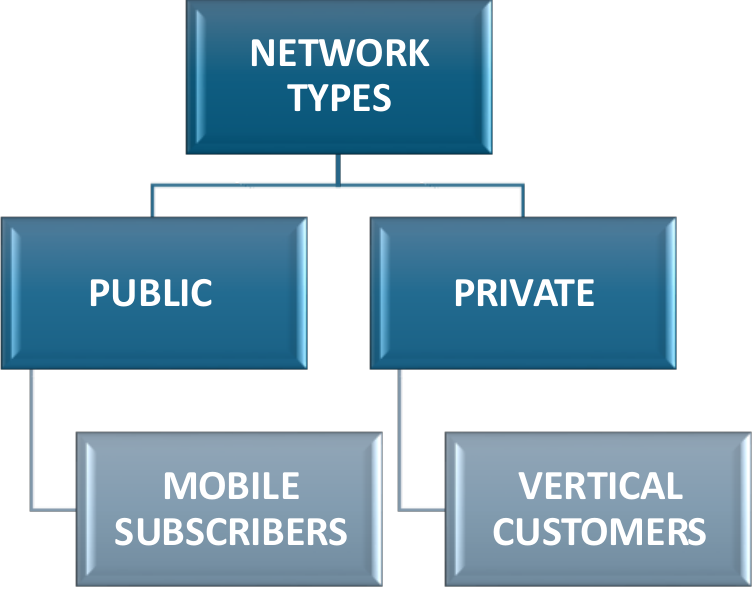}}
\caption{\label{Fig_PN}Two main types of mobile networks.}
\end{figure}

\subsubsection{PRIVATE NETWORKS}

\begin{itemize}
\item {\textbf{Vision}}:
In the 6G era, private networks are expected to become a key technology for extending the applicability of mobile networks to vertical industries by providing tailored high-quality wireless services. 

\item {\textbf{Introduction}}:
While public networks serve traditional mobile subscribers, private networks are tailored exclusively to vertical customers, as summarized in Figure \ref{Fig_PN}. Typical customers vary from private enterprises to public sector organizations, requiring secure non-public network services. Hybrid networks also exist, i.e., a mix of private and public networks, where a part of the network functions are private and other parts public. Private networks are expected to expand the service opportunities of mobile networks in the future, thereby enabling novel applications in diverse vertical industries. 

\item {\textbf{Past and Present}}:
Traditionally, mobile networks have been mostly public, offering services to conventional mobile subscribers. Private networks were first introduced in 4G. However, the usage has been rather marginal, partly due to the limitations of 4G. 5G private networks, also known as non-public networks in 3GPP, were introduced in Release 16 \cite{Prados-21}. Compared to 4G, 5G provides significantly better performance and capabilities for private networks, opening up new service opportunities, especially in industrial environments \cite{Aijaz-20}. In the 5G NR standards, private networks are supported in two categories, i.e., public network integrated and stand-alone non-public networks \cite{Prados-21}. The former relies on a PLMN and the latter performs independently. Full control and independence to customize the stand-alone network according to the needs of the customer comes at the price of higher capital and operational costs \cite{Prados-21}. Further details on the standardization of 5G non-public networks can be found in \cite{Prados-21}. In the big picture, 5G is paving the way for a paradigm shift where mobile networks are not only used for serving public subscribers but also providing versatile services for a wide range of vertical industries. 

\item {\textbf{Opportunities and Challenges}}: 
Private networks are expected to significantly expand the service opportunities of mobile networks in the 6G era, enabling novel applications in diverse vertical industries. This will generate new business opportunities for mobile network vendors and operators, as well as customer industries. Potential application domains include versatile industries and utilities \cite{Prados-21}, such as manufacturing, healthcare, transportation, retailing, agriculture, energy, education, sports, and tourism, to list a few. In particular, private mobile networks are seen as a key enabler for industry 4.0 \cite{Prados-21}. From a wider perspective, private networks will integrate mobile networks more deeply into society. In this regard, 6G is expected to penetrate all levels of future society, impacting all walks of life, utilities, businesses, and industries. 

Although private networks have been included in the 5G standardization and commercial solutions exist around the world, they are still at a rather early evolutionary stage and far from their true potential. There are diverse technological obstacles on the way toward freeing the full potential of private networks. First, private networks must fulfill the requirements of customer verticals, such as tailored services, guaranteed QoS, accurate coverage, protected data, and secure networking \cite{Prados-21}. Additionally, there is a constant need to further enhance the performance and capabilities of private networks. Whereas 5G evolution can provide rather limited enhancements, 6G is expected to take private networks to the next level, with extreme performance, pervasive AI/ML, and integrated beyond-communication technologies. In addition to opportunities, 6G will introduce new challenges as well. In this picture, the main issues are related to resource management, spectrum use, AI/ML-assisted network optimization, integrated computation, control, and sensing, fronthaul/backhaul, data privacy, and network security. In \cite{Prados-21, Wen-22, Guo-22b, Eswaran-23}, recent surveys provided detailed discussions on the main challenges of private networks.   

\begin{table*}[htb!]
\begin{center}
\caption{Summary of security technologies for 6G}
\label{Table_Security}
\centering
\begin{tabularx}{\textwidth}{| >{\centering\arraybackslash}X | >{\centering\arraybackslash}X |
>{\centering\arraybackslash}X | 
>{\centering\arraybackslash}X |
>{\centering\arraybackslash}X |
>{\centering\arraybackslash}X |
>{\centering\arraybackslash}X |
>{\centering\arraybackslash}X |}
\hline
\centering
\vspace{3mm} \textbf{Security Technologies} \vspace{3mm} & \centering \textbf{Vision} & \centering \textbf{Description} & \centering \textbf{Opportunities} & \centering \textbf{Challenges} & \centering \textbf{Past} & \vspace{1.5mm} \begin{center} \textbf{Present} \end{center} \\
\hline
\vspace{3mm} Holistic Network Security Architecture \vspace{3mm}  & Trustworthy $\&$ secure 6G & Protection against outside threats & Safe use of 6G networks & Evolving threat landscape & 4G security architecture & 5G security architecture \\
\hline
\end{tabularx}
\end{center}
\end{table*}

\item {\textbf{Literature and Future Directions}}:
In the literature, the first studies on private mobile networks date back to the early 2010s, with the main focus on 4G LTE. However, research interest in private networks remained relatively mild until the late 2010s. Then, the research started to gain more attention since the focus shifted to more promising 5G private networks. Currently, the latest works look beyond 5G. Commonly studied topics in the private network research include industry 4.0, industrial IoT, spectrum operation, network architecture, edge computing, network slicing, and security. Private networks have been thoroughly reviewed in \cite{Aijaz-20, Maman-21, Prados-21, Wen-22, Guo-22b, Eswaran-23, Collin-24}. 

The paper \cite{Aijaz-20} discussed private 5G networks in an industrial context. The main topics included industrial networking demands, 5G opportunities for industrial wireless, functional architecture of 5G private networks, industrial usage scenarios, licensed/unlicensed spectrum operation, network design challenges, and standardization efforts. In \cite{Maman-21}, 5G and beyond private networks were explored in terms of application scenarios, standardization aspects, operator models, and technological enablers. The authors of \cite{Prados-21} reviewed 5G non-public networks from a standardization perspective, focusing on enterprise customer requirements, enabling technology solutions, 3GPP standardization, single- and multi-site network scenarios, and implementation challenges. 

The work \cite{Wen-22} provided a survey on the research of private 5G networks, with discussions on the basic architectures, implementation issues, technology enablers, application scenarios, real-world field trials, and open problems. In \cite{Guo-22b}, 5G and beyond private networks were considered to deliver tailored services for vertical industries with integrated eMBB, URLLC, mMTC, and positioning. A thorough survey of 5G private networks was given in \cite{Eswaran-23}. Numerous topics were covered, including requirements, enablers, deployment modes, spectrum operations, network slicing, services, mobile operators' roles, applications, security, and future research. In \cite{Collin-24}, private 5G networks were discussed in terms of vertical industries, providing design and research guidelines. 

As mentioned earlier, private networks are still at a rather early evolutionary stage, requiring major research and development efforts in the future. Since 6G is expected to provide the tools needed for the next major leap in the private network paradigm, it is pivotal to consider the unique characteristics of 6G in research and design. Some fruitful future topics on private networks include operation in the 6G spectrum, advanced fronthaul/backhaul networks, AI/ML-based network optimization, ultra-massive IoT support, integration of beyond-communication technologies, realistic channel models, end-to-end performance evaluation, and security/privacy. Future research directions have been discussed in surveys \cite{Prados-21, Wen-22, Guo-22b, Eswaran-23}. 
\end{itemize}

\subsection{SECURITY TECHNOLOGIES FOR 6G}
Since future society will be highly dependent on 6G, it is of utmost importance to make 6G networks trustworthy and secure. A special attention needs to be paid to privacy as well, as data is becoming more abundant. Therefore, a comprehensive network security architecture is required, as discussed below and summarized in Table \ref{Table_Security}. 

\begin{figure}[!htb]
\center{\includegraphics[width=0.75\columnwidth]
{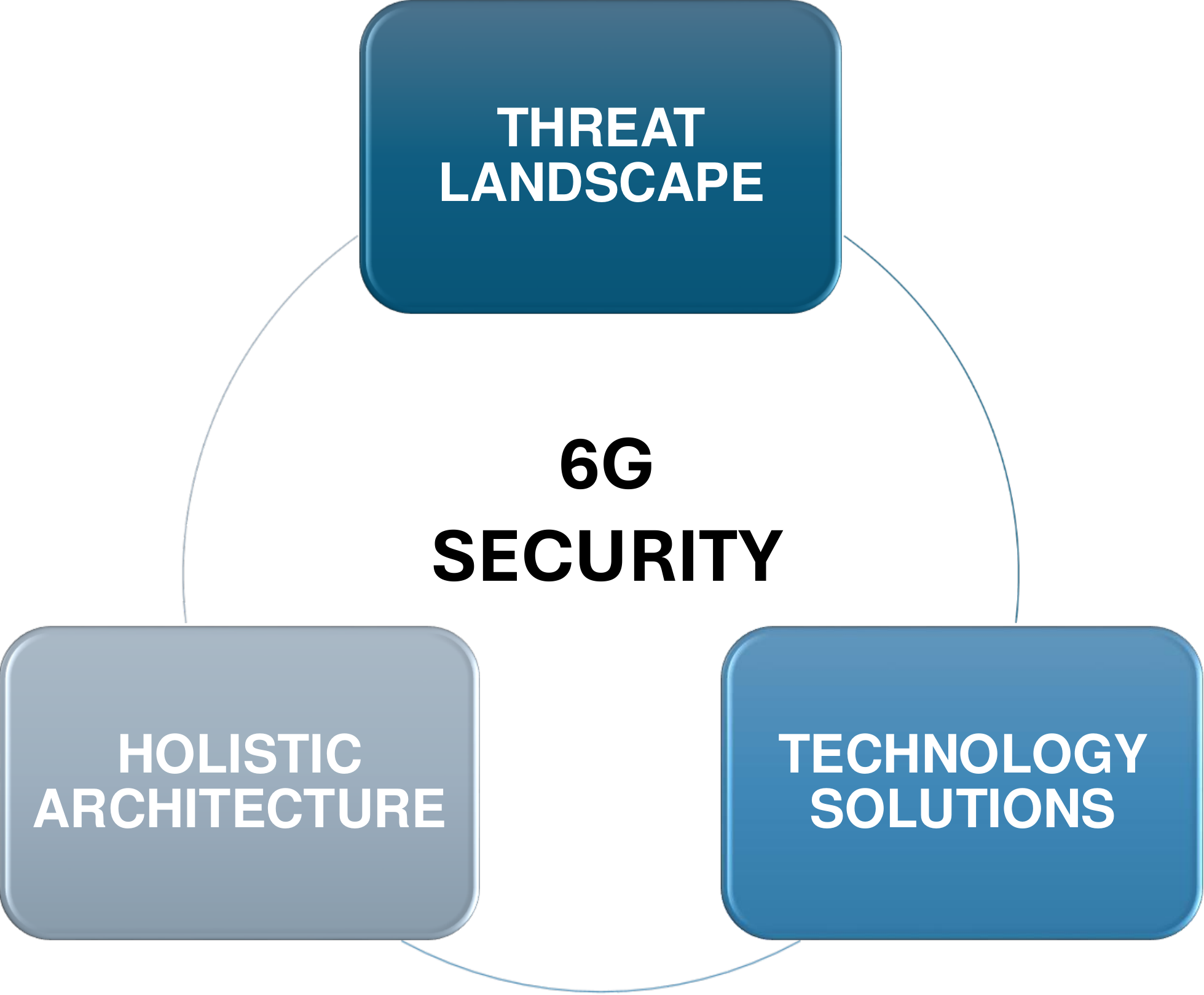}}
\caption{\label{Fig_Security}Three main phases to build 6G network security.}
\end{figure}

\subsubsection{NETWORK SECURITY ARCHITECTURE}

\begin{itemize}
\item {\textbf{Vision}}: 
It is expected that 6G networks will be secured by a comprehensive security architecture, advancing trust and privacy as well. 

\item {\textbf{Introduction}}: 
Mobile network security protects against external threats and attacks, aiming to ensure the safe and trustworthy use of the network for all customers. To achieve this target, a holistic security architecture is required, which secures all network layers from the PHY and MAC to the network and application layers. Network security will become increasingly important in the future since mobile networks will be more integrated into different levels of society. For example, security is particularly critical in healthcare, transportation, and industrial automation scenarios. In addition, data privacy must be protected. Since the use of AI/ML is constantly increasing, data is becoming more abundant, making it more vulnerable to attacks. 

\item {\textbf{Past and Present}}:
Since each new mobile generation introduces novel technologies, services, and applications, it encounters new security threats as well. Therefore, novel security solutions are required to address these threats. The first three mobile generations faced more traditional types of threats, mainly related to authentication, authorization, encryption, eavesdropping, physical attacks, and cloning \cite{Porambage-21}. Since mobile internet and its applications became mainstream in the 4G era, the security and privacy threat landscape expanded to malware applications and the denial of service attacks \cite{Porambage-21}. 

New types of security and privacy threats were introduced for 5G due to its expanded capabilities and a broad range of novel applications. In this respect, 3GPP Release 15 introduced an advanced security architecture, with features such as access-agnostic authentication, enhanced subscription privacy, user plane integrity protection, network slice-specific authentication/authorization, and advanced authentication/authorization between network functions \cite{Ziegler-21}. During the entire evolution of 5G, its security architecture will be constantly updated, considering newly added technology features and extended capabilities. Currently, a security and privacy threat paradigm is studied for 6G in academia and industry, while paying attention to the emerging 6G technologies, services, and applications.  

\item {\textbf{Opportunities and Challenges}}: 
The aim of a holistic security architecture is to ensure the trustworthy, secure, and privacy-protected use of 6G for all users. This is one of the cornerstones of 6G, and is vital for its success. Secure 6G enables the safe evolution of mobile networks and a bloom of novel applications. Developing a secure 6G ecosystem is a major challenge, which requires a holistic approach, covering security, privacy, and trust from the perspective of technology and regulations. 
To tackle diverse threats at different levels of the network, a comprehensive 6G network security architecture needs to be developed, considering privacy protection as well. At a high level, building a thorough 6G network security framework requires three (interrelated) phases, as summarized in Figure \ref{Fig_Security}. First, the entire 6G threat vector landscape must be identified and regularly updated, considering all network layers. Second, technological solutions must be developed for each identified threat. Third, a holistic network security architecture has to be built based on the identified threats and proposed solutions. 

New threat vectors arise from the emerging 6G technologies (e.g., pervasive AI/ML, NTNs, V2X, integrated beyond-communication technologies, and heterogeneous network architecture) and diverse applications (e.g., smart healthcare, smart factories, smart cities, connected autonomous vehicles, XR, and digital twins). Moreover, data will become increasingly important and abundant in the future, making 6G networks more vulnerable to attacks, and raising concerns on data privacy. The threat landscape of these technologies and applications need to be carefully studied, potential solutions developed, and a holistic security architecture designed, covering all network layers. In this framework, AI/ML has been recognized as a powerful technology for protecting against diverse security threats. On the other hand, integrated AI/ML is also a target of novel attacks. Thus, AI/ML can be seen as a double-edge sword \cite{Mao-23b}, calling for secured AI/ML solutions. The work on 6G security/privacy threats, solutions, and architecture is currently ongoing, requiring constant development efforts. 6G security has been thoroughly discussed in \cite{Porambage-21, Je-21, Ziegler-21, Nguyen-21c, Mao-23b}. 

\item {\textbf{Literature and Future Directions}}: 
Recent surveys have provided comprehensive discussions on the security, privacy, and trust issues for 6G \cite{Ylianttila-20, Sun-20, Porambage-21, Mucchi-21, Je-21, Ziegler-21, Nguyen-21c, Guo-22, Chorti-22, Mao-23b, Mitev-23, Sedjelmaci-23, Zhang-24, Tripi-24, Scalise-24, Kaur-24b}. The white paper \cite{Ylianttila-20} presented a thorough review of the trust networking, network security, PHY layer security, and privacy protection aspects for 6G. In \cite{Sun-20}, a detailed survey was provided on the privacy violation and protection in 6G by exploiting ML. In \cite{Porambage-21}, the authors discussed the security and privacy issues in the context of 6G requirements, technologies, applications, and standardization. In \cite{Mucchi-21}, PHY layer security was considered for 6G networks, focusing on challenges, solutions, and visions. The authors in \cite{Je-21} explored new security threats, arising from the introduction of promising 6G technologies and possible solutions against them. 

In \cite{Ziegler-21}, the focus was on the technology enablers for 6G security, including automated software creation, automated closed-loop security operation, privacy-preserving technologies, hardware and cloud embedded trust, quantum-safe security, jamming protection, physical layer security, and distributed ledger technologies. In \cite{Nguyen-21c}, a systematic survey was presented on the 6G security and privacy issues in the physical, connection, and service layers of the network. The authors in \cite{Guo-22} presented a detailed review of recent research progress on the security threats, attack methodologies, and defense countermeasures for 6G space-air-ground-sea network architecture. In \cite{Chorti-22}, 6G security was studied in terms of context-awareness, focusing on the PHY layer security aspects. The paper \cite{Mao-23b} provided a comprehensive review of the security and privacy of the network edge toward 6G, examining edge computing, edge caching, and EI as the targets of attacks and the sources of protection. 

In \cite{Mitev-23}, the authors discussed the opportunities that PHY layer security can offer to the security of 6G. The work \cite{Sedjelmaci-23} explored the combination of AI/ML and a zero-trust architecture as an enabler of 6G network security. In \cite{Zhang-24}, AI-enhanced PHY layer security was studied for 6G networks. The authors of \cite{Tripi-24} examined 6G security in the physical, connection, and application layers. The paper \cite{Scalise-24} reviewed the research landscape of 5G security, and discussed promising security mechanisms for 6G. In \cite{Kaur-24b}, the authors explored AI-based security approaches for 6G networks. Although 6G security has been actively researched in academia and industry for years, major efforts are still required to make 6G truly secure and trustworthy. Further research and development work is needed to identify the 6G threat vector landscape, develop AI/ML-based security and privacy solutions, and build a comprehensive architectural security framework. 
\end{itemize}

\begin{figure}[!htb]
\center{\includegraphics[width=0.85\columnwidth]
{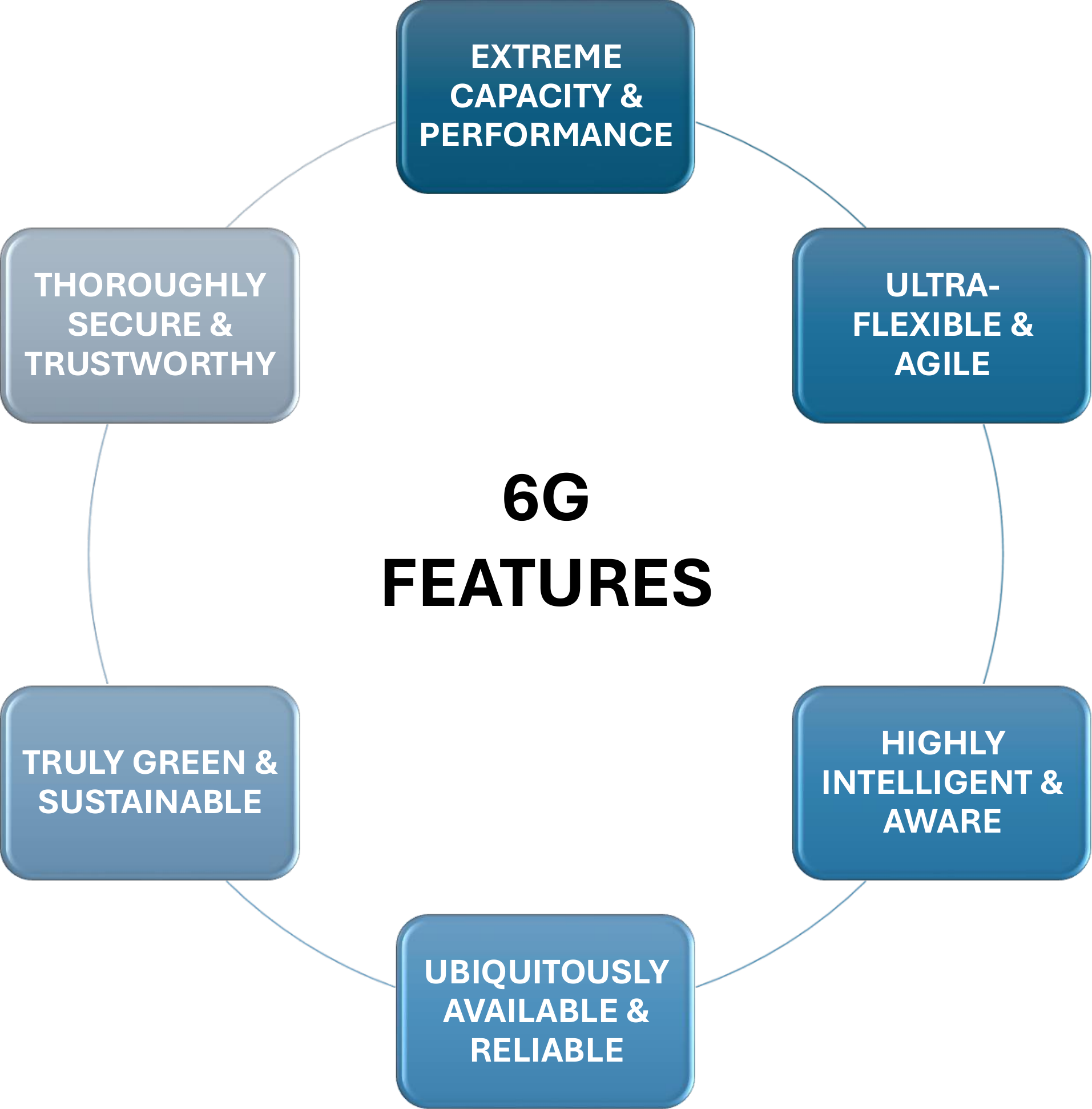}}
\caption{\label{Fig_Features}Defining features for 6G.}
\end{figure} 

\section{DEFINING FEATURES FOR 6G} 
\label{Features}
In this section, we identify 12 main features that define the essence of 6G. These features are summarized in Figure \ref{Fig_Features}. 

\begin{itemize}
\item \textbf{Extreme Capacity and Performance:}
Extreme capacity and performance will be the cornerstones of 6G. There are six main performance dimensions that need to be pushed to the extreme levels, i.e., capacity, latency, reliability, density, coverage, and mobility. Other important dimensions are energy efficiency and positioning accuracy. By pushing the performance into its limits, 6G will form a fruitful platform for a broad range of demanding services and applications. 

\item \textbf{Ultra-Flexible and Agile:}
To adapt itself to a wide range of wireless environments and application scenarios, 6G will be designed to be ultra-flexible and agile. Flexibility and agility need to be optimized at all levels of the network, from device and communication to network and service levels. 

\item \textbf{Highly Intelligent and Aware:}
One of the most revolutionizing features of 6G will be network intelligence based on pervasive AI/ML. The extensive use of AI/ML will make 6G highly intelligent, enhancing network design, operation, and management. AI/ML must be exploited at all levels of the network, including the core, edges, and air interface. The most promising technologies to realize AI/ML are DL, FL, and TL. In addition, 6G is expected to become aware of the surrounding environment through network sensing. This extends the capabilities of 6G, opening up new beyond-communication service and application opportunities. 

\item \textbf{Ubiquitously Available and Reliable:}
To enable a vast variety of robust services, 6G will be designed to be highly available and reliable. The key to ubiquitous availability is broad coverage, which is enabled by the integration of terrestrial and non-terrestrial networks. Reliability has to be maximized at different network levels, ranging from robust PHY layer schemes to network-level mechanisms. 

\item \textbf{Truly Green and Sustainable:}
Since the total energy consumption of mobile networks is constantly increasing, 6G networks will be designed for greenness and sustainability by enhancing the energy efficiency at every level of the network, from the core and edges to the MAC and PHY layer protocols. This will provide significant ecological and economic benefits. 

\item \textbf{Thoroughly Secure and Trustworthy:}
As future society will be profoundly dependent on 6G networks, it is of utmost importance to make 6G secure and trustworthy. A holistic network security architecture needs to be developed for 6G, taking into account also privacy and trust issues. Security and trustworthiness are vital elements to ensure the safe use of 6G. 
\end{itemize}

\section{6G IN A NUTSHELL}
\label{Summary}
This section provides a compact summary of our 6G vision in bullet points. 

\noindent \textbf{Vision}:
\begin{itemize}
\item {Smart Wireless World via 6G-Enabled Mobile Intelligence}
\end{itemize}

\noindent \textbf{Elements}: 
\begin{itemize}
\item {{Wireless $\circ$ Artificial Intelligence $\circ$ Internet of Everything}}
\end{itemize}

\noindent \textbf{Applications}: 
\begin{itemize}
\item Human-Machine Interactions: {{Metaverse $\circ$ Extended Reality $\circ$ Holographic-Type Communications $\circ$ Digital Twins $\circ$ Tactile Internet}}
\item Smart Environments: {{Smart Society $\circ$ Smart City $\circ$ Smart Factory $\circ$ Smart Home}}
\item Connected Autonomous Systems: {{Connected Autonomous Vehicle Systems $\circ$ Connected Autonomous Aerial Vehicle Systems $\circ$ Connected Autonomous Robotic Systems}}
\end{itemize}

\noindent  \textbf{Use Cases}: 
\begin{itemize}
\item Communication-Oriented: {{Ultra-Broadband Multimedia Communications $\circ$ Extreme Time-Sensitive and Mission-Critical Communications $\circ$ Ultra-Massive Communications $\circ$ Global-Scale Communications $\circ$ Hyper-Mobility Communications}}
\item Beyond-Communication-Oriented: {{Network Intelligence $\circ$ Network Sensing $\circ$ Network Energy}}
\end{itemize}

\noindent \textbf{Requirements}: 
\begin{itemize}
\item{{Peak Data Rate: 50/100/200 Gbit/s $\circ$ User Experienced Data Rate: 300/500 Mbit/s $\circ$ Peak Spectral Efficiency: 1.5X/3X $\circ$ Area Traffic Capacity: 30/50 Mbit/s/m$^2$ $\circ$ User Plane Latency: 0.1--1 ms $\circ$ Reliability: 1-10$^{\text{-}5}$--1-10$^{\text{-}7}$ $\circ$ Connection Density: 10$^{6}$--10$^{8}$/km$^2$ $\circ$ Maximum Mobility: 500-1000 km/h $\circ$ Positioning Accuracy: 1-10 cm}}
\end{itemize}

\noindent \textbf{Technologies}: 
\begin{itemize}
\item Spectrum-Level Technologies: {THz Communications $\circ$ Optical Wireless Communications}
\item Antenna System Technologies: {{Ultra-Massive MIMO $\circ$ Reconfigurable Intelligent Surfaces}} $\circ$ Holographic MIMO 
\item Transmission Scheme Technologies: {{Multi-Waveform Scheme $\circ$ Advanced Modulation and Coding Methods $\circ$ Non-Orthogonal Multiple Access $\circ$ Grant-Free Medium Access}}
\item Network Architectural Technologies: {{Integrated Non-Terrestrial and Terrestrial Networks $\circ$ Ultra-Dense Networks $\circ$ Cell-Free massive MIMO $\circ$ Integrated Access and Backhaul}}
\item Network Intelligence Technologies: {{Intelligent Core $\circ$ Intelligent Edge $\circ$ Intelligent Air Interface}}
\item Beyond-Communication Technologies: {{Integrated Communication, Computation, and Caching $\circ$ Integrated Sensing and Communication $\circ$ Wireless Energy Transfer}}
\item Energy-Aware Technologies: {{Green Networks $\circ$ Energy Harvesting $\circ$ Backscatter Communications}}
\item End-Device-Oriented Technologies: {{D2D Communications $\circ$ V2X Communications $\circ$ Cellular-Connected UAV Communications}}
\item Service-Oriented Technologies: {{Private Networks}}
\item Security Technologies: {{Network Security Architecture}}
\end{itemize}

\noindent \textbf{Features}: 
\begin{itemize}
\item {{Extreme Capacity $\&$ Performance $\circ$ Ultra-Flexible $\&$ Agile $\circ$ Highly Intelligence $\&$ Aware $\circ$ Ubiquitously Available $\&$ Reliable $\circ$ Truly Green $\&$ Sustainable $\circ$ Thoroughly Secure $\&$ Trustworthy}}
\end{itemize}

\section{7G VISION: A HIGH-LEVEL SKETCH}
\label{7G}
In this section, we place 6G into a wider perspective by discussing the post-6G era and sketching the first high-level vision of 7G. As each generation, 6G will eventually reach its limits and the next generation (7G) will gradually take over. Most likely, this evolution will follow the same ten year cycle seen in the past. While 6G is expected to dominate in the 2030s, we envision that the 2040s will be the 7G era. Since 7G will be significantly more complicated than 6G, it will take more time to develop it. The sooner we begin visioning 7G, the more prepared we are and the more time we have to develop it. Although 6G is in the development process and many years from the deployment phase, we can already see some major trends that evolve beyond 6G. While wireless will evolve toward unlimited, AI will expand vertically (deeper) and horizontally (broader). In addition, the range of connected devices will become wider and wider. Based on these trends, we can envision the next major disruption after 6G. 

What comes to the disruptions provided by mobile networks, it seems that the major disruptions occur in the cycles of two generations, i.e., 1G/2G, 3G/4G, and 5G/6G. In this cycle, the first generation introduces the new disruption, whereas the second one makes it boom. For example, 1G introduced mobile telephony, while 2G freed its potential. Mobile internet was first enabled by 3G and became mainstream in the 4G era. Currently, 5G is introducing mobile and intelligence and it is expected that 6G will take mobile intelligence to the next level. This evolution will most likely continue after 6G. We envisage that the next major disruption in the post-6G era will be the 7G/8G-enabled mobile hyperverse. Mobile hyperverse will rely on the cornerstones of 7G. We identify these three fundamental pillars as unlimited wireless, deep intelligence (DI), and the Internet of Intelligence (IoI). These elements are vital to enable 7G-level applications. We identify some possible applications in three categories, i.e., human-DI interactions, deeply intelligent world, and interconnected intelligent ecosystems. In the following, we provide further details on the aforementioned disruption, elements, and applications. 

\begin{table}[tb!]
\caption{A speculative evolution from 6G to 7G}
\label{Table_7G}
\begin{center}
\begin{tabular}{ |c|c| } 
 \hline
 & \\
\textbf{6G} & \textbf{7G} \\
 & \\
\hline
 \multicolumn{2}{|c|}{Disruption} \\
 \hline
 & \\
Mobile Intelligence & Mobile Hyperverse \\
 & \\
\hline
 \multicolumn{2}{|c|}{Elements} \\
 \hline
 & \\
Wireless (Extreme) & Wireless (Unlimited) \\ 
Artificial Intelligence & Deep Intelligence \\
Internet of Everything & Internet of Intelligence \\
 & \\
\hline
 \multicolumn{2}{|c|}{Applications} \\
 \hline
 & \\
Human-Machine Interactions & Human-DI Interactions \\
Smart Environments & Deeply Intelligent World \\
Autonomous Systems & Intelligent Ecosystems \\
 & \\
\hline
\end{tabular}
\end{center}
\end{table}

We envision that mobile hyperverse will be the next evolution phase after mobile intelligence, creating real-time intelligent, interconnected, interactive, and immersive entities that can ultimately integrate the digital, physical, biological, and even abstract worlds. The scale of integration can vary from nano to global range. Early signs of this kind of evolution can already be seen in Nokia's 6G vision, which aims to connect the digital, physical, and human worlds \cite{Nokia-6G}. Roughly speaking, the hyperverse can be seen as an extension of the metaverse. Mobile hyperverse can take many forms and serve different purposes. Ultimately, this evolution might lead to the Internet of Hyperverses. At a high level, mobile hyperverses can be divided into three categories: public, non-public, and hybrid. Public hyperverses are intended for consumers, whereas the non-public ones are for industrial, organizational, and enterprise purposes. Hybrid hyperverses are a mix of both, being partly public and partly non-public. From a human perspective, mobile hyperverse will enrich human interactions with other people, devices, systems, and applications, revolutionizing the human experience of the world. We envisage that the mobile hyperverse will become mainstream when all seamlessly interconnected parts of the entire ecosystem have reached a certain level of maturity, including networks, devices, user interfaces, applications, and intelligence. Mobile hyperverse requires ubiquitous, super-quality, and intelligent wireless network infrastructure; smart, lightweight, and affordable XR device entities; smooth and easy to use super-precision user interfaces; fully immersive applications exploiting all human senses; and advanced AI deeply integrated into all parts of the ecosystem. The 7G-enabled mobile hyperverse will create countless application and business opportunities at all levels of the future society. 

Unlimited wireless refers to virtually unlimited wireless resources in diverse dimensions, including connectivity, information, positioning, sensing, and energy. As AI will expand vertically and horizontally in the future, we refer to this advanced level of AI as DI. Due to DI, IoE will evolve toward IoI, which can be seen as a network of deeply intelligent entities. What comes to the 7G-level applications, human-DI interactions refer to diverse types of interactions between people and deeply intelligent entities. Examples include an augmented human, brain-DI interfaces, personal DI, telepathic communications, and holographic worlds. A deeply intelligent world refers to a world enhanced by DI at all levels and scales. Example applications in this domain include global intelligent systems, global monitoring/management/prediction entities, and a deeply intelligent society. Interconnected intelligent ecosystems refer to large-scale entities on the ground, in the air, and at the sea, exploiting wireless, DI, and IoI technologies. Overall, our 7G vision can be summarized in a few words: "Deeply Intelligent Network of Hyperverses". The evolution from 6G to 7G is summarized in Table \ref{Table_7G}. 

\section{CONCLUSION}
\label{Conclusion}
This article provided a comprehensive vision, survey, and tutorial on 6G. First, we presented the evolution road from 1G to 6G, overview of 5G, 6G development process, 6G research activities, and a literature review on 6G visions and surveys. We then introduced our overall 6G vision. After painting the big picture, the focus was shifted to the main details of 6G by identifying the fundamental elements, disruptive applications, key use cases, target performance requirements, potential technologies, and defining features. A special focus was given to a comprehensive set of potential 6G technologies, which were reviewed in a tutorial presentation style. Finally, a high-level vision was introduced for 7G so that the current 6G vision could be placed into a wider perspective. This article aims to provide a comprehensive guide to 6G to inspire future research and development work in academia, industry, and standardization bodies. 

\section{APPENDIX}
\label{Appendix}
List of Abbreviations

\newlist{abbrv}{itemize}{1}
\setlist[abbrv,1]{label=,labelwidth=1in,align=parleft,itemsep=0.1\baselineskip,leftmargin=!}
\begin{abbrv}
\item[1G] 1st generation  
\item[2D] two-dimensional 
\item[2G] 2nd generation
\item[3C] communication, computation, and caching
\item[3D] three-dimensional
\item[3G] 3rd generation
\item[3GPP] 3rd Generation Partnership Project
\item[4G] 4th generation
\item[5G] 5th generation
\item[6G] 6th generation
\item[6G-IA] 6G Smart Networks and Services Industry Association
\item[7G] 7th generation
\item[A$\&$B] access and backhaul
\item[AI] artificial intelligence
\item[AMPS] Advanced Mobile Phone Service
\item[AP] access point
\item[AR] augmented reality
\item[ATIS] Alliance for Telecommunications Industry Solutions
\item[B5G] beyond 5G
\item[BF] beamforming
\item[BPSK] binary PSK
\item[BS] base station
\item[CAAVS] connected autonomous aerial vehicle system
\item[CARS] connected autonomous robotic system
\item[CAV] connected autonomous vehicle
\item[CAVS] connected autonomous vehicle system
\item[CDMA] code-division multiple access
\item[CDMA2000] family of 3G technology standards
\item[cdmaOne] original IS-95 standard
\item[CD-NOMA] code-domain NOMA
\item[CF-mMIMO] cell-free massive MIMO
\item[CoMP] coordinated multi-point
\item[COVID-19] coronavirus disease 2019
\item[CP] cyclic prefix
\item[CPU] central processing unit
\item[CSI] channel state information
\item[C-UAV] cellular UAV
\item[D2D] device-to-device
\item[DFT] discrete Fourier transform
\item[DFT-s-OFDM] DFT-spread-OFDM
\item[DI] deep intelligence
\item[DL] deep learning
\item[DSRC] dedicated short-range communications
\item[DTL] deep transfer learning
\item[E2E] end-to-end
\item[EE] energy efficiency
\item[EH] energy harvesting
\item[EI] edge intelligence
\item[eMMB] enhanced mobile broadband
\item[ETSI] European Telecommunications Standards Institute
\item[EU] European Union
\item[FBMC] filter bank multi-carrier
\item[F-OFDM] filtered OFDM
\item[FR1] frequency range 1
\item[FR2] frequency range 2
\item[FSO] free-space optical
\item[FTL] federated transfer learning
\item[FTT] future technology trends
\item[GEO] geostationary orbit
\item[GFDM] generalized frequency division multiplexing
\item[GFMA] grant-free medium access
\item[GHz] gigahertz
\item[GMSK] Gaussian minimum-shift keying
\item[GSM] Global System for Mobile Communications
\item[HAPS] high-altitude platform station
\item[HMD] head-mounted display
\item[HMIMO] holographic MIMO
\item[HMIMOS] holographic MIMO surfaces
\item[HTC] holographic-type communications
\item[HW] hardware
\item[i3C] integrated 3C
\item[i4C] integrated 3C and control
\item[IAB] integrated access and backhaul
\item[ICC] International Conference on Communications
\item[ICT] information and communication technology
\item[IEC] International Electrotechnical Commission
\item[IEEE] Institute of Electrical and Electronics Engineers
\item[IMT] International Mobile Telecommunications
\item[IMT-2020] IMT for 2020 and beyond
\item[IMT-2030] IMT for 2030 and beyond
\item[INTNs] integrated non-terrestrial and terrestrial networks
\item[IoE] Internet of Everything
\item[IoI] Internet of Intelligence
\item[IoT] Internet of Things
\item[IoU] Internet of UAVs 
\item[IoV] Internet of Vehicles
\item[IR] infrared
\item[IRS] intelligent reflecting surface
\item[ISAC] integrated sensing and communication
\item[ISO] International Organization for Standardization
\item[ITU] International Telecommunication Union
\item[ITU-R] ITU Radiocommunication Sector
\item[KAIST] Korea Advanced Institute of Science and Technology
\item[KHz] kilohertz
\item[LD] laser diode
\item[LDPC] low density parity check
\item[LED] light-emitting diode
\item[LEO] low Earth orbit
\item[LIS] large intelligent surface
\item[LOS] line-of-sight
\item[LTE] Long-Term Evolution
\item[LTE-A] Long-Term Evolution Advanced
\item[M2M] machine-to-machine
\item[MAC] medium access control
\item[MDT] minimization of drive tests
\item[MEC] multi-access edge computing
\item[MEO] medium Earth orbit
\item[MIMO] multiple-input multiple-output
\item[MISO] multiple-input single-output
\item[mMIMO] massive MIMO
\item[mmWave] millimeter wave
\item[ML] machine learning
\item[mMTC] massive MTC
\item[ModCod] modulation and coding
\item[MoU] memorandum of understanding
\item[MR] mixed reality
\item[MTC] machine-type communications
\item[MWC] Mobile World Congress
\item[NFV] network function virtualization
\item[NG-RAN] next-generation RAN
\item[NGMN] Next-Generation Mobile Networks
\item[NICT] Japanese National Institute of Communication and Technology
\item[NMT] Nordic Mobile Telephony
\item[NOMA] non-orthogonal multiple access
\item[NR] New Radio
\item[NSA] non-stand alone
\item[NTNs] non-terrestrial networks
\item[NTT] Nippon Telegraph and Telephone
\item[OFDM] orthogonal frequency-division multiplexing
\item[OMA] orthogonal multiple access
\item[O-RAN] open RAN
\item[OWC] optical wireless communications
\item[PAPR] peak-to-average-power ratio
\item[PDC] Personal Digital Cellular
\item[PD-NOMA] power-domain NOMA
\item[PHY] physical
\item[PHz] petahertz
\item[PIN] positive-intrinsic-negative
\item[PLMN] public land mobile network
\item[PN] private network
\item[PSK] phase-shift keying
\item[QAM] quadrature amplitude modulation
\item[QoS] quality of service
\item[QPSK] quadrature phase-shift keying
\item[RAN] radio access network
\item[R$\&$D] research and development
\item[RFID] radio frequency identification
\item[RIS] reconfigurable intelligent surfaces
\item[RIT] radio interface technology
\item[S$\&$C] sensing and communication
\item[SA] stand alone
\item[SAGIN] space-air-ground integrated network
\item[SBA] service-based architecture
\item[SCS] subcarrier spacing
\item[SDN] software-defined networking
\item[SIC] successive interference cancellation
\item[SIM] subscriber identity module
\item[SNS JU] European Smart Networks and Services Joint Undertaking
\item[SON] self-organized network
\item[SRIT] set of RITs
\item[SWIPT] simultaneous wireless information and power transfer
\item[THz] terahertz
\item[TDD] time-division duplex
\item[TR] technical report
\item[TSG] technical specifications group
\item[TSG RAN] TSG radio access network
\item[TSG SA] TSG service and system aspects
\item[TX] transmitter
\item[U2D] UAV-to-device
\item[U2N] UAV-to-network
\item[U2S] UAV-to-satellite
\item[U2U] UAV-to-UAV
\item[U2V] UAV-to-vehicle
\item[U2X] UAV-to-everything
\item[UAS] unmanned aerial system
\item[UAV] unmanned aerial vehicle
\item[UDN] ultra-dense network
\item[UE] user equipment
\item[UFMC] universal filtered multi-carrier
\item[UK] United Kingdom
\item[UL] uplink
\item[umMIMO] ultra-massive MIMO
\item[UMTS] Universal Mobile Telecommunications System
\item[URLLC] ultra-reliable low-latency communications
\item[US] United States
\item[UTM] UAS traffic management
\item[UV] ultraviolet
\item[V2I] vehicle-to-infrastructure 
\item[V2N] vehicle-to-network
\item[V2P] vehicle-to-pedestrian
\item[V2V] vehicle-to-vehicle
\item[V2X] vehicle-to-everything
\item[VL] visible light
\item[VLC] visible light communications
\item[VR] virtual reality
\item[WET] wireless energy transfer
\item[WiFi] wireless fidelity
\item[WIPT] wireless information and power transfer
\item[W-OFDM] windowed OFDM
\item[WP] working party
\item[WPT] wireless power transfer
\item[WRC] World Radio Conference
\item[XL-MIMO] extremely large-scale MIMO
\item[XR] extended reality

\end{abbrv}

\section{ACKNOWLEDGMENT}
\label{Ack}
The authors would like to thank graphical designer Sallamaari Syrj{\"a} for helping with the figures. The language of the article has been checked using Paperpal, as recommended in the submission guidelines of the IEEE Access. 

\bibliographystyle{IEEEtran}
\bibliography{IEEEabrv,jour_short,conf_short,references}

\begin{IEEEbiography}
[{\includegraphics[width=1in,height=1.25in,clip,keepaspectratio]{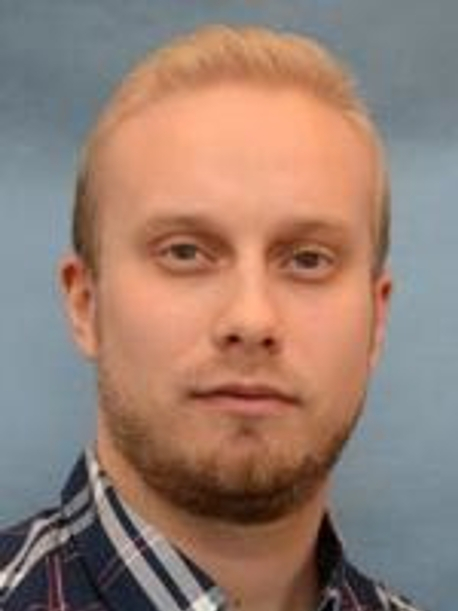}}]
{Harri Pennanen} (M'16) received the D.Sc. (Tech.) degree in telecommunications from the Centre for Wireless Communications (CWC), University of Oulu, Oulu, Finland, in 2015 with distinction. In 2015, he was a Research Associate in the Interdisciplinary Centre for Security, Reliability, and Trust, University of Luxembourg, Luxembourg. He is currently a Senior Researcher in CWC at the University of Oulu. His research interests include 6G and beyond networks.  
\end{IEEEbiography}

\begin{IEEEbiography}
[{\includegraphics[width=1in,height=1.25in,clip,keepaspectratio]{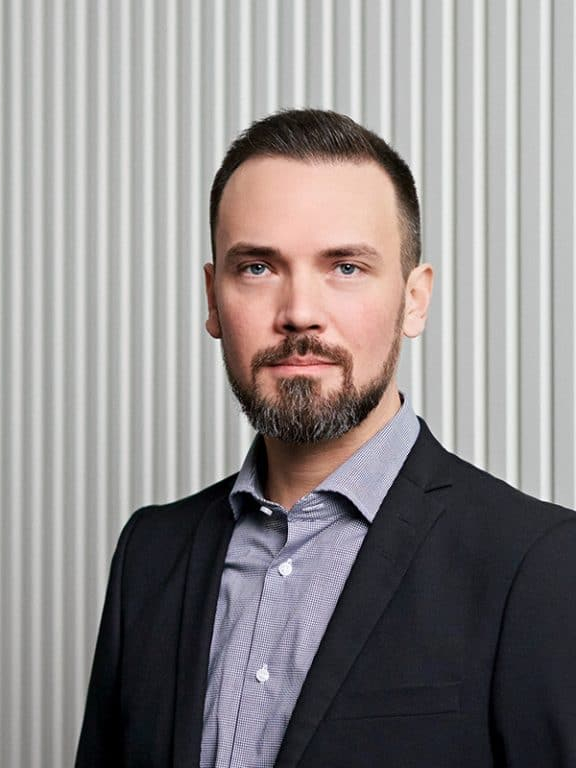}}]
{Tuomo H{\"a}nninen} is a Research Director at the Centre for Wireless Communications (CWC) in the University of Oulu. He holds a M.Sc. degree in electronics and D.Sc. degree in telecommunications. He has been coordinating several applied research projects and activities in the field of Spectrum Sharing, Critical Communication, IoT, Energy, Unmanned Aerial Vehicles (UAVs) and 5G/6G Test Network. Currently, he is coordinating one of the four Strategic Research Areas of 6G Flagship research programme: Human-centric Wireless Services.
\end{IEEEbiography}

\begin{IEEEbiography}
[{\includegraphics[width=1in,height=1.25in,clip,keepaspectratio]{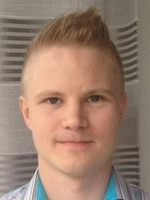}}]
{Oskari Tervo} received the D.Sc. (Tech.) degree from the University of Oulu, Finland, in 2018 with distinction. In 2014 and 2016, he was a Visiting Researcher at Kyung Hee University, Seoul, South Korea, and the Interdisciplinary Centre for Security, Reliability and Trust, University of Luxembourg, Luxembourg City, Luxembourg, respectively. He has numerous peer-reviewed articles and patent applications on the area of signal processing for communications and 3GPP physical layer cellular standardization. He has also been recognized as the best and exemplary reviewer for IEEE Transactions on Wireless Communications in 2017 and 2018, respectively. He has also co-authored a conference article receiving Best Student Paper Award in 2022. He is currently working at Nokia Standards, Oulu, where he is working as Senior Staff Research Specialist in the area of 3GPP physical layer standardization. His research interests include wide range of physical layer aspects for cellular standardization. 
\end{IEEEbiography}

\begin{IEEEbiography}
[{\includegraphics[width=1in,height=1.25in,clip,keepaspectratio]{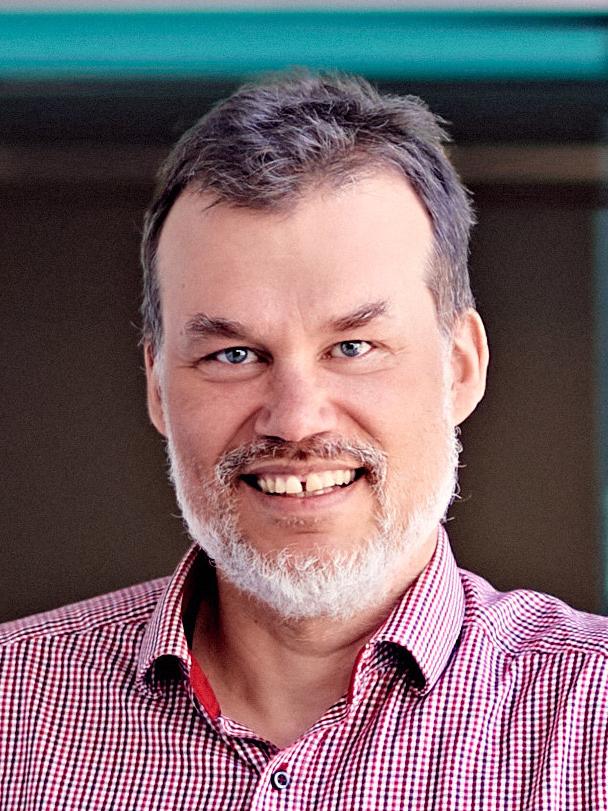}}]
{Antti T{\"o}lli} (M'08--SM'14) is a Professor with the Centre for Wireless Communications (CWC), University of Oulu. He received the Dr.Sc. (Tech.) degree in electrical engineering from the University of Oulu, Oulu, Finland, in 2008. From 1998 to 2003, he worked at Nokia Networks as a Research Engineer and Project Manager both in Finland and Spain. In May 2014, he was granted a five year (2014-2019) Academy Research Fellow post by the Academy of Finland. During the academic year 2015-2016, he visited at EURECOM, Sophia Antipolis, France, while from August 2018 till June 2019 he was visiting at the University of California Santa Barbara, USA. He has authored numerous papers in peer-reviewed international journals and conferences and several patents all in the area of signal processing and wireless communications. His research interests include radio resource management and transceiver design for broadband wireless communications with a special emphasis on distributed interference management in heterogeneous networks. From 2017 to 2021 he served as an Associate Editor for IEEE Transactions on Signal Processing. 
\end{IEEEbiography}

\begin{IEEEbiography}
[{\includegraphics[width=1in,height=1.25in,clip,keepaspectratio]{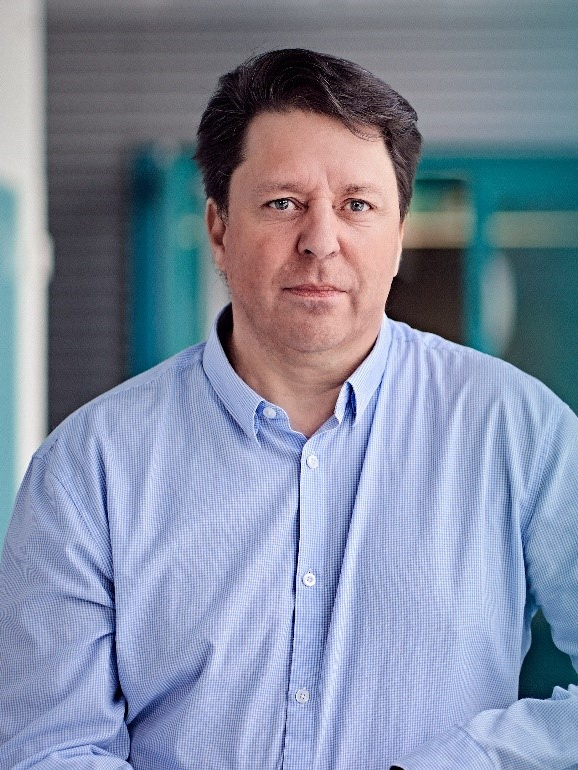}}]
{Matti Latva-aho} (IEEE Fellow) received the M.Sc., Lic.Tech. and Dr. Tech (Hons.) degrees in Electrical Engineering from the University of Oulu, Finland in 1992, 1996 and 1998, respectively. From 1992 to 1993, he was a Research Engineer at Nokia Mobile Phones, Oulu, Finland, after which he joined the Centre for Wireless Communications (CWC) at the University of Oulu. Prof. Latva-aho was Director of CWC during the years 1998-2006 and Head of the Department for Communication Engineering until August 2014. Currently, he is a Professor at the University of Oulu on wireless communications and Director for the National 6G Flagship Programme. He is also a Global Fellow at Tokyo University. Prof. Latva-aho has published over 500 conference or journal papers in the field of wireless communications. He received the Nokia Foundation Award in 2015 for his achievements in mobile communications research. 
\end{IEEEbiography}

\EOD

\end{document}